\PassOptionsToPackage{usenames,dvipsnames,table}{xcolor}
\documentclass[twocolumn,aps,prd,longbibliography]{revtex4-1}
\usepackage{graphicx,epsfig,epstopdf}
\usepackage{amsmath,amssymb,latexsym,bm}
\usepackage[english]{babel}
\usepackage{wrapfig}
\usepackage{animate}
\usepackage{times}
\usepackage[T1]{fontenc}
\usepackage{color, colortbl}
\usepackage{tikz}
\usepackage[usenames, dvipsnames]{xcolor}
\usetikzlibrary{arrows,shapes}
\usepackage{url}
\usepackage{hyphenat}
\usepackage{makecell}
\usepackage[Conny]{fncychap}
\usepackage{float}
\usepackage{lipsum}
\usepackage{mathptmx}
\usepackage{mathtools}
\usepackage[T1]{fontenc}
\usepackage[%
  colorlinks=true,
  urlcolor=blue,
  linkcolor=blue,
  citecolor=blue
]{hyperref}
\DeclareUnicodeCharacter{2212}{-}
\begin{document}
\title{Atom-field dynamics in curved spacetime}

\author{Syed Masood A. S. Bukhari$^{1}$} 
\email{masood@zju.edu.cn} \email{thamersyed@gmail.com} 
\author{Li-Gang Wang$^{1}$} \email{lgwang@zju.edu.cn}

\affiliation{$^{1}$School of Physics, Zhejiang University, Hangzhou 310027, China.}
\date{\today}

\begin{abstract}
Some aspects of atom-field interactions in curved spacetime are reviewed. Of great interest are  quantum radiative and entanglement processes  arising out of Rindler and black hole spacetimes, which involve the role of Hawking-Unruh and dynamical Casimir effects. Most of the discussion surrounds the radiative part of interactions. For this, we specifically reassess the conventional understandings of atomic  radiative transitions and energy level shifts  in curved spacetime. We also briefly outline the status quo of entanglement dynamics study in curved spacetime, and highlight literature related to some novel insights, like \textit{entanglement harvesting}.  On one hand, the study of the role played by spacetime curvature in quantum radiative and informational  phenomena has implications for fundamental physics, notably the gravity-quantum interface. In particular, one examines the viability of the Equivalence Principle, which  is at the heart of Einstein's general theory of relativity. On the other hand, it can be instructive for manipulating quantum information and light propagation in arbitrary geometries. Some issues related to nonthermal effects of acceleration are also discussed.
\end{abstract}
\maketitle
\tableofcontents
\section{Introduction}
Atom-field or light-matter interactions are at the very foundations of quantum optics, which has witnessed great progress in the past decades, opening windows to the development of many  ground-breaking theoretical tools and unprecedented experimental techniques,  having great bearing on fundamental physics and various technological applications. Some  prominent advancements include ultraintense laser systems \cite{RevModPhys.78.309}, cavity QED \cite{Walther_2006}, techniques ushering to novel quantum phases and phenomena arising out of  strong atom-photon interactions \cite{RevModPhys.90.031002},  topological photonics \cite{RevModPhys.91.015006},  gravitational wave interferometry \cite{Harry_2010, PhysRevD.88.043007, Dooley_2016, LIGOScientific:2020luc, Virgo:2020xlu}, atom interferometric precision tests for general relativity and gravity \cite{RevModPhys.90.025008},  to name a few of the recent ones. While the conventional quantum optical processes are based on flat background space(-time) with an inertial description, recent years have seen intense efforts in understanding the impact of   curved  background geometry on the optical phenomena, either in classical regime \cite{PhysRevLett.105.143901,
Schultheiss:2020wzx}, helping  emulate general relativity and black hole effects in optical structures \cite{Leonhardt_2006, Philbin:2007ji, Bekenstein2016CurvedSN,
PhysRevX.8.011001} and analogue spacetime models \cite{Faccio:2013kpa,Viermann:2022wgw}, or in quantum regime within accelerated frames and curved geometries  \cite{Lopp2018RelativityAQ, Scully:2017utk, PhysRevD.101.045017, Zhang:2014ndf,2010eqo..book.....L}.  The motivation for pursuing these directions is manifold. Firstly, one expects to  achieve possible ways to create test beds for probing cutting edge theoretical problems in fundamental physics that arise in non-Euclidean geometry which, among many include, for example, issues related to quantum gravity  \cite{PhysRevA.102.032208,
Garcia:2020slt},  Hawking-Unruh radiation \cite{Steinhauer:2015saa, Leonhardt:2016qdi, Hu:2018psq,
Sheng:2021iky}, cosmological expansion and particle creation \cite{Parker:1969au,
Parker:1971pt,
Parker:2012at,Eckel:2017uqx,
Schmit:2018hvy}, which are otherwise notoriously difficult to observe. Secondly, these efforts could be potentially helpful in designing  novel structures that may manipulate and control light propagation in arbitrary surfaces and complex media. It could also boost the progress in  quantum computation and information  phenomena by incorporating the effects of acceleration and curvature, which has been actively pursued for last two decades or so \cite{Alsing:2003es, PhysRevLett.95.120404,PhysRevLett.106.210502}, paving way for a rapidly developing field of \textit{relativistic quantum information} \cite{RevModPhys.76.93,
Mann_2012}. It has also sparked intense debates about the role of acceleration in quantum information processes \cite{PhysRevA.74.032326, Wang:2010qq, Friis:2012tb, Bruschi:2012uf,
Liu:2021dnl} and  quantum optical phenomena \cite{Lopp:2018lxl,
Martin-Martinez:2020pss}.
Similar ideas have also been explored to quantify the role of gravity in the dynamics of  Bose-Einstein condensates \cite{Sabin:2014bua, Ratzel:2018srb,PhysRevD.98.105019, Howl:2016ryt} and  Dirac equation \cite{Collas:2018jfx}.

 Quantum optical phenomena are deeply grounded in the theory of light-matter interactions, or atom-field dynamics \cite{scully_zubairy_1997,
compagno_passante_persico_1995} that occur in flat spacetime. Many novel phenomena emerge when a transition is made to curved geometries. On a thorough survey of literature, it turns out that there are many ways to go forward.

One major line of investigation is to include the contributions from acceleration radiation or celebrated Unruh effect \cite{PhysRevD.14.870}, also known as  Fulling-Davies-Unruh effect \cite{Fulling:1972md,Davies:1974th,RevModPhys.80.787}, which  posits that a Minkowski vacuum appears as a thermal state to an accelerating observer (Rindler observer), and    alludes to the idea that radiation is not a local covariant phenomenon \cite{osti_4818173,1961NCim...21..811R,BOULWARE1980169}. Likewise, we have contributions from Hawking effect \cite{Hawking1975ParticleCB}, which is the thermal radiation detected by a static observer (e.g. an atom) in a black hole geometry. Both of these effects exploit causal horizons of spacetime and use same procedures for the description of quantum fields on curved spacetime \cite{Frodden:2018mdm}. It is expected that when atoms and fields undergo acceleration in Rindler and  curved spacetimes, Hawking-Unruh effect does contribute to the atomic transition and energy level shift phenomena. Being concerned with the vacuum physics, this naturally connects Hawking-Unruh effect with the particle creation in quantum vacuum via Casimir 
\cite{Casimir:1948dh, Bordag:2009zz} or dynamical Casimir effect \cite{10.1063/1.1665432, Dodonov:2020eto}, and moving mirror models \cite{Scully:2017utk, Fulling1976RadiationFA, Davies1977QuantumVE, Anderson:2015iga, Good:2015jwa,PhysRevA.74.023807,Scully2019LaserEF,
Lock:2016rmg}. In the past three decades, this has been thoroughly worked out and forms the thrust of this review.

We also  make a brief mention of few other directions that seek to  introduce curved geometry in the description of quantum systems. The seminal work by Chandrasekhar  on solving Dirac equation in Kerr spacetime \cite{Chandrasekhar1976TheSO}  has flourished into a major activity of analyzing influence of curved spacetime on quantum mechanical behavior of particles \cite{1979PhRvD..19.1093C,Shishkin:1991ma,Finster:2009ah, Collas:2018wcc}.  In a series of papers by Parker  \cite{PhysRevLett.44.1559,Parker1981TheAA,PhysRevD.22.1922,Parker:1981wt}, the  possibility of using atoms as a probe of classical spacetime geometry was considered, where the spacetime curvature manifests itself in the atomic spectrum, having dependencies on Ricci curvature. This  has spread out into a  flurry of research activities and extended to many systems (see e.g. \cite{1993PhRvL..70.3839P,Parker:1996hc,deAMarques:2002hbv,Zhao:2007xj,Carvalho:2011krd,Roura:2021fvd}). 
 
Other considerations are based on a geometric approach applied to quantum mechanics \cite{Caianiello:1981jq}, the maximal acceleration hypothesis and its connection to Lamb shift  \cite{Lambiase:1998tn} and Unruh effect (see \cite{Benedetto:2015fta}  and the relevant references therein). Furthermore, many efforts have been devoted to the study of emission \cite{PhysRevD.56.R6071},  scattering \cite{ PhysRevLett.102.231103} and absorption
 \cite{ Macedo:2013afa, Cardoso:2019dte} of electromagnetic and other fields near black holes, which find connections to black hole superradiance\cite{Brito:2015oca} and  probing black hole geometries \cite{Bambi:2015kza}. However,  this later \textit{field-geometry} coupling is a much bigger paradigm that extends beyond the scope of present discussion. 

By introducing acceleration or spacetime curvature in optical phenomena, it necessarily involves tools from quantum field theory and differential geometry, these studies somehow lie at the boundary between quantum optics and gravity. Though the topics have sparsely been covered in some reviews
 \cite{RevModPhys.80.787, Dodonov:2020eto, Passante:2018qzj}, we believe that a comprehensive and up to date review  that could assemble all relevant works in one piece and  provide  a thorough introduction to this emerging field is still lacking at the moment. By focusing on atom-field dynamics in curved geometries,  we thus hope this short review  provides a glimpse of  this area and thus becomes handy for beginners.   \\ 
\indent We organize the work as follows. In Sec.\ref{secii}, we introduce the necessary mathematical tools including Rindler motion and techniques for quantizing fields in curved geometries. This is essential for description of Hawking-Unruh effects and is needed in subsequent discussions. Here, we also introduce the basic theory behind so-called dynamical Casimir effect.  Sec. \ref{seciii} is  devoted to the atomic radiative transition processes and Lamb shift in Rindler and black hole spacetimes, followed by discussions on dispersion and resonant interactions in curved spacetimes. In Sec. \ref{seciv}, we  discuss some aspects of particle and energy production in moving mirror models, followed by atom-mirror systems in black hole spacetime. A brief discussion is made about relativistic quantum information in Sec.\ref{secv}. Conclusions are drawn in Sec. \ref{secvi}.

\bigskip
\section{Field quantization and particle emission in curved spacetime}\label{secii}
\subsection{Accelerated observers in flat spacetime: Rindler motion}
\begin{figure}[h!]
 \begin{center}
\includegraphics[height=7cm,width=9.5cm]{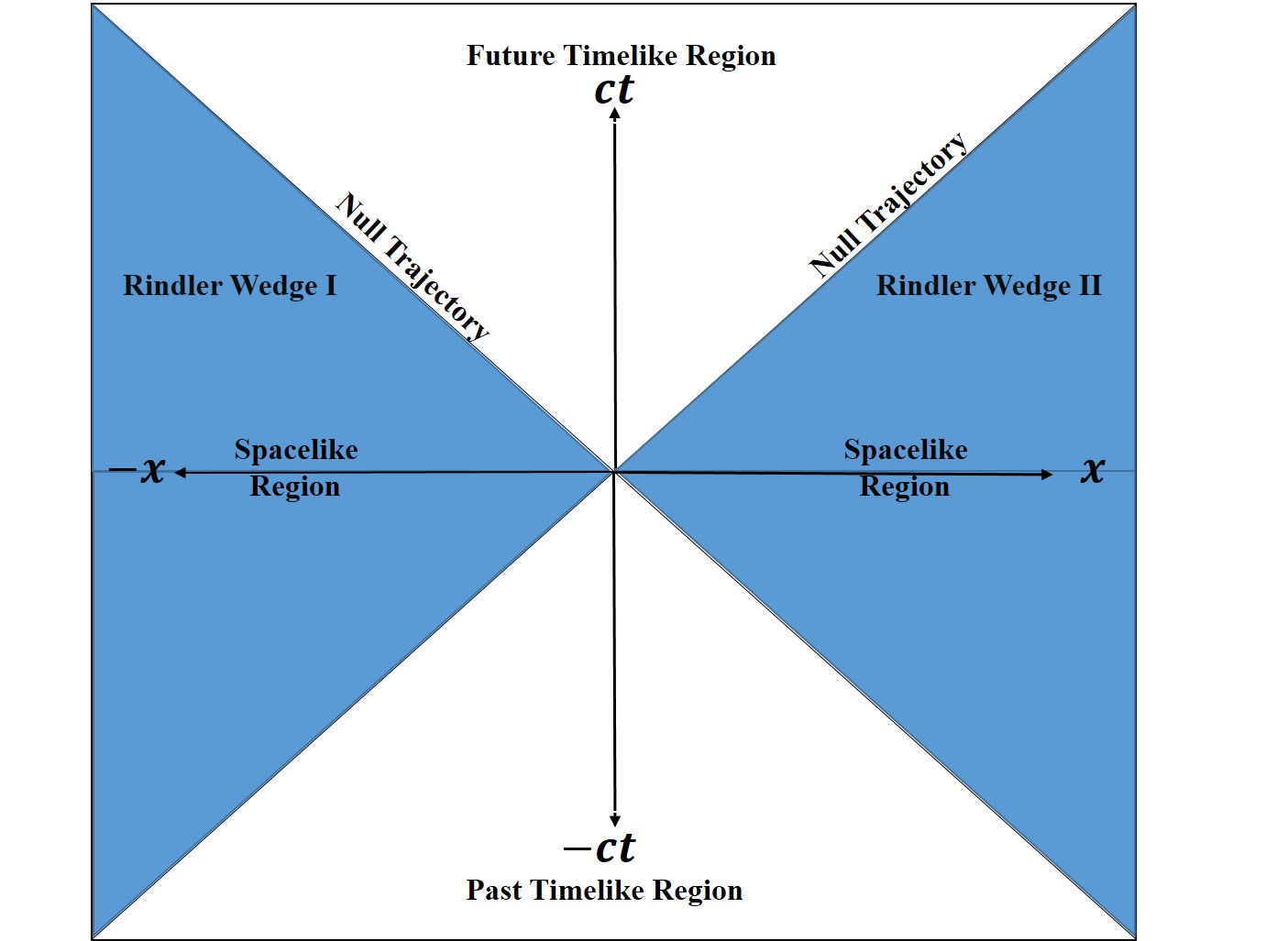}
 \end{center}
\caption{Lightcone showing different regions of spacetime. Shaded region pertains to accelerated observer.}
 \label{figC}
\end{figure}

We begin from Minkowski spacetime metric 
\begin{eqnarray}\label{interval}
 ds^2=c^2 dt^2-dx^2-dy^2-dz^2,
\end{eqnarray}
 which characterizes a  four-dimensional continuous spacetime and  is an invariant quantity under Lorentz transformation. It represents a non-Euclidean geometry, and sometimes also known as \textit{pseudo-Euclidean} spacetime. However, for a constant $t$, spatial part of the geometry remains Euclidean \cite{Hobson:2006se}. Based on the sign of $ds^2$ in Eq. (\ref{interval}), the interval can timelike $(ds^2 > 0)$, null or lightlike $(ds^2 =0)$ and spacelike $(ds^2 <0)$. Intuitively, one can view this as if a body moving along its trajectory can have three ways to go. In the first case, body is at rest and time flows. So it moves in time and does not move in space. Second case would mean that it catches up exactly with a  ray of light. And in the third case, it moves only in space and not in time, which is impossible. Hence the blue region below the light's trajectory is not causally connected to the body.  
 The situation is shown in the spacetime diagram of Fig. \ref{figC}, called \textit{lightcone}. 

Note that the shaded region on left and right sides (wedges) of the vertical axis is very important for describing accelerated observers in flat spacetime--helping in the description of Unruh effect.  A symbolic way of representing a spacetime metric is \textit{signature}. For Minkowski spacetime in Eq. (\ref{interval}), it is given by  $(+,-,-,-)$.  By using indexed coordinates $x^{\mu}(\mu=0,1,2,3)$ such that 
\begin{eqnarray}
 x^0 \equiv ct,\ \ x^1\equiv x\ \ x^2\equiv y,\ \ x^3\equiv z.\  
\end{eqnarray}
 The metric given in Eq. (\ref{interval}) can be represented in Einstein summation convention for tensors as 
 \begin{eqnarray}\label{mink}
  ds^2=\eta_{\mu\nu}dx^{\mu}dx^{\nu},
 \end{eqnarray}
where $\eta_{\mu\nu}$ is the Minkowski metric tensor. Closely related to these underlying transformations in spacetime are Lorentz and Poincar\'e groups. The Lorentz transformation is also sometimes written as \cite{Misner:1973prb}
 $x'^{\mu}=\Lambda^{\mu}_{\nu}x^{\nu}$,
where $x^{\nu}$ is any four-vector and  $\Lambda$ is the Lorentz tensor. The corresponding group comprises three rotations and three boosts and hence a six parameter (also called generators) group. Its extension by adding four spacetime translations, $
  x'^{\mu}=\Lambda^{\mu}_{\nu}x^{\nu}+ a^{\mu}$ ($a^{\mu}$ is a constant tensor)
constitutes  Poincar\'e group, which obviously has ten generators in total  \cite{1972gcpa.book.....W}. 

The inertial motion  in special relativity is very well described by the above considerations and can be found in almost every elementary textbook on relativity. So it gives an impression that only objects in uniform motion can be dealt with special theory of relativity. However, this statement is not quite complete.   So then, is it possible to incorporate acceleration in a special relativistic decription of motion? The answer is yes. Such accelerated motion is called Rindler motion. The best possible example is a constantly accelerating rocket in space.
Recall that we  mentioned in the previous section about the shaded wedges of Minkowski spacetime diagram in Fig. \ref{figC} on right and left sides of particle worldline. Rindler motion is very well described in that part of Minkowski spacetime as shown below in Fig. \ref{rind}.

 \begin{figure}[h!]
 \begin{center}
\includegraphics[width=70 mm]{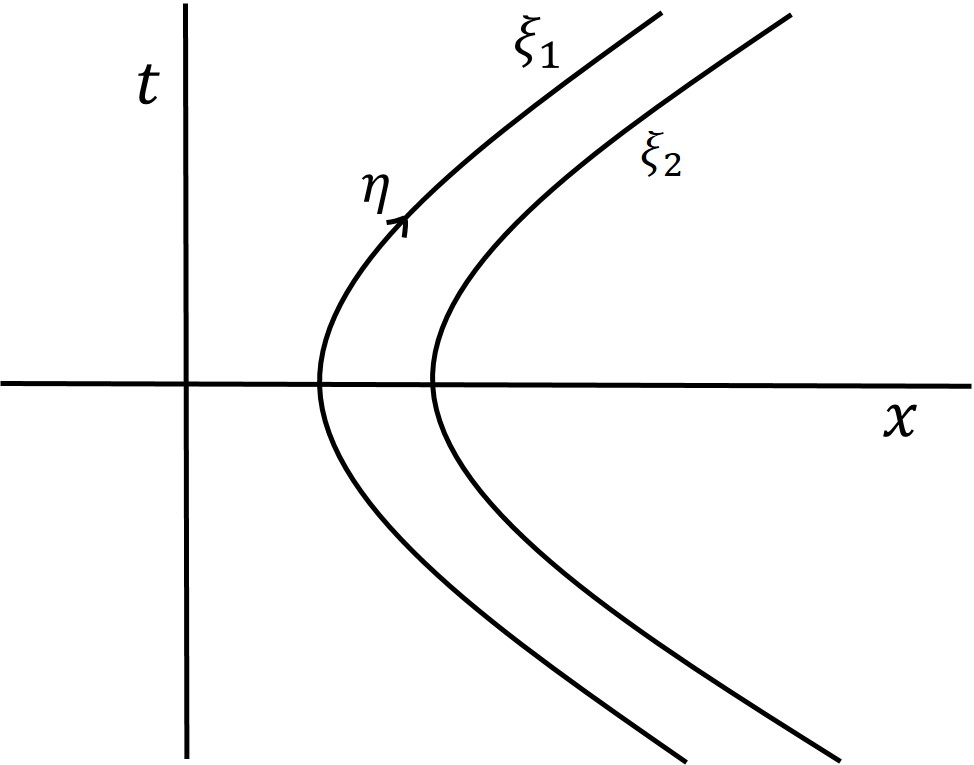}
 \end{center}
\caption{Rindler motion. Reproduced from Ref. \cite{Frodden:2018mdm}.}
 \label{rind}
\end{figure}
In Fig. \ref{rind}, $t$ and $x$ are time and space coordinates respectively in an inertial reference frame. For simplicity,  we only consider motion in right Rindler wedge as shown.
We make transformations between an inertial observer and accelerated observer as follows.

Let's consider an inertial observer $S$ and accelerated observer $B$ having his own frame in which he feels constant acceleration $\alpha$ (his proper acceleration). Also we mark the time  of $B$ in his own frame as  $\tau$ respectively ($B$'s proper time). One can approximate this accelerated motion by considering it as a succession of another inertial frame $S'$ (with $t'$ as time) having velocity $v$ with respect to  $S$. So the accelerated observer is momentarily at rest in $S'$. One can easily make a Lorentz transformation between $S$ and $S'$ which would be moving at a constant velocity $v$ with respect to $S$ at that particular instant of time. If $S$ measures the acceleration of $B$ as $a$, then acceleration of $B$ in $S'$, denoted by $\alpha'$  is given by simple transformation of special relativity $a'=\gamma^3 a$,
where $\gamma=1/\sqrt{1-v^2/c^2}$ is the usual Lorentz factor. Remember $v$ is the velocity of $S'$ w.r.t $S$ at that particular instant of time. Our aim is to get the description of $B$'s own observations in terms of $S$'s. 
By considering a particular instant of time when $B$ is at rest in $S'$, its acceleration is also constant in $S'$. At that time,

\begin{eqnarray}
 \alpha=\gamma^3 a=\Bigg[\frac{1}{\sqrt{1-\frac{v^2}{c^2}}}\Bigg]^3 a=\ const. 
\end{eqnarray}
We can also write the above equation as 
\begin{eqnarray}
 \alpha=\frac{d}{dt}(\gamma v)=\ const.
\end{eqnarray}
By considering the boundary condition $v=0$ at $t=0$, we solve the above equation to get 
\begin{eqnarray}
 v(t)=\frac{\alpha t}{\sqrt{1+({\alpha t}/c)^2}},
\end{eqnarray}
and the position $x(t)$ as 
\begin{eqnarray}
 x(t)=\alpha \int_{0}^{t}\frac{dt'}{\sqrt{1+({\alpha t'}/c)^2}}=\frac{c^2}{\alpha} \sqrt{1+({\alpha t}/c)^2}.
\end{eqnarray}
This gives rise to the following equation of motion
\begin{eqnarray}\label{motion}
 x^2-t^2=\Big(\frac{c^2}{\alpha}\Big)^2.
\end{eqnarray}
This is clearly an equation of hyperbola. So the worldline of $B$ is a hyperbola as seen from Fig. \ref{rind}. In terms of proper time $\tau$, we have \cite{Socolovsky:2013rga}
\begin{align}
 t(\tau) &=\int_{0}^{t}\sqrt{1-{v(t')^2}/c^2}\\
 &=\int_{0}^{t} \frac{dt'}{\sqrt{1+({a t'}/c)^2}}\\
 &=\big(\frac{c}{\alpha}\big)\sinh \Big(\frac{\alpha \tau}{c}\Big),
\end{align}
and 
\begin{eqnarray}
 x(\tau)=\Big( \frac{c^2}{\alpha}\Big)\cosh \Big(\frac{\alpha \tau}{c}\Big).
\end{eqnarray}
In natural units, $c=1$, we have 
\begin{eqnarray}\label{time}
 t(\tau)=\frac{1}{\alpha}\sinh \Big(\alpha \tau\Big), \ \ \label{position} x(\tau)= \frac{1}{\alpha}\cosh \Big(\alpha \tau).
\end{eqnarray}
This is called the Rindler transformation which connects an inertial and accelerating observer. Here  $\alpha$ and $\tau$ are called Rindler acceleration and Rindler time respectively \cite{1966AmJPh..34.1174R}. From Eq. (\ref{time}), one can easily identify the  hyperbolic polar coordinate representation of $x$ and $t$ as follows
\begin{eqnarray}\label{18}
 t=\xi \sinh \eta,\ x=\xi \cosh \eta,
\end{eqnarray}
where $\xi=1/\alpha$ and $\eta=\alpha\tau$. These two parameters $\xi$ and $\eta$ are called Rindler coordinates, and this  leaves the metric as 
\begin{eqnarray}\label{19}
 ds^2=\xi^2 d\eta^2-d\xi^2,
\end{eqnarray}
which is called Rindler metric. Rindler metric can be very well used to describe most the physics of uniform gravitational fields where acceleration is constant,  i.e. weak-field limit of general relativity. By using Eqs. (\ref{18}) and (\ref{19}), Rindler line element in $(1+1)$ dimensions can also be written as \cite{1966AmJPh..34.1174R} 
\begin{eqnarray}\label{rindlerl}
 ds^2=e^{2\alpha\xi}(d\tau^2-d\xi^2). 
\end{eqnarray}
To cover the whole Rindler wedge, we can accommodate many accelerating observers (infinite in principle!) in Rindler formulation with  constant accelerations parameterized by $\xi_{1}$ and $\xi_{2}$ as shown in Fig \ref{rind}. For each observer's motion, $\xi$ would be constant along the whole trajectory and only $\eta$ would change. A more general transformation would be \cite{Frodden:2018mdm}
\begin{eqnarray}
 t=\frac{1}{\alpha} f(\xi) \sinh(\eta),\ \  x=\frac{1}{\alpha}f(\xi)\cosh(\eta),
\end{eqnarray}
where $f(\xi)$ is a function of $\xi$.  
Since more the acceleration, more the observer gets close to the origin, this hints to take $f(\xi)$ an increasing function. Choosing $f(\xi)=e^{a\xi}$, the above transformation becomes
\begin{eqnarray}\label{RRR}
 t=\frac{1}{\alpha} e^{\alpha\xi} \sinh(\eta),\ \  x=\frac{1}{\alpha} e^{\alpha \xi}\cosh(\eta), 
\end{eqnarray}
which are so-called conformal Rindler coordinates \cite{Martin-Martinez:2014gra}, and very often used in  literature. 
The above derivation will be very useful in describing Unruh effect, as follows in the subsequent sections.

 \subsection{Recipes for field quantization}
Next, we discuss  some results  from quantization of fields on curved spacetimes. We begin by Minkowski description of quantum fields and develop the framework for curved cases. The simplest one is to consider a free Klein-Gordon field which possesses zero spin $(s=0)$, hence a scalar field. The flat space metric signature is $(+,-,-,-)$ and we adopt units in $c=\hbar=1$, unless stated otherwise. The detailed method can be found in many textbooks (see for example, \cite{Birrell:1982ix,Jacobson2005, Parker:2009uva}). 
In canonical quantization scheme, a scalar Klein-Gordon field $\phi(x^{\mu})$ and its associated momentum $\pi(x^{\mu})$ appear in a Legendre transformation between Hamiltonian density $\mathcal{H}$ and Lagrangian density $\mathcal{L}$
\begin{eqnarray}
\mathcal{H}=\pi\dot{\phi}-\mathcal{L}, 
\end{eqnarray}
which gives the Hamiltonian $H=\int d^{n-1}x \mathcal{H}$ for an $n$-dimensional spacetime. These two are then promoted to field operators $\hat{\phi}$ and $\hat{\pi}$, such that  they possess values at each point $x^{\mu}$. For simplicity, we drop the hat symbols on operators. The equal time commutation relations are imposed as 
\begin{align}
 [\phi(t,x),\phi(t,y)]&=[\pi(t,x),\pi(t,y)]=0,\\
 [\phi(t,x),\pi(t,y)]&=i\delta^{n-1}(x-y).
\end{align}
Also, the Lagrangian density, which is Lorentz-invariant, is given by 
\begin{eqnarray}
 \mathcal{L}=\frac{1}{2}\eta^{\mu\nu}\partial_{\mu}\phi\partial_{\nu}\phi-\frac{1}{2}m^2\phi^2,
\end{eqnarray}
where $m$ characterizes the field mass. This gives us the Klein-Gordon equation
\begin{eqnarray}\label{KGEQN}
 \eta^{\mu\nu}\partial_{\mu}\partial_{\nu}\phi+m^2\phi=0,
\end{eqnarray}
or 
\begin{eqnarray}\label{BOXEQN}
 (\Box+m^2)\phi=0,
\end{eqnarray}
where $\Box=\eta^{\mu\nu}\partial_{\mu}\partial_{\nu}$.
Thus, in a similar way to harmonic oscillator, we expand field $\phi(t,x)$ as follows
\begin{eqnarray}
 \phi(t,x)=\int d^{n-1}k\big[\hat{a}_{k}\varphi_{k}(t,x)+\hat{a^{\dagger}_{k}}{{\varphi^*}_{k}}(t,x)\big],
\end{eqnarray}
where $\varphi_{k}$ are complete set of solutions to Eq. (\ref{KGEQN}) characterized by vector $k$ (assuming $\varphi$ and $\varphi^*$ as positive and negative  frequency modes respectively), and $\hat{a}_{k}$ and $\hat{a^{\dagger}_{k}}$ are creation 
and annihilation operators respectively which follow the commutation relations as mentioned before.  By considering some explicit solutions to Eq. (\ref{KGEQN}) as $\varphi_{k}=A_{k}e^{i(k.x-\omega_{k}t)}$, with $A_{k}$ as some constant, the normalized modes are given by 
\begin{eqnarray}
 \phi_{k}(x,t)=\frac{1}{\sqrt{2\omega_{k}(2\pi)^{n-1}}}e^{i(k.x-\omega_{k}t)}.
\end{eqnarray}
In the above consideration, $\omega_{k}$ is either a positive or negative frequency satisfying the dispersion relation, $\omega_{k}^2=k^2+m^2$. With these things in hand, we can write 
\begin{eqnarray}
 H=\int d^{n-1}k\big[\hat{n}_{k}+\frac{1}{2}\delta^{(n-1)(0)}\big],
\end{eqnarray}
which makes sure that $H$ be a Hermitian operator such that $\hat{a}_{k}|n_{k}\rangle$ is its eigenstate. This operator $H$ possesses a ground state $|0\rangle$, such that $\forall\ k, \hat{a}_{k}|0\rangle=0$, which is the Fock vacuum. Thus, for Fock vacuum, the average particle number is
\begin{eqnarray}
 \langle 0|\hat{n}_{k}|0\rangle=0,\ \forall\ k. 
\end{eqnarray}
For the foregoing discussion, one important point must be made here. By considering positive frequency modes which are Lorentz-invariant, we can conclude that the definition of vacuum constructed thereby is also Lorentz-invariant. This situation is changed when we consider the background spacetime as curved.

While making transition to curved spacetime, we replace flat signature $\eta_{\mu\nu}$ by more generic signature $g_{\mu\nu} (+,-,-,...,-)$ and the derivative $\partial_{\mu}$ by covariant derivative  $\nabla_{\mu}$, which   gives the following Lagrangian for scalar field
\begin{eqnarray}
 \mathcal{L}=\frac{1}{2}|g|^{1/2}\Big(\frac{1}{2}g^{\mu\nu}\partial_{\mu}\phi\partial_{\nu}\phi-\frac{1}{2}m^2\phi^2\Big),
\end{eqnarray}
where $g$ is the determinant of metric tensor $g_{\mu\nu}$. It should be mentioned here that there is a possibility of coupling between the scalar field and gravitational background described by Ricci scalar curvature $R$. This results in the  following equation for field $\phi$
\begin{eqnarray}\nonumber
 (\nabla^{\mu}\nabla_{\mu}+m^2+\lambda R)\phi=0,
\end{eqnarray}
 where $\lambda$ is the coupling constant. We consider the case of minimal coupling where $\lambda=0$, which reduces above equation to 
 \begin{eqnarray}\label{CKG}
  (\nabla^{\mu}\nabla_{\mu}+m^2)\phi=0.
 \end{eqnarray}

We take a brief pause here and will return  to the above result. In what follows, we discuss Klein-Gordon inner product, which is very important to appreciate difference between flat and curved space quantization.  Corresponding to the harmonic oscillator case, if a function $f_{A}$  helps to  solve the oscillator equation 
\begin{eqnarray}\label{QHO}
 \ddot{x}+\omega^2 x=0,
\end{eqnarray}
by the substitution $\hat{x}=\hat{a}f_{A}(t)+\hat{a^{\dagger}}f_{A}^{*}(t)$, then it is possible to define an inner product on the space of solutions to Eq. (\ref{QHO}) as 
\begin{eqnarray}
 \langle f_{A},f_{B}\rangle=\frac{im}{\hbar}\Big(f_{A}^{*}\partial_{t}f_{B}-(\partial_{t}f^{*})f_{B}\Big),
\end{eqnarray}
where $f_{B}$ is just another function like $f_{A}$ and $\partial_{t}=\frac{\partial}{\partial t}$. With this new notation, the anticommutation relation is  $\langle f,f\rangle[\hat{a},\hat{a^{\dagger}}]=1$. For flat space field quantization with two solutions $\phi_{1}$ and $\phi_{2}$ corresponding to Eq. (\ref{KGEQN}), we define this inner product as 
\begin{eqnarray}
 \langle \phi_{1},\phi_{2}\rangle=-i\int_{\Sigma_{t}} d^{n-1}x\Big(\phi_{1}\partial_{t}\phi_{2}^*-\phi_{2}^{*}\partial_{t}\phi_{1}\Big),
\end{eqnarray}
where the integral measure is taken over a constant time hypersurface $\Sigma_{t}$ which represents Cauchy surfaces for Klein-Gordon equation given in Eq.  (\ref{KGEQN}). A Cauchy surface is a closed hypersurface intersected by every timelike curve only once, if the curve is inextendible. A spacetime is globally hyperbolic, if it has a Cauchy surface. Similarly, for the curved spacetime quantization, the inner product is given by 
\begin{eqnarray}\nonumber
 \langle \phi_{1}| \phi_{2}\rangle=i\int_{\Sigma} d^{n-1}x\ n^{\mu} \sqrt{\gamma}\Big(\phi_{1}^*\nabla_{\mu}\phi_{2}-\phi_{2}\nabla_{\mu}\phi_{1}^*\Big). 
\end{eqnarray}
Here the integral measure is defined over spacelike hypersurface $\Sigma$ with normal vector $n^{\mu}$ and induced metric $\gamma_{ij}(\gamma)$. Now it is often helpful to consider two modes of positive- and negative-frequency solutions to Eq. (\ref{CKG}) forming a complete basis  and then expanding the field operator $\phi$ as a combination of these modes. Thus, for a set of modes $f_{i}$ used by an observer, we write
\begin{eqnarray}
 \phi=\sum_{i}(\hat{a}_{i}f_{i}+\hat{a}_{i}^{\dagger}f^*_{i}),
\end{eqnarray}
where $\hat{a}_{i}$ and $\hat{a}_{i}^{\dagger}$ are identified as annihilation and creation operators following the commutation relations
\begin{align}\nonumber
 [\hat{a}_{i}, \hat{a}_{j}]=[\hat{a}_{i}^{\dagger}, \hat{a}_{j}^{\dagger}]=[\hat{a}_{i},\hat{a}_{j}^{\dagger}]=0,
\end{align}
and corresponding vacuum state is $|0_{f}\rangle$, such that $\hat{a}|0_{f}\rangle=0_{f}$.  It is very important to ascribe a timelike Killing vector for flat spacetime quantization such that one is able to classify the solutions in terms of  positive- and negative-frequency modes. 
In curved spacetime, such a Killing vector does not exist generally, hence our procedure for identifying positive- and negative-frequency mode solutions no more works. 
For the sake of clarity, we consider another set of modes $g_{i}$ with vacuum $0_{g}$ and corresponding annihilation and creation operators $\hat{b}_{i}$ and $\hat{b}_{i}^{\dagger}$ ( following same commutation relations as that of $\hat{a}_{i}$ and $\hat{a}^{\dagger}_{i}$) respectively by another observer such that 
\begin{eqnarray}
 \phi=\sum_{i}\Big(\hat{a}_{i}g_{i}+\hat{a}_{i}^{\dagger}g^*_{i}\Big).
\end{eqnarray}
It is possible to define a transformation between $f_{i}$ and $g_{i}$ such that 
\begin{eqnarray}\nonumber
 g_{i}=\sum_{i}\Big(\alpha_{ij}f_{j}+\beta_{ij}f_{j}^*\Big),\ \ 
 f_{i}=\sum_{j}\Big(\alpha^*_{ji}g_{j}-\beta_{ji}g^*_{j}\Big),
\end{eqnarray}
which characterizes the scenario where one observer expresses his results in terms of other's basis modes. This is the famous \textit{Bogoliubov transformation} and  helps to write transformation between the operators $\hat{a}_{i}$ and $\hat{b}_{i}$ as follows
\begin{eqnarray}\label{bog}
\hat{a}_{i}=\sum_{j}\Big(\alpha_{ji}\hat{b}_{j}+\beta^*_{ji}\hat{b}^{\dagger}_{j}\Big),\ \ \hat{b}_{i}=\sum_{j}\Big(\alpha^*_{ij}\hat{a}_{j}-\beta^*_{ij}\hat{a}^{\dagger}\Big),
\end{eqnarray}
with $\alpha_{ij}$ and $\beta_{ij}$ as Bogoliubov coefficients which follow the orthonormalization relations
\begin{eqnarray}\nonumber
 \sum_{j}\Big(\alpha_{ik}\alpha_{jk}^*-\beta_{ik}\beta_{jk}^*\Big)&=\delta_{ij},\\
 \sum_{j}\Big(\alpha_{ik}\beta_{jk}-\beta_{ik}\alpha_{jk}\Big)&=0.
\end{eqnarray}

Here comes the crucial step. If an observer  sees a field in $f$-vacuum while using $f$-modes, in which case there are no particles, the same system as seen by other observer  using $g$-modes would be such that expectation value of $g$-number operator is given by
\begin{eqnarray}\nonumber
 \langle 0_{f}|\hat{n}_{gi}|0_{f}\rangle&=\langle 0_{f} |\hat{b}_{i}^{\dagger}\hat{b}_{i}|0_{f}\rangle\\
 \nonumber
 &=\sum_{j}\beta_{ij}\beta^*_{ij}\\
\label{gf}
 &=\sum_{j}|\beta_{ij}|^2,
\end{eqnarray}
while we made use of Eq. (\ref{bog}). $\beta_{ij}$ is a non-zero quantity, which clearly manifests that annihilation operator of one observer is a combination of annihilation and creation operators of other one.
Eq. (\ref{gf}) signifies a remarkable result. It demonstrates that,  while one observer sees field in a vacuum state, the same field appears to other observer to be in a non-vacuum state. Thus, the uniqueness  of vacuum state in  Minkowski spacetime is broken completely in a curved spacetime. It is only for the inertial case, where 
$\beta_{ij}=0$, that the two observers agree on the definition of vacuum. It is this non-uniqueness of field states between the two observers which leads to the phenomenon of Hawking-Unruh effect.

\subsection{Thermality of the Minkowski vacuum: Unruh effect}
Variously known as Fulling-Davies-Unruh effect or  Unruh effect in short, in its simplest form, refers to the detection of thermal radiation by an accelerating observer (Rindler observer) in  Minkowski vacuum. Unruh originally discovered it while attempting to understand the underlying mechanism of black hole-originating  Hawking radiation \cite{PhysRevD.14.870}. The mathematical machinery of both these effects is same but the difference lies in the underlying spacetime geometry. While Unruh effect is observed in flat spacetime by an accelerating observer, the underlying spacetime in Hawking radiation is that of a black hole, which obviously is curved. For the derivation, we  follow books by Parker and Toms \cite{Parker:2009uva}, and Carroll \cite{Carroll:2004st}. \\
Recall from the previous section, the Rindler metric in $(1+1)$-dimensions reads \big(by using Eqs. (\ref{rindlerl}) and (\ref{RRR})\big)
\begin{eqnarray}\label{RU}
 ds^2=e^{2\alpha\xi}(-d\tau^2+d\xi^2).
\end{eqnarray}
It is not difficult to recognize that, since metric components in the line element of Eq. (\ref{RU}) are independent of $\tau$,  the vector 
\begin{align}\nonumber
 \partial_{\tau} &=\frac{\partial t}{\partial \tau}\partial_{t}+\frac{\partial x}{\partial \tau}\partial_{x}\\
 \nonumber
 &=e^{a\xi}\big[\cosh(a\tau)\partial_{t}+\sinh(a\tau)\partial_{x}\big]\\
 &=a(x\partial_{t}+t\partial_{x}),
\end{align}
is a Killing field associated with the boost in $x$-direction.\\
The Klein-Gordon equation for a massless particle $\left[m=0,\ \text{in}\ \text{Eq.} (\ref{BOXEQN}))\right]$ in Rindler metric of Eq. (\ref{RU}) can be expressed as 
\begin{eqnarray}
 \Box\phi=e^{-2\alpha\xi}(-\partial_{\tau}^2+\partial_{\xi}^2).
\end{eqnarray}
Solving this equation leads us to consider two sets of normalized plane waves corresponding to  left and right Rindler wedges respectively 
\begin{eqnarray}
 g_{k}^{(1)}=\frac{1}{\sqrt{4\pi\omega}}e^{-i(\omega\tau-k\xi)},\ 
 g_{k}^{(2)}=\frac{1}{\sqrt{4\pi\omega}}e^{i(\omega\tau+k\xi)},
\end{eqnarray}

 Here, we are considering  Minkowski  and  Rindler observers for the description of Unruh effect. For Minkowski observer, the field $\phi$ can be expressed in terms of his choice of annihilation and creation operators ($\hat{a}_{k}$ and $\hat{a}_{k}^{\dagger}$\ respectively) as 
\begin{eqnarray}
 \phi=\int dk \big(\hat{a}_{k}f_{k}+\hat{a}_{k}^{\dagger}f_{k}^*\big),
\end{eqnarray}
where $f_{k}$ are the Minkowski plane wave modes. For the Rindler observer with  $\hat{b}_{k}$ and  $\hat{b}_{k}^{\dagger}$ as annihilation and creation operators respectively, the same reads as 
\begin{eqnarray*}
 \phi =\int dk \Big(\hat{b}_{k}^{(1)}g_{k}^{(1)} +   \hat{b}_{k}^{(1)\dagger}g_{k}^{(1)*}+   \hat{b}_{k}^{(2)}g_{k}^{(2)} 
  +\hat{b}_{k}^{(2)\dagger}g_{k}^{(2)*}\Big).
\end{eqnarray*}
  Minkowski vacuum represented by $|0_{M}\rangle$ is defined here as follows
\begin{eqnarray}
 \hat{a}_{k}|0_{M}\rangle=0,
\end{eqnarray}
and the Rindler vacuum as 
\begin{eqnarray}
 \hat{b}_{k}^{(1)}|0_{R}\rangle=\hat{b}_{k}^{(2)}|0_{R}\rangle=0.
\end{eqnarray}
We emphasize here that, even though the Hilbert space for the theory is same for both observers, they however differ in Fock space description. This is because Rindler vacuum can be described as many-particle state in Minkowski representation, which arises due to fact that a Rindler mode can be written as an admixture of creation and annihilation operators of Minkowski representation. As part of the ansatz outlined in the previous section, we need to compute Bogoliubov coefficients that relate Minkowski and Rindler descriptions of the field. For a consistent formulation of field mode description in Rindler frame, we need to define a new set of functions comprising positive and negative frequency modes as 
\begin{eqnarray}\nonumber
 h_{k}^{(1)}=\frac{1}{\sqrt{2\sinh(\frac{\pi\omega}{\alpha})}}\Big(e^{\pi \omega/2\alpha}g_{k}^{(1)} + e^{-\pi \omega/2\alpha}g_{-k}^{(2)*}\Big),
\end{eqnarray}
and 
\begin{eqnarray}\nonumber
 h_{k}^{(2)}=\frac{1}{\sqrt{2\sinh(\frac{\pi\omega}{\alpha})}}\Big(e^{\pi \omega/2\alpha}g_{k}^{(2)} + e^{-\pi \omega/2\alpha}g_{-k}^{(1)*}\Big),
\end{eqnarray}
with the inner product $\Big(h_{1}^{(1)}, h_{k}^{(2)}\Big)=\delta(k_{1}-k_{2})$.
Also,by employing $h_{k}$ modes,  positive and negative frequency Minkowski modes have associated annihilation and creation operators, respectively, such that  $\forall k$
\begin{eqnarray}\nonumber
 \hat{c}_{k}^{(1)}|0_{M}\rangle= \hat{c}_{k}^{(2)}|0_{M}\rangle=0,
\end{eqnarray}
which indicates a description of Minkowski vacuum using $h_{k}$ modes. With these new modes, we write field as 
\begin{eqnarray*}
\phi =\int dk \Big(\hat{c}_{k}^{(1)}h_{k}^{(1)} +   \hat{c}_{k}^{(1)\dagger}h_{k}^{(1)*}+   \hat{c}_{k}^{(2)}h_{k}^{(2)} 
  +\hat{c}_{k}^{(2)\dagger}h_{k}^{(2)*}\Big).
\end{eqnarray*}
Now, the Rindler annihilation operators $\hat{b}_{k}$'s can be written in terms of $\hat{c}_{k}$'s as 
\begin{eqnarray}\label{NOP1}
  \hat{b}_{k}^{(1)}=\frac{1}{\sqrt{2\sinh(\frac{\pi\omega}{\alpha})}}\Big(e^{\pi \omega/2\alpha}\hat{c}_{k}^{(1)} + e^{-\pi \omega/2\alpha}\hat{c}_{-k}^{(2)\dagger}\Big).
\end{eqnarray}
Similarly, we have 
\begin{eqnarray}\label{NOP2}
  \hat{b}_{k}^{(2)}=\frac{1}{\sqrt{2\sinh(\frac{\pi\omega}{\alpha})}}\Big(e^{\pi \omega/2\alpha}\hat{c}_{k}^{(2)} + e^{-\pi \omega/2\alpha}\hat{c}_{-k}^{(1)\dagger}\Big).
\end{eqnarray}
By making use of Eqs. (\ref{NOP1}) and (\ref{NOP2}), one is able to construct number operator for Rindler observer in terms of $\hat{c}_{k}$'s, which for region I is 
\begin{eqnarray}
 \hat{n_{R}}^{(1)}(k)=\hat{b}_{1}^{(1)\dagger}\hat{b}_{k}^{(1)}.
\end{eqnarray}
Our aim is to compute the particle number in Rindler frame using Minkowski modes.  Hence, we have
\begin{eqnarray}\nonumber
 \langle 0_{M}| \hat{n_{R}}^{(1)}(k)|0_{M}\rangle= \langle \hat{b}_{1}^{(1)\dagger}\hat{b}_{k}^{(1)}|0_{M}\rangle.
\end{eqnarray}
Following Eqs.  (\ref{NOP1}) and (\ref{NOP2}), we get 
\begin{align}\nonumber
 \langle 0_{M}| \hat{n_{R}}^{(1)}(k)|0_{M}\rangle&=\frac{1}{2\sinh(\frac{\pi\omega}{\alpha})}\langle 0_{M}|e^{-\pi \omega/\alpha}\hat{c}_{-k}^{(1)} \hat{c}_{-k}^{(1)\dagger}|0_{M}\rangle\\
 \nonumber
 &=\frac{e^{-\pi\omega/\alpha}}{2\sinh(\frac{\pi\omega}{\alpha})}\delta(0)\\
 \label{FINAL}
 &=\frac{1}{e^{2\pi\omega/\alpha}-1}\delta(0).
\end{align}
Here the factor $\delta(0)$ arises out of the  use of non-square-integrable plave wave modes. In fact,  it is the expectation value of particle number 
\begin{eqnarray}
 \langle 0_{M}|\hat{c}_{-k}^{(1)} \hat{c}_{-k}^{(1)\dagger}|0_{M}\rangle=\delta(0),
\end{eqnarray}
indicating that $\hat{c}_{-k}^{(1)\dagger}|0_{M}\rangle$ is a normalized one-particle state. Eq. (\ref{FINAL}) shows that  Minkowski vacuum looks to the Rindler observer like  a thermal state with non-zero particle content. This makes notion of vacuum and particle  frame-dependent in curved spacetime, which means  that they do not represent some fundamental elements in non-inertial frames. It is thus argued that this thermal nature of vaccum in the form of Unruh effect is necessary for the consistency of quantum field theory  and needs no more experimental verification than the field theory itself \cite{RevModPhys.80.787}. Finally, taking a look at Eq. (\ref{FINAL}) shows that this is a Planck spectrum with a characteristic  temperature
\begin{eqnarray}\nonumber
 T_{U}=\frac{\alpha}{2\pi},
\end{eqnarray}
which upon putting the constants back furnishes 
\begin{eqnarray}\label{TEMP2}
 T_{U}=\frac{\hbar \alpha}{2\pi c k_{B}}.
\end{eqnarray}
This is the famous relation for Unruh temperature. The weakness of Unruh effect can be readily seen from above relation. To get a feel of it, one can put the value of constants in Eq. (\ref{TEMP2}) and this gives an estimate of required acceleration as $10^{20} m/s^{2}$, which is incredibly large.   This makes Unruh effect very difficult to observe in the lab. However, some simulation experiments have revealed the consistency of Unruh's prediction (see \cite{Hu:2018psq} as a  recent work).

\subsection{Black holes aren't black: Hawking radiation}
Though Hawking effect bears close similarity to Unruh effect in the mathematical formulation, they only differ in underlying spacetime. Unruh effect occurs in accelerated frames in Minkowski spacetime, while Hawking effect occurs for accelerated observers in curved spacetime. Einstein's equivalence principle guarantees their consistency. Since the original derivation by Hawking is tedious, so we follow simpler and brief approach by Jacobson 
 \cite{Jacobson2005} and Caroll \cite{Carroll:2004st}.
 
We write down the Schwarzschild metric of a black hole as
\begin{eqnarray}\nonumber
 ds^2=-\left(1-\frac{2GM}{r}\right)dt^2+ \frac{1}{\left(1-\frac{2GM}{r}\right)}dr^2+r^2 d\Omega^2,
\end{eqnarray}
where $2GM=R_{S}$ is the Schwarzschild radius. In Schwarzschild spacetime, given an observer with four acceleration $U^{\mu}$, we define a Killing field $K^{\mu}=V(x)U^{\mu}$, such that its magnitude given by
\begin{eqnarray}\nonumber
V=\sqrt{-K_{\mu}K^{\nu}},
\end{eqnarray}
which interestingly gives us a redshift factor $V$, that relates the emitted and observed frequencies of photons by static observers as $E=-p_{\mu}U^{\mu}$. An observer at a distance $r$ from the black hole such that $r>2GM$, has a geodesic with timelike Killing vector $\partial_t$ such that 
\begin{eqnarray}\label{RS}
 V=\sqrt{1-\frac{2GM}{r}}.
\end{eqnarray}
For this observer, the magnitude of four-acceleration is given by 
\begin{eqnarray}
\alpha=\frac{GM}{r\sqrt{r-2GM}}.
\end{eqnarray}
To put the things in perspective, we consider two observers here, one close to event horizon at a  distance $r_{1}>2GM$ and other far away at distance $r_{2}>>2GM$. Now, evidently for the one near the horizon, acceleration is very large, i.e
\begin{eqnarray}
 \alpha_{1}>>\frac{1}{2GM}.
\end{eqnarray}

Therefore, this much of acceleration sets a time and length scales such that the spacetime looks essentially flat. Here, for a (third) \textit{freely-falling} observer,  nothing unusual happens  (see also the associated \textit{"firewall"} argument \cite{Almheiri:2012rt} or \textit{"fuzzball"} scenario \cite{Mathur:2009hf}),  which implies that the  falling observer sees a vacuum state field as Minkowski vacuum, which to the observer at $r_{1}$ distance produces an Unruh effect with radiation having temperature
\begin{eqnarray}
 T_{1}=\frac{\alpha_{1}}{2\pi}.
\end{eqnarray}
Now, for the observer far away from the black hole at $r_{2}$ distance (in principle $r_{2}\rightarrow \infty $),  length and time scales set by the acceleration $\alpha_{2}^{-1}>>2GM$ are large and thus one can not ignore the  curvature effects. The overall impact of  this curvature is the redshift of radiation frequency that comes from the black hole with the temperature
\begin{eqnarray}\label{HT}
 T_{2}=\frac{V_{1}}{V_{2}}\frac{\alpha_{1}}{2\pi}.
\end{eqnarray}
One can readily see from Eq. (\ref{RS}), as $r_{2}\rightarrow \infty $, $V_{1}\rightarrow 1$, Eq. (\ref{HT}) yields 
\begin{eqnarray}
T=\lim _{r_{1}\rightarrow 2GM} \frac{V_{1}\alpha_{1}}{2\pi}.
\end{eqnarray}
In general, for a black hole, one can write
\begin{eqnarray}\label{HA}
 T_{H}=\frac{\kappa}{2\pi}=\frac{\hbar \kappa}{2\pi},
\end{eqnarray}
where $\kappa=\lim (V\alpha)$  represents the surface gravity of the black hole. Eq. (\ref{HA}) is the celebrated Hawking effect  with $T_{H}$ as Hawking temperature--the observed temperature of the thermal radiation felt by static observer at a far off distance from black hole--Schwarzschild observer.

\subsection{Dynamical Casimir effect}
A related effect to Hawking-Unruh effect which bears same underlying principle of particle generation from quantum vacuum is the famous dynamical Casimir effect (DCE) of Moore (also known as Moore-Casimir effect) \cite{10.1063/1.1665432,Dodonov:2020eto}, which has been successfully  verified in some direct  \cite{Wilson_2011, Lahteenmaki:2011cwo} and analogue \cite{Jaskula:2012ab} experiments. It is a prime example of field quanta generation due to moving or oscillating  boundaries. In fact, study of fields in presence of  boundaries is an old  subject, starting all the way from classical fields (see Refs.\cite{Jaekel:1997hr,Dodonov:2009zza}).
Here, our motive is to provide a brief introduction to DCE, most of which is based on the Ref. \cite{Dodonov:2020eto}, wherein the following has been stated as a standard definition of DCE:\\
\indent \textit{``Macroscopic phenomena caused by changes of vacuum quantum
states of fields due to fast time variations of positions (or properties) of boundaries confining the fields (or other
parameters)".}

From a classical field point of view, one may consider the wave equation for a field $A$ in units of $(c=1)$,
\begin{eqnarray}
    \left(\frac{\partial^2}{\partial t^2}-\frac{\partial^2 }{\partial x^2}\right)A=0,
\end{eqnarray}
which for a time-dependent domain $0<x<L(t)$ satisfying the boundary conditions, $A(0,t)=A(L(t),t)=0$ provides the solution 
\begin{eqnarray}
L(t)=L_{0}(1+\alpha t),
\end{eqnarray}
where $\alpha$ is a parameter signifying the scale of changes in the string length. This solution by Nicolai \cite{doi:10.1080/14786442508634593} is the one of the maiden attempts of the study of behaviour fields in presence of boundaries.  For quantum fields, the basic considerations stem from zero-point fluctuations governed by Heisenberg uncertainty principle in quantum mechanics. We assume a double wall cavity, with one wall  fixed at $x=0$ and other is movable, in such a way that the second wall is at rest for $t\leq 0$. For a field potential $A(x,t)$ perpendicular to axis $x$, satisfying the boundary conditions as stated above, one can write down
\begin{eqnarray}
    A(x,t<0)=\sum_{n=0}^{\infty} c_{n} \sin{\left(\frac{n\pi x}{L_{0}}\right)}e^{-i \omega_{n}t},
\end{eqnarray}
which is the initial state of the field. Here, $c_{n}$ are the coefficients being complex numbers and quantum operators in classical and quantum cases, respectively. $L_{0}$ is the equilibrium length of the cavity when both walls are at rest. For obtaining the solution to field $A$ for $t>0$, one has the freedom to adopt many approaches. We follow here the one by Moore \cite{10.1063/1.1665432}. Moore's solution  assumes the following form
\begin{eqnarray}
    A_{n}(x,t)=C_{n}[\exp{(-i\pi n R(t-x))}-\exp{(-i\pi n R(t+x))}],
\end{eqnarray}
where $R(\zeta)$ satisfies the equation $R(t+L(t))-R(t-L(t))=0$. The functional for of $R(\zeta)$ is given by 
\begin{eqnarray}
    R_{\alpha}(\zeta)=\frac{2\ln{|1+\alpha \zeta|}}{\ln{|(1+v)/(1-v)|}}.
\end{eqnarray}
Following some standard operational procedures \cite{10.1063/1.1665432,Dodonov:2009zza,Dodonov:2020eto}, one obtains 
\begin{multline}
    A_{n}(x,t)=C_{n}[\exp{(-i\pi n F(t-x))}
    -\exp{(-i\pi n G(t+x))}],
\end{multline}
with the following conditions:
\begin{align}
    G(t+R(t))-F(t-R(t))=2, \\
    G(t+L(t))-F(t-L(t))=0.
\end{align}
When the wall is at rest, one gets static Casimir force between the plates, $F=-\pi \hbar c/(24L_{0}^2)$. On contrary, when the distance between walls varies in such a way that $\left|dL/dt <<c\right|$, we have
\begin{multline}
    F=-\frac{\pi \hbar c}{24 L^2(t)}\left[1+\left(\frac{\dot{L}}{c}\right)^2\left(\frac{7}{3}-\frac{1}{\pi^2}\right)-\frac{L\ddot{L}}{c^2}\left(\frac{2}{3}-\frac{2}{\pi^2}\right)\right].
\end{multline}
This attractive force between the walls originates from fluctuating field modes due to oscillating boundaries. This process generates particles, which for $m$th mode,  happens at a rate \cite{Jaekel:1997hr,Dodonov:2020eto}
\begin{eqnarray}
    \frac{dN_{m}}{dt}\equiv |\epsilon|\frac{(1-(-1)^m)}{m\pi},
\end{eqnarray}
where $\epsilon$ is related to eigenmode limit $\epsilon t\ll 1$. We will see in Sec \ref{seciv}. C, the particle generation profile due to an oscillating mirror has  a Planckian factor in it and hence bears close similarity to that of Hawking-Unruh effect.

\section{Radiative atoms in Rindler and black hole spacetimes}\label{seciii}
Having discussed the prerequisites,   we now discuss atom-field interactions in Rindler and  black hole spacetimes, with contributions from Hawking-Unruh effect. 
\subsection{Unruh thermality in the atomic spectrum}
Spontaneous excitation of atoms is one of the foremost phenomena that captures the physics of atom-field interactions. It drives its origin from zero point fluctuations of electromagnetic field \cite{Lamb:1947zz,Welton:1948zz}, or the QED radiative reactions for atomic transition frequencies \cite{Ackerhalt:1973fk}, or the combination of them \cite{Milonni:1973zz}. We consider a field here that is decomposed in terms of creation and annihilation operators 
\begin{eqnarray}\nonumber
 \phi(x,t)=\int d^3k\ g_{k}[a_{k}(t)e^{ikx}+a_{k}^{\dagger}(t)e^{-ikx}],
\end{eqnarray}
where $g_{k}=[2\omega_{k}(2\pi)^3]^{-1/2}$, that interacts with an atom moving in Minkowski spacetime along the worldline  $x(\tau)$.   Here, $k$ is the wave vector of the field with the frequency $\omega_{k}$, and $\tau$ is the proper time of the atom. The Hamiltonian for atom-field interaction  in the atom's proper frame  is given by 
\begin{eqnarray}\label{SPONT}
  H(\tau)=\mu \sigma (\tau)\phi(x(\tau)).
 \end{eqnarray}
 Here $\mu$ is the atom-field coupling constant, and   $\sigma(\tau)$ depicts the internal structure of the atom and is a certain combination of  atomic lowering and raising operators.  As a part of the standard solution procedure for Heisenberg equations of motion, field $\phi$ in the interaction Eq. (\ref{SPONT}) is split into free and source parts  \cite{PhysRevA.50.1755}
\begin{eqnarray}\nonumber
  \phi^f (x(\tau))&=\int d^3k\ g_{k}[a_{k}(0)e^{i(kx-\omega_{k} t)}\\
  \label{A}
  &+a_{k}^{\dagger}(0)e^{-i(kx-\omega_{k} t)}],
 \end{eqnarray}
and 
\begin{eqnarray}
  \phi^s (x(\tau))=i\mu \int_{\tau_{0}}^{\tau} d\tau' \sigma^f(\tau')[\phi^f(x(\tau')), \phi^f(x(\tau))],
 \end{eqnarray}
respectively.
Next, an atomic observable $O(\tau)$ is defined, for which  the Heisenberg equation of motion
\begin{eqnarray}
  \frac{d O(\tau)}{d\tau}=i[H(\tau), O(\tau)],
 \end{eqnarray}
 yields  
 \begin{eqnarray}\label{D1}
\frac{d O(\tau)}{d\tau}= i\mu\phi(x(\tau))[\sigma(\tau), O(\tau)].
\end{eqnarray}
To account for contributions of vacuum fluctuations and radiation reaction for atom excitation, Dalibard, Dupont-Roc and Cohen-Tannoudji (DDC) \cite{refId0,refId1} proposed that a preferred symmetric ordering between field operator and atomic observable in Eq. (\ref{D1}) leads to a meaningful physical description of radiation reaction and vacuum fluctuations. This consideration leads to the following two equations
\begin{eqnarray}\nonumber
  \Big[\frac{d O(\tau)}{d\tau}\Big]_{VF}=\frac{1}{2}i\mu\phi^f (x(\tau))[\sigma(\tau), O(\tau)]\\
  \label{E}
  +\Big[\sigma(\tau), O(\tau)\Big]\phi^f(x(\tau)),
 \end{eqnarray}
which pertains to $\phi^f$, the free part of the field, and 
\begin{eqnarray}\nonumber
  \Big[\frac{d O(\tau)}{d\tau}\Big]_{RR}=\frac{1}{2}i\mu\phi^s (x(\tau))[\sigma(\tau), O(\tau)]\\
  \label{F}
  +\Big[\sigma(\tau), O(\tau)\Big]\phi^s(x(\tau)),
 \end{eqnarray}
 which pertains to $\phi^s$, the source part of it.
 
With regard to Unruh effect, an atom is a typical example of point-like Unruh-DeWitt detector \cite{2010grae.book.....H} that provides means to probe the Unruh thermality of vacuum.  In our case, the observable associated with the accelerated atom is its Hamiltonian, $H_{A}$, and we consider its interaction with a scalar field.  It is found that the rate at which the total energy of  the atom with two given states  $(|a\rangle$  and $|b \rangle)$  changes is given by adding Eqs.  (\ref{E}) and (\ref{F}), which  yields 
\begin{align}\nonumber
\Big\langle\frac{dH_{A}}{d\tau}\Big\rangle_{tot}&=\frac{\mu^2}{2\pi}\Bigg[-\sum_{\omega_{a}>\omega_{b}}\omega_{ab}^2|\langle a|\sigma^f (0)
|b|\rangle|^2  \\
\nonumber
& \times  \Bigg(1+\frac{1}{e^{2\pi\omega_{ab}/\alpha}-1}\Bigg)+\sum_{\omega_{a}<\omega_{b}} \omega_{ab}^2|\langle a|\sigma^f(0)
|b|\rangle|^2\\
\label{Z}
&  \times \Bigg(\frac{1}{e^{2\pi\omega_{ab}/\alpha}-1}\Bigg)\Bigg],
\end{align}
where $\omega_{ab}=\omega_{a}-\omega_{b}$ is the transition frequency of the atom \cite{PhysRevA.50.1755}. From  Eq.  (\ref{Z}), one can  see that if the atom is initially in the excited state, the term $\omega_{a}>\omega_{b} $ contributes only, leading to spontaneous emission. On the other hand, if the atom is initially in the ground state, the term $\omega_{a}<\omega_{b}$ contributes, which indicates that there is no balance between vacuum fluctuations and radiation reaction, leading to spontaneous excitation of atom even in vacuum. Note the acceleration ($\alpha$) dependence of the energy rate change, which in the limit $\alpha\rightarrow 0$ agrees with that of inertial case.  The above result shows a complete conformity with Unruh effect  for a scalar field \cite{PhysRevD.14.870}. In contrast to the scalar field, there arise some nonthermal contributions to the excitation rate of   Eq. (\ref{Z}) for linear acceleration in electromagnetic \cite{Zhu:2006wt} and circular accelerations in Dirac fields \cite{2015AnPhy.353..317C}.  This alludes to the loss of equivalence between uniform acceleration and thermality, and also affects  the contribution of radiation reaction to the atomic energy.

 It turns out that,  this thermality of Minkowski vacuum is a must for the consistency of inertial perspective which has been vindicated from the perspective of a co-accelerated observer as well, both with and without boundary \cite{Zhou:2016zsf}. For Dirac field, by considering a non-linear coupling case, a cross term appears in the rate of mean energy change given by \cite{Zhou:2012gu}.

\begin{align}\nonumber
\Big\langle\frac{dH_{A}}{d\tau}\Big\rangle_{tot}&=-\frac{\mu^2}{60\pi^3}\sum_{\omega_{a}>\omega_{b}}\omega_{ab}^6|\langle a|\sigma^f(0)
|b|\rangle|^2  \\
\nonumber
& \times \Bigg( 1+\frac{5\alpha^2 }{\omega_{ab}^2}+\frac{4\alpha^4}{\omega_{ab}^4}\Bigg) \Bigg(1+\frac{1}{e^{2\pi\omega_{ab}/\alpha}-1}\Bigg)\\
\nonumber
&+\frac{\mu^2}{60\pi^3}\sum_{\omega_{a}<\omega_{b}} \omega_{ab}^2|\langle a|\sigma^f(0)
|b|\rangle|^2\\ 
\label{D}
& \times \Bigg( 1+\frac{5\alpha^2 }{\omega_{ab}^2}+\frac{4\alpha^4}{\omega_{ab}^4}\Bigg) \Bigg(\frac{1}{e^{2\pi\omega_{ab}/\alpha}-1}\Bigg).
\end{align}
Though this response of the atom with Dirac field consistent with typical Dirac particle detector \cite{Langlois:2005nf}, however the additional term $\alpha^4$ in Eq. (\ref{D})  is absent both in case of scalar and electromagnetic fields, and the contribution of $\alpha^4$ term becomes very dominant for
$\alpha >>\omega$ \cite{Zhou:2012gu}, where $\omega$ is the  transition frequency of hydrogen atom. A more comprehensive way is to consider Einstein coefficients, since they are a must for understanding these spontaneous transition processes of atoms. In the accelerated atom case, corresponding to a ground state $1$ and an excited state $2$, we find spontaneous emission to be enhanced by a thermal factor as 
\begin{eqnarray}\label{COFA}
  A_{21}=\frac{\mu^2}{8\pi}\omega \Big(1+\frac{1}{e^{2\pi\omega/\alpha}-1}\Big),
 \end{eqnarray}
compared to the inertial case, $A_{21}=\frac{\mu^2}{8\pi}\omega$,  while we get a weighted thermal correction to the  spontaneous excitation rate 
\begin{eqnarray}\label{COFB}
 A_{12}=\frac{\mu^2}{8\pi}\omega\Big(\frac{1}{e^{2\pi\omega/\alpha}-1}\Big).
\end{eqnarray}
 It can be seen that excitation rate given by Eq. (\ref{COFB}) vanishes as $\alpha \rightarrow 0$ in agreement with Unruh effect \cite{PhysRevA.50.1755} (see also \cite{Rizzuto_2011, Zhang:2016gtp} for the inclusion of a boundary). Interestingly, based on a time-dependent perturbative  method, similar results to that of Eq. (\ref{COFB}) have been obtained  in
\cite{2005JOptB...7S..21B,2008JPhA...41p4030B,CALOGERACOS2016377}, indicating a great agreement with that of DDC formalism considered here.  

\begin{figure}[h!]
 \begin{center}
\includegraphics[width=70 mm]{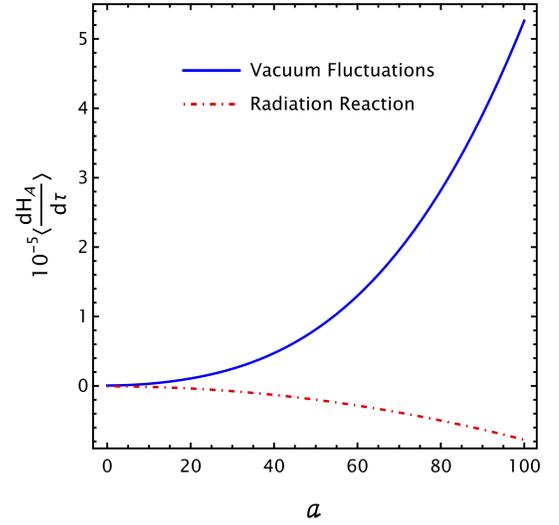}
 \end{center}
\caption{Contributions of vacuum fluctuations (blue solid line) and radiation reaction (red dashed line) for an entangled accelerated two-atom system for the variation of $a$, where $\alpha=ae^{-a\xi}$. The chosen parameters include $\omega=5, \mu=1, \mu^1=\mu^2=\mu^3=1$ and  Rindler coordinates $\xi_{1}=\xi_{2}=0$ in natural units ( Menezes \textit{et al.} \cite{Menezes:2015iva}).}
 \label{RR}
\end{figure}

Since entanglement is a key ingredient of quantum information, cryptography and computation \cite{RevModPhys.73.565}, it would be reasonable to study the impact of acceleration on radiative transitions of entangled atoms. Once again, the investigations have been done via DDC formalism (see, e.g.,  Ref. \cite{Menezes:2015iva}). Some of important treatments on the subject concern the link between spontaneous emission and entangled states in Minkowski vacuum \cite{PhysRevLett.93.140404}. Also boundaries have been shown to play decisive role in transition probabilities of entangled atoms \cite{Arias:2015moa, sym11121515}. For the inertial case, both radiation reaction and vacuum fluctuations cancel at asymptotic times, while they are responsible for decay of entanglement between the atoms. This means that entanglement could serve as an interplay between radiation reaction (a classical concept) and vacuum fluctuations (quantum concept) \cite{Menezes:2015uaa}. It is found that if the acceleration of atoms in this case is very large, the contributions of radiation reaction and vacuum fluctuations are distinct \cite{Menezes:2015iva}, as given by
\begin{eqnarray}
 \Bigg\langle \frac{d H_{A}}{d\tau}\Bigg\rangle_{VF}=\frac{\omega^4\mu^2}{3\pi}\Bigg(\frac{\alpha^2}{\omega^2}+1\Bigg)\Bigg(1+\frac{2}{e^{2\pi \omega/\alpha}-1}\Bigg),
\end{eqnarray}
and 
\begin{eqnarray}
 \Bigg\langle \frac{d H_{A}}{d\tau}\Bigg\rangle_{RR}=-\frac{\omega^4\mu^2}{3\pi}\Bigg(\frac{\alpha^2}{\omega^2}+1\Bigg)
\end{eqnarray}
which is depicted in Fig. \ref{RR}.

The situation is further modified if one considers the system interacting with a scalar field near a boundary. By introducing a perfectly reflecting boundary near the entangled system, the transition rates become dependent upon atom-boundary separation, separation between the atoms and the acceleration     $\alpha$ \cite{sym11121515}. For some cases of radiative processes of entangled atoms, the acceleration produces a nonthermal behaviour in the atomic energy changes  and  could also possibly pave way for manipulating the radiative behaviour of entangled 
systems \cite{sym11121515, Zhou:2020oqa} (see also \cite{PhysRevD.101.085009} for the case of co-accelerated observer). We also point out here that Ref.\cite{Zhou:2020oqa} contains a detailed investigation of this issue and some of the results reported in Refs. \cite{PhysRevA.93.052117, Menezes:2015iva,Menezes:2015uaa} have been challenged.

Spontaneous transitions are often associated  with radiative energy shifts in atoms, which includes Lamb shift.  However, there exists a  distinction between the behaviour of  scalar and electromagnetic fields in contributing to the energy shift \cite{PhysRevA.57.1590,PhysRevA.79.062110}. Since for inertial case, it is very well known that vacuum fluctuations, and not the radiation reaction, contribute to Lamb shift, it turns out that this situation still holds in accelerated case as well. Once again, consider a uniformly accelerated  two-level atom interacting with a massless scalar field  which undergoes an energy shift. Since in this case, radiation reaction generally 
does not contribute anything \cite{Audretsch:1995iw}, we have the following  contributions to Lamb shift 
\begin{eqnarray}
 \Delta =\Delta_{0}+ \frac{\alpha^2\mu^2}{192\pi^2}\frac{1}{\omega},
\end{eqnarray}
where $\Delta_{0}$ is the Lamb shift for inertial case. The extra factor here comes from acceleration and is purely contributed by  vacuum fluctuations \cite{1995PhRvA..52..629A}. For arbitrary stationary spacetimes, we have 
\begin{eqnarray}
 \Delta=\Delta_{0}+\mathcal{D},
\end{eqnarray}
where $\mathcal{D}$ is an extra term that can be evaluated for particular type of acceleration. For the circular acceleration case, the correction term comes out be \cite{Audretsch:1995iw}
\begin{eqnarray}\nonumber
 \mathcal{D}&=\frac{\alpha \mu^2}{64\sqrt{3}\pi^2}\Bigg[e^{-2\sqrt{3}B\omega/\alpha}\bar{Ei}\bigg(2\sqrt{3}B\frac{\omega}{\alpha}\bigg)\\
 &-e^{2\sqrt{3}B\omega/\alpha}\bar{Ei}\bigg(-2\sqrt{3}B\frac{\omega}{\alpha}\bigg)\Bigg],
\end{eqnarray}
where $\bar{Ei}$ is the principal value of exponential integral function. Here the factor $B$ is related to Lorentz factor ($v$ is the velocity of atom) as
\begin{eqnarray}
B=1-\frac{1}{5}(v\gamma)^{-2}.
\end{eqnarray}
The correction is graphically shown below in Fig. \ref{CORR} against the inertial emission rate  $\Gamma_{inertial}=\mu^2 \omega/8\pi$.

\begin{figure}[h!]
 \begin{center}
\includegraphics[width=7cm]{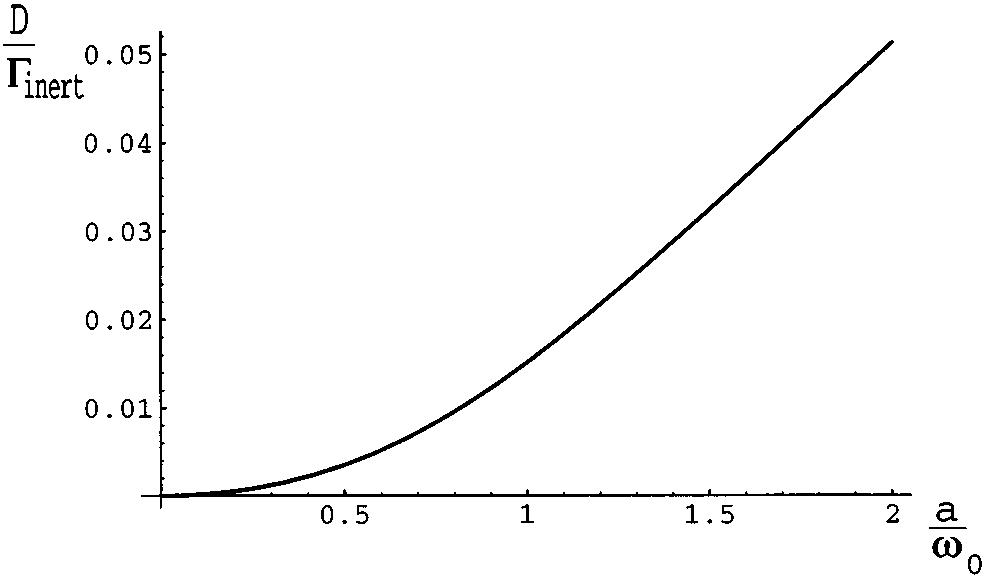}
 \end{center}
\caption{Enhancement of atomic energy shift $\mathcal{D}$ due to acceleration (from Audretsch \textit{et al.} \cite{Audretsch:1995iw}).}
 \label{CORR}
\end{figure}

The above plot shows clearly the enhancement of energy shifting compared to the  inertial emission rate of the atom. Beyond Unruh contribution, in some multilevel atoms, acceleration can induce Raman-like transitions. By virtue of these transitions, some states show no contribution from Unruh effect due to acceleration \cite{Marzlin:1997gx}.  It is worthwhile to mention here that the impact of acceleration or gravity on the atomic spectrum and level shifting has been considered in other formalisms as well, quite different from the one considered here.  The implication is that gravity plays a great role  in quantum phenomena by using atom to probe the spacetime curvature effects \cite{Parker:1969au, Audretsch:1993uc}. This has been extended to probe various gravity models \cite{Olmo:2008ye, Singh:2011cq, Wong:2017jer}. Very recently, this line of work also finds connections to some table-top experimental setups with interesting cosmological implications \cite{PhysRevD.97.084050,PhysRevLett.123.061102}.

\subsection{The case of  black holes}
\subsubsection{Hawking contributions to the atomic spectrum}
In this section, we discuss the atomic transition processes and energy level shifting occurring in different curved spacetimes of black holes.  Since the underlying space is curved,  one must expect Hawking radiation to play role , since equivalence principle guarantees the correspondence between Unruh and Hawking effect. But before jumping to the radiative processes, it is necessary to mention  the defining properties of black holes. No-hair theorem posits that all  black hole solutions of Einstein-Maxwell equations of general relativity are characterized by only three parameters \cite{Hobson:2006se}: Mass ($M$), angular momentum ($J$) and charge $(Q)$, related by  
\begin{eqnarray}
 M^2-\big(\frac{J}{M}\big)^2-Q^2\geq 0,
\end{eqnarray}
in natural units $c=G=1$. The most general case, when all these three parameters are present, corresponds to Kerr-Newmann black holes (charged, spinning), which reduces to Kerr black hole (uncharged, spinning) for $J\neq 0,Q=0$, Reissner-Nordstr\"om black hole (charged, non-spinning) for $J=0,Q\neq0$ and Schwarzschild (uncharged, non-spinning) black hole for $J=Q=0$. So these parameters are expected to play great role in the atomic radiative phenomena. The choice of vacuum state in a black hole spacetime is also of paramount importance. Contrary to Minkowski spacetime,  where vacuum is associated with non-occupation of positive frequency field modes  corresponding to a unique definition of time and time-like Killing vector, no such definition exists in curved spacetime. 

One can choose vacuum with respect to Schwarzschild time $t$ i.e. for static observer far from black hole, which happens to be  the natural definition of time. For this choice of time, Boulware vacuum is the vacuum state defined with respect to existence of normal modes for Killing vector $\partial/\partial t$, but the expectation value of renormalized stress-energy tensor in freely-falling frame diverges at horizon \cite{Sciama:1981hr}. However, this divergence is removed on the future horizon  in Unruh vacuum \cite{PhysRevD.14.870}, which is  the  most suitable choice of vacuum state of gravitational collapse of a massive body. In Unruh vacuum, the frequency modes coming from past horizon are defined with respect to Kruskal coordinates, the canonical affine parameter at past horizon. In addition to this, Hartle-Hawking vacuum \cite{Hartle:1976tp} is not empty at infinity and has rather thermal particles coming from infinity when black hole and thermal radiation are in equilibrium.  

 The effects of acceleration or curvature on atomic spectrum and radiative phenomena have been dealt in many paradigms. One of the possible ways is to incorporate Maximal Acceleration of Caianiello \cite{Papini:2015fha} into the radiative processes and Lamb shift of atoms. The basic idea is to consider an accelerating atom of mass $m$ following a worldline in the metric 
\begin{eqnarray}\label{MAM}
\bar{ds}^2=g_{\mu\nu}dx^{\mu}dx^{\nu}\Big[1-\frac{|\ddot{x}|^2}{\mathcal{A}_{m}^2}\Big],
\end{eqnarray}
where $g_{\mu\nu}$ is metric of some background gravitational field, $\ddot{x}^\mu$ its acceleration and $\mathcal{A}_{m}=(\frac{2mc^3}{\hbar})$ is the maximum acceleration limit. 
Note that the effective geometry given by Eq.  (\ref{MAM}) is a mass-dependent correction and induces curvature. With that spacetime configuration, the correction to $2S$-$2P$ Lamb shift in hydrogen atom has been demonstrated to be compatible with experiment \cite{Lambiase:1998tn}. Many other issues have been analyzed within this Maximal Acceleration model including   its relevance to Unruh effect \cite{Benedetto:2015fta} and some aspects of radiative phenomena \cite{Papini:2015fha}. However, here we continue to discuss the  curvature effects on atomic radiative and level shift phenomena using DDC formalism as done in the Rindler case. This will maintain the continuity of our discussion in a natural way.

We begin by highlighting the work by Higuchi \textit{et al.}\cite{Higuchi:1998qc}, who considered the possible excitation and emission processes of a static source outside a Schwarzschild black hole due to Hawking radiation. A point-like detector interacting with a \textit{massless}  scalar field in Unruh vacuum gives the following response rate 
\begin{eqnarray}
 R_{tot}=\frac{q^2\alpha}{4\pi^2},
\end{eqnarray}
where $q$ is the coupling constant between source and field and $\alpha$ is the proper acceleration of source held fixed in black hole spacetime. In terms of response of static sources, this result establishes an equality between  Schwarzschild spacetime (with Unruh vaccum) and Minkowski spacetime (with Minkowski vaccum), which is kind of unexpected result, since all classical formulations of equivalence principle are locally valid, while the quantum states are defined golablly \cite{Higuchi:1998qc}. However, no such  equality is found for electromagnetic \cite{Crispino:2000jx} and \textit{massive} scalar  fields \cite{Castineiras:2002fn}. Using DDC formalism, which is different than one used by Higuchi \textit{et al.} \cite{Higuchi:1998qc}, we discuss spontaneous excitation and Lamb shift of an atom interacting with scalar and electromagentic fields in black hole spacetime. However, it is interesting to see that the  essence of both of these approcahes is the vindication of Hawking's prediction \cite{Hawking1975ParticleCB}. 

Consider an atom as a point-like detector held static in Schwarzschild black hole spacetime. The metric, in the units $c=\hbar=G=k_{B}=1$, is given by 
\begin{eqnarray}\nonumber
ds^2&=\Big(1-\frac{2M}{r}\Big)dt^2 -\Big(1-\frac{2M}{r}\Big)^{-1}dr^2\\
&-r^2(d^2\theta+\sin^2\theta d\phi^2),
\end{eqnarray}
where $M$ is black hole mass and $r$ is the radial distance of atom from black the hole center. For such an atom, we explore how radiation reaction and vacuum fluctuations of massless scalar field contribute to the rate of change of  average atomic energy, when the atom-field Hamiltonian in proper frame of atom is $H(\tau)=\mu \sigma(\tau)\phi(x(\tau))$. We discuss the cases for different choices of vacua.

\textit{Boulware Vacuum.---}
For Boulware vacuum, the Wightman function for scalar field is given by \cite{Christensen:1977jc,Candelas:1980zt}
\begin{align}\nonumber
 D_{B}^{+} (x, x')&=\frac{1}{4\pi}\sum_{lm}|Y_{lm}(\theta, \phi)|^2\\
 & \times \int_{0}^{+\infty}\frac{d\omega_{k}}{\omega_{k}}e^{-i\Delta t}\big[|\overrightarrow{R}_{l}(\omega_{k}|r)|^2 + |\overleftarrow{R}_{l}(\omega_{k}|r)|^2\big],
\end{align}
where $Y_{lm}(\theta, \phi)$ are spherical harmonics, $R_{l}(\omega_{k}|r)$'s are radial functions pertaining to solution of Klein-Gordon equation, and $\omega_{k}$ the frequency of the field mode. 
The corresponding Hadamard function of the field reads as 
\begin{align}\nonumber
 C^{F}(x(\tau), x(\tau'))&=\frac{1}{8\pi}\sum_{lm}|Y_{lm}(\theta, \phi)|^2\\
 \nonumber
 & \times \int_{0}^{+\infty}\frac{d\omega_{k}}{\omega_{k}}\bigg[e^{\big(i\omega_{k}\Delta t/\sqrt{1-2M/r}\big)}\\
 \nonumber
 & + e^{-\big(i\omega\Delta t/\sqrt{1-2M/r}\big)}
 \bigg]\\
 \label{C}
 & \times \big[|\overrightarrow{R}_{l}(\omega_{k}|r)|^2 + |\overleftarrow{R}_{l}(\omega_{k}|r)|^2\big],
\end{align}
and the corresponding Pauli-Jordan (or Schwinger) function of the field are 

 \begin{align}\nonumber
 \chi^{F}(x(\tau), x(\tau'))&=\frac{1}{8\pi}\sum_{lm}|Y_{lm}(\theta, \phi)|^2\\
 \nonumber
 & \times \int_{0}^{+\infty}\frac{d\omega_{k}}{\omega}\bigg[e^{\big(i\omega_{k}\Delta t/\sqrt{1-2M/r}\big)}\\
 \nonumber
 & - e^{-\big(i\omega_{k}\Delta t/\sqrt{1-2M/r}\big)}
 \bigg]\\
 \label{CHI}
 & \times \big[|\overrightarrow{R}_{l}(\omega_{k}|r)|^2 + |\overleftarrow{R}_{l}(\omega_{k}|r)|^2\big].
\end{align}
We now make use of Eqs.  (\ref{C}) and (\ref{CHI}) in the equations for vacuum fluction and radiation reaction contributions to atomic energy in DDC formalism \cite{refId0,refId1,Audretsch:1993uc}, which is  given by 
\begin{eqnarray}\nonumber
  \Bigg\langle \frac{d H_{A}}{d\tau}\Bigg\rangle_{VF}=2i\mu^2\int_{\tau_{0}}^{\tau}d\tau' C^{F}(x(\tau), x(\tau'))\frac{d}{d\tau}\chi^A(\tau, \tau'),
\end{eqnarray}
and 
\begin{eqnarray}\nonumber
  \Bigg\langle \frac{d H_{A}}{d\tau}\Bigg\rangle_{RR}=2i\mu^2\int_{\tau_{0}}^{\tau}d\tau' \chi^{F}(x(\tau), x(\tau'))\frac{d}{d\tau}C^A(\tau, \tau'),
\end{eqnarray}
respectively, to yield the  mean rate of change of total atomic energy as \cite{Yu:2007wv,Zhou:2012eb}
\begin{align}\nonumber
  \Bigg\langle \frac{d H_{A}}{d\tau}\Bigg\rangle_{tot}&=-\frac{\mu^2}{2\pi}\sum_{\omega_{a}>\omega_{b}}\omega_{ab}^{2}|\langle a|\sigma^f(0)|b\rangle|^2 \\
  \label{ATOM}
  & \times \Big[1+ \frac{1}{16M^2\omega_{ab}^2}\sum_{l=0}^{\infty}(2l+1)|B_{l}(0)|^2\Big],
\end{align}
with $B_{l}(0)$ as reflection coefficient with the property $\overrightarrow{B_{l}}(\omega)=\overleftarrow{B_{l}}(\omega)=B_{l}(\omega)$, while the relation in Eq.  (\ref{ATOM}) is derived using Regge-Wheeler tortoise coordinate+ $r_{*}=r+2M \ln(r/2M-1)$. 
 It indicates that in Boulware vacuum, the mean atomic energy is enhanced compared to Minkowski case and also behaves normally at the horizon which is in sharp contrast to response rate of an Unruh-DeWitt detector \cite{Candelas:1980zt}. If one considers electromagnetic field,  one must first define a quantization rule before calculating the contribution from the field. Crispino \textit{et al.} \cite{Crispino:2000jx}  have carried out the quantization of electromagnetic field in exterior region of Schwarzschild black hole using Gupta-Bleuler condition in a modified Feynman gauge. This helps one to define correlation functions and vacuum states of the field. In this case, the mean rate of total energy becomes    \cite{Zhou:2012eb}
\begin{align}\nonumber
 \Bigg\langle \frac{d H_{A}}{d\tau}\Bigg\rangle_{tot}&=-\frac{\mu^2}{3\pi}\sum_{\omega_{a}>\omega_{b}}|\langle a |\sigma^f(0)|b\rangle|^2\omega_{ab}
 \\
 \label{EMP}
 & \times \Big[1+\frac{\alpha^2}{\omega_{ab}^2}+f(\omega_{ab},r)\Big] ,
\end{align}
with the proper acceleration 
\begin{eqnarray}\label{ACC}
 \alpha=\frac{M}{r^2\sqrt{g_{00}}}=\frac{M}{r^2\sqrt{1-2M/r}}.
\end{eqnarray}

Thus, it is clear from the Eq. (\ref{EMP}),  that the response of atom to  electromagnetic field is different compared to that of   scalar field with the appearance of gray-body factor $f(\omega_{ab},r)$. The energy change rate diverges as the event horizon is reached and approaches zero at asymptotic infinity where the spacetime is flat i.e. $\alpha\rightarrow 0$,  while for scalar field, it is finite \cite{Yu:2007wv,Yu:2007qu} . This indicates the equivalence of Boulware vacuum for static observer in Schwarzschild spacetime and Rindler vacuum in flat spacetime. For a  two-atom system, the mean rate change demonstrates the  possibility of  generating entanglement between them, even if they are initially prepared in a separable state in Boulware vacuum, which coincides with the conclusion in \cite{Cliche:2010fi}. It is claimed in Ref.  \cite{Menezes:2015veo} that for a pair of entangled atoms, the atomic energy rate grows rapidly as the atoms approach the black hole horizon, and this stems from the large proper accelerations with Hawking radiation consequently being a  pivotal  factor. However, in subsequent works \cite{Zhou:2020oqa,Chen:2023xbc}, these results have been challenged.

\textit{Unruh Vacuum.---}
\bigskip
In this case, we first consider the Wightman function as 
\begin{align}\nonumber
 D_{U}^{+}(x,x')&=\frac{1}{4\pi}\sum_{lm}|Y_{lm}(\theta, \phi)|^2 \times \\
 \nonumber
& \int_{-\infty}^{+\infty}\frac{d\omega_{k}}{\omega_{k}}
  \Bigg[\frac{e^{-i\omega_{k}\Delta t}}{1-e^{-2\pi\omega_{k}/\kappa}}|\overrightarrow{R_{l}}(\omega_{k})|^2\\
  &+\theta(\omega_{k})e^{-i\omega_{k} \Delta t}\overleftarrow{R_{l}}(\omega_{k})|^2\Bigg],
\end{align}
which clearly shows how surface gravity of black hole $\kappa$ enters the field quantization. This leads us to the following  mean rate of  atomic energy change for scalar field case \cite{Yu:2007wv}
\begin{align}\nonumber
 \Bigg\langle \frac{d H_{A}}{d\tau}\Bigg\rangle_{tot}&=-\frac{\mu^2}{3\pi}\sum_{\omega_{a}>\omega_{b}}|\langle a |\sigma^f (0)|b\rangle|^2\omega_{ab}^2 \\
 \nonumber
 & \times \Bigg[\bigg(1+\frac{1}{e^{2\pi\omega_{ab/\kappa_{r}}}-1}\bigg)\overrightarrow{P}(\omega_{ab})+\overleftarrow{P}(\omega_{ab})\Bigg] \\
 \label{UN}
 & + \frac{\mu^2}{3\pi}\sum_{\omega_{a}<\omega_{b}}|\langle a |\sigma^f (0)|b\rangle|^2\omega_{ab}^2\Bigg[\frac{P(\omega_{ab})}{e^{2\pi\omega_{ab/\kappa_{r}}}-1}\Bigg],
\end{align}
where we have substituted $\kappa_{r}=\kappa/(\sqrt{1-2M/r})$, $\overrightarrow{P}(\omega_{ab})$ and $\overleftarrow{P}(\omega_{ab})$  are factors that arise out radial functions $\overrightarrow{R}_{l}(\omega_{k})$ and $\overleftarrow{R}_{l}(\omega_{k})$ respectively, with $P(\omega_{ab})=\overrightarrow{P}(\omega_{ab})+ \overleftarrow{P}(\omega_{ab})$,  and this relation is same for electromagnetic case \cite{Zhou:2012eb}. One can see that  in Unruh vacuum, the balance between radiation reaction and vacuum fluctuations is broken and leads to spontaneous excitation of atom due to positive contribution from second term in Eq. (\ref{UN}). Thus, Unruh vacuum tends to destabilize the atom as compared to Boulware case [see Eq.(\ref{ACC})]. In the limiting case of $r\rightarrow 2M$, atom gets some thermal contributions at Tolman temperature, $T=(g_{00})^{-1/2}T_{H}$ ($T_{H}$ is usual Hawking temperature) from black hole appear as seen from Eq. (\ref{UN}), while for the case $r \rightarrow \infty$, some thermal contributions occur from the back-scattering of outgoing flux from black hole off the curvature \cite{Yu:2007wv,Zhou:2012eb}.

\textit{Hartle-Hawking Vacuum.---}
 We briefly mention here that in this case, the appearance of thermal 
contributions from surface gravity via the factors $1/(e^{2\pi\omega_{ab}/k_{r}}-1)$ similar to Unruh vacuum case, tends to increase the atomic energy changes and excitation occurs as if the atom is in a thermal bath at some temperature which tends to approach  Hawking temperature in the asymptotically far off distance. Some additional thermal factors in the form of  $f(\omega_{ab})$ and quadratic acceleration $\alpha^2$ appear in Eq. (\ref{UN}).
This makes situation similar to that of electromagnetic  field fluctuations in flat space in presence of boundaries \cite{Yu:2006kp}. This demonstrates that electromagnetic field fluctuations scatter off the spacetime curvature, as field modes do in flat spacetime \cite{Zhou:2012eb}.  For the entangled case, Hartle-Hawking vacuum has similar behaviour to that of  Unruh vacuum \cite{Menezes:2015veo}.

\textit{Kerr spacetime.---}
We point out here, that the Schwarzschild geometry is the simplest case of spherically symmetric spacetime.  One interesting case would be to   consider  rotating spacetime  of  Kerr black holes \cite{Visser:2007fj}, which  in Boyer-Lindquist coordinates is  given by the metric 
\begin{align}\nonumber
 ds^2&=-\big(1-\frac{2Mr}{\rho^2}\big)dt^2-\frac{4M a r \sin^2\theta}{\rho^2}dt d\phi\\
 \nonumber
 & +\frac{\rho^2}{\Delta}dr^2+\rho^2d\theta^2+\frac{\sum}{\rho^2}\sin^2\theta d\phi^2,
\end{align}
where $\rho^2=r^2+a^2\cos^2\theta$, $\Delta=r^2-2Mr+a^2$ and $\sum=(r^2+a^2)^2-a^2\Delta \sin^2\theta$. Such a metric characterizes a stationary and axially symmetric rotating black hole and has two commuting Killing vectors. In this case, some difficulties arise in defining equivalent vacuum state to Hartle-Hawking vacuum \cite{Jacobson:1994fp}, and many new vacuum states can  be considered, viz. Candelas-Chrzanowski vacuum and Frolov-Thorne vacuum. A noteworthy feature of  rotating geometry is its ability to harbour some crucial phenomena like black hole superradiance and energy extraction \cite{Brito:2015oca}. 

The problem of  atom-field interaction in Kerr geometry has been  worked by Menezes \cite{Menezes:2016quu}. The analysis has indicated that if  one considers a  static atom in Boulware vacuum, the   total rate in energy change is such that a fine  tuning exists between vacuum fluctuations and radiation reaction for atom in ground state stablizes it against spontaneous excitation. For an excited atom, they contribute equally to spontaneous excitation. This result is analogous   to atom interacting with a quantum field in Minkowski spacetime. Importantly, for asymptotic limit, $r\rightarrow \infty$, decay rate goes arbitrarily high due to occurrence of superradiance \cite{Brito:2015oca}. 

 For Unruh vacuum,  the excitation of the atom   manifests Hawking effect. For this to occur, the condition to be satisfied is  $|(\omega_{ab}/\sqrt{g_{00}})|>m\Omega_{H}$, where $\Omega_{H}$ is black hole angular velocity 
\begin{eqnarray}
 \Omega_{H}=\frac{a}{r_{+}^2+a^2},
\end{eqnarray}
with $r_{+}=M+\sqrt{M^2-a^2}$ and $a=J/M$ as black hole parameter and  $g_{00}$ is the component of metric tensor.  Also, the energy gap between the states and the angular velocity of the black hole have a great role in setting the behaviour of the total energy rate. It has also been shown that  the thermal radiation at temperature
 \begin{eqnarray}
  T=\frac{\kappa_{H}}{2\pi\sqrt{g_{00}}},
 \end{eqnarray}
allows non-superradiant energy gaps to be spontaneously excited. The Kerr black hole rotation enters the energy rate as chemical potential as expected \cite{Birrell:1982ix}.

For Candelas-Chrzanowski-Howard vacuum state, a thermal contribution for both cases, $|(\omega_{ab}/\sqrt{g_{00}})|>m\Omega_{H}$ and $|(\omega_{ab}/\sqrt{g_{00}})|<m\Omega_{H}$ is verified, unlike Unruh vacuum. For some terms proportional to radial functions, rotation doesn't affect the rate which means vacuum does not follow the $t-\phi$ reversal symmetry of Kerr geometry.   \\
\indent Other possible candidate for equivalent Hartle-Hawking vacuum for Kerr black hole is  Frolov-Thorne vacuum \cite{Frolov:1989jh,Ottewill:2000qh}. As angular velocity of black hole acts as chemical potential \cite{Birrell:1982ix}, for the atomic transition, it is greater than the energy gap. It is well known that the emission depends upon azimuthal quantum number $m$ with respect to  rotation axis through the thermal factor
\begin{eqnarray}
 \frac{1}{\exp\left({\frac{2\pi(\omega_{ab}/\sqrt{g_{00}}-m\Omega_{H})}{\kappa}}\right)-1},
\end{eqnarray}
which therefore enhances the excitation probability of the atom if its angular momentum is oriented towards that of the black hole. Furthermore, in the the limit $M\rightarrow \infty$, which corresponds to $T\rightarrow 0$, a negative flux is produced which means black hole stimulates the emission process. This is the very essence of Unruh-Starobinskii effect, the quantum analogue of superradiance \cite{Starobinskii:1973hgd,Unruh:1974bw,Matacz:1992gtm}. The case of a stationary atom with a zero angular momentum has also been studied in \cite{Menezes:2016quu}.  Superradiance also modifies the two-atom entanglement dynamics very significantly compared to Schwarzschild case, if the system is probed via Born-Markov approximation \cite{Menezes:2017oeb}. 
For the case of entangled atoms in de Sitter spacetime, the energy variation is dependent on a certain characteristic length scale. If the distance is smaller than this scale, energy change rate is same as that of thermal Minkowski vacuum. Beyond that scale, both spacetimes have distinct behaviour \cite{Liu:2018zod}.

\subsubsection{Lamb Shift: vacuum fluctuations, not the radiation reaction}

We return to Lamb shift and related radiative shifts now, citing some noted  results. 
For the Lamb shift in black hole spacetime, we continue the analysis based upon the DDC formalism, similar to what we did in Rindler case. First, we consider the Schwarzschild metric, where the atom is interacting with a massless scalar field, given by 
\begin{eqnarray}\label{SBH}
ds^2&=\Big(1-\frac{2M}{r}\Big)dt^2 -\frac{1}{\big(1-\frac{2M}{r}\big)}dr^2\\
\nonumber
&-r^2(d^2\theta+\sin^2\theta d\phi^2).
\end{eqnarray}
It has been explicitly shown that  we get the following total contribution for relative level shift in Boulware vacuum  \cite{Zhou:2010wn}
\begin{align}\nonumber
 \Delta_{B}&= \frac{\mu^2}{16\pi^2}\int_{0}^{\infty} d\omega_{k} P\Big(\frac{\omega_{k}}{\omega_{k}+\omega}-\frac{\omega_{k}}{\omega_{k}-\omega}\Big)\\
 \label{LSB}
 & + \frac{\mu^2}{16\pi^2}\int_{0}^{\infty} d\omega_{k} P\Big(\frac{\omega_{k}}{\omega_{k}+\omega}-\frac{\omega_{k}}{\omega_{k}-\omega}\Big)f(\omega_{k}, r),
\end{align}
where $\omega$ is energy level difference between levels,  $P$ represents principal value, and the grey-body factor is 
\begin{eqnarray}\nonumber
 f(r, \omega_{k})=\frac{1}{4r^2\omega_{k}^2}\sum_{l=0}^{\infty}(2l+1)|P(\sqrt{g_{00}} \omega_{k})|^2.
\end{eqnarray}
In Eq. (\ref{LSB}), the first term is just term representing Lamb shift in boundary-less Minkowski spacetime, and second one is a finite correction in  flat spacetime with no boundaries. 
This correction arises due to backscattering of field modes off the spacetime curvature and is analogous to Lamb shift of atom in presence of reflecting boundaries \cite{Audretsch:1995iw,PhysRevA.41.1587}. In Unruh vacuum, for atom close to event horizon$(r\rightarrow \infty)$,  we get corrections as 
\begin{eqnarray}\label{LB}
 \Delta_{U}= [1+f(r)]\Delta_{M}+\Delta_{T},
\end{eqnarray}
where $\Delta_{M}$ is the Minkowski term for shift given by 
\begin{eqnarray}
 \Delta_{M}=\frac{\mu^2}{16\pi^2}\int_{0}^{\infty} d\omega_{k} P\Big(\frac{\omega_{k}}{\omega_{k}+\omega}-\frac{\omega_{k}}{\omega_{k}-\omega}\Big).
\end{eqnarray}
Here, $\Delta_{T}$ is the additional thermal contribution in Unruh vacuum given by 
\begin{align}\nonumber
 \Delta_{T}=\frac{\mu^2}{16\pi^2}\int_{0}^{\infty} d\omega_{k} & P\Big(\frac{\omega_{k}}{\omega_{k}+\omega}-\frac{\omega_{k}}{\omega_{k}-\omega}\Big)\\
 \label{THERM}
& \times \Big(\frac{1}{e^{2\pi\omega_{k}}/\kappa_{r}-1}\Big),
\end{align}
where $\kappa_{r}=\kappa/\sqrt{g_{00}}$.
It is clear from the above, that the atom close to event horizon gets correction to Lamb shift as if it immersed in a thermal bath at temperature given by Tolman relation 
$T=(1/\sqrt{g_{00}})T_{H}$. Now, in the asymptotic limit , $r\rightarrow \infty$, Eq. (\ref{LB}) gets additional factor with $\Delta_{T}$ as 
\begin{eqnarray}\label{LBU}
 \Delta_{U}&= [1+f(r)]\Delta_{M}+f(r)\Delta_{T}.
\end{eqnarray}
The correction in Eq. (\ref{LBU}) appears due to backscattering of outgoing thermal flux from event horizon off the curvature of black hole spacetime. Considering same situations in Hartle-Hawking vacuum state, 
\begin{eqnarray}\label{LBH}
 \Delta_{H}&= [1+f(r)]\Delta_{M}+[1+f(r)]\Delta_{T},
\end{eqnarray}
for the asymptotic limit $r\rightarrow \infty$, which is consistent with the behaviour of Hawking-Hartle vacuum, and 
\begin{eqnarray}
 \Delta_{H}&= [1+f(r)]\Delta_{M}+[1+f(r)]\Delta_{T},
\end{eqnarray}
near the event horizon as $r\rightarrow 2M$. In both Unruh and Hartle-Hawking vacuum, thermal term $\Delta_{T}$ is due to the origin of thermal radiation from the horizon, in agreement with Hawking radiation. It is also noteworthy that this thermal term has the form
\begin{eqnarray}
 \Delta_{T}=\frac{\mu^2\omega}{4\pi^2}[1-\gamma+\ln\Big(\frac{T}{\omega}\Big)+\mathcal{O}(x_{0}^2)],
\end{eqnarray}
where $\gamma=0.577216$ is Euler constant, $x=\omega_{k}/T$ and $x_{0}=\omega/T$. 
As the temperature increases, thermal contribution to Lamb shift increases logarithmically  \cite{Zhou:2010wn}. Furthermore, in  de Sitter spacetime, the overall contribution for static atom  is just a thermal contribution of the form  similar to that of  Unruh and Hawking-Hartle case, but the thermal factor in Eq. (\ref{THERM}) includes cosmological constant $\Lambda$ as follows
\begin{eqnarray}
 \frac{1}{\exp\big({\frac{{2\pi\sqrt{3}\omega}}{\sqrt{\Lambda}}\big)-1}},
\end{eqnarray}
while for freely falling atom, the thermality occurs at the Gibbons-Hawking temperature $T_{f}=\frac{\Lambda}{2\sqrt{3}\pi}$ \cite{Gibbons:1977mu, Zhou:2010nb}. For the observational signatures of black hole-induced corrections to Lamb shift, one can look into \cite{Zhou:2012ba}.   So far, we have dealt with classical geometry of spacetime in all of  the above discussion. However, one can see in the literature,  these analyses have also been extended to quantum spacetime \cite{Cheng:2019tnk} and topological defects \cite{Cai:2019pnw}.

\indent In addition to the above radiative phenomena associated with the atoms, a considerable attention has been paid to the behavior of entanglement between atoms in non-inertial frames via DDC formalism and open quantum systems approach.  Some notable studies cover  Rindler \cite{Hu:2015lda, Chen:2021evr,Zhou:2021nyv, Soares:2022sqc}, Schwarzschild  \cite{Hu:2011pd}, Kerr \cite{Menezes:2017oeb}, cosmic string \cite{He:2020xhz,Huang:2020gha},  $\kappa$-deformed \cite{Liu:2023lok},  de Sitter \cite{Kukita:2017tpa}, and higher-dimensional \cite{Yan:2022xgg,Yan:2023ruj} spacetimes. We point out that there is another formalism concerning entanglement dynamics in curved spaces that forms the bedrock of relativistic quantum information. We will discuss that in Sec.\ref{secv}.

\subsection{Acceleration with a nonthermal character?}
When neutral atoms or molecules interact with a common electromagnetic field, a kind of interaction develops between them. We identify these interactions as dispersion and resonant interactions \cite{10.5555/1822415}.  While dispersion interactions are generally witnessed between atoms in ground state, resonant interaction generally occurs when one or both of the atoms are in excited state and possesses a long range character. Resonant interaction is also involved in resonant energy transfer in molecules \cite{2008IRPC...27..405S}, which has  relevance in biological photosynthesis \cite{Fassioli_2010} and interaction between macro-molecules \cite{Preto2013ResonantLI}. For both of these two processes, there is  detailed literature available either in books or in review papers, most of which concerns flat spacetime case or inertial atoms; see \cite{Galego_2019} for most recent analysis concerning its connection to chemical reactivity of molecules and \cite{Fiscelli:2019ywl} for analysis beyond perturbative approximation.  For a recent generic account on this topic, we refer the reader to a nice review article by Passante \cite{Passante:2018qzj}. However, we see though recent years have seen significant activity in extending these phenomena to accelerated frames and curved spacetimes including outer regions of black holes.  It is with this motivation that we intend to briefly review these interactions in curved geometries, beginning from dispersion forces followed by resonant interactions. This will also include discussions on nonthermal aspects of Hawking-Unruh effect. 

\subsubsection{Dispersion Interactions}
The general interaction energy of dispersion interaction between two atoms $A$ and $B$ is given by \cite{10.5555/1822415,doi:10.1080/01442358909353233}
\begin{align}\nonumber
 \Delta E &=-\frac{\hbar c}{\pi}\int_{0}^{\infty} du\ \alpha_{A}(iu)\ \alpha_{B}(iu)u^6\ e^{-2ur}\\
 \label{DIS}
 & \times \Bigg(\frac{1}{u^2r^2}+ \frac{3}{u^3r^3}+\frac{5}{u^4r^4}+\frac{6}{u^5r^5}+\frac{3}{u^6r^6}\Bigg),
\end{align}
where $k=iu$ is the imaginary wavenumber, $r=r_{B}-r_{A}$ is the distance between two atoms, $\alpha_{A}$ and $\alpha_{B}$ are polarizabilities of atom $A$ and $B$ respectively.
The relation in Eq. (\ref{DIS}) is valid for regions outside the overlap of two wave functions with $r$ having its dependence on relevant atomic transition wavelengths from the ground state, $\lambda_{rg}=2\pi k_{rg}^{-1}$. From (\ref{DIS}), one can study two limiting cases.
One is the  near zone limit, characterized by $r< \lambda_{rg}$
\begin{eqnarray}\label{near}
  \delta E_{near}=-\frac{2}{3}\sum_{ps}\frac{|\mu^{pg}_{A}|^2 |\mu^{sg}_{B}|^2}{E_{sg}+E_{pg}}\frac{1}{r^6},
 \end{eqnarray}
 where $p,s$ denote arbitrary atomic states, $g$ unperturbed ground state.
 Eq. (\ref{near}) is Van der Waals (nonretarded regime) relation and clearly scales as $1/r^6$. Another is  the far zone limit, characterized by $r> \lambda_{rg}$
 \begin{eqnarray}
  \delta E_{far}=-\frac{23\hbar c}{4\pi}\alpha_{A}\alpha_{B}\frac{1}{r^7},
 \end{eqnarray}
which is Casimir-Polder (or retarded) regime of dispersion interaction and scales as $1/r^7$ \cite{Passante:2018qzj, 1948PhRv...73..360C, Babb_2010}. 

As mentioned earlier, our focus is on acceleration or curved geometry effects on dispersion interactions. It turns out that the contribution from Hawking-Unruh effect has a great role to play. In fact these accelerated dispersion interactions further establish the consistency of predictions made by Hawking-Unruh effect 
\cite{Zhang:2011vsa,Zhang:2013txe,Noto:2013ona}.  Here, we consider a pair of atoms $A$ and $B$, each with frequency $\omega$, which move  with same proper acceleration $\alpha$, having constant distance $r$ between them. Following the procedure in  \cite{Marino:2014rfa}, we get the expression that shows effect of acceleration on near zone limit of dispersion energy as       
\begin{eqnarray}\nonumber
 \langle \delta E_{near}\rangle&=-\Big(1-\frac{4\alpha^2 t^2}{9c^2}\Big)\frac{3\hbar c}{2\pi r^6}\int_{0}^{\infty}\alpha_{A}\alpha_{B} du\\
 \label{NEAR}
 &+ \frac{\alpha^2 t \hbar}{\pi c^2 r^5}\int_{0}^{\infty} \alpha_{A}\alpha_{B} du,
\end{eqnarray}
which shows that the energy is time $t$ and acceleration $a$ dependent in the form of $1/r^5$ term. Compared to inertial case, this additional term decreases slowly with inter-atomic separation $r$.   Similarly, for far zone limit , we have 
\begin{eqnarray}\nonumber
 \langle \delta E_{far}\rangle &= -\frac{\hbar c}{\pi}\frac{\alpha_{A}\alpha_{B}}{r^7}\Big(\frac{23}{4}-\frac{7\alpha^2 t^2}{24c^2}\Big)\\
 \label{FAR}
 &+\frac{11\hbar \alpha^2 t}{8\pi c^2}\frac{\alpha_{A}\alpha_{B}}{r^6}, 
\end{eqnarray}
which shows that additional contributions from acceleration induce a $1/r^6$ behaviour in the energy, which is longer than usual $1/r^7$ behaviour in far zone regime for atoms at rest.  For this case, if one uses DDC formalism for calculating contributions of vacuum fluctuations assuming   a scalar field, the result turns out to yield an acceleration-dependent  length scale, $r_{0}$ characterized by the relation 
\begin{eqnarray}\label{SCALE}
 r_{0}=\frac{c^2}{\alpha},
\end{eqnarray}
which helps to identify two regimes : one where the inter-atomic distance $r\ll r_{0}$ and other $r\gg r_{0}$. If the atoms satisfy the condition $r\ll r_{0}$, it turns out that Casimir-Polder interactions  can display typical Unruh-type behaviour with temperature $T_{U}= \alpha/2\pi$, that corresponds to thermality as already discussed in \cite{Marino:2014rfa}.
In this way,   Casimir-Polder energy for a typical length scale associated with onset of quantum effects, $r>>\lambda_{th} $, where $\lambda_{th}$ is the thermal wavelength, displays the following dependence on separation 
\begin{eqnarray}\label{TH}
 E_{CP}^{th}=-\frac{1}{512}\frac{\mu^4 T}{\omega^2 r^2},
\end{eqnarray}
with $\mu$ as the atom-field coupling constant, which is a classical thermal character similar to that of  electromagnetic field \cite{Marino:2014rfa, 2001PhRvA..64c2102B}. 
However, if the separation is very large such that  $r\gg r_{0}$, the interaction energy becomes \cite{Marino:2014rfa}
\begin{eqnarray}\label{NOTH}
 E_{CP}^{acc}=-\frac{1}{512 \pi^4}\frac{\mu^4 r_{0}}{\omega^2 r^4}.
\end{eqnarray}
A comparison of Eqs. (\ref{TH}) and (\ref{NOTH}) indicates  that the interaction energy decreases faster with the mutual separation $r$ in accelerated atoms compared to both far and near zone limits. Eq. (\ref{NOTH}) signals the breakdown of Unruh thermality and is a consequence of absence of local inertial frame approximation associated with non-Minkowskian geometry over large regions of spacetime. Here it was relevant for Rindler spacetime where the atoms are in accelerated frames and background spacetime is flat. If one includes a black hole geometry, the behaviour of dispersion energy depends on choice of vacuum state: Boulware, Unruh and Hawking vacuum. Using open quantum system framework, Zhang \textit{et al.} \cite{Zhang:2011vsa} have considered  an atom interacting with a massless scalar field outside a Schwarzschild black hole of mass $M$ given by the metric in Eq. (\ref{SBH}), where the atom is at distance $r$ from the black hole center. For Boulware vacuum, a Casimir-Polder like force acts on the atom given by
\begin{eqnarray}\label{CPB}
 F^{B}=\frac{27\mu^2M^2\omega}{4\pi^2r^4}(r-3M)\ln \Big[\frac{m}{\omega}\Big],
\end{eqnarray}
where $m$ is the electron mass. Eq. (\ref{CPB}) shows that close to event horizon where $r=2M$, the force starts becoming attractive and repulsive at far off distance dropping as $1/r^3$. The turning point where $r=3M$, the vacuum modes are scattered the most. For Unruh vaccum, the force is attractive and varies as $1/(r-2M)$ and thereby diverging close to the horizon , while in the far off region, it is attractive(repulsive) for $T_{H}\gg m$ ($T_{H}\ll m$), where $T_{H}$ is Hawking temperature. If, on the other hand,   a pair of atoms is considered interacting now with electromagnetic field vacuum, the Boulware and the Unruh vacuum behaviour for Casimir-Polder interaction is like that of Minkowski spacetime with a typical $1/r^7$ behaviour while Hawking-Hartle vacuum produces a thermal effect on the interaction at temperature $T_{H}$ \cite{Zhang:2013txe}. This kind equivalence between Boulware and Unruh vacuum with that of Minkowski has been previously shown in many works(see e.g. \cite{Birrell:1982ix,Singleton:2011vh,Smerlak:2013sga,Hodgkinson:2014iua}), which raises an interesting question of distinguishability of  Minkowski and Schwarzschild geometries. However, by considering Resonance Casimir-Polder interaction (RCPI) \cite{10.5555/1822415} between entangled atoms, interaction in a Schwarzschild geometry has two distinct regimes corresponding to a characteristic length scale, that shows dependence on surface gravity $\kappa$ of black hole. If the interatomic separation $r$ is greater than that length scale, the power law behaviour of RCPI is $1/r^2$ compared to Minkowski case where it varies as $1/r$ \cite{Singha:2018vaj}. This helps to distinguish between two spacetimes. Moreover, another characteristic length scale, which distinguishes thermal and nonthermal nature of Casimir-Polder  interaction in Rindler case, is again witnessed in Schwarzschild black hole spacetime; again  the length scale is proportional to surface gravity $\kappa$ of the black hole \cite{Menezes:2017akp}.

\indent As noted, all these geometries considered so far potentially yield a plethora of physical insights into the behaviour of quantum vacuum. An interesting aspect is to analyze the behaviour with regards to metric fluctuations of spacetime in a quantum gravitational framework. It is found that if quantum corrections to classical gravitational force between two atoms are considered, the new interaction turns out to display similar behaviour to usual dispersion relations in the sense of ``near'' and ``far'' zone limits. In addition to this,  the interaction also depicts its dependency  on the material properties of the object, which is through  gravitational quadrupole polarizabilities \cite{Ford:2015wls} (see also \cite{Wu:2016esf,Hu:2016lev,Huang:2019tuf,Hu:2020cvd} for the related discussions).

\indent 
It is pertinent to mention that the interatomic energy  can offer long range behavior of quantum vacuum if one assumes the fourth-order of the coupling constant in DDC formalism \cite{Zhou:2020kvi}. Some  recent  treatments on the problem concern the coupled Unruh-De Witt detectors in Minkowski spacetime \cite{Cheng:2022xkk}, wherein near zone has been shown to be drastically amplified by  acceleration,  and investigations related to some subtleties surrounding the thermal nature of the de Sitter spacetime background \cite{Zhou:2022jkg}. 

\subsubsection{Resonant Interactions}

Like dispersion interactions, resonant interactions are also radiation-mediated interactions between neutral molecules or atoms when one or more of them are in their excited states. It involves exchange of real photons between the atoms \cite{10.5555/1822415, doi:10.1080/01442358909353233}. Resonant interaction is potentially involved in many optical phenomena,
like collective spontaneous emission \cite{PhysRev.93.99, PhysRevLett.96.010501, 2007LaPhy..17..956R}, level shifts in atoms\cite{PhysRevLett.102.143601}, resonant energy transfer between molecules \cite{doi:https://doi.org/10.1002/9780470141717.ch4} and numerous optical applications including e.g. laser cooling \cite{RevModPhys.70.721}, entanglement generation \cite{PhysRevA.61.062309} and this has been pursued rigorously in the recent decades \cite{PhysRevA.91.042127,PhysRevA.92.062711,PhysRevLett.115.033201,PhysRevLett.118.123001}. In this discussion, we briefly review the 
progress in deciphering the role of acceleration and curved spacetime in the behaviour of resonant interactions. \\
For two atoms $R$ distant apart and prepared in a correlated state, the resonant energy varies as $1/R$ in the far zone and thus are long range interactions when compared to dispersion interactions. Here we first consider two atoms $A$ and $B$ interacting with a scalar field and prepared in the following \textit{Bell-type} correlated state
\begin{eqnarray}
 |\psi_{\pm}\rangle=\frac{1}{\sqrt{2}}\Big(|g_{A},e_{B}\rangle \pm |g_{B},e_{A}\rangle \Big),
\end{eqnarray}
where $g$ and $e$ denote ground and excited states respectively. 
The Hamiltonian in this case can be written \cite{Audretsch:1993uc, Audretsch:1995iw, PhysRevA.94.012121}
\begin{align}\nonumber
 H(\tau)&=\hbar \omega \sigma_{1}^{A}(\tau)+\hbar \omega \sigma_{1}^{B}(\tau)+\sum_{k}\hbar \omega_{k}a_{k}^{\dagger}a_{k}\frac{dt}{d\tau}\\
& +\mu [\sigma_{2}^{A}\phi(x_{A}(\tau))+ \sigma_{2}^{B}\phi(x_{B}(\tau))],
\end{align}
where first two terms denote Hamiltonians of free atoms, $x_{A}(\tau)$($x_{B}(\tau)$) is the worldline of atom $A$ ($B$) and  can be inspired from Eq. (\ref{SPONT}).  By carrying out the mathematical calculations in DDC formalism, we get the contribution from radiation reactions only, given by 
\begin{align}\label{RIE}
 \delta E=\mp \frac{\mu^2}{16\pi c^2 R\sqrt{N(R,\alpha)}}\cos\Bigg[\frac{2\omega c}{\alpha}
 \sinh^{-1}\Big(\frac{R \alpha}{2c^2}\Big)\Bigg],
\end{align}
where $N(R,\alpha)$ is normalization factor and $\alpha$ is the acceleration \cite{PhysRevA.94.012121}. Eq. (\ref{RIE}) is a clear indication that acceleration does not produce any Unruh-like thermal contributions for resonant interactions. However, similar to the Casimir-Polder interaction \cite{Marino:2014rfa}, a characteristic length scale emerges as seen from relation (\ref{RIE}), given by 
\begin{eqnarray}\nonumber
 R=\frac{c^2}{\alpha}=R_{\alpha}.
\end{eqnarray}
It can be argued that resonant interaction scaling is different for interatomic distance $R$ versus $R_{\alpha}$. For $R\ll R_{\alpha}$, it is possible to find some inertial description for linear susceptibility of the field i.e. it  can be well approximated by its static counterpart \cite{PhysRevA.94.012121,Zhou:2016urt}. In this limit
\begin{eqnarray}
 \delta E\simeq \mp \frac{\mu^2}{8\pi R^2 \alpha}\cos\Bigg[\frac{2\omega c}{a}\ln\Big(\frac{R\alpha}{c^2}\Big)\Bigg].
\end{eqnarray}
However, for the $R\gg R_{\alpha}$, the acceleration affects the resonant interactions very significantly, given by 
\begin{eqnarray}
 \delta E\simeq \mp \frac{\mu^2}{16\pi \alpha}\frac{1}{R}\cos \Big(\frac{\omega}{R}\Big).
\end{eqnarray}
For this limiting value of distance, resonant energy is insensitive to Unruh effect or thermal effects of acceleration. By considering electromagnetic field, the scaling occurs either as $1/R^2$ or $1/R^4$, depending upon the orientation of dipole relative to the orthogonal directions to $R$ and also putting the system in the vicinity of a boundary, which eventually  makes it possible to control and manipulate resonant energy  by dipole orientation \cite{PhysRevA.94.012121, Zhou:2018igi}. Unlike other phenomena considered before, resonant energy in a Schwarzschild black hole does not show any distinct behavior for Boulware, Unruh or Hartle-Hawking vacuum, since the acceleration produces thermal effects only for vacuum fluctuations and resonant interactions occur due to radiation reaction. However,  like Rindler case, the manipulation of interaction strength has been shown to be possible \cite{Zhou:2017axh,Zhou:2018gqf}.

\section{Atoms and the accelerating mirrors}\label{seciv}
Quantum vacuum is full of fluctuating field modes. The feeble effects of vacuum  are normally challenging to probe as evident from the foregoing discussions; however, the amplification by various means can enhance the strength of the signatures.  In addition to Hawking-Unruh effects, this gives rise to large class of non-stationary QED effects including, previously discussed, dynamical  Casimir   effect (DCE)    \cite{10.1063/1.1665432,Dodonov:2020eto, Nation:2011dka}.  One possible way is to employ moving boundaries.   A moving boundary (mirror) thus potentially affects the structure of quantum vacuum, which results in the creation and annihilation of field quanta \cite{Haro:2006zz}.  In general, a moving mirror model in quantum field theory takes into account the impact of moving surfaces which eventually constrains the field modes.  Although much consideration has been given to single moving mirrors, two-mirror models have also received significant attention in the recent years; see e.g. \cite{Mundarain:1998kz,Dalvit:1998qs,Alves:2009ev,Fosco:2017jjf}. Moving mirror models have spanned wide area of research  activities, which makes it difficult to bring all of them under one roof, for  one of its manifestations in a flat spacetime version viz. DCE,  has already been worked out in many aspects (see e.g,\cite{PhysRevA.97.032514,PhysRevA.98.063807,PhysRevA.102.033703} for some recent investigations and \cite{Dodonov:2020eto} for a most updated review).  Most  importantly, these models are relevant in studying  the particle production in various cosmological models and radiation from collapsing black holes    \cite{Birrell:1982ix,Brevik:2000zb} (see also \cite{Wittemer:2019agm} for a recent analogue experimental setup), quantum decoherence\cite{Dalvit:1999sg}, Entanglement dynamics \cite{Andreata2005DynamicsOE} and harvesting \cite{Cong:2018vqx}. Furthermore, it has also been a successful model for shedding new light on the deep workings of Hawking-Unruh effect \cite{Scully:2017utk,Ben-Benjamin:2019opz, Svidzinsky:2018jkp, Svidzinsky:2019jqr} and equivalence principle of relativity   \cite{Fulling:2018lez}. In this section, we touch some of the aspects of accelerated mirrors that are very relevant for our discussion, viz atoms and accelerated mirrors on curved spacetimes and Hawking-Unruh effect.  In particular, we first review the works related to the general principles governing energy and  particle creation under different boundary conditions from accelerated mirrors, and later discuss the relevant scenario of curved spacetime extension of  DCE.  Afterwords, we discuss   atom-moving (accelerating) mirror physics in black hole spacetimes, and discuss some of the recent issues concerning acceleration radiation.  Although the usual reference during the analysis is to massless scalar fields for simplicity, however the allusion to quantum radiation (light photons) and optical phenomena is naturally implied and can be worked out.

\subsection{Parameterizing energy and particle production in moving mirrors}
Under general conditions, a $(1+1)$-dimensional moving mirror comprises a massless scalar field $\phi$ obeying  Dirichlet boundary conditions on a perfectly reflecting boundary with the wave equation $(c=\hbar=1)$
\begin{eqnarray}\label{KGEMM}
  \Big(\frac{\partial}{\partial t^2}-\frac{\partial}{\partial x^2}\Bigg)\phi=0
\end{eqnarray}
By introducing conformal (null) coordinates, 
\begin{eqnarray}\nonumber
 u=t-x,\ \  v=t+x,
\end{eqnarray}
the solution to Eq. (\ref{KGEMM}) is generally written  as 
\begin{eqnarray}
 \phi_{\omega_{k}}=g(v)+h(u),
\end{eqnarray}
where $\phi_{\omega}$ are the mode functions, and $g$ and $h$ are arbitrary functions. The inner product is defined as
\begin{eqnarray}\label{IP}
 (\phi_{1}, \phi_{2})=-i\int_{\Sigma} d\Sigma n^{\mu}[\phi_{1}(x)\partial_{\mu}\phi_{2}^*],
\end{eqnarray}
where $\Sigma$ is some Cauchy surface for spacetime and $n^{\mu}$ is future-directed unit normal \cite{Birrell:1982ix}. Without boundaries in Minkowski spacetime, the normalized modes are
\begin{eqnarray}\nonumber
 \phi_{\omega_{k} u}=\frac{1}{\sqrt{4\pi\omega_{k}}}e^{-i\omega_{k} u},\ \ 
 \phi_{\omega_{k} v}=\frac{1}{\sqrt{4\pi\omega_{k}}}e^{-i\omega_{k} v},
\end{eqnarray}
which gives the solution to Klein-Gordon equation of in Eq. (\ref{KGEMM}) as 
\begin{align}\nonumber
 \phi=\frac{1}{\sqrt{4\pi\omega_{k}}} & \int_{0}^{\infty} d\omega_{k} \big[a_{\omega_{k} u}e^{-i\omega_{k} u}+a_{\omega_{k} v}e^{-i\omega_{k}}\\
 & +a_{\omega_{k} u}^{\dagger}e^{+i\omega_{k} u}+a_{\omega_{k} v}^{\dagger}e^{+i\omega_{k} v}\big],
\end{align}
where $a_{\omega_{k} u} (a_{\omega_{k} u}^{\dagger})$ and $a_{\omega_{k} v} (a_{\omega_{k} v}^{\dagger})$ are creation (annihilation) operators. For the sake of brevity and less mathematical rigor, we avoid detailed mathematical calculations, for which interested reader can look into \cite{Birrell:1982ix}. The inner product in Eq. (\ref{IP}) is to be evaluated for a particular mirror trajectory. If  past and  future null infinities for mirror trajectory $x=z(t)$ are denoted by $\mathcal{I}^{-}$ and   $\mathcal{I}^{+}$ respectively, the scalar product of Eq. (\ref{IP}) for $\mathcal{I}^-$ gives  \cite{Good:2011nue} 
\begin{align}\label{IMINUS}
 (\phi_{1}, \phi_{2})=-i\int_{-\infty}^{+\infty}[\phi_{1}(u=-\infty,v)\overleftrightarrow{\partial_{v}}\phi_{2}^*(u=-\infty),v],
\end{align}
and for $\mathcal{I}^+$ 
\begin{align}\nonumber
 (\phi_{1}, \phi_{2})&=-i\int_{-\infty}^{+\infty}[\phi_{1}(u,v=\infty) \overleftrightarrow{\partial_{u}}\phi_{2}^*(u,v=\infty)]\\
 & -i\int_{v_{0}}^{+\infty}[\phi_{1}(u,v=\infty) \overleftrightarrow{\partial_{u}}\phi_{2}^*(u,v=\infty)].
\end{align}
When Dirichlet boundary conditions \cite{Fulling1976RadiationFA} are imposed, $\phi_{\omega_{k}}$ must vanish at the mirror's location, we get the value of two functions, $u=t-z(t)$ and $v=t+z(t)$. We can choose the mode functions either to be positive w.r.t $\mathcal{I}^-$ becoming an  \textit{in} vacuum state with frequency $\omega_{l}'$, or to be positive w.r.t $\mathcal{I}^{+}$ becoming an \textit{out} vacuum state with frequency $\omega_{k}$. Thus, we write 
\begin{eqnarray}\label{OUTVACUUM}
 \phi_{\omega_{l}'}^{in}=\frac{1}{\sqrt{4\pi\omega_{l}'}}\big(e^{-i\omega_{l}'v}-e^{-i\omega_{l}'p(u)}\big),
\end{eqnarray}
where $p(u)=v$ is some function of $u$, which implies field mode vanishes at mirror's location. For $\mathcal{I}^{+}$, there are two sets of mode functions. One set given by
\begin{align}\label{PHIOUT}
 \phi_{\omega_{k}}^{R,\text{out}}=\frac{1}{\sqrt{4\pi\omega_{k}}}\big(e^{-i\omega_{k} f(v)}-e^{-i\omega_{k} u}\big),\ \ \text{for}\  v< v_{0},
\end{align}
which is non-zero for right $\mathcal{I}_{R}^+$ and zero for left $\mathcal{I}_{L}^+$ and other set which is positive w.r.t $\mathcal{I}^{-}$ is denoted by $\phi_{\omega_{k}}^L$ which are only included if the mirror trajectory is asymptotic to the null surface $v=v_{0}$, and these modes don't interact with the mirror \cite{Carlitz:1986nh,Haro:2008zza,Nicolaevici:2009zz}. It is important to mention canonical relation for mode functions here, given by
\begin{align}\nonumber
 \Big(\phi_{\omega_{k}}(x),\phi_{\omega_{l}'}(x)\Big)&=-\Big(\phi_{\omega_{k}}^*(x),
 \phi_{\omega_{l}'}^*(x)\Big)=\delta(\omega_{k}-\omega_{l}')\\
 \label{INNERP}
 & \Big(\phi_{\omega_{k}}(x),\phi_{\omega_{l}'}^*(x)\Big)=0,
\end{align}
which for the modes $\phi_{\omega_{k}}^L$ gives the inner product
\begin{eqnarray}\label{INNERL}
 \Big(\phi_{\omega_{k}}^L,\phi_{\omega_{l}'}^L\Big)=-i\int_{v_{0}}^{\infty} dv \phi_{\omega_{k}}^L\overleftrightarrow{\partial_{v}}\phi_{\omega_{l}'}^{L*}=\delta(\omega_{k}-\omega_{l}').
\end{eqnarray}

The much deeper analysis of the above mode functions needs a specific choice of trajectory for the mirror. In fact, this has been carried out in many works. Few of  such trajectories include: Carley and Willey trajectory \cite{Carlitz:1986nh}, Walker-Davies trajectory \cite{1982JPhA...15L.477W}, and some new types of trajectories, recently introduced in a series of papers  by Good \textit{et al.} \cite{Good:2013lca,Good:2016oey,Good:2015jwa,Good:2017kjr,Good:2020rmk}. By virtue of  choosing a particular trajectory, it is possible to calculate a physical observable like particle number, energy etc. In addition to choice of trajectories, several other ways to \textit{parameterize} the behaviour of moving mirror models vis-\`a-vis  energy or particle production include the simple Dirichlet and Newman \cite{Fulling1976RadiationFA,Davies1977QuantumVE} 
or Robin \cite{Mintz:2006yz} boundary conditions. Some more sophisticated boundary situations include the one studied by Barton \textit{et al.} \cite{Barton:1995he,Calogeracos:1995he}, which includes a mass term  for the field at the position of the mirror and which acts as delta-function type potential. In another model, Golestan \textit{et al.} \cite{Golestanian:1997ks,
Golestanian:1998bx} constrain the field amplitude around the position of the mirror by utilizing an auxiliary field, while the proposal by Sopova \textit{et al.} \cite{Sopova:2002cs}  replaces the mirror by a dispersive dielectric.  In another of the very recent models, Galley \textit{et al.} \cite{Galley:2012qz} have introduced a mirror-oscillator-field (MOF) model, where a new internal degree of freedom associated with the mirror mimics the mirror-field microscopic interaction by minimally coupling to the field modes present at the position of the mirror. Later, similar coupling was considered by Wang \textit{et al.} \cite{Wang:2013lex} to calculate the force on mirror due to vacuum fluctuations, which however produces some divergent effective mass.  In a later model, this was thoroughly worked out and removed by considering the minimal coupling between internal oscillator and a massive scalar field \cite{Wang:2015axa}. \\
Having defined the inner products in Eqs. (\ref{INNERP}) and (\ref{INNERL}), one of the standard methods to describe quanta production include Bogoliubov transformation between modes at $\mathcal{I}^-$ and $\mathcal{I}^+$.  We expand modes at $\mathcal{I}^+$ as
\begin{eqnarray}
\phi_{\omega_{k}}^{J}=\int_{0}^{\infty}d\omega_{l}' \big(\alpha_{\omega_{k}\omega_{l}'}^J\phi_{\omega_{l}'}^{in}+\beta_{\omega_{k}\omega_{l}'}^J\phi_{\omega_{k}'}^{in*}\big).
\end{eqnarray}
The Bogoliubov coefficients are 
\begin{align}
 \alpha_{\omega_{k}\omega_{l}'}^J&=(\phi_{\omega_{k}}^J,\phi_{\omega_{l}'}^{in}),\\
 \beta_{\omega_{k}\omega_{l}'}^J&=-(\phi_{\omega_{k}}^J,\phi_{\omega_{l}'}^{in*}),
\end{align}
where $J$ denotes either of the right $\mathcal{I}_{R}^+$ or left $\mathcal{I}_{L}^+$
modes. The average particle number associated $\mathcal{I}^+$  is given by
\begin{align}
 \langle N^J\rangle=\langle 0_{in}| N^J|0_{in}\rangle=\int_{0}^{\infty}d\omega_{k} \int_{0}^{\infty}d\omega_{l}'|\beta_{\omega_{k} \omega_{l}'}^J|^2.
\end{align}
By using cauchy surface $\mathcal{I}^-$ and making use of  Eqs. (\ref{IMINUS}), (\ref{OUTVACUUM}) and (\ref{PHIOUT}), we get the $\beta_{\omega_{k} \omega_{l}'}^R$ corresponding to Cauchy surface  $\mathcal{I}_{L}^+$ as 
\begin{align}\label{BETA1}
 \beta_{\omega_{k}\omega_{l}'}^{R}=\frac{1}{4\pi\sqrt{\omega_{k}\omega_{l}'}}\int_{-\infty}^{v_{0}}dv\ e^{-i\omega_{l}'v-i\omega_{k} f(v)}\Big[\omega_{l}'-\omega_{k}\frac{df(v)}{dv}\Big],
\end{align}
and we get an equivalent expression if one uses $\mathcal{I}_{R}^+$ Cauchy surface given by
\begin{align}\label{BETA2}
  \beta_{\omega_{k}\omega_{l}'}^{R}=\frac{1}{4\pi\sqrt{\omega_{k}\omega_{l}'}}\int_{-\infty}^{\infty}du\ e^{-i\omega_{l}'u-i\omega_{k} p(u)}\Big[\omega_{l}'-\omega_{k}\frac{dp(u)}{du}\Big].
\end{align}
It is interesting to write Eq. (\ref{BETA2}) in terms of a time integral over the trajectory $z(t)$ by substituting $u=t-z(t)$, where $t-z(t)$ gives the values of $u$ at a given location of mirror along the trajectory $z(t)$. Thus we write Eq. (\ref{BETA2}) as follows
\begin{align}
  \beta_{\omega_{k}\omega_{l}'}^{R}=\frac{1}{4\pi\sqrt{\omega_{k}\omega_{l}'}}\int_{-\infty}^{\infty}dt\ e^{-i\omega_{+}t+i\omega_{-}z(t)}[\omega_{+}+\dot{z}(t)-\omega_{-}], 
\end{align}
where we defined $\omega_{+}=\omega_{k}+\omega_{l}'$ and $\omega_{-}=\omega_{k}-\omega_{l}'$ and it is most suitable when the trajectory is asympotically inertial. For an inertial trajectory, $\beta_{\omega_{k}\omega_{l}'}^{R}=0$, which obviously is the case since the mirror is not accelerating. For a trajectory that is initially inertial and characterized a 
finite time accelaration, the total energy by summing over the modes is given by the relation \cite{Walker:1984vj}  
\begin{align}
 E=\int_{0}^{\infty}\omega_{k} \langle N_{\omega_{k}}\rangle d\omega_{k}.
\end{align}
There are two other  popular methods to quantify particle creation or energy content. One is to employ the formalism by Davies and Fulling \cite{Fulling1976RadiationFA}, which calculates the expectation value of stress-energy tensor for massless minimally coupled field, which gives the energy flux as 
\begin{eqnarray}
 \langle T_{uu}\rangle=\frac{1}{24\pi}\Big[\frac{3}{2}\bigg(\frac{p''}{p'}\bigg)^2-\frac{p'''}{p'}\Big],
\end{eqnarray}
where $p'$ denotes derivatives of $p$ with respect to $u$. For inertial trajectory, $\langle T_{uu}\rangle=0$, and it only survives with acceleration.  In terms of trajectory $z(t)$, the energy flux is given by 
\begin{eqnarray}
 E=\frac{1}{12\pi}\int_{-\infty}^{\infty}\frac{\ddot{z}^2}{(1+\dot{z})^2(1-\dot{z})^3}dt,
\end{eqnarray}
where time derivatives of $z$ are with respect to lab frame, not the proper frame of mirror. The other framework uses wave packets to calculate particle number. One famous example is by Hawking's proposal for particle creation in black holes \cite{Hawking1975ParticleCB}, which enables one to study time dependence of particle creation. Detailed aspects of this method can be found in \cite{Fabbri:2005mw}. A wave packet $\phi_{jn}$ is constructed from the $\phi_{\omega_{k}}$ as
\begin{eqnarray}
 \phi_{jn}=\frac{1}{\sqrt(\epsilon)}\int_{j\epsilon}^{(j+1)\epsilon}d\omega_{k} e^{\frac{2\pi n i\omega_{k}}{\epsilon}}\phi_{\omega_{k}},
\end{eqnarray}
where $n$ is an integer and $j$ is non-negative integer, $\epsilon$ is the width of frequency range for each packet $(j+\frac{1}{2})\epsilon$. The application of these methods for different mirror trajectories has been  worked out for some cases, such as Carlitz-Willey   \cite{Carlitz:1986nh} and Walker-Davies \cite{Walker:1984vj} trajectories. Here, we only write down the expresion for energy flux
\begin{eqnarray}
 E=\sum_{j,n}\Big(j+\frac{1}{2}\Big)\epsilon\langle N_{jn}\rangle,
\end{eqnarray}
where $N_{jn}$ is the average particle number that reaches 
$\mathcal{I}_{R}^+$ in the frequency range $j\epsilon\leq \omega_{k} \leq (j+1)\epsilon$ within an approximate time range $(\frac{2\pi n-\pi}{\epsilon})\leq u\leq (\frac{2\pi n+\pi}{\epsilon})$.  

 Good \textit{et al.} \cite{Good:2013lca} have solved mirror problem for a class of trajectories with \textit{time-dependent} particle production. We find, for example, a trajectory called Arctx (from arctangent exponential) given by 
\begin{eqnarray}
 z(t)=-\frac{1}{\mu}\tan^{-1}(e^{\mu t}), 
\end{eqnarray}
where $\mu$ is a positive constant, gives an estimate of energy produced as 
\begin{eqnarray}
 E=\frac{\mu}{2592\pi}(13\sqrt{13}-36).
\end{eqnarray}
A good time resolution of particle production can be obtained by these trajectories (see also \cite{Good:2017kjr}). 

\subsection{Dynamical Casimir effect in black holes}
We pointed out earlier that DCE is one of the primal examples of moving mirror problem. We  discuss here the application of accelerating mirror models to the description of DCE in curved spacetime.  The impact of gravity on static Casimir  effect has been considered in some works by Sorge   \textit{et al.} \cite{Sorge:2005ed,Sorge:2019ldb} and on DCE by C\'eleri \textit{et al.} \cite{Celeri:2008ui}. A similar related  study by R\"atzel \textit{et al.} \cite{Ratzel:2017etl} discusses frequency spectrum of optical resonator in a curved geometry.   Lock \textit{et al.} \cite{Lock:2016rmg} have recently presented a general formalism for incorporating spacetime curvature effects into DCE,  when  one cavity boundary is fixed. In a similar vein, detailed investigations for a Casimir apparatus in free fall  in a Schwarzschild black hole have been carried out in some very recent works \cite{Sorge:2019ecb,Wilson:2019ago}, that further substantiate the points we discuss here.  

It has been demonstrated that a massless scalar field inside a cavity can describe phononic excitations in relativistic BEC system \cite{Fagnocchi:2010sn} or  suitably approximate an electromagnetic field when the polarization effects are very small \cite{Friis:2013eva}. By considering such a system  with inertial coordinates $(x,t)$, the Klein-Gordon equation yields normalized solutions given by 
\begin{eqnarray}\label{DCEKG}
 \phi_{m}(t)=\frac{1}{\sqrt{m\pi}}e^{-i\omega_{m}t}\sin[\omega_{m}(x-x_{1})],
\end{eqnarray}
where $\omega_{m}=m\pi/L$ are mode frequencies with $L=(x_{2}-x_{1})$ as the cavity length with $m$ being an integer. A column vector from the mode solutions to the Eq.(\ref{DCEKG}) is given by
$\Psi=[\phi_{1}, \phi_{2},...,\phi_{1}^*, \phi_{2}^*,...]^T$, where $T$ is the matrix transpose operation, can be related to another set of mode solutions $\bar{\Psi}$ by
$\bar{\Psi}=S\Psi$ where 
\[S=\begin{bmatrix}\alpha&\beta\\ \beta^*&\alpha^*\\ \end{bmatrix}.\]
Here $\ SKS^{\dagger}=K$ and $K$ is $2\times2$ identity matrix. $\alpha_{mn}$ and $\beta_{mn}$ are Bogoliubov coefficients and help to calculate the particle number in an initial vacuum state, given by $\mathcal{N}=\sum_{n}|\beta_{mn}|^2$. For our case, the coefficients for finite time range $t=0$ to $t=T$ turn out to be 
\begin{align}\nonumber
 \alpha_{mn}=e^{i\int_{0}^{T}dt \omega_{m}(t)}&\Bigg[\delta_{mn}+ \sum_{j=1}^{2}\int_{0}^{T}dt A_{mn}^{(j)}\\
 &\times e^{-i\int_{0}^{t}dt'[\omega_{m}(t')-\omega_{n}(t')]}\frac{dx_{j}}{dt}\Bigg],
 \end{align}
and
\begin{align}\nonumber
 \beta_{mn}=e^{i\int_{0}^{T}dt \omega_{m}(t)}& \sum_{j=1}^{2}\int_{0}^{T}dt B_{mn}^{(j)}\\
 &\times e^{-i\int_{0}^{t}dt'[\omega_{m}(t')+\omega_{n}(t')]}\frac{dx_{j}}{dt},
\end{align}
where $j=1,2$. 

Consider now a spacetime curvature characterized by the Schwarzschild metric 
\begin{eqnarray}
 ds^2=-f(r)dt^2+\frac{1}{f(r)}dr^2,
\end{eqnarray}
where $f(r)=1-2GM/r$ with $2GM$ as Schwarzschild radius. We also assume one boundary at $r=r_{0}$ fixed radial distance and other movable boundary $r=(r_{0}+L_{0})[1+\delta(t)]$.  Following  detailed calculations as given in \cite{Lock:2016rmg}, in this case, the Bogoliubov coefficients are modified by spacetime curvature as follows 
\begin{align}\nonumber
 \beta_{mn}&=e^{i\omega_{n}T}[\epsilon\upsilon]\sqrt{\omega_{m}\omega_{n}}\frac{f(r_{0})}{f(r_{0}+L_{0})}\Bigg[i\frac{(-1)^p-e^{i(\omega_{m}+\omega_{n})T}}{(\omega_{m}+\omega_{n})^2-\upsilon^2}\\
 \label{BDCE}
  & + \frac{A 2GM}{(r_{0}+L_{0})^2}\frac{\upsilon}{\omega_{m}+\omega_{n}}\frac{e^{i(\omega_{m}+\omega_{n})T}-1}{\Big(\frac{\omega_{m}+\omega_{n}}{2}\Big)^2-\upsilon^2}\Bigg].
\end{align}
Here $\epsilon=A/L_{0}$, where $A$ is related to $r_{2}(t)=r_{0}+L_{0}+A\sin(\upsilon t)$ and $\upsilon$ is the oscillation frequency of proper length of cavity. A close look at Eq. (\ref{BDCE}) reveals some interesting features. The first term contributes when curvature is zero and corresponds to resonant frequency $\upsilon=\omega_{m}+\omega_{n}$. The second term shows a novel contribution due to black hole spacetime curvature, which depicts a resonance at subharmonic $\upsilon=\frac{\omega_{m}+\omega_{n}}{2}$. Furthermore, for a particular frequency, $\upsilon=\omega_{q}+\omega_{r}$ for some $q,r$ in the regime $\upsilon T\gg 1$, we get
\begin{align}
 |\beta_{mn}|^2=\frac{1}{4}\Big[1-4GM\frac{L_{0}}{r_{0}^2}\Big]\Big[\epsilon\frac{f(r_{0})\pi T}{L_{0}}\delta_{m+n,q+r}\Big],
\end{align}
which indicates particle reduction that can be attributed to curvature and agrees with the results in \cite{Celeri:2008ui} (see also \cite{Lima:2019pbo} for  comparative results). 

\subsection{Atomic excitation with accelerated mirrors}
\subsubsection{The Rindler case}
 It has been observed that a moving mirror or boundary potentially reflects virtual particles into real ones \cite{10.1063/1.1665432}, like an atom that jumps to an excited state with the emission of a photon in Unruh-type virtual processes  \cite{PhysRevA.74.023807,Scully:2003zz}. The preceding discussion dealt with mirrors alone that accelerate in Minkowski or black hole spacetime. This section is an extension of that  in the sense that we now include an atom in the vicinity of a mirror, the combined system in a Minkowski or a black hole spacetime. In such configuration, a relative acceleration between atom and mirror renders virtual photons into real ones leading to the atomic transition.  Such a construction has been pursued for many years now and it yields a myriad of phenomena  with connections to Hawking-Unruh effects \cite{Scully:2017utk,Scully:2003zz, Dolan:2020hzm}, causality in  acceleration radiation \cite{PhysRevResearch.1.033115} equivalence principle \cite{Fulling:2018lez},  Fano interference 
 \cite{Svidzinsky:2018jkp} and  Cherenkov radiation \cite{Svidzinsky:2019jqr}. 
 
 Consider a two-level atom uniformly accelerated with respect to a fixed mirror in flat spacetime, moving along the trajectory
\begin{eqnarray}\label{AMT}
 t(\tau)=\frac{c}{\alpha}\sinh \bigg(\frac{\alpha\tau}{c}\bigg),\ \ z(\tau)=\frac{c^2}{\alpha}\cosh \bigg(\frac{\alpha\tau}{c}\bigg), 
\end{eqnarray}
where as usual $t$ stands for lab time, $\alpha$ is proper acceleration of the atom and $\tau$ is proper time of atom.   Denoting field mode by $\phi_{\nu}[t(\tau),z(\tau)]$ and  atomic lowering operator by  $\sigma$, interaction between atom and the photon gives rise to following Hamiltonian
\begin{eqnarray}
 H(\tau)=\hbar \mu\big(a_{\nu}\phi_{\nu}+ H.c\big)\big(\sigma e^{-i\omega\tau}+H.c\big),
\end{eqnarray}
where $a_{\nu}$ is photon annihilation operator and $\mu$ is atom-field coupling constant and $\omega$ is transition frequency of atom. Let's assume that mirror is fixed at $z=z_{0}\ll c^2/\alpha$. The probability of excitation of atom along  with the emission of a photon is given by \cite{Svidzinsky:2018jkp}
\begin{eqnarray}\label{PEX}
 P=\frac{8\pi c\mu^2}{\alpha \omega}\frac{\sin^2\big[\frac{\nu z_{0}}{c}+\varphi\big]}{e^{2\pi c \omega/\alpha}-1},
\end{eqnarray}
where $\varphi$ does not depend on $z_{0}$. The probability as given by Eq. (\ref{PEX}), clearly is an oscillates as function of mirror position $z_{0}$ and is has Plank-type thermal behaviour at Unruh temperature $T_{U}$ with the factor
$[e^{\hbar \omega/k_{B}T_{U}}-1]^{-1}$. The corresponding average photon occupation number in the mode frequency $\nu$ turns out to be 
\begin{eqnarray}\label{PDA}
 \bar{n}_{\nu}=\frac{1}{(e^{2\pi c \omega/\alpha}-1)}.
\end{eqnarray}
Alternatively, if the atom is fixed and mirror moves according to  the trajectory given in Eq. (\ref{AMT}), the probability of the event is given by
\begin{eqnarray}\label{PDN}
 P=\frac{8\pi c \nu \mu^2}{\alpha \omega^2}\frac{\sin^2\big[\frac{\omega z_{0}}{c}+\varphi\big]}{e^{(2\pi c \nu/\alpha}-1)},
\end{eqnarray}
which shows now the atomic excitation probability with the generation of photon depends on photon frequency $\nu$ and not atomic frequency $\omega$.  Also note that the oscillatory behaviour of probability is determined by atomic wave number $\omega/c$ unlike earlier case where it is governed by photon wavelength. In this case, the photon distribution gotten from Eq. (\ref{PDN})  
\begin{eqnarray}\label{PDM}
 \bar{n}_{\nu}=\frac{1}{(e^{2\pi c \nu/\alpha}-1)}.
\end{eqnarray}
From Eqs. (\ref{PDA}) and (\ref{PDM}), we see different photon distributions arise depending on whether atom or mirror is accelerated. If instead of an accelerating atom and mirror system, one considers a uniformly moving atom in the vicinity of a medium like a flat metal surface or an optical cavity, the excitation of atom of atom is followed by the emission of a surface plasmon, which is found to be connected to Cherenkov and Unruh effects \cite{Svidzinsky:2019jqr}. The field quantum is emitted at the expense of kinetic energy of atom through vacuum fluctuations.

\subsubsection{Black hole case: Horizon Brightened Acceleration Radiation (HBAR)}
The relative acceleration between atom and field modes can be envisioned in a different scenario as well: a mirror held fixed against the gravitational pull of a black hole, while the atom falls freely in the black hole. This scheme has been recently considered in a work by Scully \textit{et al.} \cite{Scully:2017utk}. Some more aspects of this problem have also been covered in \cite{PhysRevA.74.023807, Scully2019LaserEF}.  The idea is to consider an atomic cloud  such that covariant acceleration of atoms is zero (which means atoms are in inertial frame as implied by equivalence principle), while the mirror is held fixed by applying a force to counter the pull of black hole's gravity and is thus accelerated. In this case, the probability of the event is given by
\begin{eqnarray}
 P=\frac{4\pi \mu^2\nu}{\omega^2(1+2\nu/\omega)^2}\Big[\frac{1}{\exp{\big(\frac{4\pi R_{s}\nu}{c}\big)}-1}\Big],
\end{eqnarray}
where $R_{s}=2GM/c^2$ is the Schwarzschild radius of the black hole. It is important to note here that Hawking radiation does not contribute here as the atoms are shielded from it by the fixed mirror.  This acceleration radiation, called \textit{horizon brightened acceleration radiation} (HBAR), emitted by freely falling atoms appears to a distant observer much like (but different from) Hawking radiation. A master equation technique yields the following relation for entropy of radiation
\begin{eqnarray}
 \dot{S_{p}}=\frac{k_{B}c^3}{4\hbar G}\dot{A_{p}},
\end{eqnarray}
where $\dot{A}_{p}=(2\dot{m}_{p}/M)A$ is the rate of change of change of black hole area. Note here $A=4\pi R_{s}^2$ is black hole surface area and $\dot{m}_{p}c^2=\sum_{\nu}\dot{\bar{n}}_{\nu}\nu$ is the power carried away by photons and results in decrease in black hole mass. The photon distribution here $\bar{n}_{\nu}$ is analogous to that of Minkowski case, worked out in \cite{Svidzinsky:2018jkp}. We believe that this problem of radiation from falling atoms  in some sense provides a typical example of a phenomena of particle creation in general correspondence between accelerated mirrors and black holes, a formalism that  has been thoroughly touched in Refs. \cite{Anderson:2015iga,Good:2015jwa, Good:2016bsq,Good:2020nmz,
Good:2020byh}.

There are other interesting aspects of HBAR radiation that are worthy of attention, as listed below:
\begin{itemize}
\item 
The origin of black hole thermodynamics has been an intriguing subject since the seminal work by Bekenstein \cite{Bekenstein:1973ur} and Hawking\cite{Hawking1975ParticleCB}.  An insightful way of looking  at it via a conformal field theory in $0+1$ dimension, so-called conformal quantum  mechanics (CQM) \cite{1976NCimA..34..569A}.  It has been previously shown that Bekenstein-Hawking entropy emerges from CQM to field as a near-horizon approximation \cite{Camblong:2004ye,
Camblong:2004ec}.  The near-horizon approximation for black hole metric coefficient $f(r)$ is achieved by using Taylor expansion while respecting the  condition $(r/r_{g}-1)\ll 1$ such that $f(r)\approx f'(r_{g})(r-r_{g})$.  These ideas have been recently applied to HBAR radiation, and can be found in Refs. \cite{Camblong:2020pme,Azizi:2021qcu,Azizi:2021yto}.  Connected to the near-horizon analysis is the very question of underlying mechanism that generates thermality in  black hole horizons as in the Bekenstein-Hawking temperature. In fact this is a vast area of investigation in black hole physics and finds many diverse ideas knitted together including chaos theory. Some important works in this direction can be found in Refs. \cite{Maldacena:2005he,Morita:2019bfr,Maitra:2019eix,Dalui:2020qpt,Dalui:2019esx,Dalui:2021tvy,Dalui:2021sme,Kane:2022zcg}.     
\item Since HBAR is essentially related to radiation emission from a freely-falling atom, it finds inevitable connection to Equivalence Principle of general relativity.  There are many aspects of this issue for timelike geodesics discussed in \cite{Scully:2017utk,Fulling:2018lez,Chatterjee:2021fue,Sen:2022tru}. Whether the thermalization of the detectors occurs along the  null geodesics is also a crucial question which can be found in Ref. \cite{Chakraborty:2019ltu}. A recent extension covers the possible enhancement or degradation of HBAR intensity in presence of dark matter \cite{Bukhari:2023yuy}.
\item  The usual notion of Planckian thermality associated with Hawking-Unruh effect is generally true on ideal grounds. The situation changes if one takes into account backreaction or scattering \cite{Parikh:1999mf,Visser:2014ypa,Ma:2017odv}. Pertinent to this   scenario, the nonthermal flux emission from black holes with multiple Killing horizons has been a subject of debate for a long time. The prime example includes black holes immersed in a positive cosmological constant \cite{Kastor:1993mj,
Bhattacharya:2018ltm, Qiu:2019qgp}. The conditions that generate nonthermal emission have been tied up to the choice of vacuum states and coordinate systems.  However, it is possible to realize the possible nonthermal particle emission in presence of dark energy \cite{Bukhari:2022wyx}. It is reasonable to expect that is nonthermality is linked to the magnitude of cosmological constant.  
\end{itemize}

\section{A note on relativistic quantum information}\label{secv}
The proceeding discussion concerned a major aspect of atom-field dynamics which involved the radiative aspect of the interactions in curved spaces. In fact, it was the main focus of our work. However, there is  another aspect of these interactions concerning the quantum informational protocols that also deserves a mention. The study of relativistic aspects of quantum information theory  is not a subject without precedent.  Many noted review works exist on these lines. We may provide a brief glimpse into the status quo of the field in a piece-wise manner below: 
\begin{itemize}
    \item 
One of the earliest and  detailed reviews is that by  Peres and Terno \cite{RevModPhys.76.93}, which contains a voluminous treatment of quantum informational phenomena,  with regards to causality in special and general theory of relativity, with connections to Hawking information paradox. Pertinent to this, the role of reference frames and superselection rules have been reviewed in Ref. \cite{Bartlett:2006tzx}.
\item 
Quantum entanglement has been shown to be different for parties involving acceleration compared with inertial observers \cite{PhysRevLett.95.120404}. This realization has given rise to information and communication protocols in accelerated frames and curved geometries, and various aspects of these results have been summarized in Ref.  \cite{Mann_2012,Martin-Martinez:2014gra,Ralph:2011hp} 
\item Entanglement in quantum field states is a well-known effect \cite{Summers:1987ze, Reznik:2003mnx}. It has been suggested that  two Unruh-DeWitt type detectors, initially uncorrelated, can become entangled after interacting locally with a quantum field, which means the field entanglement has been swapped to the detectors, a process termed as \textit{entanglement harvesting} \cite{Salton:2014jaa,Pozas-Kerstjens:2015gta}.
\item Entanglement harvesting protocol is a phenomenon that has been shown to be highly dependent on various factors, such as detector motion type \cite{Zhou:2022nur,Liu:2022uhf,Bozanic:2023okm,Zhang:2020xvo}, presence of boundaries \cite{Liu:2020jaj,Ye:2021muj,Liu:2023zro,Li:2024dvs,Barman:2023wkr}, spacetime topology \cite{Ji:2024fcq,
Martin-Martinez:2015qwa}, internal structure of the detectors \cite{Hu:2022nxc}, external environment \cite{Cong:2020nec,Chen:2021evr}, spacetime dimensions \cite{Yan:2023ruj}, and curvature \cite{Henderson:2017yuv,Caribe:2023fhr,Liu:2023lok,Hu:2013ypa}. 
\item Many intriguing phenomena have emerged with regard  to entanglement extraction from black hole vacuum. For example, it has been found  \cite{Henderson:2017yuv} that two static detectors hovering outside a Bañados-
Teitelboim-Zanelli (BTZ)  black hole can not harvest entanglement in a certain region, a process termed as \textit{entanglement shadow}. This phenomenon has been further investigated in Refs. \cite{Bueley:2022ple,Gallock-Yoshimura:2021yok} 
\end{itemize}

Above facts could provide brief insights into some of the major advances in the field of relativistic quantum information. We however believe that this field has witnessed an explosive growth in the past few decades and demands a separate review work.

\section{Conclusive Remarks}\label{secvi}
Atom-field interactions are at the heart of quantum optics which have been greatly studied in Minkowski spacetime with flat a background geometry. In recent times, following the theoretical and experimental progress of Einstein's general theory of  relativity and other gravitational theories, considerable attention has been paid to the impact of curved geometries on radiative phenomena, which on one side could help binding gravity with quantum theory within a theory of quantum gravity, and on other side help designing of novel systems for manipulation of light and quantum communication signals in arbitrary frames of references and complicated geometries. In this brief review, we   attempted to highlight the ongoing progress in studying quantum radiative  and entanglement  phenomena in curved spacetime that involve contributions from the well-known Hawking-Unruh effect. For atomic radiative transitions and Lamb shift, we observed that Hawking-Unruh effect enhance the probability and strength of  radiation emission with an explicit dependence on atom's acceleration. This in turn validates the existence of Hawking-Unruh  effect. This was discussed in Sec \ref{seciii}.A and \ref{seciii}.B.    In Sec.\ref{seciii}.C, we discussed the dispersion and resonant interactions and pointed out how Unruh thermality can break down beyond a certain acceleration-dependent distance scale between the atoms.    In Sec. \ref{seciii}.D, we discussed role of accelerated mirrors in particle production, as in dynamical Casimir effect, and in the excitation of atoms in their vicinity, both in flat and curved spacetime. Finally, in Sec.\ref{secv}, we touched upon briefly on quantum entanglement aspects in curved spacetimes.   We hope this short piece of review  is useful for some beginners in the field.

{\section* {Acknowledgments}}
\setlength{\parskip}{0cm}
    \setlength{\parindent}{1em}
This research is supported by the National Natural Science Foundation of China (NSFC) (Grant No. 11974309). SMASB acknowledges financial support from China Scholarship Council at Zhejiang University.

\bibliographystyle{apsrev4-1}
\bibliography{masood.bib}

\begin{thebibliography}{343}%
\makeatletter
\providecommand \@ifxundefined [1]{%
 \@ifx{#1\undefined}
}%
\providecommand \@ifnum [1]{%
 \ifnum #1\expandafter \@firstoftwo
 \else \expandafter \@secondoftwo
 \fi
}%
\providecommand \@ifx [1]{%
 \ifx #1\expandafter \@firstoftwo
 \else \expandafter \@secondoftwo
 \fi
}%
\providecommand \natexlab [1]{#1}%
\providecommand \enquote  [1]{``#1''}%
\providecommand \bibnamefont  [1]{#1}%
\providecommand \bibfnamefont [1]{#1}%
\providecommand \citenamefont [1]{#1}%
\providecommand \href@noop [0]{\@secondoftwo}%
\providecommand \href [0]{\begingroup \@sanitize@url \@href}%
\providecommand \@href[1]{\@@startlink{#1}\@@href}%
\providecommand \@@href[1]{\endgroup#1\@@endlink}%
\providecommand \@sanitize@url [0]{\catcode `\\12\catcode `\$12\catcode `\&12\catcode `\#12\catcode `\^12\catcode `\_12\catcode `\%12\relax}%
\providecommand \@@startlink[1]{}%
\providecommand \@@endlink[0]{}%
\providecommand \url  [0]{\begingroup\@sanitize@url \@url }%
\providecommand \@url [1]{\endgroup\@href {#1}{\urlprefix }}%
\providecommand \urlprefix  [0]{URL }%
\providecommand \Eprint [0]{\href }%
\providecommand \doibase [0]{http://dx.doi.org/}%
\providecommand \selectlanguage [0]{\@gobble}%
\providecommand \bibinfo  [0]{\@secondoftwo}%
\providecommand \bibfield  [0]{\@secondoftwo}%
\providecommand \translation [1]{[#1]}%
\providecommand \BibitemOpen [0]{}%
\providecommand \bibitemStop [0]{}%
\providecommand \bibitemNoStop [0]{.\EOS\space}%
\providecommand \EOS [0]{\spacefactor3000\relax}%
\providecommand \BibitemShut  [1]{\csname bibitem#1\endcsname}%
\let\auto@bib@innerbib\@empty
\bibitem [{\citenamefont {Mourou}\ \emph {et~al.}(2006)\citenamefont {Mourou}, \citenamefont {Tajima},\ and\ \citenamefont {Bulanov}}]{RevModPhys.78.309}%
  \BibitemOpen
  \bibfield  {author} {\bibinfo {author} {\bibfnamefont {G.~A.}\ \bibnamefont {Mourou}}, \bibinfo {author} {\bibfnamefont {T.}~\bibnamefont {Tajima}}, \ and\ \bibinfo {author} {\bibfnamefont {S.~V.}\ \bibnamefont {Bulanov}},\ }\href {\doibase 10.1103/RevModPhys.78.309} {\bibfield  {journal} {\bibinfo  {journal} {Rev. Mod. Phys.}\ }\textbf {\bibinfo {volume} {78}},\ \bibinfo {pages} {309} (\bibinfo {year} {2006})}\BibitemShut {NoStop}%
\bibitem [{\citenamefont {Walther}\ \emph {et~al.}(2006)\citenamefont {Walther}, \citenamefont {Varcoe}, \citenamefont {Englert},\ and\ \citenamefont {Becker}}]{Walther_2006}%
  \BibitemOpen
  \bibfield  {author} {\bibinfo {author} {\bibfnamefont {H.}~\bibnamefont {Walther}}, \bibinfo {author} {\bibfnamefont {B.~T.~H.}\ \bibnamefont {Varcoe}}, \bibinfo {author} {\bibfnamefont {B.-G.}\ \bibnamefont {Englert}}, \ and\ \bibinfo {author} {\bibfnamefont {T.}~\bibnamefont {Becker}},\ }\href {\doibase 10.1088/0034-4885/69/5/R02} {\bibfield  {journal} {\bibinfo  {journal} {Reports on Progress in Physics}\ }\textbf {\bibinfo {volume} {69}},\ \bibinfo {pages} {1325} (\bibinfo {year} {2006})}\BibitemShut {NoStop}%
\bibitem [{\citenamefont {Chang}\ \emph {et~al.}(2018)\citenamefont {Chang}, \citenamefont {Douglas}, \citenamefont {Gonz\'alez-Tudela}, \citenamefont {Hung},\ and\ \citenamefont {Kimble}}]{RevModPhys.90.031002}%
  \BibitemOpen
  \bibfield  {author} {\bibinfo {author} {\bibfnamefont {D.~E.}\ \bibnamefont {Chang}}, \bibinfo {author} {\bibfnamefont {J.~S.}\ \bibnamefont {Douglas}}, \bibinfo {author} {\bibfnamefont {A.}~\bibnamefont {Gonz\'alez-Tudela}}, \bibinfo {author} {\bibfnamefont {C.-L.}\ \bibnamefont {Hung}}, \ and\ \bibinfo {author} {\bibfnamefont {H.~J.}\ \bibnamefont {Kimble}},\ }\href {\doibase 10.1103/RevModPhys.90.031002} {\bibfield  {journal} {\bibinfo  {journal} {Rev. Mod. Phys.}\ }\textbf {\bibinfo {volume} {90}},\ \bibinfo {pages} {031002} (\bibinfo {year} {2018})}\BibitemShut {NoStop}%
\bibitem [{\citenamefont {Ozawa}\ \emph {et~al.}(2019)\citenamefont {Ozawa}, \citenamefont {Price}, \citenamefont {Amo}, \citenamefont {Goldman}, \citenamefont {Hafezi}, \citenamefont {Lu}, \citenamefont {Rechtsman}, \citenamefont {Schuster}, \citenamefont {Simon}, \citenamefont {Zilberberg},\ and\ \citenamefont {Carusotto}}]{RevModPhys.91.015006}%
  \BibitemOpen
  \bibfield  {author} {\bibinfo {author} {\bibfnamefont {T.}~\bibnamefont {Ozawa}}, \bibinfo {author} {\bibfnamefont {H.~M.}\ \bibnamefont {Price}}, \bibinfo {author} {\bibfnamefont {A.}~\bibnamefont {Amo}}, \bibinfo {author} {\bibfnamefont {N.}~\bibnamefont {Goldman}}, \bibinfo {author} {\bibfnamefont {M.}~\bibnamefont {Hafezi}}, \bibinfo {author} {\bibfnamefont {L.}~\bibnamefont {Lu}}, \bibinfo {author} {\bibfnamefont {M.~C.}\ \bibnamefont {Rechtsman}}, \bibinfo {author} {\bibfnamefont {D.}~\bibnamefont {Schuster}}, \bibinfo {author} {\bibfnamefont {J.}~\bibnamefont {Simon}}, \bibinfo {author} {\bibfnamefont {O.}~\bibnamefont {Zilberberg}}, \ and\ \bibinfo {author} {\bibfnamefont {I.}~\bibnamefont {Carusotto}},\ }\href {\doibase 10.1103/RevModPhys.91.015006} {\bibfield  {journal} {\bibinfo  {journal} {Rev. Mod. Phys.}\ }\textbf {\bibinfo {volume} {91}},\ \bibinfo {pages} {015006} (\bibinfo {year} {2019})}\BibitemShut {NoStop}%
\bibitem [{\citenamefont {Harry}\ and\ \citenamefont {(forthe LIGO Scientific~Collaboration)}(2010)}]{Harry_2010}%
  \BibitemOpen
  \bibfield  {author} {\bibinfo {author} {\bibfnamefont {G.~M.}\ \bibnamefont {Harry}}\ and\ \bibinfo {author} {\bibnamefont {(forthe LIGO Scientific~Collaboration)}},\ }\href {\doibase 10.1088/0264-9381/27/8/084006} {\bibfield  {journal} {\bibinfo  {journal} {Classical and Quantum Gravity}\ }\textbf {\bibinfo {volume} {27}},\ \bibinfo {pages} {084006} (\bibinfo {year} {2010})}\BibitemShut {NoStop}%
\bibitem [{\citenamefont {Aso}\ \emph {et~al.}(2013)\citenamefont {Aso}, \citenamefont {Michimura}, \citenamefont {Somiya}, \citenamefont {Ando}, \citenamefont {Miyakawa}, \citenamefont {Sekiguchi}, \citenamefont {Tatsumi},\ and\ \citenamefont {Yamamoto}}]{PhysRevD.88.043007}%
  \BibitemOpen
  \bibfield  {author} {\bibinfo {author} {\bibfnamefont {Y.}~\bibnamefont {Aso}}, \bibinfo {author} {\bibfnamefont {Y.}~\bibnamefont {Michimura}}, \bibinfo {author} {\bibfnamefont {K.}~\bibnamefont {Somiya}}, \bibinfo {author} {\bibfnamefont {M.}~\bibnamefont {Ando}}, \bibinfo {author} {\bibfnamefont {O.}~\bibnamefont {Miyakawa}}, \bibinfo {author} {\bibfnamefont {T.}~\bibnamefont {Sekiguchi}}, \bibinfo {author} {\bibfnamefont {D.}~\bibnamefont {Tatsumi}}, \ and\ \bibinfo {author} {\bibfnamefont {H.}~\bibnamefont {Yamamoto}} (\bibinfo {collaboration} {The KAGRA Collaboration}),\ }\href {\doibase 10.1103/PhysRevD.88.043007} {\bibfield  {journal} {\bibinfo  {journal} {Phys. Rev. D}\ }\textbf {\bibinfo {volume} {88}},\ \bibinfo {pages} {043007} (\bibinfo {year} {2013})}\BibitemShut {NoStop}%
\bibitem [{\citenamefont {Dooley}\ \emph {et~al.}(2016)\citenamefont {Dooley}, \citenamefont {Leong}, \citenamefont {Adams}, \citenamefont {Affeldt}, \citenamefont {Bisht}, \citenamefont {Bogan}, \citenamefont {Degallaix}, \citenamefont {Gräf}, \citenamefont {Hild}, \citenamefont {Hough}, \citenamefont {Khalaidovski}, \citenamefont {Lastzka}, \citenamefont {Lough}, \citenamefont {Lück}, \citenamefont {Macleod}, \citenamefont {Nuttall}, \citenamefont {Prijatelj}, \citenamefont {Schnabel}, \citenamefont {Schreiber}, \citenamefont {Slutsky}, \citenamefont {Sorazu}, \citenamefont {Strain}, \citenamefont {Vahlbruch}, \citenamefont {Wąs}, \citenamefont {Willke}, \citenamefont {Wittel}, \citenamefont {Danzmann},\ and\ \citenamefont {Grote}}]{Dooley_2016}%
  \BibitemOpen
  \bibfield  {author} {\bibinfo {author} {\bibfnamefont {K.~L.}\ \bibnamefont {Dooley}}, \bibinfo {author} {\bibfnamefont {J.~R.}\ \bibnamefont {Leong}}, \bibinfo {author} {\bibfnamefont {T.}~\bibnamefont {Adams}}, \bibinfo {author} {\bibfnamefont {C.}~\bibnamefont {Affeldt}}, \bibinfo {author} {\bibfnamefont {A.}~\bibnamefont {Bisht}}, \bibinfo {author} {\bibfnamefont {C.}~\bibnamefont {Bogan}}, \bibinfo {author} {\bibfnamefont {J.}~\bibnamefont {Degallaix}}, \bibinfo {author} {\bibfnamefont {C.}~\bibnamefont {Gräf}}, \bibinfo {author} {\bibfnamefont {S.}~\bibnamefont {Hild}}, \bibinfo {author} {\bibfnamefont {J.}~\bibnamefont {Hough}}, \bibinfo {author} {\bibfnamefont {A.}~\bibnamefont {Khalaidovski}}, \bibinfo {author} {\bibfnamefont {N.}~\bibnamefont {Lastzka}}, \bibinfo {author} {\bibfnamefont {J.}~\bibnamefont {Lough}}, \bibinfo {author} {\bibfnamefont {H.}~\bibnamefont {Lück}}, \bibinfo {author} {\bibfnamefont {D.}~\bibnamefont {Macleod}}, \bibinfo {author} {\bibfnamefont {L.}~\bibnamefont
  {Nuttall}}, \bibinfo {author} {\bibfnamefont {M.}~\bibnamefont {Prijatelj}}, \bibinfo {author} {\bibfnamefont {R.}~\bibnamefont {Schnabel}}, \bibinfo {author} {\bibfnamefont {E.}~\bibnamefont {Schreiber}}, \bibinfo {author} {\bibfnamefont {J.}~\bibnamefont {Slutsky}}, \bibinfo {author} {\bibfnamefont {B.}~\bibnamefont {Sorazu}}, \bibinfo {author} {\bibfnamefont {K.~A.}\ \bibnamefont {Strain}}, \bibinfo {author} {\bibfnamefont {H.}~\bibnamefont {Vahlbruch}}, \bibinfo {author} {\bibfnamefont {M.}~\bibnamefont {Wąs}}, \bibinfo {author} {\bibfnamefont {B.}~\bibnamefont {Willke}}, \bibinfo {author} {\bibfnamefont {H.}~\bibnamefont {Wittel}}, \bibinfo {author} {\bibfnamefont {K.}~\bibnamefont {Danzmann}}, \ and\ \bibinfo {author} {\bibfnamefont {H.}~\bibnamefont {Grote}},\ }\href {\doibase 10.1088/0264-9381/33/7/075009} {\bibfield  {journal} {\bibinfo  {journal} {Classical and Quantum Gravity}\ }\textbf {\bibinfo {volume} {33}},\ \bibinfo {pages} {075009} (\bibinfo {year} {2016})}\BibitemShut {NoStop}%
\bibitem [{\citenamefont {Yu}\ \emph {et~al.}(2020)\citenamefont {Yu} \emph {et~al.}}]{LIGOScientific:2020luc}%
  \BibitemOpen
  \bibfield  {author} {\bibinfo {author} {\bibfnamefont {H.}~\bibnamefont {Yu}} \emph {et~al.} (\bibinfo {collaboration} {LIGO Scientific}),\ }\href {\doibase 10.1038/s41586-020-2420-8} {\bibfield  {journal} {\bibinfo  {journal} {Nature}\ }\textbf {\bibinfo {volume} {583}},\ \bibinfo {pages} {43} (\bibinfo {year} {2020})},\ \Eprint {http://arxiv.org/abs/2002.01519} {arXiv:2002.01519 [quant-ph]} \BibitemShut {NoStop}%
\bibitem [{\citenamefont {Acernese}\ \emph {et~al.}(2020)\citenamefont {Acernese} \emph {et~al.}}]{Virgo:2020xlu}%
  \BibitemOpen
  \bibfield  {author} {\bibinfo {author} {\bibfnamefont {F.}~\bibnamefont {Acernese}} \emph {et~al.} (\bibinfo {collaboration} {Virgo}),\ }\href {\doibase 10.1103/PhysRevLett.125.131101} {\bibfield  {journal} {\bibinfo  {journal} {Phys. Rev. Lett.}\ }\textbf {\bibinfo {volume} {125}},\ \bibinfo {pages} {131101} (\bibinfo {year} {2020})}\BibitemShut {NoStop}%
\bibitem [{\citenamefont {Safronova}\ \emph {et~al.}(2018)\citenamefont {Safronova}, \citenamefont {Budker}, \citenamefont {DeMille}, \citenamefont {Kimball}, \citenamefont {Derevianko},\ and\ \citenamefont {Clark}}]{RevModPhys.90.025008}%
  \BibitemOpen
  \bibfield  {author} {\bibinfo {author} {\bibfnamefont {M.~S.}\ \bibnamefont {Safronova}}, \bibinfo {author} {\bibfnamefont {D.}~\bibnamefont {Budker}}, \bibinfo {author} {\bibfnamefont {D.}~\bibnamefont {DeMille}}, \bibinfo {author} {\bibfnamefont {D.~F.~J.}\ \bibnamefont {Kimball}}, \bibinfo {author} {\bibfnamefont {A.}~\bibnamefont {Derevianko}}, \ and\ \bibinfo {author} {\bibfnamefont {C.~W.}\ \bibnamefont {Clark}},\ }\href {\doibase 10.1103/RevModPhys.90.025008} {\bibfield  {journal} {\bibinfo  {journal} {Rev. Mod. Phys.}\ }\textbf {\bibinfo {volume} {90}},\ \bibinfo {pages} {025008} (\bibinfo {year} {2018})}\BibitemShut {NoStop}%
\bibitem [{\citenamefont {Schultheiss}\ \emph {et~al.}(2010)\citenamefont {Schultheiss}, \citenamefont {Batz}, \citenamefont {Szameit}, \citenamefont {Dreisow}, \citenamefont {Nolte}, \citenamefont {T\"unnermann}, \citenamefont {Longhi},\ and\ \citenamefont {Peschel}}]{PhysRevLett.105.143901}%
  \BibitemOpen
  \bibfield  {author} {\bibinfo {author} {\bibfnamefont {V.~H.}\ \bibnamefont {Schultheiss}}, \bibinfo {author} {\bibfnamefont {S.}~\bibnamefont {Batz}}, \bibinfo {author} {\bibfnamefont {A.}~\bibnamefont {Szameit}}, \bibinfo {author} {\bibfnamefont {F.}~\bibnamefont {Dreisow}}, \bibinfo {author} {\bibfnamefont {S.}~\bibnamefont {Nolte}}, \bibinfo {author} {\bibfnamefont {A.}~\bibnamefont {T\"unnermann}}, \bibinfo {author} {\bibfnamefont {S.}~\bibnamefont {Longhi}}, \ and\ \bibinfo {author} {\bibfnamefont {U.}~\bibnamefont {Peschel}},\ }\href {\doibase 10.1103/PhysRevLett.105.143901} {\bibfield  {journal} {\bibinfo  {journal} {Phys. Rev. Lett.}\ }\textbf {\bibinfo {volume} {105}},\ \bibinfo {pages} {143901} (\bibinfo {year} {2010})}\BibitemShut {NoStop}%
\bibitem [{\citenamefont {Schultheiss}\ \emph {et~al.}(2020)\citenamefont {Schultheiss}, \citenamefont {Batz},\ and\ \citenamefont {Peschel}}]{Schultheiss:2020wzx}%
  \BibitemOpen
  \bibfield  {author} {\bibinfo {author} {\bibfnamefont {V.~H.}\ \bibnamefont {Schultheiss}}, \bibinfo {author} {\bibfnamefont {S.}~\bibnamefont {Batz}}, \ and\ \bibinfo {author} {\bibfnamefont {U.}~\bibnamefont {Peschel}},\ }\href {\doibase 10.1080/23746149.2020.1759451} {\bibfield  {journal} {\bibinfo  {journal} {Adv. Phys. X}\ }\textbf {\bibinfo {volume} {5}},\ \bibinfo {pages} {1759451} (\bibinfo {year} {2020})}\BibitemShut {NoStop}%
\bibitem [{\citenamefont {Leonhardt}\ and\ \citenamefont {Philbin}(2006)}]{Leonhardt_2006}%
  \BibitemOpen
  \bibfield  {author} {\bibinfo {author} {\bibfnamefont {U.}~\bibnamefont {Leonhardt}}\ and\ \bibinfo {author} {\bibfnamefont {T.~G.}\ \bibnamefont {Philbin}},\ }\href {\doibase 10.1088/1367-2630/8/10/247} {\bibfield  {journal} {\bibinfo  {journal} {New Journal of Physics}\ }\textbf {\bibinfo {volume} {8}},\ \bibinfo {pages} {247} (\bibinfo {year} {2006})}\BibitemShut {NoStop}%
\bibitem [{\citenamefont {Philbin}\ \emph {et~al.}(2008)\citenamefont {Philbin}, \citenamefont {Kuklewicz}, \citenamefont {Robertson}, \citenamefont {Hill}, \citenamefont {Konig},\ and\ \citenamefont {Leonhardt}}]{Philbin:2007ji}%
  \BibitemOpen
  \bibfield  {author} {\bibinfo {author} {\bibfnamefont {T.~G.}\ \bibnamefont {Philbin}}, \bibinfo {author} {\bibfnamefont {C.}~\bibnamefont {Kuklewicz}}, \bibinfo {author} {\bibfnamefont {S.}~\bibnamefont {Robertson}}, \bibinfo {author} {\bibfnamefont {S.}~\bibnamefont {Hill}}, \bibinfo {author} {\bibfnamefont {F.}~\bibnamefont {Konig}}, \ and\ \bibinfo {author} {\bibfnamefont {U.}~\bibnamefont {Leonhardt}},\ }\href {\doibase 10.1126/science.1153625} {\bibfield  {journal} {\bibinfo  {journal} {Science}\ }\textbf {\bibinfo {volume} {319}},\ \bibinfo {pages} {1367} (\bibinfo {year} {2008})},\ \Eprint {http://arxiv.org/abs/0711.4796} {arXiv:0711.4796 [gr-qc]} \BibitemShut {NoStop}%
\bibitem [{\citenamefont {Bekenstein}\ \emph {et~al.}(2016)\citenamefont {Bekenstein}, \citenamefont {Kabessa}, \citenamefont {Sharabi}, \citenamefont {Tal}, \citenamefont {Engheta}, \citenamefont {Eisenstein}, \citenamefont {Agranat},\ and\ \citenamefont {Segev}}]{Bekenstein2016CurvedSN}%
  \BibitemOpen
  \bibfield  {author} {\bibinfo {author} {\bibfnamefont {R.}~\bibnamefont {Bekenstein}}, \bibinfo {author} {\bibfnamefont {Y.}~\bibnamefont {Kabessa}}, \bibinfo {author} {\bibfnamefont {Y.}~\bibnamefont {Sharabi}}, \bibinfo {author} {\bibfnamefont {O.}~\bibnamefont {Tal}}, \bibinfo {author} {\bibfnamefont {N.}~\bibnamefont {Engheta}}, \bibinfo {author} {\bibfnamefont {G.}~\bibnamefont {Eisenstein}}, \bibinfo {author} {\bibfnamefont {A.~J.}\ \bibnamefont {Agranat}}, \ and\ \bibinfo {author} {\bibfnamefont {M.}~\bibnamefont {Segev}},\ }\href@noop {} {\bibfield  {journal} {\bibinfo  {journal} {2016 Conference on Lasers and Electro-Optics (CLEO)}\ ,\ \bibinfo {pages} {1}} (\bibinfo {year} {2016})}\BibitemShut {NoStop}%
\bibitem [{\citenamefont {Patsyk}\ \emph {et~al.}(2018)\citenamefont {Patsyk}, \citenamefont {Bandres}, \citenamefont {Bekenstein},\ and\ \citenamefont {Segev}}]{PhysRevX.8.011001}%
  \BibitemOpen
  \bibfield  {author} {\bibinfo {author} {\bibfnamefont {A.}~\bibnamefont {Patsyk}}, \bibinfo {author} {\bibfnamefont {M.~A.}\ \bibnamefont {Bandres}}, \bibinfo {author} {\bibfnamefont {R.}~\bibnamefont {Bekenstein}}, \ and\ \bibinfo {author} {\bibfnamefont {M.}~\bibnamefont {Segev}},\ }\href {\doibase 10.1103/PhysRevX.8.011001} {\bibfield  {journal} {\bibinfo  {journal} {Phys. Rev. X}\ }\textbf {\bibinfo {volume} {8}},\ \bibinfo {pages} {011001} (\bibinfo {year} {2018})}\BibitemShut {NoStop}%
\bibitem [{\citenamefont {Faccio}\ \emph {et~al.}(2013)\citenamefont {Faccio}, \citenamefont {Belgiorno}, \citenamefont {Cacciatori}, \citenamefont {Gorini}, \citenamefont {Liberati},\ and\ \citenamefont {Moschella}}]{Faccio:2013kpa}%
  \BibitemOpen
  \bibinfo {editor} {\bibfnamefont {D.}~\bibnamefont {Faccio}}, \bibinfo {editor} {\bibfnamefont {F.}~\bibnamefont {Belgiorno}}, \bibinfo {editor} {\bibfnamefont {S.}~\bibnamefont {Cacciatori}}, \bibinfo {editor} {\bibfnamefont {V.}~\bibnamefont {Gorini}}, \bibinfo {editor} {\bibfnamefont {S.}~\bibnamefont {Liberati}}, \ and\ \bibinfo {editor} {\bibfnamefont {U.}~\bibnamefont {Moschella}},\ eds.,\ \href {\doibase 10.1007/978-3-319-00266-8} {\emph {\bibinfo {title} {{Analogue Gravity Phenomenology}}}},\ Vol.\ \bibinfo {volume} {870}\ (\bibinfo {year} {2013})\BibitemShut {NoStop}%
\bibitem [{\citenamefont {Viermann}\ \emph {et~al.}(2022)\citenamefont {Viermann} \emph {et~al.}}]{Viermann:2022wgw}%
  \BibitemOpen
  \bibfield  {author} {\bibinfo {author} {\bibfnamefont {C.}~\bibnamefont {Viermann}} \emph {et~al.},\ }\href {\doibase 10.1038/s41586-022-05313-9} {\bibfield  {journal} {\bibinfo  {journal} {Nature}\ }\textbf {\bibinfo {volume} {611}},\ \bibinfo {pages} {260} (\bibinfo {year} {2022})},\ \Eprint {http://arxiv.org/abs/2202.10399} {arXiv:2202.10399 [cond-mat.quant-gas]} \BibitemShut {NoStop}%
\bibitem [{\citenamefont {Lopp}\ \emph {et~al.}(2018{\natexlab{a}})\citenamefont {Lopp}, \citenamefont {Mart{\'i}n-Martinez},\ and\ \citenamefont {Page}}]{Lopp2018RelativityAQ}%
  \BibitemOpen
  \bibfield  {author} {\bibinfo {author} {\bibfnamefont {R.}~\bibnamefont {Lopp}}, \bibinfo {author} {\bibfnamefont {E.~S.}\ \bibnamefont {Mart{\'i}n-Martinez}}, \ and\ \bibinfo {author} {\bibfnamefont {D.~N.}\ \bibnamefont {Page}},\ }\href@noop {} {\bibfield  {journal} {\bibinfo  {journal} {Classical and Quantum Gravity}\ }\textbf {\bibinfo {volume} {35}} (\bibinfo {year} {2018}{\natexlab{a}})}\BibitemShut {NoStop}%
\bibitem [{\citenamefont {Scully}\ \emph {et~al.}(2018)\citenamefont {Scully}, \citenamefont {Fulling}, \citenamefont {Lee}, \citenamefont {Page}, \citenamefont {Schleich},\ and\ \citenamefont {Svidzinsky}}]{Scully:2017utk}%
  \BibitemOpen
  \bibfield  {author} {\bibinfo {author} {\bibfnamefont {M.~O.}\ \bibnamefont {Scully}}, \bibinfo {author} {\bibfnamefont {S.}~\bibnamefont {Fulling}}, \bibinfo {author} {\bibfnamefont {D.}~\bibnamefont {Lee}}, \bibinfo {author} {\bibfnamefont {D.~N.}\ \bibnamefont {Page}}, \bibinfo {author} {\bibfnamefont {W.}~\bibnamefont {Schleich}}, \ and\ \bibinfo {author} {\bibfnamefont {A.}~\bibnamefont {Svidzinsky}},\ }\href {\doibase 10.1073/pnas.1807703115} {\bibfield  {journal} {\bibinfo  {journal} {Proc. Nat. Acad. Sci.}\ }\textbf {\bibinfo {volume} {115}},\ \bibinfo {pages} {8131} (\bibinfo {year} {2018})},\ \Eprint {http://arxiv.org/abs/1709.00481} {arXiv:1709.00481 [quant-ph]} \BibitemShut {NoStop}%
\bibitem [{\citenamefont {Mart\'{\i}n-Mart\'{\i}nez}\ \emph {et~al.}(2020)\citenamefont {Mart\'{\i}n-Mart\'{\i}nez}, \citenamefont {Perche},\ and\ \citenamefont {de~S.~L.~Torres}}]{PhysRevD.101.045017}%
  \BibitemOpen
  \bibfield  {author} {\bibinfo {author} {\bibfnamefont {E.}~\bibnamefont {Mart\'{\i}n-Mart\'{\i}nez}}, \bibinfo {author} {\bibfnamefont {T.~R.}\ \bibnamefont {Perche}}, \ and\ \bibinfo {author} {\bibfnamefont {B.}~\bibnamefont {de~S.~L.~Torres}},\ }\href {\doibase 10.1103/PhysRevD.101.045017} {\bibfield  {journal} {\bibinfo  {journal} {Phys. Rev. D}\ }\textbf {\bibinfo {volume} {101}},\ \bibinfo {pages} {045017} (\bibinfo {year} {2020})}\BibitemShut {NoStop}%
\bibitem [{\citenamefont {Zhang}\ \emph {et~al.}(2014)\citenamefont {Zhang}, \citenamefont {Cai},\ and\ \citenamefont {Zhan}}]{Zhang:2014ndf}%
  \BibitemOpen
  \bibfield  {author} {\bibinfo {author} {\bibfnamefont {B.~C.}\ \bibnamefont {Zhang}}, \bibinfo {author} {\bibfnamefont {Q.~Y.}\ \bibnamefont {Cai}}, \ and\ \bibinfo {author} {\bibfnamefont {M.~S.}\ \bibnamefont {Zhan}},\ }\href {\doibase 10.1360/SSPMA-2013-00095} {\bibfield  {journal} {\bibinfo  {journal} {Sci. Sin. Phys. Mech. Astron.}\ }\textbf {\bibinfo {volume} {44}},\ \bibinfo {pages} {879} (\bibinfo {year} {2014})}\BibitemShut {NoStop}%
\bibitem [{\citenamefont {{Leonhardt}}(2010)}]{2010eqo..book.....L}%
  \BibitemOpen
  \bibfield  {author} {\bibinfo {author} {\bibfnamefont {U.}~\bibnamefont {{Leonhardt}}},\ }\href@noop {} {\emph {\bibinfo {title} {{Essential Quantum Optics}}}}\ (\bibinfo {year} {2010})\BibitemShut {NoStop}%
\bibitem [{\citenamefont {Boettcher}\ \emph {et~al.}(2020)\citenamefont {Boettcher}, \citenamefont {Bienias}, \citenamefont {Belyansky}, \citenamefont {Koll\'ar},\ and\ \citenamefont {Gorshkov}}]{PhysRevA.102.032208}%
  \BibitemOpen
  \bibfield  {author} {\bibinfo {author} {\bibfnamefont {I.}~\bibnamefont {Boettcher}}, \bibinfo {author} {\bibfnamefont {P.}~\bibnamefont {Bienias}}, \bibinfo {author} {\bibfnamefont {R.}~\bibnamefont {Belyansky}}, \bibinfo {author} {\bibfnamefont {A.~J.}\ \bibnamefont {Koll\'ar}}, \ and\ \bibinfo {author} {\bibfnamefont {A.~V.}\ \bibnamefont {Gorshkov}},\ }\href {\doibase 10.1103/PhysRevA.102.032208} {\bibfield  {journal} {\bibinfo  {journal} {Phys. Rev. A}\ }\textbf {\bibinfo {volume} {102}},\ \bibinfo {pages} {032208} (\bibinfo {year} {2020})}\BibitemShut {NoStop}%
\bibitem [{\citenamefont {Garcia}\ \emph {et~al.}(2020)\citenamefont {Garcia}, \citenamefont {Chaplain}, \citenamefont {B\v{e}l\'\i{}n}, \citenamefont {Tyc}, \citenamefont {Englert},\ and\ \citenamefont {Courtial}}]{Garcia:2020slt}%
  \BibitemOpen
  \bibfield  {author} {\bibinfo {author} {\bibfnamefont {D.~G.}\ \bibnamefont {Garcia}}, \bibinfo {author} {\bibfnamefont {G.~J.}\ \bibnamefont {Chaplain}}, \bibinfo {author} {\bibfnamefont {J.}~\bibnamefont {B\v{e}l\'\i{}n}}, \bibinfo {author} {\bibfnamefont {T.}~\bibnamefont {Tyc}}, \bibinfo {author} {\bibfnamefont {C.}~\bibnamefont {Englert}}, \ and\ \bibinfo {author} {\bibfnamefont {J.}~\bibnamefont {Courtial}},\ }\href {\doibase 10.1364/OPTICA.378357} {\bibfield  {journal} {\bibinfo  {journal} {Optica}\ }\textbf {\bibinfo {volume} {7}},\ \bibinfo {pages} {142} (\bibinfo {year} {2020})}\BibitemShut {NoStop}%
\bibitem [{\citenamefont {Steinhauer}(2016)}]{Steinhauer:2015saa}%
  \BibitemOpen
  \bibfield  {author} {\bibinfo {author} {\bibfnamefont {J.}~\bibnamefont {Steinhauer}},\ }\href {\doibase 10.1038/nphys3863} {\bibfield  {journal} {\bibinfo  {journal} {Nature Phys.}\ }\textbf {\bibinfo {volume} {12}},\ \bibinfo {pages} {959} (\bibinfo {year} {2016})},\ \Eprint {http://arxiv.org/abs/1510.00621} {arXiv:1510.00621 [gr-qc]} \BibitemShut {NoStop}%
\bibitem [{\citenamefont {Leonhardt}(2018)}]{Leonhardt:2016qdi}%
  \BibitemOpen
  \bibfield  {author} {\bibinfo {author} {\bibfnamefont {U.}~\bibnamefont {Leonhardt}},\ }\href {\doibase 10.1002/andp.201700114} {\bibfield  {journal} {\bibinfo  {journal} {Annalen Phys.}\ }\textbf {\bibinfo {volume} {530}},\ \bibinfo {pages} {1700114} (\bibinfo {year} {2018})},\ \Eprint {http://arxiv.org/abs/1609.03803} {arXiv:1609.03803 [gr-qc]} \BibitemShut {NoStop}%
\bibitem [{\citenamefont {Hu}\ \emph {et~al.}(2019)\citenamefont {Hu}, \citenamefont {Feng}, \citenamefont {Zhang},\ and\ \citenamefont {Chin}}]{Hu:2018psq}%
  \BibitemOpen
  \bibfield  {author} {\bibinfo {author} {\bibfnamefont {J.}~\bibnamefont {Hu}}, \bibinfo {author} {\bibfnamefont {L.}~\bibnamefont {Feng}}, \bibinfo {author} {\bibfnamefont {Z.}~\bibnamefont {Zhang}}, \ and\ \bibinfo {author} {\bibfnamefont {C.}~\bibnamefont {Chin}},\ }\href {\doibase 10.1038/s41567-019-0537-1} {\bibfield  {journal} {\bibinfo  {journal} {Nature Phys.}\ }\textbf {\bibinfo {volume} {15}},\ \bibinfo {pages} {785} (\bibinfo {year} {2019})},\ \Eprint {http://arxiv.org/abs/1807.07504} {arXiv:1807.07504 [physics.atom-ph]} \BibitemShut {NoStop}%
\bibitem [{\citenamefont {Sheng}\ \emph {et~al.}(2021)\citenamefont {Sheng}, \citenamefont {Qian}, \citenamefont {Li}, \citenamefont {Niu},\ and\ \citenamefont {Gong}}]{Sheng:2021iky}%
  \BibitemOpen
  \bibfield  {author} {\bibinfo {author} {\bibfnamefont {T.}~\bibnamefont {Sheng}}, \bibinfo {author} {\bibfnamefont {J.}~\bibnamefont {Qian}}, \bibinfo {author} {\bibfnamefont {X.}~\bibnamefont {Li}}, \bibinfo {author} {\bibfnamefont {Y.}~\bibnamefont {Niu}}, \ and\ \bibinfo {author} {\bibfnamefont {S.}~\bibnamefont {Gong}},\ }\href {\doibase 10.1103/PhysRevA.103.013301} {\bibfield  {journal} {\bibinfo  {journal} {Phys. Rev. A}\ }\textbf {\bibinfo {volume} {103}},\ \bibinfo {pages} {013301} (\bibinfo {year} {2021})}\BibitemShut {NoStop}%
\bibitem [{\citenamefont {Parker}(1969)}]{Parker:1969au}%
  \BibitemOpen
  \bibfield  {author} {\bibinfo {author} {\bibfnamefont {L.}~\bibnamefont {Parker}},\ }\href {\doibase 10.1103/PhysRev.183.1057} {\bibfield  {journal} {\bibinfo  {journal} {Phys. Rev.}\ }\textbf {\bibinfo {volume} {183}},\ \bibinfo {pages} {1057} (\bibinfo {year} {1969})}\BibitemShut {NoStop}%
\bibitem [{\citenamefont {Parker}(1971)}]{Parker:1971pt}%
  \BibitemOpen
  \bibfield  {author} {\bibinfo {author} {\bibfnamefont {L.}~\bibnamefont {Parker}},\ }\href {\doibase 10.1103/PhysRevD.3.346} {\bibfield  {journal} {\bibinfo  {journal} {Phys. Rev. D}\ }\textbf {\bibinfo {volume} {3}},\ \bibinfo {pages} {346} (\bibinfo {year} {1971})},\ \bibinfo {note} {[Erratum: Phys.Rev.D 3, 2546--2546 (1971)]}\BibitemShut {NoStop}%
\bibitem [{\citenamefont {Parker}(2012)}]{Parker:2012at}%
  \BibitemOpen
  \bibfield  {author} {\bibinfo {author} {\bibfnamefont {L.}~\bibnamefont {Parker}},\ }\href {\doibase 10.1088/1751-8113/45/37/374023} {\bibfield  {journal} {\bibinfo  {journal} {J. Phys. A}\ }\textbf {\bibinfo {volume} {45}},\ \bibinfo {pages} {374023} (\bibinfo {year} {2012})},\ \Eprint {http://arxiv.org/abs/1205.5616} {arXiv:1205.5616 [astro-ph.CO]} \BibitemShut {NoStop}%
\bibitem [{\citenamefont {Eckel}\ \emph {et~al.}(2018)\citenamefont {Eckel}, \citenamefont {Kumar}, \citenamefont {Jacobson}, \citenamefont {Spielman},\ and\ \citenamefont {Campbell}}]{Eckel:2017uqx}%
  \BibitemOpen
  \bibfield  {author} {\bibinfo {author} {\bibfnamefont {S.}~\bibnamefont {Eckel}}, \bibinfo {author} {\bibfnamefont {A.}~\bibnamefont {Kumar}}, \bibinfo {author} {\bibfnamefont {T.}~\bibnamefont {Jacobson}}, \bibinfo {author} {\bibfnamefont {I.~B.}\ \bibnamefont {Spielman}}, \ and\ \bibinfo {author} {\bibfnamefont {G.~K.}\ \bibnamefont {Campbell}},\ }\href {\doibase 10.1103/PhysRevX.8.021021} {\bibfield  {journal} {\bibinfo  {journal} {Phys. Rev. X}\ }\textbf {\bibinfo {volume} {8}},\ \bibinfo {pages} {021021} (\bibinfo {year} {2018})},\ \Eprint {http://arxiv.org/abs/1710.05800} {arXiv:1710.05800 [cond-mat.quant-gas]} \BibitemShut {NoStop}%
\bibitem [{\citenamefont {Schmit}\ \emph {et~al.}(2020)\citenamefont {Schmit}, \citenamefont {Taketani},\ and\ \citenamefont {Wilhelm}}]{Schmit:2018hvy}%
  \BibitemOpen
  \bibfield  {author} {\bibinfo {author} {\bibfnamefont {R.~P.}\ \bibnamefont {Schmit}}, \bibinfo {author} {\bibfnamefont {B.~G.}\ \bibnamefont {Taketani}}, \ and\ \bibinfo {author} {\bibfnamefont {F.~K.}\ \bibnamefont {Wilhelm}},\ }\href {\doibase 10.1371/journal.pone.0229382} {\bibfield  {journal} {\bibinfo  {journal} {PLoS One}\ }\textbf {\bibinfo {volume} {15}},\ \bibinfo {pages} {e0229382} (\bibinfo {year} {2020})},\ \Eprint {http://arxiv.org/abs/1804.04092} {arXiv:1804.04092 [quant-ph]} \BibitemShut {NoStop}%
\bibitem [{\citenamefont {Alsing}\ and\ \citenamefont {Milburn}(2003)}]{Alsing:2003es}%
  \BibitemOpen
  \bibfield  {author} {\bibinfo {author} {\bibfnamefont {P.~M.}\ \bibnamefont {Alsing}}\ and\ \bibinfo {author} {\bibfnamefont {G.~J.}\ \bibnamefont {Milburn}},\ }\href {\doibase 10.1103/PhysRevLett.91.180404} {\bibfield  {journal} {\bibinfo  {journal} {Phys. Rev. Lett.}\ }\textbf {\bibinfo {volume} {91}},\ \bibinfo {pages} {180404} (\bibinfo {year} {2003})},\ \Eprint {http://arxiv.org/abs/quant-ph/0302179} {arXiv:quant-ph/0302179} \BibitemShut {NoStop}%
\bibitem [{\citenamefont {Fuentes-Schuller}\ and\ \citenamefont {Mann}(2005)}]{PhysRevLett.95.120404}%
  \BibitemOpen
  \bibfield  {author} {\bibinfo {author} {\bibfnamefont {I.}~\bibnamefont {Fuentes-Schuller}}\ and\ \bibinfo {author} {\bibfnamefont {R.~B.}\ \bibnamefont {Mann}},\ }\href {\doibase 10.1103/PhysRevLett.95.120404} {\bibfield  {journal} {\bibinfo  {journal} {Phys. Rev. Lett.}\ }\textbf {\bibinfo {volume} {95}},\ \bibinfo {pages} {120404} (\bibinfo {year} {2005})}\BibitemShut {NoStop}%
\bibitem [{\citenamefont {Downes}\ \emph {et~al.}(2011)\citenamefont {Downes}, \citenamefont {Fuentes},\ and\ \citenamefont {Ralph}}]{PhysRevLett.106.210502}%
  \BibitemOpen
  \bibfield  {author} {\bibinfo {author} {\bibfnamefont {T.~G.}\ \bibnamefont {Downes}}, \bibinfo {author} {\bibfnamefont {I.}~\bibnamefont {Fuentes}}, \ and\ \bibinfo {author} {\bibfnamefont {T.~C.}\ \bibnamefont {Ralph}},\ }\href {\doibase 10.1103/PhysRevLett.106.210502} {\bibfield  {journal} {\bibinfo  {journal} {Phys. Rev. Lett.}\ }\textbf {\bibinfo {volume} {106}},\ \bibinfo {pages} {210502} (\bibinfo {year} {2011})}\BibitemShut {NoStop}%
\bibitem [{\citenamefont {Peres}\ and\ \citenamefont {Terno}(2004)}]{RevModPhys.76.93}%
  \BibitemOpen
  \bibfield  {author} {\bibinfo {author} {\bibfnamefont {A.}~\bibnamefont {Peres}}\ and\ \bibinfo {author} {\bibfnamefont {D.~R.}\ \bibnamefont {Terno}},\ }\href {\doibase 10.1103/RevModPhys.76.93} {\bibfield  {journal} {\bibinfo  {journal} {Rev. Mod. Phys.}\ }\textbf {\bibinfo {volume} {76}},\ \bibinfo {pages} {93} (\bibinfo {year} {2004})}\BibitemShut {NoStop}%
\bibitem [{\citenamefont {Mann}\ and\ \citenamefont {Ralph}(2012)}]{Mann_2012}%
  \BibitemOpen
  \bibfield  {author} {\bibinfo {author} {\bibfnamefont {R.~B.}\ \bibnamefont {Mann}}\ and\ \bibinfo {author} {\bibfnamefont {T.~C.}\ \bibnamefont {Ralph}},\ }\href {\doibase 10.1088/0264-9381/29/22/220301} {\bibfield  {journal} {\bibinfo  {journal} {Classical and Quantum Gravity}\ }\textbf {\bibinfo {volume} {29}},\ \bibinfo {pages} {220301} (\bibinfo {year} {2012})}\BibitemShut {NoStop}%
\bibitem [{\citenamefont {Alsing}\ \emph {et~al.}(2006)\citenamefont {Alsing}, \citenamefont {Fuentes-Schuller}, \citenamefont {Mann},\ and\ \citenamefont {Tessier}}]{PhysRevA.74.032326}%
  \BibitemOpen
  \bibfield  {author} {\bibinfo {author} {\bibfnamefont {P.~M.}\ \bibnamefont {Alsing}}, \bibinfo {author} {\bibfnamefont {I.}~\bibnamefont {Fuentes-Schuller}}, \bibinfo {author} {\bibfnamefont {R.~B.}\ \bibnamefont {Mann}}, \ and\ \bibinfo {author} {\bibfnamefont {T.~E.}\ \bibnamefont {Tessier}},\ }\href {\doibase 10.1103/PhysRevA.74.032326} {\bibfield  {journal} {\bibinfo  {journal} {Phys. Rev. A}\ }\textbf {\bibinfo {volume} {74}},\ \bibinfo {pages} {032326} (\bibinfo {year} {2006})}\BibitemShut {NoStop}%
\bibitem [{\citenamefont {Wang}\ and\ \citenamefont {Jing}(2011)}]{Wang:2010qq}%
  \BibitemOpen
  \bibfield  {author} {\bibinfo {author} {\bibfnamefont {J.}~\bibnamefont {Wang}}\ and\ \bibinfo {author} {\bibfnamefont {J.}~\bibnamefont {Jing}},\ }\href {\doibase 10.1103/PhysRevA.83.022314} {\bibfield  {journal} {\bibinfo  {journal} {Phys. Rev. A}\ }\textbf {\bibinfo {volume} {83}},\ \bibinfo {pages} {022314} (\bibinfo {year} {2011})},\ \bibinfo {note} {[Erratum: Phys.Rev.A 97, 029902 (2018)]},\ \Eprint {http://arxiv.org/abs/1012.4268} {arXiv:1012.4268 [quant-ph]} \BibitemShut {NoStop}%
\bibitem [{\citenamefont {Friis}\ \emph {et~al.}(2012)\citenamefont {Friis}, \citenamefont {Bruschi}, \citenamefont {Louko},\ and\ \citenamefont {Fuentes}}]{Friis:2012tb}%
  \BibitemOpen
  \bibfield  {author} {\bibinfo {author} {\bibfnamefont {N.}~\bibnamefont {Friis}}, \bibinfo {author} {\bibfnamefont {D.~E.}\ \bibnamefont {Bruschi}}, \bibinfo {author} {\bibfnamefont {J.}~\bibnamefont {Louko}}, \ and\ \bibinfo {author} {\bibfnamefont {I.}~\bibnamefont {Fuentes}},\ }\href {\doibase 10.1103/PhysRevD.85.081701} {\bibfield  {journal} {\bibinfo  {journal} {Phys. Rev. D}\ }\textbf {\bibinfo {volume} {85}},\ \bibinfo {pages} {081701} (\bibinfo {year} {2012})},\ \Eprint {http://arxiv.org/abs/1201.0549} {arXiv:1201.0549 [quant-ph]} \BibitemShut {NoStop}%
\bibitem [{\citenamefont {Bruschi}\ \emph {et~al.}(2013)\citenamefont {Bruschi}, \citenamefont {Dragan}, \citenamefont {Lee}, \citenamefont {Fuentes},\ and\ \citenamefont {Louko}}]{Bruschi:2012uf}%
  \BibitemOpen
  \bibfield  {author} {\bibinfo {author} {\bibfnamefont {D.~E.}\ \bibnamefont {Bruschi}}, \bibinfo {author} {\bibfnamefont {A.}~\bibnamefont {Dragan}}, \bibinfo {author} {\bibfnamefont {A.~R.}\ \bibnamefont {Lee}}, \bibinfo {author} {\bibfnamefont {I.}~\bibnamefont {Fuentes}}, \ and\ \bibinfo {author} {\bibfnamefont {J.}~\bibnamefont {Louko}},\ }\href {\doibase 10.1103/PhysRevLett.111.090504} {\bibfield  {journal} {\bibinfo  {journal} {Phys. Rev. Lett.}\ }\textbf {\bibinfo {volume} {111}},\ \bibinfo {pages} {090504} (\bibinfo {year} {2013})},\ \Eprint {http://arxiv.org/abs/1201.0663} {arXiv:1201.0663 [quant-ph]} \BibitemShut {NoStop}%
\bibitem [{\citenamefont {Liu}\ \emph {et~al.}(2022)\citenamefont {Liu}, \citenamefont {Zhang}, \citenamefont {Mann},\ and\ \citenamefont {Yu}}]{Liu:2021dnl}%
  \BibitemOpen
  \bibfield  {author} {\bibinfo {author} {\bibfnamefont {Z.}~\bibnamefont {Liu}}, \bibinfo {author} {\bibfnamefont {J.}~\bibnamefont {Zhang}}, \bibinfo {author} {\bibfnamefont {R.~B.}\ \bibnamefont {Mann}}, \ and\ \bibinfo {author} {\bibfnamefont {H.}~\bibnamefont {Yu}},\ }\href {\doibase 10.1103/PhysRevD.105.085012} {\bibfield  {journal} {\bibinfo  {journal} {Phys. Rev. D}\ }\textbf {\bibinfo {volume} {105}},\ \bibinfo {pages} {085012} (\bibinfo {year} {2022})},\ \Eprint {http://arxiv.org/abs/2111.04392} {arXiv:2111.04392 [quant-ph]} \BibitemShut {NoStop}%
\bibitem [{\citenamefont {Lopp}\ \emph {et~al.}(2018{\natexlab{b}})\citenamefont {Lopp}, \citenamefont {Martin-Martinez},\ and\ \citenamefont {Page}}]{Lopp:2018lxl}%
  \BibitemOpen
  \bibfield  {author} {\bibinfo {author} {\bibfnamefont {R.}~\bibnamefont {Lopp}}, \bibinfo {author} {\bibfnamefont {E.}~\bibnamefont {Martin-Martinez}}, \ and\ \bibinfo {author} {\bibfnamefont {D.~N.}\ \bibnamefont {Page}},\ }\href {\doibase 10.1088/1361-6382/aae750} {\bibfield  {journal} {\bibinfo  {journal} {Class. Quant. Grav.}\ }\textbf {\bibinfo {volume} {35}},\ \bibinfo {pages} {224001} (\bibinfo {year} {2018}{\natexlab{b}})},\ \Eprint {http://arxiv.org/abs/1806.10158} {arXiv:1806.10158 [quant-ph]} \BibitemShut {NoStop}%
\bibitem [{\citenamefont {Mart\'\i{}n-Mart\'\i{}nez}\ \emph {et~al.}(2020)\citenamefont {Mart\'\i{}n-Mart\'\i{}nez}, \citenamefont {Perche},\ and\ \citenamefont {de~S.~L.~Torres}}]{Martin-Martinez:2020pss}%
  \BibitemOpen
  \bibfield  {author} {\bibinfo {author} {\bibfnamefont {E.}~\bibnamefont {Mart\'\i{}n-Mart\'\i{}nez}}, \bibinfo {author} {\bibfnamefont {T.~R.}\ \bibnamefont {Perche}}, \ and\ \bibinfo {author} {\bibfnamefont {B.}~\bibnamefont {de~S.~L.~Torres}},\ }\href {\doibase 10.1103/PhysRevD.101.045017} {\bibfield  {journal} {\bibinfo  {journal} {Phys. Rev. D}\ }\textbf {\bibinfo {volume} {101}},\ \bibinfo {pages} {045017} (\bibinfo {year} {2020})},\ \Eprint {http://arxiv.org/abs/2001.10010} {arXiv:2001.10010 [quant-ph]} \BibitemShut {NoStop}%
\bibitem [{\citenamefont {Sabin}\ \emph {et~al.}(2014)\citenamefont {Sabin}, \citenamefont {Bruschi}, \citenamefont {Ahmadi},\ and\ \citenamefont {Fuentes}}]{Sabin:2014bua}%
  \BibitemOpen
  \bibfield  {author} {\bibinfo {author} {\bibfnamefont {C.}~\bibnamefont {Sabin}}, \bibinfo {author} {\bibfnamefont {D.~E.}\ \bibnamefont {Bruschi}}, \bibinfo {author} {\bibfnamefont {M.}~\bibnamefont {Ahmadi}}, \ and\ \bibinfo {author} {\bibfnamefont {I.}~\bibnamefont {Fuentes}},\ }\href {\doibase 10.1088/1367-2630/16/8/085003} {\bibfield  {journal} {\bibinfo  {journal} {New J. Phys.}\ }\textbf {\bibinfo {volume} {16}},\ \bibinfo {pages} {085003} (\bibinfo {year} {2014})},\ \Eprint {http://arxiv.org/abs/1402.7009} {arXiv:1402.7009 [quant-ph]} \BibitemShut {NoStop}%
\bibitem [{\citenamefont {R\"atzel}\ \emph {et~al.}(2018{\natexlab{a}})\citenamefont {R\"atzel}, \citenamefont {Howl}, \citenamefont {Lindkvist},\ and\ \citenamefont {Fuentes}}]{Ratzel:2018srb}%
  \BibitemOpen
  \bibfield  {author} {\bibinfo {author} {\bibfnamefont {D.}~\bibnamefont {R\"atzel}}, \bibinfo {author} {\bibfnamefont {R.}~\bibnamefont {Howl}}, \bibinfo {author} {\bibfnamefont {J.}~\bibnamefont {Lindkvist}}, \ and\ \bibinfo {author} {\bibfnamefont {I.}~\bibnamefont {Fuentes}},\ }\href {\doibase 10.1088/1367-2630/aad272} {\bibfield  {journal} {\bibinfo  {journal} {New J. Phys.}\ }\textbf {\bibinfo {volume} {20}},\ \bibinfo {pages} {073044} (\bibinfo {year} {2018}{\natexlab{a}})},\ \Eprint {http://arxiv.org/abs/1804.11122} {arXiv:1804.11122 [quant-ph]} \BibitemShut {NoStop}%
\bibitem [{\citenamefont {Sch\"utzhold}(2018)}]{PhysRevD.98.105019}%
  \BibitemOpen
  \bibfield  {author} {\bibinfo {author} {\bibfnamefont {R.}~\bibnamefont {Sch\"utzhold}},\ }\href {\doibase 10.1103/PhysRevD.98.105019} {\bibfield  {journal} {\bibinfo  {journal} {Phys. Rev. D}\ }\textbf {\bibinfo {volume} {98}},\ \bibinfo {pages} {105019} (\bibinfo {year} {2018})}\BibitemShut {NoStop}%
\bibitem [{\citenamefont {Howl}\ \emph {et~al.}(2018)\citenamefont {Howl}, \citenamefont {Hackerm\"uller}, \citenamefont {Bruschi},\ and\ \citenamefont {Fuentes}}]{Howl:2016ryt}%
  \BibitemOpen
  \bibfield  {author} {\bibinfo {author} {\bibfnamefont {R.}~\bibnamefont {Howl}}, \bibinfo {author} {\bibfnamefont {L.}~\bibnamefont {Hackerm\"uller}}, \bibinfo {author} {\bibfnamefont {D.~E.}\ \bibnamefont {Bruschi}}, \ and\ \bibinfo {author} {\bibfnamefont {I.}~\bibnamefont {Fuentes}},\ }\href {\doibase 10.1080/23746149.2017.1383184} {\bibfield  {journal} {\bibinfo  {journal} {Adv. Phys. X}\ }\textbf {\bibinfo {volume} {3}},\ \bibinfo {pages} {1383184} (\bibinfo {year} {2018})},\ \Eprint {http://arxiv.org/abs/1607.06666} {arXiv:1607.06666 [quant-ph]} \BibitemShut {NoStop}%
\bibitem [{\citenamefont {Collas}\ and\ \citenamefont {Klein}(2019)}]{Collas:2018jfx}%
  \BibitemOpen
  \bibfield  {author} {\bibinfo {author} {\bibfnamefont {P.}~\bibnamefont {Collas}}\ and\ \bibinfo {author} {\bibfnamefont {D.}~\bibnamefont {Klein}},\ }\href {\doibase 10.1007/978-3-030-14825-6} {\emph {\bibinfo {title} {{The Dirac Equation in Curved Spacetime}: {A Guide for Calculations}}}},\ SpringerBriefs in Physics\ (\bibinfo  {publisher} {Springer},\ \bibinfo {year} {2019})\ \Eprint {http://arxiv.org/abs/1809.02764} {arXiv:1809.02764 [gr-qc]} \BibitemShut {NoStop}%
\bibitem [{\citenamefont {Scully}\ and\ \citenamefont {Zubairy}(1997)}]{scully_zubairy_1997}%
  \BibitemOpen
  \bibfield  {author} {\bibinfo {author} {\bibfnamefont {M.~O.}\ \bibnamefont {Scully}}\ and\ \bibinfo {author} {\bibfnamefont {M.~S.}\ \bibnamefont {Zubairy}},\ }\href {\doibase 10.1017/CBO9780511813993} {\emph {\bibinfo {title} {Quantum Optics}}}\ (\bibinfo  {publisher} {Cambridge University Press},\ \bibinfo {year} {1997})\BibitemShut {NoStop}%
\bibitem [{\citenamefont {Compagno}\ \emph {et~al.}(1995)\citenamefont {Compagno}, \citenamefont {Passante},\ and\ \citenamefont {Persico}}]{compagno_passante_persico_1995}%
  \BibitemOpen
  \bibfield  {author} {\bibinfo {author} {\bibfnamefont {G.}~\bibnamefont {Compagno}}, \bibinfo {author} {\bibfnamefont {R.}~\bibnamefont {Passante}}, \ and\ \bibinfo {author} {\bibfnamefont {F.}~\bibnamefont {Persico}},\ }\href {\doibase 10.1017/CBO9780511599774} {\emph {\bibinfo {title} {Atom-Field Interactions and Dressed Atoms}}},\ Cambridge Studies in Modern Optics\ (\bibinfo  {publisher} {Cambridge University Press},\ \bibinfo {year} {1995})\BibitemShut {NoStop}%
\bibitem [{\citenamefont {Unruh}(1976)}]{PhysRevD.14.870}%
  \BibitemOpen
  \bibfield  {author} {\bibinfo {author} {\bibfnamefont {W.~G.}\ \bibnamefont {Unruh}},\ }\href {\doibase 10.1103/PhysRevD.14.870} {\bibfield  {journal} {\bibinfo  {journal} {Phys. Rev. D}\ }\textbf {\bibinfo {volume} {14}},\ \bibinfo {pages} {870} (\bibinfo {year} {1976})}\BibitemShut {NoStop}%
\bibitem [{\citenamefont {Fulling}(1973)}]{Fulling:1972md}%
  \BibitemOpen
  \bibfield  {author} {\bibinfo {author} {\bibfnamefont {S.~A.}\ \bibnamefont {Fulling}},\ }\href {\doibase 10.1103/PhysRevD.7.2850} {\bibfield  {journal} {\bibinfo  {journal} {Phys. Rev. D}\ }\textbf {\bibinfo {volume} {7}},\ \bibinfo {pages} {2850} (\bibinfo {year} {1973})}\BibitemShut {NoStop}%
\bibitem [{\citenamefont {Davies}(1975)}]{Davies:1974th}%
  \BibitemOpen
  \bibfield  {author} {\bibinfo {author} {\bibfnamefont {P.~C.~W.}\ \bibnamefont {Davies}},\ }\href {\doibase 10.1088/0305-4470/8/4/022} {\bibfield  {journal} {\bibinfo  {journal} {J. Phys. A}\ }\textbf {\bibinfo {volume} {8}},\ \bibinfo {pages} {609} (\bibinfo {year} {1975})}\BibitemShut {NoStop}%
\bibitem [{\citenamefont {Crispino}\ \emph {et~al.}(2008)\citenamefont {Crispino}, \citenamefont {Higuchi},\ and\ \citenamefont {Matsas}}]{RevModPhys.80.787}%
  \BibitemOpen
  \bibfield  {author} {\bibinfo {author} {\bibfnamefont {L.~C.~B.}\ \bibnamefont {Crispino}}, \bibinfo {author} {\bibfnamefont {A.}~\bibnamefont {Higuchi}}, \ and\ \bibinfo {author} {\bibfnamefont {G.~E.~A.}\ \bibnamefont {Matsas}},\ }\href {\doibase 10.1103/RevModPhys.80.787} {\bibfield  {journal} {\bibinfo  {journal} {Rev. Mod. Phys.}\ }\textbf {\bibinfo {volume} {80}},\ \bibinfo {pages} {787} (\bibinfo {year} {2008})}\BibitemShut {NoStop}%
\bibitem [{\citenamefont {Rohrlich}(1961)}]{osti_4818173}%
  \BibitemOpen
  \bibfield  {author} {\bibinfo {author} {\bibfnamefont {F.}~\bibnamefont {Rohrlich}},\ }\href {\doibase 10.1007/BF02785607} {\bibfield  {journal} {\bibinfo  {journal} {Nuovo Cimento}\ }\textbf {\bibinfo {volume} {21}} (\bibinfo {year} {1961}),\ 10.1007/BF02785607}\BibitemShut {NoStop}%
\bibitem [{\citenamefont {{Rohrlich}}(1961)}]{1961NCim...21..811R}%
  \BibitemOpen
  \bibfield  {author} {\bibinfo {author} {\bibfnamefont {F.}~\bibnamefont {{Rohrlich}}},\ }\href {\doibase 10.1007/BF02785607} {\bibfield  {journal} {\bibinfo  {journal} {Il Nuovo Cimento}\ }\textbf {\bibinfo {volume} {21}},\ \bibinfo {pages} {811} (\bibinfo {year} {1961})}\BibitemShut {NoStop}%
\bibitem [{\citenamefont {Boulware}(1980)}]{BOULWARE1980169}%
  \BibitemOpen
  \bibfield  {author} {\bibinfo {author} {\bibfnamefont {D.~G.}\ \bibnamefont {Boulware}},\ }\href {\doibase https://doi.org/10.1016/0003-4916(80)90360-7} {\bibfield  {journal} {\bibinfo  {journal} {Annals of Physics}\ }\textbf {\bibinfo {volume} {124}},\ \bibinfo {pages} {169} (\bibinfo {year} {1980})}\BibitemShut {NoStop}%
\bibitem [{\citenamefont {Hawking}(1975)}]{Hawking1975ParticleCB}%
  \BibitemOpen
  \bibfield  {author} {\bibinfo {author} {\bibfnamefont {S.~W.}\ \bibnamefont {Hawking}},\ }\href@noop {} {\bibfield  {journal} {\bibinfo  {journal} {Communications in Mathematical Physics}\ }\textbf {\bibinfo {volume} {43}},\ \bibinfo {pages} {199} (\bibinfo {year} {1975})}\BibitemShut {NoStop}%
\bibitem [{\citenamefont {Frodden}\ and\ \citenamefont {Vald\'es}(2018)}]{Frodden:2018mdm}%
  \BibitemOpen
  \bibfield  {author} {\bibinfo {author} {\bibfnamefont {E.}~\bibnamefont {Frodden}}\ and\ \bibinfo {author} {\bibfnamefont {N.}~\bibnamefont {Vald\'es}},\ }\href {\doibase 10.1142/S0217751X18300260} {\bibfield  {journal} {\bibinfo  {journal} {Int. J. Mod. Phys. A}\ }\textbf {\bibinfo {volume} {33}},\ \bibinfo {pages} {1830026} (\bibinfo {year} {2018})},\ \Eprint {http://arxiv.org/abs/1806.11157} {arXiv:1806.11157 [gr-qc]} \BibitemShut {NoStop}%
\bibitem [{\citenamefont {Casimir}(1948)}]{Casimir:1948dh}%
  \BibitemOpen
  \bibfield  {author} {\bibinfo {author} {\bibfnamefont {H.~B.~G.}\ \bibnamefont {Casimir}},\ }\href@noop {} {\bibfield  {journal} {\bibinfo  {journal} {Indag. Math.}\ }\textbf {\bibinfo {volume} {10}},\ \bibinfo {pages} {261} (\bibinfo {year} {1948})}\BibitemShut {NoStop}%
\bibitem [{\citenamefont {Bordag}\ \emph {et~al.}(2009)\citenamefont {Bordag}, \citenamefont {Klimchitskaya}, \citenamefont {Mohideen},\ and\ \citenamefont {Mostepanenko}}]{Bordag:2009zz}%
  \BibitemOpen
  \bibfield  {author} {\bibinfo {author} {\bibfnamefont {M.}~\bibnamefont {Bordag}}, \bibinfo {author} {\bibfnamefont {G.~L.}\ \bibnamefont {Klimchitskaya}}, \bibinfo {author} {\bibfnamefont {U.}~\bibnamefont {Mohideen}}, \ and\ \bibinfo {author} {\bibfnamefont {V.~M.}\ \bibnamefont {Mostepanenko}},\ }\href@noop {} {\emph {\bibinfo {title} {{Advances in the Casimir effect}}}},\ Vol.\ \bibinfo {volume} {145}\ (\bibinfo  {publisher} {Oxford University Press},\ \bibinfo {year} {2009})\BibitemShut {NoStop}%
\bibitem [{\citenamefont {Moore}(2003)}]{10.1063/1.1665432}%
  \BibitemOpen
  \bibfield  {author} {\bibinfo {author} {\bibfnamefont {G.~T.}\ \bibnamefont {Moore}},\ }\href {\doibase 10.1063/1.1665432} {\bibfield  {journal} {\bibinfo  {journal} {Journal of Mathematical Physics}\ }\textbf {\bibinfo {volume} {11}},\ \bibinfo {pages} {2679} (\bibinfo {year} {2003})},\ \Eprint {http://arxiv.org/abs/https://pubs.aip.org/aip/jmp/article-pdf/11/9/2679/8144916/2679\_1\_online.pdf} {https://pubs.aip.org/aip/jmp/article-pdf/11/9/2679/8144916/2679\_1\_online.pdf} \BibitemShut {NoStop}%
\bibitem [{\citenamefont {Dodonov}(2020)}]{Dodonov:2020eto}%
  \BibitemOpen
  \bibfield  {author} {\bibinfo {author} {\bibfnamefont {V.~V.}\ \bibnamefont {Dodonov}},\ }\href {\doibase 10.3390/physics2010007} {\bibfield  {journal} {\bibinfo  {journal} {MDPI Physics}\ }\textbf {\bibinfo {volume} {2}},\ \bibinfo {pages} {67} (\bibinfo {year} {2020})}\BibitemShut {NoStop}%
\bibitem [{\citenamefont {Fulling}\ and\ \citenamefont {Davies}(1976)}]{Fulling1976RadiationFA}%
  \BibitemOpen
  \bibfield  {author} {\bibinfo {author} {\bibfnamefont {S.~A.}\ \bibnamefont {Fulling}}\ and\ \bibinfo {author} {\bibfnamefont {P.~C.~W.}\ \bibnamefont {Davies}},\ }\href@noop {} {\bibfield  {journal} {\bibinfo  {journal} {Proceedings of the Royal Society of London. A. Mathematical and Physical Sciences}\ }\textbf {\bibinfo {volume} {348}},\ \bibinfo {pages} {393 } (\bibinfo {year} {1976})}\BibitemShut {NoStop}%
\bibitem [{\citenamefont {Davies}\ and\ \citenamefont {Fulling}(1977)}]{Davies1977QuantumVE}%
  \BibitemOpen
  \bibfield  {author} {\bibinfo {author} {\bibfnamefont {P.~C.~W.}\ \bibnamefont {Davies}}\ and\ \bibinfo {author} {\bibfnamefont {S.~A.}\ \bibnamefont {Fulling}},\ }\href@noop {} {\bibfield  {journal} {\bibinfo  {journal} {Proceedings of the Royal Society of London. A. Mathematical and Physical Sciences}\ }\textbf {\bibinfo {volume} {354}},\ \bibinfo {pages} {59 } (\bibinfo {year} {1977})}\BibitemShut {NoStop}%
\bibitem [{\citenamefont {Anderson}\ \emph {et~al.}(2017)\citenamefont {Anderson}, \citenamefont {Good},\ and\ \citenamefont {Evans}}]{Anderson:2015iga}%
  \BibitemOpen
  \bibfield  {author} {\bibinfo {author} {\bibfnamefont {P.~R.}\ \bibnamefont {Anderson}}, \bibinfo {author} {\bibfnamefont {M.~R.~R.}\ \bibnamefont {Good}}, \ and\ \bibinfo {author} {\bibfnamefont {C.~R.}\ \bibnamefont {Evans}},\ }in\ \href {\doibase 10.1142/9789813226609_0171} {\emph {\bibinfo {booktitle} {{14th Marcel Grossmann Meeting on Recent Developments in Theoretical and Experimental General Relativity, Astrophysics, and Relativistic Field Theories}}}},\ Vol.~\bibinfo {volume} {2}\ (\bibinfo {year} {2017})\ pp.\ \bibinfo {pages} {1701--1704},\ \Eprint {http://arxiv.org/abs/1507.03489} {arXiv:1507.03489 [gr-qc]} \BibitemShut {NoStop}%
\bibitem [{\citenamefont {Good}\ \emph {et~al.}(2017)\citenamefont {Good}, \citenamefont {Anderson},\ and\ \citenamefont {Evans}}]{Good:2015jwa}%
  \BibitemOpen
  \bibfield  {author} {\bibinfo {author} {\bibfnamefont {M.~R.~R.}\ \bibnamefont {Good}}, \bibinfo {author} {\bibfnamefont {P.~R.}\ \bibnamefont {Anderson}}, \ and\ \bibinfo {author} {\bibfnamefont {C.~R.}\ \bibnamefont {Evans}},\ }in\ \href {\doibase 10.1142/9789813226609_0172} {\emph {\bibinfo {booktitle} {{14th Marcel Grossmann Meeting on Recent Developments in Theoretical and Experimental General Relativity, Astrophysics, and Relativistic Field Theories}}}},\ Vol.~\bibinfo {volume} {2}\ (\bibinfo {year} {2017})\ pp.\ \bibinfo {pages} {1705--1708},\ \Eprint {http://arxiv.org/abs/1507.05048} {arXiv:1507.05048 [gr-qc]} \BibitemShut {NoStop}%
\bibitem [{\citenamefont {Belyanin}\ \emph {et~al.}(2006)\citenamefont {Belyanin}, \citenamefont {Kocharovsky}, \citenamefont {Capasso}, \citenamefont {Fry}, \citenamefont {Zubairy},\ and\ \citenamefont {Scully}}]{PhysRevA.74.023807}%
  \BibitemOpen
  \bibfield  {author} {\bibinfo {author} {\bibfnamefont {A.}~\bibnamefont {Belyanin}}, \bibinfo {author} {\bibfnamefont {V.~V.}\ \bibnamefont {Kocharovsky}}, \bibinfo {author} {\bibfnamefont {F.}~\bibnamefont {Capasso}}, \bibinfo {author} {\bibfnamefont {E.}~\bibnamefont {Fry}}, \bibinfo {author} {\bibfnamefont {M.~S.}\ \bibnamefont {Zubairy}}, \ and\ \bibinfo {author} {\bibfnamefont {M.~O.}\ \bibnamefont {Scully}},\ }\href {\doibase 10.1103/PhysRevA.74.023807} {\bibfield  {journal} {\bibinfo  {journal} {Phys. Rev. A}\ }\textbf {\bibinfo {volume} {74}},\ \bibinfo {pages} {023807} (\bibinfo {year} {2006})}\BibitemShut {NoStop}%
\bibitem [{\citenamefont {Scully}(2019)}]{Scully2019LaserEF}%
  \BibitemOpen
  \bibfield  {author} {\bibinfo {author} {\bibfnamefont {M.~O.}\ \bibnamefont {Scully}},\ }\href@noop {} {\bibfield  {journal} {\bibinfo  {journal} {Physica Scripta}\ }\textbf {\bibinfo {volume} {95}} (\bibinfo {year} {2019})}\BibitemShut {NoStop}%
\bibitem [{\citenamefont {Lock}\ and\ \citenamefont {Fuentes}(2017)}]{Lock:2016rmg}%
  \BibitemOpen
  \bibfield  {author} {\bibinfo {author} {\bibfnamefont {M.~P.~E.}\ \bibnamefont {Lock}}\ and\ \bibinfo {author} {\bibfnamefont {I.}~\bibnamefont {Fuentes}},\ }\href {\doibase 10.1088/1367-2630/aa7651} {\bibfield  {journal} {\bibinfo  {journal} {New J. Phys.}\ }\textbf {\bibinfo {volume} {19}},\ \bibinfo {pages} {073005} (\bibinfo {year} {2017})},\ \Eprint {http://arxiv.org/abs/1607.05444} {arXiv:1607.05444 [quant-ph]} \BibitemShut {NoStop}%
\bibitem [{\citenamefont {Chandrasekhar}(1976)}]{Chandrasekhar1976TheSO}%
  \BibitemOpen
  \bibfield  {author} {\bibinfo {author} {\bibfnamefont {S.}~\bibnamefont {Chandrasekhar}},\ }\href@noop {} {\bibfield  {journal} {\bibinfo  {journal} {Proceedings of the Royal Society of London. A. Mathematical and Physical Sciences}\ }\textbf {\bibinfo {volume} {349}},\ \bibinfo {pages} {571 } (\bibinfo {year} {1976})}\BibitemShut {NoStop}%
\bibitem [{\citenamefont {{Carter}}\ and\ \citenamefont {{McLenaghan}}(1979)}]{1979PhRvD..19.1093C}%
  \BibitemOpen
  \bibfield  {author} {\bibinfo {author} {\bibfnamefont {B.}~\bibnamefont {{Carter}}}\ and\ \bibinfo {author} {\bibfnamefont {R.~G.}\ \bibnamefont {{McLenaghan}}},\ }\href {\doibase 10.1103/PhysRevD.19.1093} {\bibfield  {journal} {\bibinfo  {journal} {\prd}\ }\textbf {\bibinfo {volume} {19}},\ \bibinfo {pages} {1093} (\bibinfo {year} {1979})}\BibitemShut {NoStop}%
\bibitem [{\citenamefont {Shishkin}(1991)}]{Shishkin:1991ma}%
  \BibitemOpen
  \bibfield  {author} {\bibinfo {author} {\bibfnamefont {G.~V.}\ \bibnamefont {Shishkin}},\ }\href {\doibase 10.1088/0264-9381/8/1/017} {\bibfield  {journal} {\bibinfo  {journal} {Class. Quant. Grav.}\ }\textbf {\bibinfo {volume} {8}},\ \bibinfo {pages} {175} (\bibinfo {year} {1991})}\BibitemShut {NoStop}%
\bibitem [{\citenamefont {Finster}\ and\ \citenamefont {Reintjes}(2009)}]{Finster:2009ah}%
  \BibitemOpen
  \bibfield  {author} {\bibinfo {author} {\bibfnamefont {F.}~\bibnamefont {Finster}}\ and\ \bibinfo {author} {\bibfnamefont {M.}~\bibnamefont {Reintjes}},\ }\href {\doibase 10.1088/0264-9381/26/10/105021} {\bibfield  {journal} {\bibinfo  {journal} {Class. Quant. Grav.}\ }\textbf {\bibinfo {volume} {26}},\ \bibinfo {pages} {105021} (\bibinfo {year} {2009})},\ \Eprint {http://arxiv.org/abs/0901.0602} {arXiv:0901.0602 [math-ph]} \BibitemShut {NoStop}%
\bibitem [{\citenamefont {Collas}\ and\ \citenamefont {Klein}(2018)}]{Collas:2018wcc}%
  \BibitemOpen
  \bibfield  {author} {\bibinfo {author} {\bibfnamefont {P.}~\bibnamefont {Collas}}\ and\ \bibinfo {author} {\bibfnamefont {D.}~\bibnamefont {Klein}},\ }\href {\doibase 10.1088/1361-6382/aac144} {\bibfield  {journal} {\bibinfo  {journal} {Class. Quant. Grav.}\ }\textbf {\bibinfo {volume} {35}},\ \bibinfo {pages} {125006} (\bibinfo {year} {2018})},\ \Eprint {http://arxiv.org/abs/1801.02756} {arXiv:1801.02756 [gr-qc]} \BibitemShut {NoStop}%
\bibitem [{\citenamefont {Parker}(1980{\natexlab{a}})}]{PhysRevLett.44.1559}%
  \BibitemOpen
  \bibfield  {author} {\bibinfo {author} {\bibfnamefont {L.}~\bibnamefont {Parker}},\ }\href {\doibase 10.1103/PhysRevLett.44.1559} {\bibfield  {journal} {\bibinfo  {journal} {Phys. Rev. Lett.}\ }\textbf {\bibinfo {volume} {44}},\ \bibinfo {pages} {1559} (\bibinfo {year} {1980}{\natexlab{a}})}\BibitemShut {NoStop}%
\bibitem [{\citenamefont {Parker}(1981{\natexlab{a}})}]{Parker1981TheAA}%
  \BibitemOpen
  \bibfield  {author} {\bibinfo {author} {\bibfnamefont {L.}~\bibnamefont {Parker}},\ }\href@noop {} {\bibfield  {journal} {\bibinfo  {journal} {General Relativity and Gravitation}\ }\textbf {\bibinfo {volume} {13}},\ \bibinfo {pages} {307} (\bibinfo {year} {1981}{\natexlab{a}})}\BibitemShut {NoStop}%
\bibitem [{\citenamefont {Parker}(1980{\natexlab{b}})}]{PhysRevD.22.1922}%
  \BibitemOpen
  \bibfield  {author} {\bibinfo {author} {\bibfnamefont {L.}~\bibnamefont {Parker}},\ }\href {\doibase 10.1103/PhysRevD.22.1922} {\bibfield  {journal} {\bibinfo  {journal} {Phys. Rev. D}\ }\textbf {\bibinfo {volume} {22}},\ \bibinfo {pages} {1922} (\bibinfo {year} {1980}{\natexlab{b}})}\BibitemShut {NoStop}%
\bibitem [{\citenamefont {Parker}(1981{\natexlab{b}})}]{Parker:1981wt}%
  \BibitemOpen
  \bibfield  {author} {\bibinfo {author} {\bibfnamefont {L.}~\bibnamefont {Parker}},\ }\href {\doibase 10.1103/PhysRevD.24.535} {\bibfield  {journal} {\bibinfo  {journal} {Phys. Rev. D}\ }\textbf {\bibinfo {volume} {24}},\ \bibinfo {pages} {535} (\bibinfo {year} {1981}{\natexlab{b}})}\BibitemShut {NoStop}%
\bibitem [{\citenamefont {{Pinto}}(1993)}]{1993PhRvL..70.3839P}%
  \BibitemOpen
  \bibfield  {author} {\bibinfo {author} {\bibfnamefont {F.}~\bibnamefont {{Pinto}}},\ }\href {\doibase 10.1103/PhysRevLett.70.3839} {\bibfield  {journal} {\bibinfo  {journal} {\prl}\ }\textbf {\bibinfo {volume} {70}},\ \bibinfo {pages} {3839} (\bibinfo {year} {1993})}\BibitemShut {NoStop}%
\bibitem [{\citenamefont {Parker}\ \emph {et~al.}(1997)\citenamefont {Parker}, \citenamefont {Vollick},\ and\ \citenamefont {Redmount}}]{Parker:1996hc}%
  \BibitemOpen
  \bibfield  {author} {\bibinfo {author} {\bibfnamefont {L.}~\bibnamefont {Parker}}, \bibinfo {author} {\bibfnamefont {D.}~\bibnamefont {Vollick}}, \ and\ \bibinfo {author} {\bibfnamefont {I.}~\bibnamefont {Redmount}},\ }\href {\doibase 10.1103/PhysRevD.56.2113} {\bibfield  {journal} {\bibinfo  {journal} {Phys. Rev. D}\ }\textbf {\bibinfo {volume} {56}},\ \bibinfo {pages} {2113} (\bibinfo {year} {1997})}\BibitemShut {NoStop}%
\bibitem [{\citenamefont {de~A.~Marques}\ and\ \citenamefont {Bezerra}(2002)}]{deAMarques:2002hbv}%
  \BibitemOpen
  \bibfield  {author} {\bibinfo {author} {\bibfnamefont {G.}~\bibnamefont {de~A.~Marques}}\ and\ \bibinfo {author} {\bibfnamefont {V.~B.}\ \bibnamefont {Bezerra}},\ }\href {\doibase 10.1103/PhysRevD.66.105011} {\bibfield  {journal} {\bibinfo  {journal} {Phys. Rev. D}\ }\textbf {\bibinfo {volume} {66}},\ \bibinfo {pages} {105011} (\bibinfo {year} {2002})},\ \Eprint {http://arxiv.org/abs/gr-qc/0209030} {arXiv:gr-qc/0209030} \BibitemShut {NoStop}%
\bibitem [{\citenamefont {Zhao}\ \emph {et~al.}(2007)\citenamefont {Zhao}, \citenamefont {Liu},\ and\ \citenamefont {Li}}]{Zhao:2007xj}%
  \BibitemOpen
  \bibfield  {author} {\bibinfo {author} {\bibfnamefont {Z.-H.}\ \bibnamefont {Zhao}}, \bibinfo {author} {\bibfnamefont {Y.-X.}\ \bibnamefont {Liu}}, \ and\ \bibinfo {author} {\bibfnamefont {X.-G.}\ \bibnamefont {Li}},\ }\href {\doibase 10.1103/PhysRevD.76.064016} {\bibfield  {journal} {\bibinfo  {journal} {Phys. Rev. D}\ }\textbf {\bibinfo {volume} {76}},\ \bibinfo {pages} {064016} (\bibinfo {year} {2007})},\ \Eprint {http://arxiv.org/abs/0705.1571} {arXiv:0705.1571 [gr-qc]} \BibitemShut {NoStop}%
\bibitem [{\citenamefont {Carvalho}\ \emph {et~al.}(2011)\citenamefont {Carvalho}, \citenamefont {Furtado},\ and\ \citenamefont {Moraes}}]{Carvalho:2011krd}%
  \BibitemOpen
  \bibfield  {author} {\bibinfo {author} {\bibfnamefont {J.}~\bibnamefont {Carvalho}}, \bibinfo {author} {\bibfnamefont {C.}~\bibnamefont {Furtado}}, \ and\ \bibinfo {author} {\bibfnamefont {F.}~\bibnamefont {Moraes}},\ }\href {\doibase 10.1103/PhysRevA.84.032109} {\bibfield  {journal} {\bibinfo  {journal} {Phys. Rev. A}\ }\textbf {\bibinfo {volume} {84}},\ \bibinfo {pages} {032109} (\bibinfo {year} {2011})}\BibitemShut {NoStop}%
\bibitem [{\citenamefont {Roura}(2021)}]{Roura:2021fvd}%
  \BibitemOpen
  \bibfield  {author} {\bibinfo {author} {\bibfnamefont {A.}~\bibnamefont {Roura}},\ }\href {\doibase 10.1126/science.abm6854} {\bibfield  {journal} {\bibinfo  {journal} {Science}\ }\textbf {\bibinfo {volume} {375}},\ \bibinfo {pages} {142} (\bibinfo {year} {2021})}\BibitemShut {NoStop}%
\bibitem [{\citenamefont {Caianiello}(1981)}]{Caianiello:1981jq}%
  \BibitemOpen
  \bibfield  {author} {\bibinfo {author} {\bibfnamefont {E.~R.}\ \bibnamefont {Caianiello}},\ }\href {\doibase 10.1007/BF02745135} {\bibfield  {journal} {\bibinfo  {journal} {Lett. Nuovo Cim.}\ }\textbf {\bibinfo {volume} {32}},\ \bibinfo {pages} {65} (\bibinfo {year} {1981})}\BibitemShut {NoStop}%
\bibitem [{\citenamefont {Lambiase}\ \emph {et~al.}(1998)\citenamefont {Lambiase}, \citenamefont {Papini},\ and\ \citenamefont {Scarpetta}}]{Lambiase:1998tn}%
  \BibitemOpen
  \bibfield  {author} {\bibinfo {author} {\bibfnamefont {G.}~\bibnamefont {Lambiase}}, \bibinfo {author} {\bibfnamefont {G.}~\bibnamefont {Papini}}, \ and\ \bibinfo {author} {\bibfnamefont {G.}~\bibnamefont {Scarpetta}},\ }\href {\doibase 10.1016/S0375-9601(98)00364-8} {\bibfield  {journal} {\bibinfo  {journal} {Phys. Lett. A}\ }\textbf {\bibinfo {volume} {244}},\ \bibinfo {pages} {349} (\bibinfo {year} {1998})},\ \Eprint {http://arxiv.org/abs/hep-ph/9804438} {arXiv:hep-ph/9804438} \BibitemShut {NoStop}%
\bibitem [{\citenamefont {Benedetto}\ and\ \citenamefont {Feoli}(2015)}]{Benedetto:2015fta}%
  \BibitemOpen
  \bibfield  {author} {\bibinfo {author} {\bibfnamefont {E.}~\bibnamefont {Benedetto}}\ and\ \bibinfo {author} {\bibfnamefont {A.}~\bibnamefont {Feoli}},\ }\href {\doibase 10.1142/S0217732315500753} {\bibfield  {journal} {\bibinfo  {journal} {Mod. Phys. Lett. A}\ }\textbf {\bibinfo {volume} {30}},\ \bibinfo {pages} {1550075} (\bibinfo {year} {2015})}\BibitemShut {NoStop}%
\bibitem [{\citenamefont {Higuchi}\ \emph {et~al.}(1997)\citenamefont {Higuchi}, \citenamefont {Matsas},\ and\ \citenamefont {Sudarsky}}]{PhysRevD.56.R6071}%
  \BibitemOpen
  \bibfield  {author} {\bibinfo {author} {\bibfnamefont {A.}~\bibnamefont {Higuchi}}, \bibinfo {author} {\bibfnamefont {G.~E.~A.}\ \bibnamefont {Matsas}}, \ and\ \bibinfo {author} {\bibfnamefont {D.}~\bibnamefont {Sudarsky}},\ }\href {\doibase 10.1103/PhysRevD.56.R6071} {\bibfield  {journal} {\bibinfo  {journal} {Phys. Rev. D}\ }\textbf {\bibinfo {volume} {56}},\ \bibinfo {pages} {R6071} (\bibinfo {year} {1997})}\BibitemShut {NoStop}%
\bibitem [{\citenamefont {Crispino}\ \emph {et~al.}(2009)\citenamefont {Crispino}, \citenamefont {Dolan},\ and\ \citenamefont {Oliveira}}]{PhysRevLett.102.231103}%
  \BibitemOpen
  \bibfield  {author} {\bibinfo {author} {\bibfnamefont {L.~C.~B.}\ \bibnamefont {Crispino}}, \bibinfo {author} {\bibfnamefont {S.~R.}\ \bibnamefont {Dolan}}, \ and\ \bibinfo {author} {\bibfnamefont {E.~S.}\ \bibnamefont {Oliveira}},\ }\href {\doibase 10.1103/PhysRevLett.102.231103} {\bibfield  {journal} {\bibinfo  {journal} {Phys. Rev. Lett.}\ }\textbf {\bibinfo {volume} {102}},\ \bibinfo {pages} {231103} (\bibinfo {year} {2009})}\BibitemShut {NoStop}%
\bibitem [{\citenamefont {Macedo}\ \emph {et~al.}(2013)\citenamefont {Macedo}, \citenamefont {Leite}, \citenamefont {Oliveira}, \citenamefont {Dolan},\ and\ \citenamefont {Crispino}}]{Macedo:2013afa}%
  \BibitemOpen
  \bibfield  {author} {\bibinfo {author} {\bibfnamefont {C.~F.~B.}\ \bibnamefont {Macedo}}, \bibinfo {author} {\bibfnamefont {L.~C.~S.}\ \bibnamefont {Leite}}, \bibinfo {author} {\bibfnamefont {E.~S.}\ \bibnamefont {Oliveira}}, \bibinfo {author} {\bibfnamefont {S.~R.}\ \bibnamefont {Dolan}}, \ and\ \bibinfo {author} {\bibfnamefont {L.~C.~B.}\ \bibnamefont {Crispino}},\ }\href {\doibase 10.1103/PhysRevD.88.064033} {\bibfield  {journal} {\bibinfo  {journal} {Phys. Rev. D}\ }\textbf {\bibinfo {volume} {88}},\ \bibinfo {pages} {064033} (\bibinfo {year} {2013})},\ \Eprint {http://arxiv.org/abs/1308.0018} {arXiv:1308.0018 [gr-qc]} \BibitemShut {NoStop}%
\bibitem [{\citenamefont {Cardoso}\ and\ \citenamefont {Vicente}(2019)}]{Cardoso:2019dte}%
  \BibitemOpen
  \bibfield  {author} {\bibinfo {author} {\bibfnamefont {V.}~\bibnamefont {Cardoso}}\ and\ \bibinfo {author} {\bibfnamefont {R.}~\bibnamefont {Vicente}},\ }\href {\doibase 10.1103/PhysRevD.100.084001} {\bibfield  {journal} {\bibinfo  {journal} {Phys. Rev. D}\ }\textbf {\bibinfo {volume} {100}},\ \bibinfo {pages} {084001} (\bibinfo {year} {2019})},\ \Eprint {http://arxiv.org/abs/1906.10140} {arXiv:1906.10140 [gr-qc]} \BibitemShut {NoStop}%
\bibitem [{\citenamefont {Brito}\ \emph {et~al.}(2015)\citenamefont {Brito}, \citenamefont {Cardoso},\ and\ \citenamefont {Pani}}]{Brito:2015oca}%
  \BibitemOpen
  \bibfield  {author} {\bibinfo {author} {\bibfnamefont {R.}~\bibnamefont {Brito}}, \bibinfo {author} {\bibfnamefont {V.}~\bibnamefont {Cardoso}}, \ and\ \bibinfo {author} {\bibfnamefont {P.}~\bibnamefont {Pani}},\ }\href {\doibase 10.1007/978-3-319-19000-6} {\bibfield  {journal} {\bibinfo  {journal} {Lect. Notes Phys.}\ }\textbf {\bibinfo {volume} {906}},\ \bibinfo {pages} {pp.1} (\bibinfo {year} {2015})},\ \Eprint {http://arxiv.org/abs/1501.06570} {arXiv:1501.06570 [gr-qc]} \BibitemShut {NoStop}%
\bibitem [{\citenamefont {Bambi}(2017)}]{Bambi:2015kza}%
  \BibitemOpen
  \bibfield  {author} {\bibinfo {author} {\bibfnamefont {C.}~\bibnamefont {Bambi}},\ }\href {\doibase 10.1103/RevModPhys.89.025001} {\bibfield  {journal} {\bibinfo  {journal} {Rev. Mod. Phys.}\ }\textbf {\bibinfo {volume} {89}},\ \bibinfo {pages} {025001} (\bibinfo {year} {2017})},\ \Eprint {http://arxiv.org/abs/1509.03884} {arXiv:1509.03884 [gr-qc]} \BibitemShut {NoStop}%
\bibitem [{\citenamefont {Passante}(2018)}]{Passante:2018qzj}%
  \BibitemOpen
  \bibfield  {author} {\bibinfo {author} {\bibfnamefont {R.}~\bibnamefont {Passante}},\ }\href {\doibase 10.3390/sym10120735} {\bibfield  {journal} {\bibinfo  {journal} {Symmetry}\ }\textbf {\bibinfo {volume} {10}},\ \bibinfo {pages} {735} (\bibinfo {year} {2018})},\ \Eprint {http://arxiv.org/abs/1812.05078} {arXiv:1812.05078 [quant-ph]} \BibitemShut {NoStop}%
\bibitem [{\citenamefont {Hobson}\ \emph {et~al.}(2006)\citenamefont {Hobson}, \citenamefont {Efstathiou},\ and\ \citenamefont {Lasenby}}]{Hobson:2006se}%
  \BibitemOpen
  \bibfield  {author} {\bibinfo {author} {\bibfnamefont {M.~P.}\ \bibnamefont {Hobson}}, \bibinfo {author} {\bibfnamefont {G.~P.}\ \bibnamefont {Efstathiou}}, \ and\ \bibinfo {author} {\bibfnamefont {A.~N.}\ \bibnamefont {Lasenby}},\ }\href@noop {} {\emph {\bibinfo {title} {{General relativity: An introduction for physicists}}}}\ (\bibinfo {year} {2006})\BibitemShut {NoStop}%
\bibitem [{\citenamefont {Misner}\ \emph {et~al.}(1973)\citenamefont {Misner}, \citenamefont {Thorne},\ and\ \citenamefont {Wheeler}}]{Misner:1973prb}%
  \BibitemOpen
  \bibfield  {author} {\bibinfo {author} {\bibfnamefont {C.~W.}\ \bibnamefont {Misner}}, \bibinfo {author} {\bibfnamefont {K.~S.}\ \bibnamefont {Thorne}}, \ and\ \bibinfo {author} {\bibfnamefont {J.~A.}\ \bibnamefont {Wheeler}},\ }\href@noop {} {\emph {\bibinfo {title} {{Gravitation}}}}\ (\bibinfo  {publisher} {W. H. Freeman},\ \bibinfo {address} {San Francisco},\ \bibinfo {year} {1973})\BibitemShut {NoStop}%
\bibitem [{\citenamefont {{Weinberg}}(1972)}]{1972gcpa.book.....W}%
  \BibitemOpen
  \bibfield  {author} {\bibinfo {author} {\bibfnamefont {S.}~\bibnamefont {{Weinberg}}},\ }\href@noop {} {\emph {\bibinfo {title} {{Gravitation and Cosmology: Principles and Applications of the General Theory of Relativity}}}}\ (\bibinfo {year} {1972})\BibitemShut {NoStop}%
\bibitem [{\citenamefont {Socolovsky}(2013)}]{Socolovsky:2013rga}%
  \BibitemOpen
  \bibfield  {author} {\bibinfo {author} {\bibfnamefont {M.}~\bibnamefont {Socolovsky}},\ }\href@noop {} {\  (\bibinfo {year} {2013})},\ \Eprint {http://arxiv.org/abs/1304.2833} {arXiv:1304.2833 [gr-qc]} \BibitemShut {NoStop}%
\bibitem [{\citenamefont {{Rindler}}(1966)}]{1966AmJPh..34.1174R}%
  \BibitemOpen
  \bibfield  {author} {\bibinfo {author} {\bibfnamefont {W.}~\bibnamefont {{Rindler}}},\ }\href {\doibase 10.1119/1.1972547} {\bibfield  {journal} {\bibinfo  {journal} {American Journal of Physics}\ }\textbf {\bibinfo {volume} {34}},\ \bibinfo {pages} {1174} (\bibinfo {year} {1966})}\BibitemShut {NoStop}%
\bibitem [{\citenamefont {Martin-Martinez}\ and\ \citenamefont {Menicucci}(2014)}]{Martin-Martinez:2014gra}%
  \BibitemOpen
  \bibfield  {author} {\bibinfo {author} {\bibfnamefont {E.}~\bibnamefont {Martin-Martinez}}\ and\ \bibinfo {author} {\bibfnamefont {N.~C.}\ \bibnamefont {Menicucci}},\ }\href {\doibase 10.1088/0264-9381/31/21/214001} {\bibfield  {journal} {\bibinfo  {journal} {Class. Quant. Grav.}\ }\textbf {\bibinfo {volume} {31}},\ \bibinfo {pages} {214001} (\bibinfo {year} {2014})},\ \Eprint {http://arxiv.org/abs/1408.3420} {arXiv:1408.3420 [quant-ph]} \BibitemShut {NoStop}%
\bibitem [{\citenamefont {Birrell}\ and\ \citenamefont {Davies}(1984)}]{Birrell:1982ix}%
  \BibitemOpen
  \bibfield  {author} {\bibinfo {author} {\bibfnamefont {N.~D.}\ \bibnamefont {Birrell}}\ and\ \bibinfo {author} {\bibfnamefont {P.~C.~W.}\ \bibnamefont {Davies}},\ }\href {\doibase 10.1017/CBO9780511622632} {\emph {\bibinfo {title} {{Quantum Fields in Curved Space}}}},\ Cambridge Monographs on Mathematical Physics\ (\bibinfo  {publisher} {Cambridge Univ. Press},\ \bibinfo {address} {Cambridge, UK},\ \bibinfo {year} {1984})\BibitemShut {NoStop}%
\bibitem [{\citenamefont {Jacobson}(2005)}]{Jacobson2005}%
  \BibitemOpen
  \bibfield  {author} {\bibinfo {author} {\bibfnamefont {T.}~\bibnamefont {Jacobson}},\ }\enquote {\bibinfo {title} {Introduction to quantum fields in curved spacetime and the hawking effect},}\ in\ \href {\doibase 10.1007/0-387-24992-3_2} {\emph {\bibinfo {booktitle} {Lectures on Quantum Gravity}}},\ \bibinfo {editor} {edited by\ \bibinfo {editor} {\bibfnamefont {A.}~\bibnamefont {Gomberoff}}\ and\ \bibinfo {editor} {\bibfnamefont {D.}~\bibnamefont {Marolf}}}\ (\bibinfo  {publisher} {Springer US},\ \bibinfo {address} {Boston, MA},\ \bibinfo {year} {2005})\ pp.\ \bibinfo {pages} {39--89}\BibitemShut {NoStop}%
\bibitem [{\citenamefont {Parker}\ and\ \citenamefont {Toms}(2009)}]{Parker:2009uva}%
  \BibitemOpen
  \bibfield  {author} {\bibinfo {author} {\bibfnamefont {L.~E.}\ \bibnamefont {Parker}}\ and\ \bibinfo {author} {\bibfnamefont {D.}~\bibnamefont {Toms}},\ }\href {\doibase 10.1017/CBO9780511813924} {\emph {\bibinfo {title} {{Quantum Field Theory in Curved Spacetime}: {Quantized Field and Gravity}}}},\ Cambridge Monographs on Mathematical Physics\ (\bibinfo  {publisher} {Cambridge University Press},\ \bibinfo {year} {2009})\BibitemShut {NoStop}%
\bibitem [{\citenamefont {Carroll}(2019)}]{Carroll:2004st}%
  \BibitemOpen
  \bibfield  {author} {\bibinfo {author} {\bibfnamefont {S.~M.}\ \bibnamefont {Carroll}},\ }\href@noop {} {\emph {\bibinfo {title} {{Spacetime and Geometry}}}}\ (\bibinfo  {publisher} {Cambridge University Press},\ \bibinfo {year} {2019})\BibitemShut {NoStop}%
\bibitem [{\citenamefont {Almheiri}\ \emph {et~al.}(2013)\citenamefont {Almheiri}, \citenamefont {Marolf}, \citenamefont {Polchinski},\ and\ \citenamefont {Sully}}]{Almheiri:2012rt}%
  \BibitemOpen
  \bibfield  {author} {\bibinfo {author} {\bibfnamefont {A.}~\bibnamefont {Almheiri}}, \bibinfo {author} {\bibfnamefont {D.}~\bibnamefont {Marolf}}, \bibinfo {author} {\bibfnamefont {J.}~\bibnamefont {Polchinski}}, \ and\ \bibinfo {author} {\bibfnamefont {J.}~\bibnamefont {Sully}},\ }\href {\doibase 10.1007/JHEP02(2013)062} {\bibfield  {journal} {\bibinfo  {journal} {JHEP}\ }\textbf {\bibinfo {volume} {02}},\ \bibinfo {pages} {062} (\bibinfo {year} {2013})},\ \Eprint {http://arxiv.org/abs/1207.3123} {arXiv:1207.3123 [hep-th]} \BibitemShut {NoStop}%
\bibitem [{\citenamefont {Mathur}(2009)}]{Mathur:2009hf}%
  \BibitemOpen
  \bibfield  {author} {\bibinfo {author} {\bibfnamefont {S.~D.}\ \bibnamefont {Mathur}},\ }\href {\doibase 10.1088/0264-9381/26/22/224001} {\bibfield  {journal} {\bibinfo  {journal} {Class. Quant. Grav.}\ }\textbf {\bibinfo {volume} {26}},\ \bibinfo {pages} {224001} (\bibinfo {year} {2009})},\ \Eprint {http://arxiv.org/abs/0909.1038} {arXiv:0909.1038 [hep-th]} \BibitemShut {NoStop}%
\bibitem [{\citenamefont {Wilson}\ \emph {et~al.}(2011)\citenamefont {Wilson}, \citenamefont {Johansson}, \citenamefont {Pourkabirian}, \citenamefont {Simoen}, \citenamefont {Johansson}, \citenamefont {Duty}, \citenamefont {Nori},\ and\ \citenamefont {Delsing}}]{Wilson_2011}%
  \BibitemOpen
  \bibfield  {author} {\bibinfo {author} {\bibfnamefont {C.~M.}\ \bibnamefont {Wilson}}, \bibinfo {author} {\bibfnamefont {G.}~\bibnamefont {Johansson}}, \bibinfo {author} {\bibfnamefont {A.}~\bibnamefont {Pourkabirian}}, \bibinfo {author} {\bibfnamefont {M.}~\bibnamefont {Simoen}}, \bibinfo {author} {\bibfnamefont {J.~R.}\ \bibnamefont {Johansson}}, \bibinfo {author} {\bibfnamefont {T.}~\bibnamefont {Duty}}, \bibinfo {author} {\bibfnamefont {F.}~\bibnamefont {Nori}}, \ and\ \bibinfo {author} {\bibfnamefont {P.}~\bibnamefont {Delsing}},\ }\href {\doibase 10.1038/nature10561} {\bibfield  {journal} {\bibinfo  {journal} {Nature}\ }\textbf {\bibinfo {volume} {479}},\ \bibinfo {pages} {376} (\bibinfo {year} {2011})}\BibitemShut {NoStop}%
\bibitem [{\citenamefont {L\"ahteenm\"aki}\ \emph {et~al.}(2013)\citenamefont {L\"ahteenm\"aki}, \citenamefont {Paraoanu}, \citenamefont {Hassel},\ and\ \citenamefont {Hakonen}}]{Lahteenmaki:2011cwo}%
  \BibitemOpen
  \bibfield  {author} {\bibinfo {author} {\bibfnamefont {P.}~\bibnamefont {L\"ahteenm\"aki}}, \bibinfo {author} {\bibfnamefont {G.~S.}\ \bibnamefont {Paraoanu}}, \bibinfo {author} {\bibfnamefont {J.}~\bibnamefont {Hassel}}, \ and\ \bibinfo {author} {\bibfnamefont {P.~J.}\ \bibnamefont {Hakonen}},\ }\href {\doibase 10.1073/pnas.1212705110} {\bibfield  {journal} {\bibinfo  {journal} {Proc. Nat. Acad. Sci.}\ }\textbf {\bibinfo {volume} {110}},\ \bibinfo {pages} {4234} (\bibinfo {year} {2013})},\ \Eprint {http://arxiv.org/abs/1111.5608} {arXiv:1111.5608 [cond-mat.mes-hall]} \BibitemShut {NoStop}%
\bibitem [{\citenamefont {Jaskula}\ \emph {et~al.}(2012)\citenamefont {Jaskula}, \citenamefont {Partridge}, \citenamefont {Bonneau}, \citenamefont {Lopes}, \citenamefont {Ruaudel}, \citenamefont {Boiron},\ and\ \citenamefont {Westbrook}}]{Jaskula:2012ab}%
  \BibitemOpen
  \bibfield  {author} {\bibinfo {author} {\bibfnamefont {J.~C.}\ \bibnamefont {Jaskula}}, \bibinfo {author} {\bibfnamefont {G.~B.}\ \bibnamefont {Partridge}}, \bibinfo {author} {\bibfnamefont {M.}~\bibnamefont {Bonneau}}, \bibinfo {author} {\bibfnamefont {R.}~\bibnamefont {Lopes}}, \bibinfo {author} {\bibfnamefont {J.}~\bibnamefont {Ruaudel}}, \bibinfo {author} {\bibfnamefont {D.}~\bibnamefont {Boiron}}, \ and\ \bibinfo {author} {\bibfnamefont {C.~I.}\ \bibnamefont {Westbrook}},\ }\href {\doibase 10.1103/PhysRevLett.109.220401} {\bibfield  {journal} {\bibinfo  {journal} {Phys. Rev. Lett.}\ }\textbf {\bibinfo {volume} {109}},\ \bibinfo {pages} {220401} (\bibinfo {year} {2012})},\ \Eprint {http://arxiv.org/abs/1207.1338} {arXiv:1207.1338 [cond-mat.quant-gas]} \BibitemShut {NoStop}%
\bibitem [{\citenamefont {Jaekel}\ and\ \citenamefont {Reynaud}(1997)}]{Jaekel:1997hr}%
  \BibitemOpen
  \bibfield  {author} {\bibinfo {author} {\bibfnamefont {M.-T.}\ \bibnamefont {Jaekel}}\ and\ \bibinfo {author} {\bibfnamefont {S.}~\bibnamefont {Reynaud}},\ }\href {\doibase 10.1088/0034-4885/60/9/001} {\bibfield  {journal} {\bibinfo  {journal} {Rept. Prog. Phys.}\ }\textbf {\bibinfo {volume} {60}},\ \bibinfo {pages} {863} (\bibinfo {year} {1997})},\ \Eprint {http://arxiv.org/abs/quant-ph/9706035} {arXiv:quant-ph/9706035} \BibitemShut {NoStop}%
\bibitem [{\citenamefont {Dodonov}(2009)}]{Dodonov:2009zza}%
  \BibitemOpen
  \bibfield  {author} {\bibinfo {author} {\bibfnamefont {V.~V.}\ \bibnamefont {Dodonov}},\ }\href {\doibase 10.1088/1742-6596/161/1/012027} {\bibfield  {journal} {\bibinfo  {journal} {J. Phys. Conf. Ser.}\ }\textbf {\bibinfo {volume} {161}},\ \bibinfo {pages} {012027} (\bibinfo {year} {2009})}\BibitemShut {NoStop}%
\bibitem [{\citenamefont {Nicolai}(1925)}]{doi:10.1080/14786442508634593}%
  \BibitemOpen
  \bibfield  {author} {\bibinfo {author} {\bibfnamefont {E.}~\bibnamefont {Nicolai}},\ }\href {\doibase 10.1080/14786442508634593} {\bibfield  {journal} {\bibinfo  {journal} {The London, Edinburgh, and Dublin Philosophical Magazine and Journal of Science}\ }\textbf {\bibinfo {volume} {49}},\ \bibinfo {pages} {171} (\bibinfo {year} {1925})},\ \Eprint {http://arxiv.org/abs/https://doi.org/10.1080/14786442508634593} {https://doi.org/10.1080/14786442508634593} \BibitemShut {NoStop}%
\bibitem [{\citenamefont {Lamb}\ and\ \citenamefont {Retherford}(1947)}]{Lamb:1947zz}%
  \BibitemOpen
  \bibfield  {author} {\bibinfo {author} {\bibfnamefont {W.~E.}\ \bibnamefont {Lamb}}\ and\ \bibinfo {author} {\bibfnamefont {R.~C.}\ \bibnamefont {Retherford}},\ }\href {\doibase 10.1103/PhysRev.72.241} {\bibfield  {journal} {\bibinfo  {journal} {Phys. Rev.}\ }\textbf {\bibinfo {volume} {72}},\ \bibinfo {pages} {241} (\bibinfo {year} {1947})}\BibitemShut {NoStop}%
\bibitem [{\citenamefont {Welton}(1948)}]{Welton:1948zz}%
  \BibitemOpen
  \bibfield  {author} {\bibinfo {author} {\bibfnamefont {T.~A.}\ \bibnamefont {Welton}},\ }\href {\doibase 10.1103/PhysRev.74.1157} {\bibfield  {journal} {\bibinfo  {journal} {Phys. Rev.}\ }\textbf {\bibinfo {volume} {74}},\ \bibinfo {pages} {1157} (\bibinfo {year} {1948})}\BibitemShut {NoStop}%
\bibitem [{\citenamefont {Ackerhalt}\ \emph {et~al.}(1973)\citenamefont {Ackerhalt}, \citenamefont {Knight},\ and\ \citenamefont {Eberly}}]{Ackerhalt:1973fk}%
  \BibitemOpen
  \bibfield  {author} {\bibinfo {author} {\bibfnamefont {J.~R.}\ \bibnamefont {Ackerhalt}}, \bibinfo {author} {\bibfnamefont {P.~L.}\ \bibnamefont {Knight}}, \ and\ \bibinfo {author} {\bibfnamefont {J.~H.}\ \bibnamefont {Eberly}},\ }\href {\doibase 10.1103/PhysRevLett.30.456} {\bibfield  {journal} {\bibinfo  {journal} {Phys. Rev. Lett.}\ }\textbf {\bibinfo {volume} {30}},\ \bibinfo {pages} {456} (\bibinfo {year} {1973})}\BibitemShut {NoStop}%
\bibitem [{\citenamefont {Milonni}\ \emph {et~al.}(1973)\citenamefont {Milonni}, \citenamefont {Ackerhalt},\ and\ \citenamefont {Smith}}]{Milonni:1973zz}%
  \BibitemOpen
  \bibfield  {author} {\bibinfo {author} {\bibfnamefont {P.~W.}\ \bibnamefont {Milonni}}, \bibinfo {author} {\bibfnamefont {J.~R.}\ \bibnamefont {Ackerhalt}}, \ and\ \bibinfo {author} {\bibfnamefont {W.~A.}\ \bibnamefont {Smith}},\ }\href {\doibase 10.1103/PhysRevLett.31.958} {\bibfield  {journal} {\bibinfo  {journal} {Phys. Rev. Lett.}\ }\textbf {\bibinfo {volume} {31}},\ \bibinfo {pages} {958} (\bibinfo {year} {1973})}\BibitemShut {NoStop}%
\bibitem [{\citenamefont {Audretsch}\ and\ \citenamefont {M\"uller}(1994)}]{PhysRevA.50.1755}%
  \BibitemOpen
  \bibfield  {author} {\bibinfo {author} {\bibfnamefont {J.}~\bibnamefont {Audretsch}}\ and\ \bibinfo {author} {\bibfnamefont {R.}~\bibnamefont {M\"uller}},\ }\href {\doibase 10.1103/PhysRevA.50.1755} {\bibfield  {journal} {\bibinfo  {journal} {Phys. Rev. A}\ }\textbf {\bibinfo {volume} {50}},\ \bibinfo {pages} {1755} (\bibinfo {year} {1994})}\BibitemShut {NoStop}%
\bibitem [{\citenamefont {{Dalibard, J.}}\ \emph {et~al.}(1982)\citenamefont {{Dalibard, J.}}, \citenamefont {{Dupont-Roc, J.}},\ and\ \citenamefont {{Cohen-Tannoudji, C.}}}]{refId0}%
  \BibitemOpen
  \bibfield  {author} {\bibinfo {author} {\bibnamefont {{Dalibard, J.}}}, \bibinfo {author} {\bibnamefont {{Dupont-Roc, J.}}}, \ and\ \bibinfo {author} {\bibnamefont {{Cohen-Tannoudji, C.}}},\ }\href {\doibase 10.1051/jphys:0198200430110161700} {\bibfield  {journal} {\bibinfo  {journal} {J. Phys. France}\ }\textbf {\bibinfo {volume} {43}},\ \bibinfo {pages} {1617} (\bibinfo {year} {1982})}\BibitemShut {NoStop}%
\bibitem [{\citenamefont {{Dalibard, J.}}\ \emph {et~al.}(1984)\citenamefont {{Dalibard, J.}}, \citenamefont {{Dupont-Roc, J.}},\ and\ \citenamefont {{Cohen-Tannoudji, C.}}}]{refId1}%
  \BibitemOpen
  \bibfield  {author} {\bibinfo {author} {\bibnamefont {{Dalibard, J.}}}, \bibinfo {author} {\bibnamefont {{Dupont-Roc, J.}}}, \ and\ \bibinfo {author} {\bibnamefont {{Cohen-Tannoudji, C.}}},\ }\href {\doibase 10.1051/jphys:01984004504063700} {\bibfield  {journal} {\bibinfo  {journal} {J. Phys. France}\ }\textbf {\bibinfo {volume} {45}},\ \bibinfo {pages} {637} (\bibinfo {year} {1984})}\BibitemShut {NoStop}%
\bibitem [{\citenamefont {{Hawking}}\ and\ \citenamefont {{Israel}}(2010)}]{2010grae.book.....H}%
  \BibitemOpen
  \bibfield  {author} {\bibinfo {author} {\bibfnamefont {S.}~\bibnamefont {{Hawking}}}\ and\ \bibinfo {author} {\bibfnamefont {W.}~\bibnamefont {{Israel}}},\ }\href@noop {} {\emph {\bibinfo {title} {{General Relativity: an Einstein Centenary Survey}}}}\ (\bibinfo {year} {2010})\BibitemShut {NoStop}%
\bibitem [{\citenamefont {Zhu}\ \emph {et~al.}(2006)\citenamefont {Zhu}, \citenamefont {Yu},\ and\ \citenamefont {Lu}}]{Zhu:2006wt}%
  \BibitemOpen
  \bibfield  {author} {\bibinfo {author} {\bibfnamefont {Z.}~\bibnamefont {Zhu}}, \bibinfo {author} {\bibfnamefont {H.~W.}\ \bibnamefont {Yu}}, \ and\ \bibinfo {author} {\bibfnamefont {S.}~\bibnamefont {Lu}},\ }\href {\doibase 10.1103/PhysRevD.73.107501} {\bibfield  {journal} {\bibinfo  {journal} {Phys. Rev. D}\ }\textbf {\bibinfo {volume} {73}},\ \bibinfo {pages} {107501} (\bibinfo {year} {2006})},\ \Eprint {http://arxiv.org/abs/gr-qc/0604116} {arXiv:gr-qc/0604116} \BibitemShut {NoStop}%
\bibitem [{\citenamefont {{Chen}}\ \emph {et~al.}(2015)\citenamefont {{Chen}}, \citenamefont {{Hu}},\ and\ \citenamefont {{Yu}}}]{2015AnPhy.353..317C}%
  \BibitemOpen
  \bibfield  {author} {\bibinfo {author} {\bibfnamefont {J.}~\bibnamefont {{Chen}}}, \bibinfo {author} {\bibfnamefont {J.}~\bibnamefont {{Hu}}}, \ and\ \bibinfo {author} {\bibfnamefont {H.}~\bibnamefont {{Yu}}},\ }\href {\doibase 10.1016/j.aop.2014.12.003} {\bibfield  {journal} {\bibinfo  {journal} {Annals of Physics}\ }\textbf {\bibinfo {volume} {353}},\ \bibinfo {pages} {317} (\bibinfo {year} {2015})}\BibitemShut {NoStop}%
\bibitem [{\citenamefont {Zhou}(2016)}]{Zhou:2016zsf}%
  \BibitemOpen
  \bibfield  {author} {\bibinfo {author} {\bibfnamefont {W.}~\bibnamefont {Zhou}},\ }\href {\doibase 10.1142/S0217732316501893} {\bibfield  {journal} {\bibinfo  {journal} {Mod. Phys. Lett. A}\ }\textbf {\bibinfo {volume} {31}},\ \bibinfo {pages} {1650189} (\bibinfo {year} {2016})}\BibitemShut {NoStop}%
\bibitem [{\citenamefont {Zhou}\ and\ \citenamefont {Yu}(2012{\natexlab{a}})}]{Zhou:2012gu}%
  \BibitemOpen
  \bibfield  {author} {\bibinfo {author} {\bibfnamefont {W.}~\bibnamefont {Zhou}}\ and\ \bibinfo {author} {\bibfnamefont {H.}~\bibnamefont {Yu}},\ }\href {\doibase 10.1103/PhysRevA.86.033841} {\bibfield  {journal} {\bibinfo  {journal} {Phys. Rev. A}\ }\textbf {\bibinfo {volume} {86}},\ \bibinfo {pages} {033841} (\bibinfo {year} {2012}{\natexlab{a}})},\ \Eprint {http://arxiv.org/abs/1209.2494} {arXiv:1209.2494 [quant-ph]} \BibitemShut {NoStop}%
\bibitem [{\citenamefont {Langlois}(2006)}]{Langlois:2005nf}%
  \BibitemOpen
  \bibfield  {author} {\bibinfo {author} {\bibfnamefont {P.}~\bibnamefont {Langlois}},\ }\href {\doibase 10.1016/j.aop.2006.01.013} {\bibfield  {journal} {\bibinfo  {journal} {Annals Phys.}\ }\textbf {\bibinfo {volume} {321}},\ \bibinfo {pages} {2027} (\bibinfo {year} {2006})},\ \Eprint {http://arxiv.org/abs/gr-qc/0510049} {arXiv:gr-qc/0510049} \BibitemShut {NoStop}%
\bibitem [{\citenamefont {Rizzuto}\ and\ \citenamefont {Spagnolo}(2011)}]{Rizzuto_2011}%
  \BibitemOpen
  \bibfield  {author} {\bibinfo {author} {\bibfnamefont {L.}~\bibnamefont {Rizzuto}}\ and\ \bibinfo {author} {\bibfnamefont {S.}~\bibnamefont {Spagnolo}},\ }\href {\doibase 10.1088/0031-8949/2011/T143/014021} {\bibfield  {journal} {\bibinfo  {journal} {Physica Scripta}\ }\textbf {\bibinfo {volume} {2011}},\ \bibinfo {pages} {014021} (\bibinfo {year} {2011})}\BibitemShut {NoStop}%
\bibitem [{\citenamefont {Zhang}(2016)}]{Zhang:2016gtp}%
  \BibitemOpen
  \bibfield  {author} {\bibinfo {author} {\bibfnamefont {A.}~\bibnamefont {Zhang}},\ }\href {\doibase 10.1007/s10701-016-0016-9} {\bibfield  {journal} {\bibinfo  {journal} {Found. Phys.}\ }\textbf {\bibinfo {volume} {46}},\ \bibinfo {pages} {1199} (\bibinfo {year} {2016})},\ \Eprint {http://arxiv.org/abs/1612.01844} {arXiv:1612.01844 [quant-ph]} \BibitemShut {NoStop}%
\bibitem [{\citenamefont {{Barton}}\ and\ \citenamefont {{Calogeracos}}(2005)}]{2005JOptB...7S..21B}%
  \BibitemOpen
  \bibfield  {author} {\bibinfo {author} {\bibfnamefont {G.}~\bibnamefont {{Barton}}}\ and\ \bibinfo {author} {\bibfnamefont {A.}~\bibnamefont {{Calogeracos}}},\ }\href {\doibase 10.1088/1464-4266/7/3/003} {\bibfield  {journal} {\bibinfo  {journal} {Journal of Optics B: Quantum and Semiclassical Optics}\ }\textbf {\bibinfo {volume} {7}},\ \bibinfo {pages} {S21} (\bibinfo {year} {2005})}\BibitemShut {NoStop}%
\bibitem [{\citenamefont {{Barton}}\ and\ \citenamefont {{Calogeracos}}(2008)}]{2008JPhA...41p4030B}%
  \BibitemOpen
  \bibfield  {author} {\bibinfo {author} {\bibfnamefont {G.}~\bibnamefont {{Barton}}}\ and\ \bibinfo {author} {\bibfnamefont {A.}~\bibnamefont {{Calogeracos}}},\ }\href {\doibase 10.1088/1751-8113/41/16/164030} {\bibfield  {journal} {\bibinfo  {journal} {Journal of Physics A Mathematical General}\ }\textbf {\bibinfo {volume} {41}},\ \bibinfo {eid} {164030} (\bibinfo {year} {2008})}\BibitemShut {NoStop}%
\bibitem [{\citenamefont {Calogeracos}(2016)}]{CALOGERACOS2016377}%
  \BibitemOpen
  \bibfield  {author} {\bibinfo {author} {\bibfnamefont {A.}~\bibnamefont {Calogeracos}},\ }\href {\doibase https://doi.org/10.1016/j.rinp.2016.05.008} {\bibfield  {journal} {\bibinfo  {journal} {Results in Physics}\ }\textbf {\bibinfo {volume} {6}},\ \bibinfo {pages} {377} (\bibinfo {year} {2016})}\BibitemShut {NoStop}%
\bibitem [{\citenamefont {Raimond}\ \emph {et~al.}(2001)\citenamefont {Raimond}, \citenamefont {Brune},\ and\ \citenamefont {Haroche}}]{RevModPhys.73.565}%
  \BibitemOpen
  \bibfield  {author} {\bibinfo {author} {\bibfnamefont {J.~M.}\ \bibnamefont {Raimond}}, \bibinfo {author} {\bibfnamefont {M.}~\bibnamefont {Brune}}, \ and\ \bibinfo {author} {\bibfnamefont {S.}~\bibnamefont {Haroche}},\ }\href {\doibase 10.1103/RevModPhys.73.565} {\bibfield  {journal} {\bibinfo  {journal} {Rev. Mod. Phys.}\ }\textbf {\bibinfo {volume} {73}},\ \bibinfo {pages} {565} (\bibinfo {year} {2001})}\BibitemShut {NoStop}%
\bibitem [{\citenamefont {Menezes}\ and\ \citenamefont {Svaiter}(2016{\natexlab{a}})}]{Menezes:2015iva}%
  \BibitemOpen
  \bibfield  {author} {\bibinfo {author} {\bibfnamefont {G.}~\bibnamefont {Menezes}}\ and\ \bibinfo {author} {\bibfnamefont {N.~F.}\ \bibnamefont {Svaiter}},\ }\href {\doibase 10.1103/PhysRevA.93.052117} {\bibfield  {journal} {\bibinfo  {journal} {Phys. Rev. A}\ }\textbf {\bibinfo {volume} {93}},\ \bibinfo {pages} {052117} (\bibinfo {year} {2016}{\natexlab{a}})},\ \Eprint {http://arxiv.org/abs/1512.02886} {arXiv:1512.02886 [hep-th]} \BibitemShut {NoStop}%
\bibitem [{\citenamefont {Yu}\ and\ \citenamefont {Eberly}(2004)}]{PhysRevLett.93.140404}%
  \BibitemOpen
  \bibfield  {author} {\bibinfo {author} {\bibfnamefont {T.}~\bibnamefont {Yu}}\ and\ \bibinfo {author} {\bibfnamefont {J.~H.}\ \bibnamefont {Eberly}},\ }\href {\doibase 10.1103/PhysRevLett.93.140404} {\bibfield  {journal} {\bibinfo  {journal} {Phys. Rev. Lett.}\ }\textbf {\bibinfo {volume} {93}},\ \bibinfo {pages} {140404} (\bibinfo {year} {2004})}\BibitemShut {NoStop}%
\bibitem [{\citenamefont {Arias}\ \emph {et~al.}(2016)\citenamefont {Arias}, \citenamefont {Due\~nas}, \citenamefont {Menezes},\ and\ \citenamefont {Svaiter}}]{Arias:2015moa}%
  \BibitemOpen
  \bibfield  {author} {\bibinfo {author} {\bibfnamefont {E.}~\bibnamefont {Arias}}, \bibinfo {author} {\bibfnamefont {J.~G.}\ \bibnamefont {Due\~nas}}, \bibinfo {author} {\bibfnamefont {G.}~\bibnamefont {Menezes}}, \ and\ \bibinfo {author} {\bibfnamefont {N.~F.}\ \bibnamefont {Svaiter}},\ }\href {\doibase 10.1007/JHEP07(2016)147} {\bibfield  {journal} {\bibinfo  {journal} {JHEP}\ }\textbf {\bibinfo {volume} {07}},\ \bibinfo {pages} {147} (\bibinfo {year} {2016})},\ \Eprint {http://arxiv.org/abs/1510.00047} {arXiv:1510.00047 [quant-ph]} \BibitemShut {NoStop}%
\bibitem [{\citenamefont {Zhang}\ and\ \citenamefont {Zhou}(2019)}]{sym11121515}%
  \BibitemOpen
  \bibfield  {author} {\bibinfo {author} {\bibfnamefont {C.}~\bibnamefont {Zhang}}\ and\ \bibinfo {author} {\bibfnamefont {W.}~\bibnamefont {Zhou}},\ }\href {\doibase 10.3390/sym11121515} {\bibfield  {journal} {\bibinfo  {journal} {Symmetry}\ }\textbf {\bibinfo {volume} {11}} (\bibinfo {year} {2019}),\ 10.3390/sym11121515}\BibitemShut {NoStop}%
\bibitem [{\citenamefont {Menezes}\ and\ \citenamefont {Svaiter}(2015)}]{Menezes:2015uaa}%
  \BibitemOpen
  \bibfield  {author} {\bibinfo {author} {\bibfnamefont {G.}~\bibnamefont {Menezes}}\ and\ \bibinfo {author} {\bibfnamefont {N.~F.}\ \bibnamefont {Svaiter}},\ }\href {\doibase 10.1103/PhysRevA.92.062131} {\bibfield  {journal} {\bibinfo  {journal} {Phys. Rev. A}\ }\textbf {\bibinfo {volume} {92}},\ \bibinfo {pages} {062131} (\bibinfo {year} {2015})},\ \Eprint {http://arxiv.org/abs/1508.04513} {arXiv:1508.04513 [hep-th]} \BibitemShut {NoStop}%
\bibitem [{\citenamefont {Zhou}\ and\ \citenamefont {Yu}(2020{\natexlab{a}})}]{Zhou:2020oqa}%
  \BibitemOpen
  \bibfield  {author} {\bibinfo {author} {\bibfnamefont {W.}~\bibnamefont {Zhou}}\ and\ \bibinfo {author} {\bibfnamefont {H.}~\bibnamefont {Yu}},\ }\href {\doibase 10.1103/PhysRevD.101.025009} {\bibfield  {journal} {\bibinfo  {journal} {Phys. Rev. D}\ }\textbf {\bibinfo {volume} {101}},\ \bibinfo {pages} {025009} (\bibinfo {year} {2020}{\natexlab{a}})},\ \Eprint {http://arxiv.org/abs/2001.00750} {arXiv:2001.00750 [quant-ph]} \BibitemShut {NoStop}%
\bibitem [{\citenamefont {Zhou}\ and\ \citenamefont {Yu}(2020{\natexlab{b}})}]{PhysRevD.101.085009}%
  \BibitemOpen
  \bibfield  {author} {\bibinfo {author} {\bibfnamefont {W.}~\bibnamefont {Zhou}}\ and\ \bibinfo {author} {\bibfnamefont {H.}~\bibnamefont {Yu}},\ }\href {\doibase 10.1103/PhysRevD.101.085009} {\bibfield  {journal} {\bibinfo  {journal} {Phys. Rev. D}\ }\textbf {\bibinfo {volume} {101}},\ \bibinfo {pages} {085009} (\bibinfo {year} {2020}{\natexlab{b}})}\BibitemShut {NoStop}%
\bibitem [{\citenamefont {Menezes}\ and\ \citenamefont {Svaiter}(2016{\natexlab{b}})}]{PhysRevA.93.052117}%
  \BibitemOpen
  \bibfield  {author} {\bibinfo {author} {\bibfnamefont {G.}~\bibnamefont {Menezes}}\ and\ \bibinfo {author} {\bibfnamefont {N.~F.}\ \bibnamefont {Svaiter}},\ }\href {\doibase 10.1103/PhysRevA.93.052117} {\bibfield  {journal} {\bibinfo  {journal} {Phys. Rev. A}\ }\textbf {\bibinfo {volume} {93}},\ \bibinfo {pages} {052117} (\bibinfo {year} {2016}{\natexlab{b}})}\BibitemShut {NoStop}%
\bibitem [{\citenamefont {Passante}(1998)}]{PhysRevA.57.1590}%
  \BibitemOpen
  \bibfield  {author} {\bibinfo {author} {\bibfnamefont {R.}~\bibnamefont {Passante}},\ }\href {\doibase 10.1103/PhysRevA.57.1590} {\bibfield  {journal} {\bibinfo  {journal} {Phys. Rev. A}\ }\textbf {\bibinfo {volume} {57}},\ \bibinfo {pages} {1590} (\bibinfo {year} {1998})}\BibitemShut {NoStop}%
\bibitem [{\citenamefont {Rizzuto}\ and\ \citenamefont {Spagnolo}(2009)}]{PhysRevA.79.062110}%
  \BibitemOpen
  \bibfield  {author} {\bibinfo {author} {\bibfnamefont {L.}~\bibnamefont {Rizzuto}}\ and\ \bibinfo {author} {\bibfnamefont {S.}~\bibnamefont {Spagnolo}},\ }\href {\doibase 10.1103/PhysRevA.79.062110} {\bibfield  {journal} {\bibinfo  {journal} {Phys. Rev. A}\ }\textbf {\bibinfo {volume} {79}},\ \bibinfo {pages} {062110} (\bibinfo {year} {2009})}\BibitemShut {NoStop}%
\bibitem [{\citenamefont {Audretsch}\ \emph {et~al.}(1995)\citenamefont {Audretsch}, \citenamefont {Mueller},\ and\ \citenamefont {Holzmann}}]{Audretsch:1995iw}%
  \BibitemOpen
  \bibfield  {author} {\bibinfo {author} {\bibfnamefont {J.}~\bibnamefont {Audretsch}}, \bibinfo {author} {\bibfnamefont {R.}~\bibnamefont {Mueller}}, \ and\ \bibinfo {author} {\bibfnamefont {M.}~\bibnamefont {Holzmann}},\ }\href {\doibase 10.1088/0264-9381/12/12/010} {\bibfield  {journal} {\bibinfo  {journal} {Class. Quant. Grav.}\ }\textbf {\bibinfo {volume} {12}},\ \bibinfo {pages} {2927} (\bibinfo {year} {1995})},\ \Eprint {http://arxiv.org/abs/quant-ph/9510025} {arXiv:quant-ph/9510025} \BibitemShut {NoStop}%
\bibitem [{\citenamefont {{Audretsch}}\ and\ \citenamefont {{M{\"u}ller}}(1995)}]{1995PhRvA..52..629A}%
  \BibitemOpen
  \bibfield  {author} {\bibinfo {author} {\bibfnamefont {J.}~\bibnamefont {{Audretsch}}}\ and\ \bibinfo {author} {\bibfnamefont {R.}~\bibnamefont {{M{\"u}ller}}},\ }\href {\doibase 10.1103/PhysRevA.52.629} {\bibfield  {journal} {\bibinfo  {journal} {\pra}\ }\textbf {\bibinfo {volume} {52}},\ \bibinfo {pages} {629} (\bibinfo {year} {1995})},\ \Eprint {http://arxiv.org/abs/gr-qc/9503058} {arXiv:gr-qc/9503058 [gr-qc]} \BibitemShut {NoStop}%
\bibitem [{\citenamefont {Marzlin}\ and\ \citenamefont {Audretsch}(1998)}]{Marzlin:1997gx}%
  \BibitemOpen
  \bibfield  {author} {\bibinfo {author} {\bibfnamefont {K.-P.}\ \bibnamefont {Marzlin}}\ and\ \bibinfo {author} {\bibfnamefont {J.}~\bibnamefont {Audretsch}},\ }\href {\doibase 10.1103/PhysRevD.57.1045} {\bibfield  {journal} {\bibinfo  {journal} {Phys. Rev. D}\ }\textbf {\bibinfo {volume} {57}},\ \bibinfo {pages} {1045} (\bibinfo {year} {1998})},\ \Eprint {http://arxiv.org/abs/gr-qc/9707058} {arXiv:gr-qc/9707058} \BibitemShut {NoStop}%
\bibitem [{\citenamefont {Audretsch}\ and\ \citenamefont {Marzlin}(1994)}]{Audretsch:1993uc}%
  \BibitemOpen
  \bibfield  {author} {\bibinfo {author} {\bibfnamefont {J.}~\bibnamefont {Audretsch}}\ and\ \bibinfo {author} {\bibfnamefont {K.~P.}\ \bibnamefont {Marzlin}},\ }\href {\doibase 10.1103/PhysRevA.50.2080} {\bibfield  {journal} {\bibinfo  {journal} {Phys. Rev. A}\ }\textbf {\bibinfo {volume} {50}},\ \bibinfo {pages} {2080} (\bibinfo {year} {1994})},\ \Eprint {http://arxiv.org/abs/gr-qc/9310029} {arXiv:gr-qc/9310029} \BibitemShut {NoStop}%
\bibitem [{\citenamefont {Olmo}(2008)}]{Olmo:2008ye}%
  \BibitemOpen
  \bibfield  {author} {\bibinfo {author} {\bibfnamefont {G.~J.}\ \bibnamefont {Olmo}},\ }\href {\doibase 10.1103/PhysRevD.77.084021} {\bibfield  {journal} {\bibinfo  {journal} {Phys. Rev. D}\ }\textbf {\bibinfo {volume} {77}},\ \bibinfo {pages} {084021} (\bibinfo {year} {2008})},\ \Eprint {http://arxiv.org/abs/0802.4038} {arXiv:0802.4038 [gr-qc]} \BibitemShut {NoStop}%
\bibitem [{\citenamefont {Singh}\ and\ \citenamefont {Mobed}(2011)}]{Singh:2011cq}%
  \BibitemOpen
  \bibfield  {author} {\bibinfo {author} {\bibfnamefont {D.}~\bibnamefont {Singh}}\ and\ \bibinfo {author} {\bibfnamefont {N.}~\bibnamefont {Mobed}},\ }\href {\doibase 10.1088/0264-9381/28/10/105024} {\bibfield  {journal} {\bibinfo  {journal} {Class. Quant. Grav.}\ }\textbf {\bibinfo {volume} {28}},\ \bibinfo {pages} {105024} (\bibinfo {year} {2011})},\ \Eprint {http://arxiv.org/abs/1101.1030} {arXiv:1101.1030 [gr-qc]} \BibitemShut {NoStop}%
\bibitem [{\citenamefont {Wong}\ and\ \citenamefont {Davis}(2017)}]{Wong:2017jer}%
  \BibitemOpen
  \bibfield  {author} {\bibinfo {author} {\bibfnamefont {L.~K.}\ \bibnamefont {Wong}}\ and\ \bibinfo {author} {\bibfnamefont {A.-C.}\ \bibnamefont {Davis}},\ }\href {\doibase 10.1103/PhysRevD.95.104010} {\bibfield  {journal} {\bibinfo  {journal} {Phys. Rev. D}\ }\textbf {\bibinfo {volume} {95}},\ \bibinfo {pages} {104010} (\bibinfo {year} {2017})},\ \Eprint {http://arxiv.org/abs/1703.05659} {arXiv:1703.05659 [astro-ph.CO]} \BibitemShut {NoStop}%
\bibitem [{\citenamefont {Brax}\ \emph {et~al.}(2018)\citenamefont {Brax}, \citenamefont {Davis}, \citenamefont {Elder},\ and\ \citenamefont {Wong}}]{PhysRevD.97.084050}%
  \BibitemOpen
  \bibfield  {author} {\bibinfo {author} {\bibfnamefont {P.}~\bibnamefont {Brax}}, \bibinfo {author} {\bibfnamefont {A.-C.}\ \bibnamefont {Davis}}, \bibinfo {author} {\bibfnamefont {B.}~\bibnamefont {Elder}}, \ and\ \bibinfo {author} {\bibfnamefont {L.~K.}\ \bibnamefont {Wong}},\ }\href {\doibase 10.1103/PhysRevD.97.084050} {\bibfield  {journal} {\bibinfo  {journal} {Phys. Rev. D}\ }\textbf {\bibinfo {volume} {97}},\ \bibinfo {pages} {084050} (\bibinfo {year} {2018})}\BibitemShut {NoStop}%
\bibitem [{\citenamefont {Sabulsky}\ \emph {et~al.}(2019)\citenamefont {Sabulsky}, \citenamefont {Dutta}, \citenamefont {Hinds}, \citenamefont {Elder}, \citenamefont {Burrage},\ and\ \citenamefont {Copeland}}]{PhysRevLett.123.061102}%
  \BibitemOpen
  \bibfield  {author} {\bibinfo {author} {\bibfnamefont {D.~O.}\ \bibnamefont {Sabulsky}}, \bibinfo {author} {\bibfnamefont {I.}~\bibnamefont {Dutta}}, \bibinfo {author} {\bibfnamefont {E.~A.}\ \bibnamefont {Hinds}}, \bibinfo {author} {\bibfnamefont {B.}~\bibnamefont {Elder}}, \bibinfo {author} {\bibfnamefont {C.}~\bibnamefont {Burrage}}, \ and\ \bibinfo {author} {\bibfnamefont {E.~J.}\ \bibnamefont {Copeland}},\ }\href {\doibase 10.1103/PhysRevLett.123.061102} {\bibfield  {journal} {\bibinfo  {journal} {Phys. Rev. Lett.}\ }\textbf {\bibinfo {volume} {123}},\ \bibinfo {pages} {061102} (\bibinfo {year} {2019})}\BibitemShut {NoStop}%
\bibitem [{\citenamefont {Sciama}\ \emph {et~al.}(1981)\citenamefont {Sciama}, \citenamefont {Candelas},\ and\ \citenamefont {Deutsch}}]{Sciama:1981hr}%
  \BibitemOpen
  \bibfield  {author} {\bibinfo {author} {\bibfnamefont {D.~W.}\ \bibnamefont {Sciama}}, \bibinfo {author} {\bibfnamefont {P.}~\bibnamefont {Candelas}}, \ and\ \bibinfo {author} {\bibfnamefont {D.}~\bibnamefont {Deutsch}},\ }\href {\doibase 10.1080/00018738100101457} {\bibfield  {journal} {\bibinfo  {journal} {Adv. Phys.}\ }\textbf {\bibinfo {volume} {30}},\ \bibinfo {pages} {327} (\bibinfo {year} {1981})}\BibitemShut {NoStop}%
\bibitem [{\citenamefont {Hartle}\ and\ \citenamefont {Hawking}(1976)}]{Hartle:1976tp}%
  \BibitemOpen
  \bibfield  {author} {\bibinfo {author} {\bibfnamefont {J.~B.}\ \bibnamefont {Hartle}}\ and\ \bibinfo {author} {\bibfnamefont {S.~W.}\ \bibnamefont {Hawking}},\ }\href {\doibase 10.1103/PhysRevD.13.2188} {\bibfield  {journal} {\bibinfo  {journal} {Phys. Rev. D}\ }\textbf {\bibinfo {volume} {13}},\ \bibinfo {pages} {2188} (\bibinfo {year} {1976})}\BibitemShut {NoStop}%
\bibitem [{\citenamefont {Papini}(2015)}]{Papini:2015fha}%
  \BibitemOpen
  \bibfield  {author} {\bibinfo {author} {\bibfnamefont {G.}~\bibnamefont {Papini}},\ }\href {\doibase 10.1142/S0217732315501667} {\bibfield  {journal} {\bibinfo  {journal} {Mod. Phys. Lett. A}\ }\textbf {\bibinfo {volume} {30}},\ \bibinfo {pages} {1550166} (\bibinfo {year} {2015})},\ \Eprint {http://arxiv.org/abs/1507.02198} {arXiv:1507.02198 [gr-qc]} \BibitemShut {NoStop}%
\bibitem [{\citenamefont {Higuchi}\ \emph {et~al.}(1998)\citenamefont {Higuchi}, \citenamefont {Matsas},\ and\ \citenamefont {Sudarsky}}]{Higuchi:1998qc}%
  \BibitemOpen
  \bibfield  {author} {\bibinfo {author} {\bibfnamefont {A.}~\bibnamefont {Higuchi}}, \bibinfo {author} {\bibfnamefont {G.~E.~A.}\ \bibnamefont {Matsas}}, \ and\ \bibinfo {author} {\bibfnamefont {D.}~\bibnamefont {Sudarsky}},\ }\href {\doibase 10.1103/PhysRevD.58.104021} {\bibfield  {journal} {\bibinfo  {journal} {Phys. Rev. D}\ }\textbf {\bibinfo {volume} {58}},\ \bibinfo {pages} {104021} (\bibinfo {year} {1998})},\ \Eprint {http://arxiv.org/abs/gr-qc/9806093} {arXiv:gr-qc/9806093} \BibitemShut {NoStop}%
\bibitem [{\citenamefont {Crispino}\ \emph {et~al.}(2001)\citenamefont {Crispino}, \citenamefont {Higuchi},\ and\ \citenamefont {Matsas}}]{Crispino:2000jx}%
  \BibitemOpen
  \bibfield  {author} {\bibinfo {author} {\bibfnamefont {L.~C.~B.}\ \bibnamefont {Crispino}}, \bibinfo {author} {\bibfnamefont {A.}~\bibnamefont {Higuchi}}, \ and\ \bibinfo {author} {\bibfnamefont {G.~E.~A.}\ \bibnamefont {Matsas}},\ }\href {\doibase 10.1103/PhysRevD.80.029906} {\bibfield  {journal} {\bibinfo  {journal} {Phys. Rev. D}\ }\textbf {\bibinfo {volume} {63}},\ \bibinfo {pages} {124008} (\bibinfo {year} {2001})},\ \bibinfo {note} {[Erratum: Phys.Rev.D 80, 029906 (2009)]},\ \Eprint {http://arxiv.org/abs/gr-qc/0011070} {arXiv:gr-qc/0011070} \BibitemShut {NoStop}%
\bibitem [{\citenamefont {Castineiras}\ \emph {et~al.}(2003)\citenamefont {Castineiras}, \citenamefont {Costa~e Silva},\ and\ \citenamefont {Matsas}}]{Castineiras:2002fn}%
  \BibitemOpen
  \bibfield  {author} {\bibinfo {author} {\bibfnamefont {J.}~\bibnamefont {Castineiras}}, \bibinfo {author} {\bibfnamefont {I.~P.}\ \bibnamefont {Costa~e Silva}}, \ and\ \bibinfo {author} {\bibfnamefont {G.~E.~A.}\ \bibnamefont {Matsas}},\ }\href {\doibase 10.1103/PhysRevD.67.067502} {\bibfield  {journal} {\bibinfo  {journal} {Phys. Rev. D}\ }\textbf {\bibinfo {volume} {67}},\ \bibinfo {pages} {067502} (\bibinfo {year} {2003})},\ \Eprint {http://arxiv.org/abs/gr-qc/0211053} {arXiv:gr-qc/0211053} \BibitemShut {NoStop}%
\bibitem [{\citenamefont {Christensen}\ and\ \citenamefont {Fulling}(1977)}]{Christensen:1977jc}%
  \BibitemOpen
  \bibfield  {author} {\bibinfo {author} {\bibfnamefont {S.~M.}\ \bibnamefont {Christensen}}\ and\ \bibinfo {author} {\bibfnamefont {S.~A.}\ \bibnamefont {Fulling}},\ }\href {\doibase 10.1103/PhysRevD.15.2088} {\bibfield  {journal} {\bibinfo  {journal} {Phys. Rev. D}\ }\textbf {\bibinfo {volume} {15}},\ \bibinfo {pages} {2088} (\bibinfo {year} {1977})}\BibitemShut {NoStop}%
\bibitem [{\citenamefont {Candelas}(1980)}]{Candelas:1980zt}%
  \BibitemOpen
  \bibfield  {author} {\bibinfo {author} {\bibfnamefont {P.}~\bibnamefont {Candelas}},\ }\href {\doibase 10.1103/PhysRevD.21.2185} {\bibfield  {journal} {\bibinfo  {journal} {Phys. Rev. D}\ }\textbf {\bibinfo {volume} {21}},\ \bibinfo {pages} {2185} (\bibinfo {year} {1980})}\BibitemShut {NoStop}%
\bibitem [{\citenamefont {Yu}\ and\ \citenamefont {Zhou}(2007{\natexlab{a}})}]{Yu:2007wv}%
  \BibitemOpen
  \bibfield  {author} {\bibinfo {author} {\bibfnamefont {H.~W.}\ \bibnamefont {Yu}}\ and\ \bibinfo {author} {\bibfnamefont {W.}~\bibnamefont {Zhou}},\ }\href {\doibase 10.1103/PhysRevD.76.044023} {\bibfield  {journal} {\bibinfo  {journal} {Phys. Rev. D}\ }\textbf {\bibinfo {volume} {76}},\ \bibinfo {pages} {044023} (\bibinfo {year} {2007}{\natexlab{a}})},\ \Eprint {http://arxiv.org/abs/0707.2613} {arXiv:0707.2613 [gr-qc]} \BibitemShut {NoStop}%
\bibitem [{\citenamefont {Zhou}\ and\ \citenamefont {Yu}(2012{\natexlab{b}})}]{Zhou:2012eb}%
  \BibitemOpen
  \bibfield  {author} {\bibinfo {author} {\bibfnamefont {W.}~\bibnamefont {Zhou}}\ and\ \bibinfo {author} {\bibfnamefont {H.}~\bibnamefont {Yu}},\ }\href {\doibase 10.1088/0264-9381/29/8/085003} {\bibfield  {journal} {\bibinfo  {journal} {Class. Quant. Grav.}\ }\textbf {\bibinfo {volume} {29}},\ \bibinfo {pages} {085003} (\bibinfo {year} {2012}{\natexlab{b}})},\ \Eprint {http://arxiv.org/abs/1203.5867} {arXiv:1203.5867 [gr-qc]} \BibitemShut {NoStop}%
\bibitem [{\citenamefont {Yu}\ and\ \citenamefont {Zhou}(2007{\natexlab{b}})}]{Yu:2007qu}%
  \BibitemOpen
  \bibfield  {author} {\bibinfo {author} {\bibfnamefont {H.~W.}\ \bibnamefont {Yu}}\ and\ \bibinfo {author} {\bibfnamefont {W.}~\bibnamefont {Zhou}},\ }\href {\doibase 10.1103/PhysRevD.76.027503} {\bibfield  {journal} {\bibinfo  {journal} {Phys. Rev. D}\ }\textbf {\bibinfo {volume} {76}},\ \bibinfo {pages} {027503} (\bibinfo {year} {2007}{\natexlab{b}})},\ \Eprint {http://arxiv.org/abs/0706.2207} {arXiv:0706.2207 [hep-th]} \BibitemShut {NoStop}%
\bibitem [{\citenamefont {Cliche}\ and\ \citenamefont {Kempf}(2011)}]{Cliche:2010fi}%
  \BibitemOpen
  \bibfield  {author} {\bibinfo {author} {\bibfnamefont {M.}~\bibnamefont {Cliche}}\ and\ \bibinfo {author} {\bibfnamefont {A.}~\bibnamefont {Kempf}},\ }\href {\doibase 10.1103/PhysRevD.83.045019} {\bibfield  {journal} {\bibinfo  {journal} {Phys. Rev. D}\ }\textbf {\bibinfo {volume} {83}},\ \bibinfo {pages} {045019} (\bibinfo {year} {2011})},\ \Eprint {http://arxiv.org/abs/1008.4926} {arXiv:1008.4926 [quant-ph]} \BibitemShut {NoStop}%
\bibitem [{\citenamefont {Menezes}(2016)}]{Menezes:2015veo}%
  \BibitemOpen
  \bibfield  {author} {\bibinfo {author} {\bibfnamefont {G.}~\bibnamefont {Menezes}},\ }\href {\doibase 10.1103/PhysRevD.94.105008} {\bibfield  {journal} {\bibinfo  {journal} {Phys. Rev. D}\ }\textbf {\bibinfo {volume} {94}},\ \bibinfo {pages} {105008} (\bibinfo {year} {2016})},\ \Eprint {http://arxiv.org/abs/1512.03636} {arXiv:1512.03636 [gr-qc]} \BibitemShut {NoStop}%
\bibitem [{\citenamefont {Chen}\ \emph {et~al.}(2023)\citenamefont {Chen}, \citenamefont {Hu},\ and\ \citenamefont {Yu}}]{Chen:2023xbc}%
  \BibitemOpen
  \bibfield  {author} {\bibinfo {author} {\bibfnamefont {Y.}~\bibnamefont {Chen}}, \bibinfo {author} {\bibfnamefont {J.}~\bibnamefont {Hu}}, \ and\ \bibinfo {author} {\bibfnamefont {H.}~\bibnamefont {Yu}},\ }\href {\doibase 10.1103/PhysRevD.107.025015} {\bibfield  {journal} {\bibinfo  {journal} {Phys. Rev. D}\ }\textbf {\bibinfo {volume} {107}},\ \bibinfo {pages} {025015} (\bibinfo {year} {2023})}\BibitemShut {NoStop}%
\bibitem [{\citenamefont {Yu}\ \emph {et~al.}(2006)\citenamefont {Yu}, \citenamefont {Yu},\ and\ \citenamefont {Zhu}}]{Yu:2006kp}%
  \BibitemOpen
  \bibfield  {author} {\bibinfo {author} {\bibfnamefont {H.}~\bibnamefont {Yu}}, \bibinfo {author} {\bibfnamefont {H.~W.}\ \bibnamefont {Yu}}, \ and\ \bibinfo {author} {\bibfnamefont {Z.}~\bibnamefont {Zhu}},\ }\href {\doibase 10.1103/PhysRevD.74.044032} {\bibfield  {journal} {\bibinfo  {journal} {Phys. Rev. D}\ }\textbf {\bibinfo {volume} {74}},\ \bibinfo {pages} {044032} (\bibinfo {year} {2006})},\ \Eprint {http://arxiv.org/abs/quant-ph/0608179} {arXiv:quant-ph/0608179} \BibitemShut {NoStop}%
\bibitem [{\citenamefont {Visser}(2007)}]{Visser:2007fj}%
  \BibitemOpen
  \bibfield  {author} {\bibinfo {author} {\bibfnamefont {M.}~\bibnamefont {Visser}},\ }in\ \href@noop {} {\emph {\bibinfo {booktitle} {{Kerr Fest: Black Holes in Astrophysics, General Relativity and Quantum Gravity}}}}\ (\bibinfo {year} {2007})\ \Eprint {http://arxiv.org/abs/0706.0622} {arXiv:0706.0622 [gr-qc]} \BibitemShut {NoStop}%
\bibitem [{\citenamefont {Jacobson}(1994)}]{Jacobson:1994fp}%
  \BibitemOpen
  \bibfield  {author} {\bibinfo {author} {\bibfnamefont {T.}~\bibnamefont {Jacobson}},\ }\href {\doibase 10.1103/PhysRevD.50.R6031} {\bibfield  {journal} {\bibinfo  {journal} {Phys. Rev. D}\ }\textbf {\bibinfo {volume} {50}},\ \bibinfo {pages} {R6031} (\bibinfo {year} {1994})},\ \Eprint {http://arxiv.org/abs/gr-qc/9407022} {arXiv:gr-qc/9407022} \BibitemShut {NoStop}%
\bibitem [{\citenamefont {Menezes}(2017)}]{Menezes:2016quu}%
  \BibitemOpen
  \bibfield  {author} {\bibinfo {author} {\bibfnamefont {G.}~\bibnamefont {Menezes}},\ }\href {\doibase 10.1103/PhysRevD.95.065015} {\bibfield  {journal} {\bibinfo  {journal} {Phys. Rev. D}\ }\textbf {\bibinfo {volume} {95}},\ \bibinfo {pages} {065015} (\bibinfo {year} {2017})},\ \bibinfo {note} {[Erratum: Phys.Rev.D 97, 029901 (2018)]},\ \Eprint {http://arxiv.org/abs/1611.00056} {arXiv:1611.00056 [gr-qc]} \BibitemShut {NoStop}%
\bibitem [{\citenamefont {Frolov}\ and\ \citenamefont {Thorne}(1989)}]{Frolov:1989jh}%
  \BibitemOpen
  \bibfield  {author} {\bibinfo {author} {\bibfnamefont {V.~P.}\ \bibnamefont {Frolov}}\ and\ \bibinfo {author} {\bibfnamefont {K.~S.}\ \bibnamefont {Thorne}},\ }\href {\doibase 10.1103/PhysRevD.39.2125} {\bibfield  {journal} {\bibinfo  {journal} {Phys. Rev. D}\ }\textbf {\bibinfo {volume} {39}},\ \bibinfo {pages} {2125} (\bibinfo {year} {1989})}\BibitemShut {NoStop}%
\bibitem [{\citenamefont {Ottewill}\ and\ \citenamefont {Winstanley}(2000)}]{Ottewill:2000qh}%
  \BibitemOpen
  \bibfield  {author} {\bibinfo {author} {\bibfnamefont {A.~C.}\ \bibnamefont {Ottewill}}\ and\ \bibinfo {author} {\bibfnamefont {E.}~\bibnamefont {Winstanley}},\ }\href {\doibase 10.1103/PhysRevD.62.084018} {\bibfield  {journal} {\bibinfo  {journal} {Phys. Rev. D}\ }\textbf {\bibinfo {volume} {62}},\ \bibinfo {pages} {084018} (\bibinfo {year} {2000})},\ \Eprint {http://arxiv.org/abs/gr-qc/0004022} {arXiv:gr-qc/0004022} \BibitemShut {NoStop}%
\bibitem [{\citenamefont {Starobinskii}(1973)}]{Starobinskii:1973hgd}%
  \BibitemOpen
  \bibfield  {author} {\bibinfo {author} {\bibfnamefont {A.~A.}\ \bibnamefont {Starobinskii}},\ }\href@noop {} {\bibfield  {journal} {\bibinfo  {journal} {Sov. Phys. JETP}\ }\textbf {\bibinfo {volume} {64}},\ \bibinfo {pages} {48} (\bibinfo {year} {1973})}\BibitemShut {NoStop}%
\bibitem [{\citenamefont {Unruh}(1974)}]{Unruh:1974bw}%
  \BibitemOpen
  \bibfield  {author} {\bibinfo {author} {\bibfnamefont {W.~G.}\ \bibnamefont {Unruh}},\ }\href {\doibase 10.1103/PhysRevD.10.3194} {\bibfield  {journal} {\bibinfo  {journal} {Phys. Rev. D}\ }\textbf {\bibinfo {volume} {10}},\ \bibinfo {pages} {3194} (\bibinfo {year} {1974})}\BibitemShut {NoStop}%
\bibitem [{\citenamefont {Matacz}\ \emph {et~al.}(1993)\citenamefont {Matacz}, \citenamefont {Davies},\ and\ \citenamefont {Ottewill}}]{Matacz:1992gtm}%
  \BibitemOpen
  \bibfield  {author} {\bibinfo {author} {\bibfnamefont {A.~L.}\ \bibnamefont {Matacz}}, \bibinfo {author} {\bibfnamefont {P.~C.~W.}\ \bibnamefont {Davies}}, \ and\ \bibinfo {author} {\bibfnamefont {A.~C.}\ \bibnamefont {Ottewill}},\ }\href {\doibase 10.1103/PhysRevD.47.1557} {\bibfield  {journal} {\bibinfo  {journal} {Phys. Rev. D}\ }\textbf {\bibinfo {volume} {47}},\ \bibinfo {pages} {1557} (\bibinfo {year} {1993})},\ \Eprint {http://arxiv.org/abs/gr-qc/9212004} {arXiv:gr-qc/9212004} \BibitemShut {NoStop}%
\bibitem [{\citenamefont {Menezes}(2018)}]{Menezes:2017oeb}%
  \BibitemOpen
  \bibfield  {author} {\bibinfo {author} {\bibfnamefont {G.}~\bibnamefont {Menezes}},\ }\href {\doibase 10.1103/PhysRevD.97.085021} {\bibfield  {journal} {\bibinfo  {journal} {Phys. Rev. D}\ }\textbf {\bibinfo {volume} {97}},\ \bibinfo {pages} {085021} (\bibinfo {year} {2018})},\ \Eprint {http://arxiv.org/abs/1712.07151} {arXiv:1712.07151 [gr-qc]} \BibitemShut {NoStop}%
\bibitem [{\citenamefont {Liu}\ \emph {et~al.}(2018)\citenamefont {Liu}, \citenamefont {Tian}, \citenamefont {Wang},\ and\ \citenamefont {Jing}}]{Liu:2018zod}%
  \BibitemOpen
  \bibfield  {author} {\bibinfo {author} {\bibfnamefont {X.}~\bibnamefont {Liu}}, \bibinfo {author} {\bibfnamefont {Z.}~\bibnamefont {Tian}}, \bibinfo {author} {\bibfnamefont {J.}~\bibnamefont {Wang}}, \ and\ \bibinfo {author} {\bibfnamefont {J.}~\bibnamefont {Jing}},\ }\href {\doibase 10.1103/PhysRevD.97.105030} {\bibfield  {journal} {\bibinfo  {journal} {Phys. Rev. D}\ }\textbf {\bibinfo {volume} {97}},\ \bibinfo {pages} {105030} (\bibinfo {year} {2018})},\ \Eprint {http://arxiv.org/abs/1805.04470} {arXiv:1805.04470 [gr-qc]} \BibitemShut {NoStop}%
\bibitem [{\citenamefont {Zhou}\ and\ \citenamefont {Yu}(2010{\natexlab{a}})}]{Zhou:2010wn}%
  \BibitemOpen
  \bibfield  {author} {\bibinfo {author} {\bibfnamefont {W.}~\bibnamefont {Zhou}}\ and\ \bibinfo {author} {\bibfnamefont {H.~W.}\ \bibnamefont {Yu}},\ }\href {\doibase 10.1103/PhysRevD.82.104030} {\bibfield  {journal} {\bibinfo  {journal} {Phys. Rev. D}\ }\textbf {\bibinfo {volume} {82}},\ \bibinfo {pages} {104030} (\bibinfo {year} {2010}{\natexlab{a}})},\ \Eprint {http://arxiv.org/abs/1011.1619} {arXiv:1011.1619 [hep-th]} \BibitemShut {NoStop}%
\bibitem [{\citenamefont {Meschede}\ \emph {et~al.}(1990)\citenamefont {Meschede}, \citenamefont {Jhe},\ and\ \citenamefont {Hinds}}]{PhysRevA.41.1587}%
  \BibitemOpen
  \bibfield  {author} {\bibinfo {author} {\bibfnamefont {D.}~\bibnamefont {Meschede}}, \bibinfo {author} {\bibfnamefont {W.}~\bibnamefont {Jhe}}, \ and\ \bibinfo {author} {\bibfnamefont {E.~A.}\ \bibnamefont {Hinds}},\ }\href {\doibase 10.1103/PhysRevA.41.1587} {\bibfield  {journal} {\bibinfo  {journal} {Phys. Rev. A}\ }\textbf {\bibinfo {volume} {41}},\ \bibinfo {pages} {1587} (\bibinfo {year} {1990})}\BibitemShut {NoStop}%
\bibitem [{\citenamefont {Gibbons}\ and\ \citenamefont {Hawking}(1977)}]{Gibbons:1977mu}%
  \BibitemOpen
  \bibfield  {author} {\bibinfo {author} {\bibfnamefont {G.~W.}\ \bibnamefont {Gibbons}}\ and\ \bibinfo {author} {\bibfnamefont {S.~W.}\ \bibnamefont {Hawking}},\ }\href {\doibase 10.1103/PhysRevD.15.2738} {\bibfield  {journal} {\bibinfo  {journal} {Phys. Rev. D}\ }\textbf {\bibinfo {volume} {15}},\ \bibinfo {pages} {2738} (\bibinfo {year} {1977})}\BibitemShut {NoStop}%
\bibitem [{\citenamefont {Zhou}\ and\ \citenamefont {Yu}(2010{\natexlab{b}})}]{Zhou:2010nb}%
  \BibitemOpen
  \bibfield  {author} {\bibinfo {author} {\bibfnamefont {W.}~\bibnamefont {Zhou}}\ and\ \bibinfo {author} {\bibfnamefont {H.~W.}\ \bibnamefont {Yu}},\ }\href {\doibase 10.1103/PhysRevD.82.124067} {\bibfield  {journal} {\bibinfo  {journal} {Phys. Rev. D}\ }\textbf {\bibinfo {volume} {82}},\ \bibinfo {pages} {124067} (\bibinfo {year} {2010}{\natexlab{b}})},\ \Eprint {http://arxiv.org/abs/1012.4055} {arXiv:1012.4055 [hep-th]} \BibitemShut {NoStop}%
\bibitem [{\citenamefont {Zhou}\ and\ \citenamefont {Yu}(2012{\natexlab{c}})}]{Zhou:2012ba}%
  \BibitemOpen
  \bibfield  {author} {\bibinfo {author} {\bibfnamefont {W.}~\bibnamefont {Zhou}}\ and\ \bibinfo {author} {\bibfnamefont {H.}~\bibnamefont {Yu}},\ }\href {\doibase 10.1007/JHEP10(2012)172} {\bibfield  {journal} {\bibinfo  {journal} {JHEP}\ }\textbf {\bibinfo {volume} {10}},\ \bibinfo {pages} {172} (\bibinfo {year} {2012}{\natexlab{c}})},\ \Eprint {http://arxiv.org/abs/1204.2015} {arXiv:1204.2015 [gr-qc]} \BibitemShut {NoStop}%
\bibitem [{\citenamefont {Cheng}\ \emph {et~al.}(2019)\citenamefont {Cheng}, \citenamefont {Hu},\ and\ \citenamefont {Yu}}]{Cheng:2019tnk}%
  \BibitemOpen
  \bibfield  {author} {\bibinfo {author} {\bibfnamefont {S.}~\bibnamefont {Cheng}}, \bibinfo {author} {\bibfnamefont {J.}~\bibnamefont {Hu}}, \ and\ \bibinfo {author} {\bibfnamefont {H.}~\bibnamefont {Yu}},\ }\href {\doibase 10.1103/PhysRevD.100.025010} {\bibfield  {journal} {\bibinfo  {journal} {Phys. Rev. D}\ }\textbf {\bibinfo {volume} {100}},\ \bibinfo {pages} {025010} (\bibinfo {year} {2019})},\ \Eprint {http://arxiv.org/abs/1907.00715} {arXiv:1907.00715 [gr-qc]} \BibitemShut {NoStop}%
\bibitem [{\citenamefont {Cai}\ and\ \citenamefont {Ren}(2019)}]{Cai:2019pnw}%
  \BibitemOpen
  \bibfield  {author} {\bibinfo {author} {\bibfnamefont {H.}~\bibnamefont {Cai}}\ and\ \bibinfo {author} {\bibfnamefont {Z.}~\bibnamefont {Ren}},\ }\href {\doibase 10.1088/1361-6382/ab30d0} {\bibfield  {journal} {\bibinfo  {journal} {Class. Quant. Grav.}\ }\textbf {\bibinfo {volume} {36}},\ \bibinfo {pages} {165001} (\bibinfo {year} {2019})}\BibitemShut {NoStop}%
\bibitem [{\citenamefont {Hu}\ and\ \citenamefont {Yu}(2015)}]{Hu:2015lda}%
  \BibitemOpen
  \bibfield  {author} {\bibinfo {author} {\bibfnamefont {J.}~\bibnamefont {Hu}}\ and\ \bibinfo {author} {\bibfnamefont {H.}~\bibnamefont {Yu}},\ }\href {\doibase 10.1103/PhysRevA.91.012327} {\bibfield  {journal} {\bibinfo  {journal} {Phys. Rev. A}\ }\textbf {\bibinfo {volume} {91}},\ \bibinfo {pages} {012327} (\bibinfo {year} {2015})},\ \Eprint {http://arxiv.org/abs/1501.03321} {arXiv:1501.03321 [quant-ph]} \BibitemShut {NoStop}%
\bibitem [{\citenamefont {Chen}\ \emph {et~al.}(2022)\citenamefont {Chen}, \citenamefont {Hu},\ and\ \citenamefont {Yu}}]{Chen:2021evr}%
  \BibitemOpen
  \bibfield  {author} {\bibinfo {author} {\bibfnamefont {Y.}~\bibnamefont {Chen}}, \bibinfo {author} {\bibfnamefont {J.}~\bibnamefont {Hu}}, \ and\ \bibinfo {author} {\bibfnamefont {H.}~\bibnamefont {Yu}},\ }\href {\doibase 10.1103/PhysRevD.105.045013} {\bibfield  {journal} {\bibinfo  {journal} {Phys. Rev. D}\ }\textbf {\bibinfo {volume} {105}},\ \bibinfo {pages} {045013} (\bibinfo {year} {2022})},\ \Eprint {http://arxiv.org/abs/2110.01780} {arXiv:2110.01780 [quant-ph]} \BibitemShut {NoStop}%
\bibitem [{\citenamefont {Zhou}\ \emph {et~al.}(2021{\natexlab{a}})\citenamefont {Zhou}, \citenamefont {Hu},\ and\ \citenamefont {Yu}}]{Zhou:2021nyv}%
  \BibitemOpen
  \bibfield  {author} {\bibinfo {author} {\bibfnamefont {Y.}~\bibnamefont {Zhou}}, \bibinfo {author} {\bibfnamefont {J.}~\bibnamefont {Hu}}, \ and\ \bibinfo {author} {\bibfnamefont {H.}~\bibnamefont {Yu}},\ }\href {\doibase 10.1007/JHEP09(2021)088} {\bibfield  {journal} {\bibinfo  {journal} {JHEP}\ }\textbf {\bibinfo {volume} {09}},\ \bibinfo {pages} {088} (\bibinfo {year} {2021}{\natexlab{a}})},\ \Eprint {http://arxiv.org/abs/2105.14735} {arXiv:2105.14735 [gr-qc]} \BibitemShut {NoStop}%
\bibitem [{\citenamefont {Soares}\ \emph {et~al.}(2022)\citenamefont {Soares}, \citenamefont {Menezes},\ and\ \citenamefont {Svaiter}}]{Soares:2022sqc}%
  \BibitemOpen
  \bibfield  {author} {\bibinfo {author} {\bibfnamefont {M.~S.}\ \bibnamefont {Soares}}, \bibinfo {author} {\bibfnamefont {G.}~\bibnamefont {Menezes}}, \ and\ \bibinfo {author} {\bibfnamefont {N.~F.}\ \bibnamefont {Svaiter}},\ }\href {\doibase 10.1103/PhysRevA.106.062440} {\bibfield  {journal} {\bibinfo  {journal} {Phys. Rev. A}\ }\textbf {\bibinfo {volume} {106}},\ \bibinfo {pages} {062440} (\bibinfo {year} {2022})},\ \Eprint {http://arxiv.org/abs/2205.11628} {arXiv:2205.11628 [hep-th]} \BibitemShut {NoStop}%
\bibitem [{\citenamefont {Hu}\ and\ \citenamefont {Yu}(2011)}]{Hu:2011pd}%
  \BibitemOpen
  \bibfield  {author} {\bibinfo {author} {\bibfnamefont {J.}~\bibnamefont {Hu}}\ and\ \bibinfo {author} {\bibfnamefont {H.}~\bibnamefont {Yu}},\ }\href {\doibase 10.1007/JHEP08(2011)137} {\bibfield  {journal} {\bibinfo  {journal} {JHEP}\ }\textbf {\bibinfo {volume} {08}},\ \bibinfo {pages} {137} (\bibinfo {year} {2011})},\ \Eprint {http://arxiv.org/abs/1109.0335} {arXiv:1109.0335 [hep-th]} \BibitemShut {NoStop}%
\bibitem [{\citenamefont {He}\ \emph {et~al.}(2020)\citenamefont {He}, \citenamefont {Yu},\ and\ \citenamefont {Hu}}]{He:2020xhz}%
  \BibitemOpen
  \bibfield  {author} {\bibinfo {author} {\bibfnamefont {P.}~\bibnamefont {He}}, \bibinfo {author} {\bibfnamefont {H.}~\bibnamefont {Yu}}, \ and\ \bibinfo {author} {\bibfnamefont {J.}~\bibnamefont {Hu}},\ }\href {\doibase 10.1140/epjc/s10052-020-7663-x} {\bibfield  {journal} {\bibinfo  {journal} {Eur. Phys. J. C}\ }\textbf {\bibinfo {volume} {80}},\ \bibinfo {pages} {134} (\bibinfo {year} {2020})}\BibitemShut {NoStop}%
\bibitem [{\citenamefont {Huang}(2021)}]{Huang:2020gha}%
  \BibitemOpen
  \bibfield  {author} {\bibinfo {author} {\bibfnamefont {Z.}~\bibnamefont {Huang}},\ }\href {\doibase 10.1007/s11128-021-03119-8} {\bibfield  {journal} {\bibinfo  {journal} {Quant. Inf. Proc.}\ }\textbf {\bibinfo {volume} {20}},\ \bibinfo {pages} {173} (\bibinfo {year} {2021})},\ \Eprint {http://arxiv.org/abs/2003.02223} {arXiv:2003.02223 [quant-ph]} \BibitemShut {NoStop}%
\bibitem [{\citenamefont {Liu}\ \emph {et~al.}(2023{\natexlab{a}})\citenamefont {Liu}, \citenamefont {Tian},\ and\ \citenamefont {Jing}}]{Liu:2023lok}%
  \BibitemOpen
  \bibfield  {author} {\bibinfo {author} {\bibfnamefont {X.}~\bibnamefont {Liu}}, \bibinfo {author} {\bibfnamefont {Z.}~\bibnamefont {Tian}}, \ and\ \bibinfo {author} {\bibfnamefont {J.}~\bibnamefont {Jing}},\ }\href@noop {} {\  (\bibinfo {year} {2023}{\natexlab{a}})},\ \Eprint {http://arxiv.org/abs/2309.08135} {arXiv:2309.08135 [hep-th]} \BibitemShut {NoStop}%
\bibitem [{\citenamefont {Kukita}\ and\ \citenamefont {Nambu}(2017)}]{Kukita:2017tpa}%
  \BibitemOpen
  \bibfield  {author} {\bibinfo {author} {\bibfnamefont {S.}~\bibnamefont {Kukita}}\ and\ \bibinfo {author} {\bibfnamefont {Y.}~\bibnamefont {Nambu}},\ }\href {\doibase 10.1088/1361-6382/aa8e31} {\bibfield  {journal} {\bibinfo  {journal} {Class. Quant. Grav.}\ }\textbf {\bibinfo {volume} {34}},\ \bibinfo {pages} {235010} (\bibinfo {year} {2017})},\ \Eprint {http://arxiv.org/abs/1706.09175} {arXiv:1706.09175 [gr-qc]} \BibitemShut {NoStop}%
\bibitem [{\citenamefont {Yan}\ and\ \citenamefont {Zhang}(2022)}]{Yan:2022xgg}%
  \BibitemOpen
  \bibfield  {author} {\bibinfo {author} {\bibfnamefont {J.}~\bibnamefont {Yan}}\ and\ \bibinfo {author} {\bibfnamefont {B.}~\bibnamefont {Zhang}},\ }\href {\doibase 10.1007/JHEP10(2022)051} {\bibfield  {journal} {\bibinfo  {journal} {JHEP}\ }\textbf {\bibinfo {volume} {10}},\ \bibinfo {pages} {051} (\bibinfo {year} {2022})},\ \Eprint {http://arxiv.org/abs/2206.13681} {arXiv:2206.13681 [gr-qc]} \BibitemShut {NoStop}%
\bibitem [{\citenamefont {Yan}\ \emph {et~al.}(2023)\citenamefont {Yan}, \citenamefont {Zhang},\ and\ \citenamefont {Cai}}]{Yan:2023ruj}%
  \BibitemOpen
  \bibfield  {author} {\bibinfo {author} {\bibfnamefont {J.}~\bibnamefont {Yan}}, \bibinfo {author} {\bibfnamefont {B.}~\bibnamefont {Zhang}}, \ and\ \bibinfo {author} {\bibfnamefont {Q.}~\bibnamefont {Cai}},\ }\href@noop {} {\  (\bibinfo {year} {2023})},\ \Eprint {http://arxiv.org/abs/2311.04610} {arXiv:2311.04610 [hep-th]} \BibitemShut {NoStop}%
\bibitem [{\citenamefont {Salam}(2009)}]{10.5555/1822415}%
  \BibitemOpen
  \bibfield  {author} {\bibinfo {author} {\bibfnamefont {A.}~\bibnamefont {Salam}},\ }\href@noop {} {\emph {\bibinfo {title} {Molecular Quantum Electrodynamics: Long-Range Intermolecular Interactions}}}\ (\bibinfo  {publisher} {Wiley Publishing},\ \bibinfo {year} {2009})\BibitemShut {NoStop}%
\bibitem [{\citenamefont {{Salam}}(2008)}]{2008IRPC...27..405S}%
  \BibitemOpen
  \bibfield  {author} {\bibinfo {author} {\bibfnamefont {A.}~\bibnamefont {{Salam}}},\ }\href {\doibase 10.1080/01442350802045206} {\bibfield  {journal} {\bibinfo  {journal} {International Reviews in Physical Chemistry}\ }\textbf {\bibinfo {volume} {27}},\ \bibinfo {pages} {405} (\bibinfo {year} {2008})}\BibitemShut {NoStop}%
\bibitem [{\citenamefont {Fassioli}\ and\ \citenamefont {Olaya-Castro}(2010)}]{Fassioli_2010}%
  \BibitemOpen
  \bibfield  {author} {\bibinfo {author} {\bibfnamefont {F.}~\bibnamefont {Fassioli}}\ and\ \bibinfo {author} {\bibfnamefont {A.}~\bibnamefont {Olaya-Castro}},\ }\href {\doibase 10.1088/1367-2630/12/8/085006} {\bibfield  {journal} {\bibinfo  {journal} {New Journal of Physics}\ }\textbf {\bibinfo {volume} {12}},\ \bibinfo {pages} {085006} (\bibinfo {year} {2010})}\BibitemShut {NoStop}%
\bibitem [{\citenamefont {Preto}\ and\ \citenamefont {Pettini}(2013)}]{Preto2013ResonantLI}%
  \BibitemOpen
  \bibfield  {author} {\bibinfo {author} {\bibfnamefont {J.}~\bibnamefont {Preto}}\ and\ \bibinfo {author} {\bibfnamefont {M.}~\bibnamefont {Pettini}},\ }\href@noop {} {\bibfield  {journal} {\bibinfo  {journal} {Physics Letters A}\ }\textbf {\bibinfo {volume} {377}},\ \bibinfo {pages} {587} (\bibinfo {year} {2013})}\BibitemShut {NoStop}%
\bibitem [{\citenamefont {Galego}\ \emph {et~al.}(2019)\citenamefont {Galego}, \citenamefont {Climent}, \citenamefont {Garcia-Vidal},\ and\ \citenamefont {Feist}}]{Galego_2019}%
  \BibitemOpen
  \bibfield  {author} {\bibinfo {author} {\bibfnamefont {J.}~\bibnamefont {Galego}}, \bibinfo {author} {\bibfnamefont {C.}~\bibnamefont {Climent}}, \bibinfo {author} {\bibfnamefont {F.~J.}\ \bibnamefont {Garcia-Vidal}}, \ and\ \bibinfo {author} {\bibfnamefont {J.}~\bibnamefont {Feist}},\ }\href {\doibase 10.1103/physrevx.9.021057} {\bibfield  {journal} {\bibinfo  {journal} {Physical Review X}\ }\textbf {\bibinfo {volume} {9}} (\bibinfo {year} {2019}),\ 10.1103/physrevx.9.021057}\BibitemShut {NoStop}%
\bibitem [{\citenamefont {Fiscelli}\ \emph {et~al.}(2020)\citenamefont {Fiscelli}, \citenamefont {Rizzuto},\ and\ \citenamefont {Passante}}]{Fiscelli:2019ywl}%
  \BibitemOpen
  \bibfield  {author} {\bibinfo {author} {\bibfnamefont {G.}~\bibnamefont {Fiscelli}}, \bibinfo {author} {\bibfnamefont {L.}~\bibnamefont {Rizzuto}}, \ and\ \bibinfo {author} {\bibfnamefont {R.}~\bibnamefont {Passante}},\ }\href {\doibase 10.1103/PhysRevLett.124.013604} {\bibfield  {journal} {\bibinfo  {journal} {Phys. Rev. Lett.}\ }\textbf {\bibinfo {volume} {124}},\ \bibinfo {pages} {013604} (\bibinfo {year} {2020})},\ \Eprint {http://arxiv.org/abs/1909.03517} {arXiv:1909.03517 [quant-ph]} \BibitemShut {NoStop}%
\bibitem [{\citenamefont {Andrews}\ \emph {et~al.}(1989)\citenamefont {Andrews}, \citenamefont {Craig},\ and\ \citenamefont {Thirunamachandran}}]{doi:10.1080/01442358909353233}%
  \BibitemOpen
  \bibfield  {author} {\bibinfo {author} {\bibfnamefont {D.~L.}\ \bibnamefont {Andrews}}, \bibinfo {author} {\bibfnamefont {D.~P.}\ \bibnamefont {Craig}}, \ and\ \bibinfo {author} {\bibfnamefont {T.}~\bibnamefont {Thirunamachandran}},\ }\href {\doibase 10.1080/01442358909353233} {\bibfield  {journal} {\bibinfo  {journal} {International Reviews in Physical Chemistry}\ }\textbf {\bibinfo {volume} {8}},\ \bibinfo {pages} {339} (\bibinfo {year} {1989})},\ \Eprint {http://arxiv.org/abs/https://doi.org/10.1080/01442358909353233} {https://doi.org/10.1080/01442358909353233} \BibitemShut {NoStop}%
\bibitem [{\citenamefont {{Casimir}}\ and\ \citenamefont {{Polder}}(1948)}]{1948PhRv...73..360C}%
  \BibitemOpen
  \bibfield  {author} {\bibinfo {author} {\bibfnamefont {H.~B.}\ \bibnamefont {{Casimir}}}\ and\ \bibinfo {author} {\bibfnamefont {D.}~\bibnamefont {{Polder}}},\ }\href {\doibase 10.1103/PhysRev.73.360} {\bibfield  {journal} {\bibinfo  {journal} {Physical Review}\ }\textbf {\bibinfo {volume} {73}},\ \bibinfo {pages} {360} (\bibinfo {year} {1948})}\BibitemShut {NoStop}%
\bibitem [{\citenamefont {Babb}(2010)}]{Babb_2010}%
  \BibitemOpen
  \bibfield  {author} {\bibinfo {author} {\bibfnamefont {J.~F.}\ \bibnamefont {Babb}},\ }in\ \href {\doibase 10.1016/s1049-250x(10)59001-3} {\emph {\bibinfo {booktitle} {Advances In Atomic, Molecular, and Optical Physics}}}\ (\bibinfo  {publisher} {Elsevier},\ \bibinfo {year} {2010})\ pp.\ \bibinfo {pages} {1--20}\BibitemShut {NoStop}%
\bibitem [{\citenamefont {Zhang}\ and\ \citenamefont {Yu}(2011)}]{Zhang:2011vsa}%
  \BibitemOpen
  \bibfield  {author} {\bibinfo {author} {\bibfnamefont {J.}~\bibnamefont {Zhang}}\ and\ \bibinfo {author} {\bibfnamefont {H.}~\bibnamefont {Yu}},\ }\href {\doibase 10.1103/PhysRevA.84.042103} {\bibfield  {journal} {\bibinfo  {journal} {Phys. Rev. A}\ }\textbf {\bibinfo {volume} {84}},\ \bibinfo {pages} {042103} (\bibinfo {year} {2011})},\ \Eprint {http://arxiv.org/abs/1109.4704} {arXiv:1109.4704 [quant-ph]} \BibitemShut {NoStop}%
\bibitem [{\citenamefont {Zhang}\ and\ \citenamefont {Yu}(2013)}]{Zhang:2013txe}%
  \BibitemOpen
  \bibfield  {author} {\bibinfo {author} {\bibfnamefont {J.}~\bibnamefont {Zhang}}\ and\ \bibinfo {author} {\bibfnamefont {H.}~\bibnamefont {Yu}},\ }\href {\doibase 10.1103/PhysRevA.88.064501} {\bibfield  {journal} {\bibinfo  {journal} {Phys. Rev. A}\ }\textbf {\bibinfo {volume} {88}},\ \bibinfo {pages} {064501} (\bibinfo {year} {2013})},\ \Eprint {http://arxiv.org/abs/1401.4838} {arXiv:1401.4838 [gr-qc]} \BibitemShut {NoStop}%
\bibitem [{\citenamefont {Noto}\ and\ \citenamefont {Passante}(2013)}]{Noto:2013ona}%
  \BibitemOpen
  \bibfield  {author} {\bibinfo {author} {\bibfnamefont {A.}~\bibnamefont {Noto}}\ and\ \bibinfo {author} {\bibfnamefont {R.}~\bibnamefont {Passante}},\ }\href {\doibase 10.1103/PhysRevD.88.025041} {\bibfield  {journal} {\bibinfo  {journal} {Phys. Rev. D}\ }\textbf {\bibinfo {volume} {88}},\ \bibinfo {pages} {025041} (\bibinfo {year} {2013})},\ \Eprint {http://arxiv.org/abs/1304.5786} {arXiv:1304.5786 [quant-ph]} \BibitemShut {NoStop}%
\bibitem [{\citenamefont {Marino}\ \emph {et~al.}(2014)\citenamefont {Marino}, \citenamefont {Noto},\ and\ \citenamefont {Passante}}]{Marino:2014rfa}%
  \BibitemOpen
  \bibfield  {author} {\bibinfo {author} {\bibfnamefont {J.}~\bibnamefont {Marino}}, \bibinfo {author} {\bibfnamefont {A.}~\bibnamefont {Noto}}, \ and\ \bibinfo {author} {\bibfnamefont {R.}~\bibnamefont {Passante}},\ }\href {\doibase 10.1103/PhysRevLett.113.020403} {\bibfield  {journal} {\bibinfo  {journal} {Phys. Rev. Lett.}\ }\textbf {\bibinfo {volume} {113}},\ \bibinfo {pages} {020403} (\bibinfo {year} {2014})},\ \Eprint {http://arxiv.org/abs/1403.2437} {arXiv:1403.2437 [quant-ph]} \BibitemShut {NoStop}%
\bibitem [{\citenamefont {{Barton}}(2001)}]{2001PhRvA..64c2102B}%
  \BibitemOpen
  \bibfield  {author} {\bibinfo {author} {\bibfnamefont {G.}~\bibnamefont {{Barton}}},\ }\href {\doibase 10.1103/PhysRevA.64.032102} {\bibfield  {journal} {\bibinfo  {journal} {\pra}\ }\textbf {\bibinfo {volume} {64}},\ \bibinfo {eid} {032102} (\bibinfo {year} {2001})}\BibitemShut {NoStop}%
\bibitem [{\citenamefont {Singleton}\ and\ \citenamefont {Wilburn}(2011)}]{Singleton:2011vh}%
  \BibitemOpen
  \bibfield  {author} {\bibinfo {author} {\bibfnamefont {D.}~\bibnamefont {Singleton}}\ and\ \bibinfo {author} {\bibfnamefont {S.}~\bibnamefont {Wilburn}},\ }\href {\doibase 10.1103/PhysRevLett.107.081102} {\bibfield  {journal} {\bibinfo  {journal} {Phys. Rev. Lett.}\ }\textbf {\bibinfo {volume} {107}},\ \bibinfo {pages} {081102} (\bibinfo {year} {2011})},\ \Eprint {http://arxiv.org/abs/1102.5564} {arXiv:1102.5564 [gr-qc]} \BibitemShut {NoStop}%
\bibitem [{\citenamefont {Smerlak}\ and\ \citenamefont {Singh}(2013)}]{Smerlak:2013sga}%
  \BibitemOpen
  \bibfield  {author} {\bibinfo {author} {\bibfnamefont {M.}~\bibnamefont {Smerlak}}\ and\ \bibinfo {author} {\bibfnamefont {S.}~\bibnamefont {Singh}},\ }\href {\doibase 10.1103/PhysRevD.88.104023} {\bibfield  {journal} {\bibinfo  {journal} {Phys. Rev. D}\ }\textbf {\bibinfo {volume} {88}},\ \bibinfo {pages} {104023} (\bibinfo {year} {2013})},\ \Eprint {http://arxiv.org/abs/1304.2858} {arXiv:1304.2858 [gr-qc]} \BibitemShut {NoStop}%
\bibitem [{\citenamefont {Hodgkinson}\ \emph {et~al.}(2014)\citenamefont {Hodgkinson}, \citenamefont {Louko},\ and\ \citenamefont {Ottewill}}]{Hodgkinson:2014iua}%
  \BibitemOpen
  \bibfield  {author} {\bibinfo {author} {\bibfnamefont {L.}~\bibnamefont {Hodgkinson}}, \bibinfo {author} {\bibfnamefont {J.}~\bibnamefont {Louko}}, \ and\ \bibinfo {author} {\bibfnamefont {A.~C.}\ \bibnamefont {Ottewill}},\ }\href {\doibase 10.1103/PhysRevD.89.104002} {\bibfield  {journal} {\bibinfo  {journal} {Phys. Rev. D}\ }\textbf {\bibinfo {volume} {89}},\ \bibinfo {pages} {104002} (\bibinfo {year} {2014})},\ \Eprint {http://arxiv.org/abs/1401.2667} {arXiv:1401.2667 [gr-qc]} \BibitemShut {NoStop}%
\bibitem [{\citenamefont {Singha}(2019)}]{Singha:2018vaj}%
  \BibitemOpen
  \bibfield  {author} {\bibinfo {author} {\bibfnamefont {C.}~\bibnamefont {Singha}},\ }\href {\doibase 10.1142/S0217732319503565} {\bibfield  {journal} {\bibinfo  {journal} {Mod. Phys. Lett. A}\ }\textbf {\bibinfo {volume} {35}},\ \bibinfo {pages} {1950356} (\bibinfo {year} {2019})},\ \Eprint {http://arxiv.org/abs/1808.07041} {arXiv:1808.07041 [gr-qc]} \BibitemShut {NoStop}%
\bibitem [{\citenamefont {Menezes}\ \emph {et~al.}(2017)\citenamefont {Menezes}, \citenamefont {Kiefer},\ and\ \citenamefont {Marino}}]{Menezes:2017akp}%
  \BibitemOpen
  \bibfield  {author} {\bibinfo {author} {\bibfnamefont {G.}~\bibnamefont {Menezes}}, \bibinfo {author} {\bibfnamefont {C.}~\bibnamefont {Kiefer}}, \ and\ \bibinfo {author} {\bibfnamefont {J.}~\bibnamefont {Marino}},\ }\href {\doibase 10.1103/PhysRevD.95.085014} {\bibfield  {journal} {\bibinfo  {journal} {Phys. Rev. D}\ }\textbf {\bibinfo {volume} {95}},\ \bibinfo {pages} {085014} (\bibinfo {year} {2017})},\ \Eprint {http://arxiv.org/abs/1703.00193} {arXiv:1703.00193 [gr-qc]} \BibitemShut {NoStop}%
\bibitem [{\citenamefont {Ford}\ \emph {et~al.}(2016)\citenamefont {Ford}, \citenamefont {Hertzberg},\ and\ \citenamefont {Karouby}}]{Ford:2015wls}%
  \BibitemOpen
  \bibfield  {author} {\bibinfo {author} {\bibfnamefont {L.~H.}\ \bibnamefont {Ford}}, \bibinfo {author} {\bibfnamefont {M.~P.}\ \bibnamefont {Hertzberg}}, \ and\ \bibinfo {author} {\bibfnamefont {J.}~\bibnamefont {Karouby}},\ }\href {\doibase 10.1103/PhysRevLett.116.151301} {\bibfield  {journal} {\bibinfo  {journal} {Phys. Rev. Lett.}\ }\textbf {\bibinfo {volume} {116}},\ \bibinfo {pages} {151301} (\bibinfo {year} {2016})},\ \Eprint {http://arxiv.org/abs/1512.07632} {arXiv:1512.07632 [hep-th]} \BibitemShut {NoStop}%
\bibitem [{\citenamefont {Wu}\ \emph {et~al.}(2016)\citenamefont {Wu}, \citenamefont {Hu},\ and\ \citenamefont {Yu}}]{Wu:2016esf}%
  \BibitemOpen
  \bibfield  {author} {\bibinfo {author} {\bibfnamefont {P.}~\bibnamefont {Wu}}, \bibinfo {author} {\bibfnamefont {J.}~\bibnamefont {Hu}}, \ and\ \bibinfo {author} {\bibfnamefont {H.}~\bibnamefont {Yu}},\ }\href {\doibase 10.1016/j.physletb.2016.10.025} {\bibfield  {journal} {\bibinfo  {journal} {Phys. Lett. B}\ }\textbf {\bibinfo {volume} {763}},\ \bibinfo {pages} {40} (\bibinfo {year} {2016})},\ \Eprint {http://arxiv.org/abs/1607.04929} {arXiv:1607.04929 [gr-qc]} \BibitemShut {NoStop}%
\bibitem [{\citenamefont {Hu}\ and\ \citenamefont {Yu}(2017)}]{Hu:2016lev}%
  \BibitemOpen
  \bibfield  {author} {\bibinfo {author} {\bibfnamefont {J.}~\bibnamefont {Hu}}\ and\ \bibinfo {author} {\bibfnamefont {H.}~\bibnamefont {Yu}},\ }\href {\doibase 10.1016/j.physletb.2017.01.038} {\bibfield  {journal} {\bibinfo  {journal} {Phys. Lett. B}\ }\textbf {\bibinfo {volume} {767}},\ \bibinfo {pages} {16} (\bibinfo {year} {2017})},\ \Eprint {http://arxiv.org/abs/1605.02193} {arXiv:1605.02193 [hep-th]} \BibitemShut {NoStop}%
\bibitem [{\citenamefont {Huang}(2019)}]{Huang:2019tuf}%
  \BibitemOpen
  \bibfield  {author} {\bibinfo {author} {\bibfnamefont {Z.}~\bibnamefont {Huang}},\ }\href {\doibase 10.1088/1361-6382/ab2e41} {\bibfield  {journal} {\bibinfo  {journal} {Class. Quant. Grav.}\ }\textbf {\bibinfo {volume} {36}},\ \bibinfo {pages} {155001} (\bibinfo {year} {2019})}\BibitemShut {NoStop}%
\bibitem [{\citenamefont {Hu}\ \emph {et~al.}(2020)\citenamefont {Hu}, \citenamefont {Hu},\ and\ \citenamefont {Yu}}]{Hu:2020cvd}%
  \BibitemOpen
  \bibfield  {author} {\bibinfo {author} {\bibfnamefont {Y.}~\bibnamefont {Hu}}, \bibinfo {author} {\bibfnamefont {J.}~\bibnamefont {Hu}}, \ and\ \bibinfo {author} {\bibfnamefont {H.}~\bibnamefont {Yu}},\ }\href {\doibase 10.1103/PhysRevD.101.066015} {\bibfield  {journal} {\bibinfo  {journal} {Phys. Rev. D}\ }\textbf {\bibinfo {volume} {101}},\ \bibinfo {pages} {066015} (\bibinfo {year} {2020})},\ \Eprint {http://arxiv.org/abs/2006.06354} {arXiv:2006.06354 [gr-qc]} \BibitemShut {NoStop}%
\bibitem [{\citenamefont {Zhou}\ \emph {et~al.}(2021{\natexlab{b}})\citenamefont {Zhou}, \citenamefont {Cheng},\ and\ \citenamefont {Yu}}]{Zhou:2020kvi}%
  \BibitemOpen
  \bibfield  {author} {\bibinfo {author} {\bibfnamefont {W.}~\bibnamefont {Zhou}}, \bibinfo {author} {\bibfnamefont {S.}~\bibnamefont {Cheng}}, \ and\ \bibinfo {author} {\bibfnamefont {H.}~\bibnamefont {Yu}},\ }\href {\doibase 10.1103/PhysRevA.103.012227} {\bibfield  {journal} {\bibinfo  {journal} {Phys. Rev. A}\ }\textbf {\bibinfo {volume} {103}},\ \bibinfo {pages} {012227} (\bibinfo {year} {2021}{\natexlab{b}})},\ \Eprint {http://arxiv.org/abs/2011.11481} {arXiv:2011.11481 [quant-ph]} \BibitemShut {NoStop}%
\bibitem [{\citenamefont {Cheng}\ \emph {et~al.}(2022)\citenamefont {Cheng}, \citenamefont {Zhou},\ and\ \citenamefont {Yu}}]{Cheng:2022xkk}%
  \BibitemOpen
  \bibfield  {author} {\bibinfo {author} {\bibfnamefont {S.}~\bibnamefont {Cheng}}, \bibinfo {author} {\bibfnamefont {W.}~\bibnamefont {Zhou}}, \ and\ \bibinfo {author} {\bibfnamefont {H.}~\bibnamefont {Yu}},\ }\href {\doibase 10.1016/j.physletb.2022.137440} {\bibfield  {journal} {\bibinfo  {journal} {Phys. Lett. B}\ }\textbf {\bibinfo {volume} {834}},\ \bibinfo {pages} {137440} (\bibinfo {year} {2022})},\ \Eprint {http://arxiv.org/abs/2205.11086} {arXiv:2205.11086 [hep-th]} \BibitemShut {NoStop}%
\bibitem [{\citenamefont {Zhou}\ \emph {et~al.}(2023)\citenamefont {Zhou}, \citenamefont {Cheng},\ and\ \citenamefont {Yu}}]{Zhou:2022jkg}%
  \BibitemOpen
  \bibfield  {author} {\bibinfo {author} {\bibfnamefont {W.}~\bibnamefont {Zhou}}, \bibinfo {author} {\bibfnamefont {S.}~\bibnamefont {Cheng}}, \ and\ \bibinfo {author} {\bibfnamefont {H.}~\bibnamefont {Yu}},\ }\href {\doibase 10.1016/j.physletb.2023.138097} {\bibfield  {journal} {\bibinfo  {journal} {Phys. Lett. B}\ }\textbf {\bibinfo {volume} {844}},\ \bibinfo {pages} {138097} (\bibinfo {year} {2023})},\ \Eprint {http://arxiv.org/abs/2211.14747} {arXiv:2211.14747 [gr-qc]} \BibitemShut {NoStop}%
\bibitem [{\citenamefont {Dicke}(1954)}]{PhysRev.93.99}%
  \BibitemOpen
  \bibfield  {author} {\bibinfo {author} {\bibfnamefont {R.~H.}\ \bibnamefont {Dicke}},\ }\href {\doibase 10.1103/PhysRev.93.99} {\bibfield  {journal} {\bibinfo  {journal} {Phys. Rev.}\ }\textbf {\bibinfo {volume} {93}},\ \bibinfo {pages} {99} (\bibinfo {year} {1954})}\BibitemShut {NoStop}%
\bibitem [{\citenamefont {Scully}\ \emph {et~al.}(2006)\citenamefont {Scully}, \citenamefont {Fry}, \citenamefont {Ooi},\ and\ \citenamefont {W\'odkiewicz}}]{PhysRevLett.96.010501}%
  \BibitemOpen
  \bibfield  {author} {\bibinfo {author} {\bibfnamefont {M.~O.}\ \bibnamefont {Scully}}, \bibinfo {author} {\bibfnamefont {E.~S.}\ \bibnamefont {Fry}}, \bibinfo {author} {\bibfnamefont {C.~H.~R.}\ \bibnamefont {Ooi}}, \ and\ \bibinfo {author} {\bibfnamefont {K.}~\bibnamefont {W\'odkiewicz}},\ }\href {\doibase 10.1103/PhysRevLett.96.010501} {\bibfield  {journal} {\bibinfo  {journal} {Phys. Rev. Lett.}\ }\textbf {\bibinfo {volume} {96}},\ \bibinfo {pages} {010501} (\bibinfo {year} {2006})}\BibitemShut {NoStop}%
\bibitem [{\citenamefont {{Raymond Ooi}}\ \emph {et~al.}(2007)\citenamefont {{Raymond Ooi}}, \citenamefont {{Rostovtsev}},\ and\ \citenamefont {{Scully}}}]{2007LaPhy..17..956R}%
  \BibitemOpen
  \bibfield  {author} {\bibinfo {author} {\bibfnamefont {C.~H.}\ \bibnamefont {{Raymond Ooi}}}, \bibinfo {author} {\bibfnamefont {Y.}~\bibnamefont {{Rostovtsev}}}, \ and\ \bibinfo {author} {\bibfnamefont {M.~O.}\ \bibnamefont {{Scully}}},\ }\href {\doibase 10.1134/S1054660X07070092} {\bibfield  {journal} {\bibinfo  {journal} {Laser Physics}\ }\textbf {\bibinfo {volume} {17}},\ \bibinfo {pages} {956} (\bibinfo {year} {2007})}\BibitemShut {NoStop}%
\bibitem [{\citenamefont {Scully}(2009)}]{PhysRevLett.102.143601}%
  \BibitemOpen
  \bibfield  {author} {\bibinfo {author} {\bibfnamefont {M.~O.}\ \bibnamefont {Scully}},\ }\href {\doibase 10.1103/PhysRevLett.102.143601} {\bibfield  {journal} {\bibinfo  {journal} {Phys. Rev. Lett.}\ }\textbf {\bibinfo {volume} {102}},\ \bibinfo {pages} {143601} (\bibinfo {year} {2009})}\BibitemShut {NoStop}%
\bibitem [{\citenamefont {Juzeliūnas}\ and\ \citenamefont {Andrews}(2000)}]{doi:https://doi.org/10.1002/9780470141717.ch4}%
  \BibitemOpen
  \bibfield  {author} {\bibinfo {author} {\bibfnamefont {G.}~\bibnamefont {Juzeliūnas}}\ and\ \bibinfo {author} {\bibfnamefont {D.~L.}\ \bibnamefont {Andrews}},\ }\enquote {\bibinfo {title} {Quantum electrodynamics of resonance energy transfer},}\ in\ \href {\doibase https://doi.org/10.1002/9780470141717.ch4} {\emph {\bibinfo {booktitle} {Advances in Chemical Physics}}}\ (\bibinfo  {publisher} {John Wiley \& Sons, Ltd},\ \bibinfo {year} {2000})\ pp.\ \bibinfo {pages} {357--410},\ \Eprint {http://arxiv.org/abs/https://onlinelibrary.wiley.com/doi/pdf/10.1002/9780470141717.ch4} {https://onlinelibrary.wiley.com/doi/pdf/10.1002/9780470141717.ch4} \BibitemShut {NoStop}%
\bibitem [{\citenamefont {Phillips}(1998)}]{RevModPhys.70.721}%
  \BibitemOpen
  \bibfield  {author} {\bibinfo {author} {\bibfnamefont {W.~D.}\ \bibnamefont {Phillips}},\ }\href {\doibase 10.1103/RevModPhys.70.721} {\bibfield  {journal} {\bibinfo  {journal} {Rev. Mod. Phys.}\ }\textbf {\bibinfo {volume} {70}},\ \bibinfo {pages} {721} (\bibinfo {year} {1998})}\BibitemShut {NoStop}%
\bibitem [{\citenamefont {Brennen}\ \emph {et~al.}(2000)\citenamefont {Brennen}, \citenamefont {Deutsch},\ and\ \citenamefont {Jessen}}]{PhysRevA.61.062309}%
  \BibitemOpen
  \bibfield  {author} {\bibinfo {author} {\bibfnamefont {G.~K.}\ \bibnamefont {Brennen}}, \bibinfo {author} {\bibfnamefont {I.~H.}\ \bibnamefont {Deutsch}}, \ and\ \bibinfo {author} {\bibfnamefont {P.~S.}\ \bibnamefont {Jessen}},\ }\href {\doibase 10.1103/PhysRevA.61.062309} {\bibfield  {journal} {\bibinfo  {journal} {Phys. Rev. A}\ }\textbf {\bibinfo {volume} {61}},\ \bibinfo {pages} {062309} (\bibinfo {year} {2000})}\BibitemShut {NoStop}%
\bibitem [{\citenamefont {Berman}(2015)}]{PhysRevA.91.042127}%
  \BibitemOpen
  \bibfield  {author} {\bibinfo {author} {\bibfnamefont {P.~R.}\ \bibnamefont {Berman}},\ }\href {\doibase 10.1103/PhysRevA.91.042127} {\bibfield  {journal} {\bibinfo  {journal} {Phys. Rev. A}\ }\textbf {\bibinfo {volume} {91}},\ \bibinfo {pages} {042127} (\bibinfo {year} {2015})}\BibitemShut {NoStop}%
\bibitem [{\citenamefont {Milonni}\ and\ \citenamefont {Rafsanjani}(2015)}]{PhysRevA.92.062711}%
  \BibitemOpen
  \bibfield  {author} {\bibinfo {author} {\bibfnamefont {P.~W.}\ \bibnamefont {Milonni}}\ and\ \bibinfo {author} {\bibfnamefont {S.~M.~H.}\ \bibnamefont {Rafsanjani}},\ }\href {\doibase 10.1103/PhysRevA.92.062711} {\bibfield  {journal} {\bibinfo  {journal} {Phys. Rev. A}\ }\textbf {\bibinfo {volume} {92}},\ \bibinfo {pages} {062711} (\bibinfo {year} {2015})}\BibitemShut {NoStop}%
\bibitem [{\citenamefont {Donaire}\ \emph {et~al.}(2015)\citenamefont {Donaire}, \citenamefont {Gu\'erout},\ and\ \citenamefont {Lambrecht}}]{PhysRevLett.115.033201}%
  \BibitemOpen
  \bibfield  {author} {\bibinfo {author} {\bibfnamefont {M.}~\bibnamefont {Donaire}}, \bibinfo {author} {\bibfnamefont {R.}~\bibnamefont {Gu\'erout}}, \ and\ \bibinfo {author} {\bibfnamefont {A.}~\bibnamefont {Lambrecht}},\ }\href {\doibase 10.1103/PhysRevLett.115.033201} {\bibfield  {journal} {\bibinfo  {journal} {Phys. Rev. Lett.}\ }\textbf {\bibinfo {volume} {115}},\ \bibinfo {pages} {033201} (\bibinfo {year} {2015})}\BibitemShut {NoStop}%
\bibitem [{\citenamefont {Jentschura}\ \emph {et~al.}(2017)\citenamefont {Jentschura}, \citenamefont {Adhikari},\ and\ \citenamefont {Debierre}}]{PhysRevLett.118.123001}%
  \BibitemOpen
  \bibfield  {author} {\bibinfo {author} {\bibfnamefont {U.~D.}\ \bibnamefont {Jentschura}}, \bibinfo {author} {\bibfnamefont {C.~M.}\ \bibnamefont {Adhikari}}, \ and\ \bibinfo {author} {\bibfnamefont {V.}~\bibnamefont {Debierre}},\ }\href {\doibase 10.1103/PhysRevLett.118.123001} {\bibfield  {journal} {\bibinfo  {journal} {Phys. Rev. Lett.}\ }\textbf {\bibinfo {volume} {118}},\ \bibinfo {pages} {123001} (\bibinfo {year} {2017})}\BibitemShut {NoStop}%
\bibitem [{\citenamefont {Rizzuto}\ \emph {et~al.}(2016)\citenamefont {Rizzuto}, \citenamefont {Lattuca}, \citenamefont {Marino}, \citenamefont {Noto}, \citenamefont {Spagnolo}, \citenamefont {Zhou},\ and\ \citenamefont {Passante}}]{PhysRevA.94.012121}%
  \BibitemOpen
  \bibfield  {author} {\bibinfo {author} {\bibfnamefont {L.}~\bibnamefont {Rizzuto}}, \bibinfo {author} {\bibfnamefont {M.}~\bibnamefont {Lattuca}}, \bibinfo {author} {\bibfnamefont {J.}~\bibnamefont {Marino}}, \bibinfo {author} {\bibfnamefont {A.}~\bibnamefont {Noto}}, \bibinfo {author} {\bibfnamefont {S.}~\bibnamefont {Spagnolo}}, \bibinfo {author} {\bibfnamefont {W.}~\bibnamefont {Zhou}}, \ and\ \bibinfo {author} {\bibfnamefont {R.}~\bibnamefont {Passante}},\ }\href {\doibase 10.1103/PhysRevA.94.012121} {\bibfield  {journal} {\bibinfo  {journal} {Phys. Rev. A}\ }\textbf {\bibinfo {volume} {94}},\ \bibinfo {pages} {012121} (\bibinfo {year} {2016})}\BibitemShut {NoStop}%
\bibitem [{\citenamefont {Zhou}\ \emph {et~al.}(2016)\citenamefont {Zhou}, \citenamefont {Passante},\ and\ \citenamefont {Rizzuto}}]{Zhou:2016urt}%
  \BibitemOpen
  \bibfield  {author} {\bibinfo {author} {\bibfnamefont {W.}~\bibnamefont {Zhou}}, \bibinfo {author} {\bibfnamefont {R.}~\bibnamefont {Passante}}, \ and\ \bibinfo {author} {\bibfnamefont {L.}~\bibnamefont {Rizzuto}},\ }\href {\doibase 10.1103/PhysRevD.94.105025} {\bibfield  {journal} {\bibinfo  {journal} {Phys. Rev. D}\ }\textbf {\bibinfo {volume} {94}},\ \bibinfo {pages} {105025} (\bibinfo {year} {2016})},\ \Eprint {http://arxiv.org/abs/1609.06931} {arXiv:1609.06931 [quant-ph]} \BibitemShut {NoStop}%
\bibitem [{\citenamefont {Zhou}\ \emph {et~al.}(2018)\citenamefont {Zhou}, \citenamefont {Passante},\ and\ \citenamefont {Rizzuto}}]{Zhou:2018igi}%
  \BibitemOpen
  \bibfield  {author} {\bibinfo {author} {\bibfnamefont {W.}~\bibnamefont {Zhou}}, \bibinfo {author} {\bibfnamefont {R.}~\bibnamefont {Passante}}, \ and\ \bibinfo {author} {\bibfnamefont {L.}~\bibnamefont {Rizzuto}},\ }\href {\doibase 10.3390/sym10060185} {\bibfield  {journal} {\bibinfo  {journal} {Symmetry}\ }\textbf {\bibinfo {volume} {10}},\ \bibinfo {pages} {185} (\bibinfo {year} {2018})}\BibitemShut {NoStop}%
\bibitem [{\citenamefont {Zhou}\ and\ \citenamefont {Yu}(2017)}]{Zhou:2017axh}%
  \BibitemOpen
  \bibfield  {author} {\bibinfo {author} {\bibfnamefont {W.}~\bibnamefont {Zhou}}\ and\ \bibinfo {author} {\bibfnamefont {H.}~\bibnamefont {Yu}},\ }\href {\doibase 10.1103/PhysRevD.96.045018} {\bibfield  {journal} {\bibinfo  {journal} {Phys. Rev. D}\ }\textbf {\bibinfo {volume} {96}},\ \bibinfo {pages} {045018} (\bibinfo {year} {2017})}\BibitemShut {NoStop}%
\bibitem [{\citenamefont {Zhou}\ and\ \citenamefont {Yu}(2018)}]{Zhou:2018gqf}%
  \BibitemOpen
  \bibfield  {author} {\bibinfo {author} {\bibfnamefont {W.}~\bibnamefont {Zhou}}\ and\ \bibinfo {author} {\bibfnamefont {H.}~\bibnamefont {Yu}},\ }\href {\doibase 10.1103/PhysRevD.97.045007} {\bibfield  {journal} {\bibinfo  {journal} {Phys. Rev. D}\ }\textbf {\bibinfo {volume} {97}},\ \bibinfo {pages} {045007} (\bibinfo {year} {2018})},\ \Eprint {http://arxiv.org/abs/1802.01699} {arXiv:1802.01699 [gr-qc]} \BibitemShut {NoStop}%
\bibitem [{\citenamefont {Nation}\ \emph {et~al.}(2012)\citenamefont {Nation}, \citenamefont {Johansson}, \citenamefont {Blencowe},\ and\ \citenamefont {Nori}}]{Nation:2011dka}%
  \BibitemOpen
  \bibfield  {author} {\bibinfo {author} {\bibfnamefont {P.~D.}\ \bibnamefont {Nation}}, \bibinfo {author} {\bibfnamefont {J.~R.}\ \bibnamefont {Johansson}}, \bibinfo {author} {\bibfnamefont {M.~P.}\ \bibnamefont {Blencowe}}, \ and\ \bibinfo {author} {\bibfnamefont {F.}~\bibnamefont {Nori}},\ }\href {\doibase 10.1103/RevModPhys.84.1} {\bibfield  {journal} {\bibinfo  {journal} {Rev. Mod. Phys.}\ }\textbf {\bibinfo {volume} {84}},\ \bibinfo {pages} {1} (\bibinfo {year} {2012})},\ \Eprint {http://arxiv.org/abs/1103.0835} {arXiv:1103.0835 [quant-ph]} \BibitemShut {NoStop}%
\bibitem [{\citenamefont {Haro}\ and\ \citenamefont {Elizalde}(2006)}]{Haro:2006zz}%
  \BibitemOpen
  \bibfield  {author} {\bibinfo {author} {\bibfnamefont {J.}~\bibnamefont {Haro}}\ and\ \bibinfo {author} {\bibfnamefont {E.}~\bibnamefont {Elizalde}},\ }\href {\doibase 10.1103/PhysRevLett.97.130401} {\bibfield  {journal} {\bibinfo  {journal} {Phys. Rev. Lett.}\ }\textbf {\bibinfo {volume} {97}},\ \bibinfo {pages} {130401} (\bibinfo {year} {2006})}\BibitemShut {NoStop}%
\bibitem [{\citenamefont {Mundarain}\ and\ \citenamefont {Maia~Neto}(1998)}]{Mundarain:1998kz}%
  \BibitemOpen
  \bibfield  {author} {\bibinfo {author} {\bibfnamefont {D.~F.}\ \bibnamefont {Mundarain}}\ and\ \bibinfo {author} {\bibfnamefont {P.~A.}\ \bibnamefont {Maia~Neto}},\ }\href {\doibase 10.1103/PhysRevA.57.1379} {\bibfield  {journal} {\bibinfo  {journal} {Phys. Rev. A}\ }\textbf {\bibinfo {volume} {57}},\ \bibinfo {pages} {1379} (\bibinfo {year} {1998})},\ \Eprint {http://arxiv.org/abs/quant-ph/9808064} {arXiv:quant-ph/9808064} \BibitemShut {NoStop}%
\bibitem [{\citenamefont {Dalvit}\ and\ \citenamefont {Mazzitelli}(1999)}]{Dalvit:1998qs}%
  \BibitemOpen
  \bibfield  {author} {\bibinfo {author} {\bibfnamefont {D.~A.~R.}\ \bibnamefont {Dalvit}}\ and\ \bibinfo {author} {\bibfnamefont {F.~D.}\ \bibnamefont {Mazzitelli}},\ }\href {\doibase 10.1103/PhysRevA.59.3049} {\bibfield  {journal} {\bibinfo  {journal} {Phys. Rev. A}\ }\textbf {\bibinfo {volume} {59}},\ \bibinfo {pages} {3049} (\bibinfo {year} {1999})},\ \Eprint {http://arxiv.org/abs/quant-ph/9810092} {arXiv:quant-ph/9810092} \BibitemShut {NoStop}%
\bibitem [{\citenamefont {Alves}\ \emph {et~al.}(2010)\citenamefont {Alves}, \citenamefont {Granhen},\ and\ \citenamefont {Pires}}]{Alves:2009ev}%
  \BibitemOpen
  \bibfield  {author} {\bibinfo {author} {\bibfnamefont {D.~T.}\ \bibnamefont {Alves}}, \bibinfo {author} {\bibfnamefont {E.~R.}\ \bibnamefont {Granhen}}, \ and\ \bibinfo {author} {\bibfnamefont {W.~P.}\ \bibnamefont {Pires}},\ }\href {\doibase 10.1103/PhysRevD.82.045028} {\bibfield  {journal} {\bibinfo  {journal} {Phys. Rev. D}\ }\textbf {\bibinfo {volume} {82}},\ \bibinfo {pages} {045028} (\bibinfo {year} {2010})},\ \Eprint {http://arxiv.org/abs/0912.1802} {arXiv:0912.1802 [hep-th]} \BibitemShut {NoStop}%
\bibitem [{\citenamefont {Fosco}\ \emph {et~al.}(2017)\citenamefont {Fosco}, \citenamefont {Giraldo},\ and\ \citenamefont {Mazzitelli}}]{Fosco:2017jjf}%
  \BibitemOpen
  \bibfield  {author} {\bibinfo {author} {\bibfnamefont {C.~D.}\ \bibnamefont {Fosco}}, \bibinfo {author} {\bibfnamefont {A.}~\bibnamefont {Giraldo}}, \ and\ \bibinfo {author} {\bibfnamefont {F.~D.}\ \bibnamefont {Mazzitelli}},\ }\href {\doibase 10.1103/PhysRevD.96.045004} {\bibfield  {journal} {\bibinfo  {journal} {Phys. Rev. D}\ }\textbf {\bibinfo {volume} {96}},\ \bibinfo {pages} {045004} (\bibinfo {year} {2017})},\ \Eprint {http://arxiv.org/abs/1704.07198} {arXiv:1704.07198 [hep-th]} \BibitemShut {NoStop}%
\bibitem [{\citenamefont {Souza}\ \emph {et~al.}(2018)\citenamefont {Souza}, \citenamefont {Impens},\ and\ \citenamefont {Neto}}]{PhysRevA.97.032514}%
  \BibitemOpen
  \bibfield  {author} {\bibinfo {author} {\bibfnamefont {R.~d. M.~e.}\ \bibnamefont {Souza}}, \bibinfo {author} {\bibfnamefont {F.~m.~c.}\ \bibnamefont {Impens}}, \ and\ \bibinfo {author} {\bibfnamefont {P.~A.~M.}\ \bibnamefont {Neto}},\ }\href {\doibase 10.1103/PhysRevA.97.032514} {\bibfield  {journal} {\bibinfo  {journal} {Phys. Rev. A}\ }\textbf {\bibinfo {volume} {97}},\ \bibinfo {pages} {032514} (\bibinfo {year} {2018})}\BibitemShut {NoStop}%
\bibitem [{\citenamefont {Lo}\ and\ \citenamefont {Law}(2018)}]{PhysRevA.98.063807}%
  \BibitemOpen
  \bibfield  {author} {\bibinfo {author} {\bibfnamefont {L.}~\bibnamefont {Lo}}\ and\ \bibinfo {author} {\bibfnamefont {C.~K.}\ \bibnamefont {Law}},\ }\href {\doibase 10.1103/PhysRevA.98.063807} {\bibfield  {journal} {\bibinfo  {journal} {Phys. Rev. A}\ }\textbf {\bibinfo {volume} {98}},\ \bibinfo {pages} {063807} (\bibinfo {year} {2018})}\BibitemShut {NoStop}%
\bibitem [{\citenamefont {Lo}\ \emph {et~al.}(2020)\citenamefont {Lo}, \citenamefont {Fong},\ and\ \citenamefont {Law}}]{PhysRevA.102.033703}%
  \BibitemOpen
  \bibfield  {author} {\bibinfo {author} {\bibfnamefont {L.}~\bibnamefont {Lo}}, \bibinfo {author} {\bibfnamefont {P.~T.}\ \bibnamefont {Fong}}, \ and\ \bibinfo {author} {\bibfnamefont {C.~K.}\ \bibnamefont {Law}},\ }\href {\doibase 10.1103/PhysRevA.102.033703} {\bibfield  {journal} {\bibinfo  {journal} {Phys. Rev. A}\ }\textbf {\bibinfo {volume} {102}},\ \bibinfo {pages} {033703} (\bibinfo {year} {2020})}\BibitemShut {NoStop}%
\bibitem [{\citenamefont {Brevik}\ \emph {et~al.}(2000)\citenamefont {Brevik}, \citenamefont {Milton}, \citenamefont {Odintsov},\ and\ \citenamefont {Osetrin}}]{Brevik:2000zb}%
  \BibitemOpen
  \bibfield  {author} {\bibinfo {author} {\bibfnamefont {I.~H.}\ \bibnamefont {Brevik}}, \bibinfo {author} {\bibfnamefont {K.~A.}\ \bibnamefont {Milton}}, \bibinfo {author} {\bibfnamefont {S.~D.}\ \bibnamefont {Odintsov}}, \ and\ \bibinfo {author} {\bibfnamefont {K.~E.}\ \bibnamefont {Osetrin}},\ }\href {\doibase 10.1103/PhysRevD.62.064005} {\bibfield  {journal} {\bibinfo  {journal} {Phys. Rev. D}\ }\textbf {\bibinfo {volume} {62}},\ \bibinfo {pages} {064005} (\bibinfo {year} {2000})},\ \Eprint {http://arxiv.org/abs/hep-th/0003158} {arXiv:hep-th/0003158} \BibitemShut {NoStop}%
\bibitem [{\citenamefont {Wittemer}\ \emph {et~al.}(2019)\citenamefont {Wittemer}, \citenamefont {Hakelberg}, \citenamefont {Kiefer}, \citenamefont {Schr\"oder}, \citenamefont {Fey}, \citenamefont {Sch\"utzhold}, \citenamefont {Warring},\ and\ \citenamefont {Schaetz}}]{Wittemer:2019agm}%
  \BibitemOpen
  \bibfield  {author} {\bibinfo {author} {\bibfnamefont {M.}~\bibnamefont {Wittemer}}, \bibinfo {author} {\bibfnamefont {F.}~\bibnamefont {Hakelberg}}, \bibinfo {author} {\bibfnamefont {P.}~\bibnamefont {Kiefer}}, \bibinfo {author} {\bibfnamefont {J.-P.}\ \bibnamefont {Schr\"oder}}, \bibinfo {author} {\bibfnamefont {C.}~\bibnamefont {Fey}}, \bibinfo {author} {\bibfnamefont {R.}~\bibnamefont {Sch\"utzhold}}, \bibinfo {author} {\bibfnamefont {U.}~\bibnamefont {Warring}}, \ and\ \bibinfo {author} {\bibfnamefont {T.}~\bibnamefont {Schaetz}},\ }\href {\doibase 10.1103/PhysRevLett.123.180502} {\bibfield  {journal} {\bibinfo  {journal} {Phys. Rev. Lett.}\ }\textbf {\bibinfo {volume} {123}},\ \bibinfo {pages} {180502} (\bibinfo {year} {2019})},\ \Eprint {http://arxiv.org/abs/1903.05523} {arXiv:1903.05523 [quant-ph]} \BibitemShut {NoStop}%
\bibitem [{\citenamefont {Dalvit}\ and\ \citenamefont {Maia~Neto}(2000)}]{Dalvit:1999sg}%
  \BibitemOpen
  \bibfield  {author} {\bibinfo {author} {\bibfnamefont {D.~A.~R.}\ \bibnamefont {Dalvit}}\ and\ \bibinfo {author} {\bibfnamefont {P.~A.}\ \bibnamefont {Maia~Neto}},\ }\href {\doibase 10.1103/PhysRevLett.84.798} {\bibfield  {journal} {\bibinfo  {journal} {Phys. Rev. Lett.}\ }\textbf {\bibinfo {volume} {84}},\ \bibinfo {pages} {798} (\bibinfo {year} {2000})},\ \Eprint {http://arxiv.org/abs/quant-ph/9910027} {arXiv:quant-ph/9910027} \BibitemShut {NoStop}%
\bibitem [{\citenamefont {Andreata}\ and\ \citenamefont {Dodonov}(2005)}]{Andreata2005DynamicsOE}%
  \BibitemOpen
  \bibfield  {author} {\bibinfo {author} {\bibfnamefont {M.~A.}\ \bibnamefont {Andreata}}\ and\ \bibinfo {author} {\bibfnamefont {V.~V.}\ \bibnamefont {Dodonov}},\ }\href@noop {} {\bibfield  {journal} {\bibinfo  {journal} {Journal of Optics B-quantum and Semiclassical Optics}\ }\textbf {\bibinfo {volume} {7}} (\bibinfo {year} {2005})}\BibitemShut {NoStop}%
\bibitem [{\citenamefont {Cong}\ \emph {et~al.}(2019)\citenamefont {Cong}, \citenamefont {Tjoa},\ and\ \citenamefont {Mann}}]{Cong:2018vqx}%
  \BibitemOpen
  \bibfield  {author} {\bibinfo {author} {\bibfnamefont {W.}~\bibnamefont {Cong}}, \bibinfo {author} {\bibfnamefont {E.}~\bibnamefont {Tjoa}}, \ and\ \bibinfo {author} {\bibfnamefont {R.~B.}\ \bibnamefont {Mann}},\ }\href {\doibase 10.1007/JHEP06(2019)021} {\bibfield  {journal} {\bibinfo  {journal} {JHEP}\ }\textbf {\bibinfo {volume} {06}},\ \bibinfo {pages} {021} (\bibinfo {year} {2019})},\ \bibinfo {note} {[Erratum: JHEP 07, 051 (2019)]},\ \Eprint {http://arxiv.org/abs/1810.07359} {arXiv:1810.07359 [quant-ph]} \BibitemShut {NoStop}%
\bibitem [{\citenamefont {Ben-Benjamin}\ \emph {et~al.}(2019)\citenamefont {Ben-Benjamin} \emph {et~al.}}]{Ben-Benjamin:2019opz}%
  \BibitemOpen
  \bibfield  {author} {\bibinfo {author} {\bibfnamefont {J.~S.}\ \bibnamefont {Ben-Benjamin}} \emph {et~al.},\ }\href {\doibase 10.1142/S0217751X19410057} {\bibfield  {journal} {\bibinfo  {journal} {Int. J. Mod. Phys. A}\ }\textbf {\bibinfo {volume} {34}},\ \bibinfo {pages} {1941005} (\bibinfo {year} {2019})},\ \Eprint {http://arxiv.org/abs/1906.01729} {arXiv:1906.01729 [quant-ph]} \BibitemShut {NoStop}%
\bibitem [{\citenamefont {Svidzinsky}\ \emph {et~al.}(2018)\citenamefont {Svidzinsky}, \citenamefont {Ben-Benjamin}, \citenamefont {Fulling},\ and\ \citenamefont {Page}}]{Svidzinsky:2018jkp}%
  \BibitemOpen
  \bibfield  {author} {\bibinfo {author} {\bibfnamefont {A.~A.}\ \bibnamefont {Svidzinsky}}, \bibinfo {author} {\bibfnamefont {J.~S.}\ \bibnamefont {Ben-Benjamin}}, \bibinfo {author} {\bibfnamefont {S.~A.}\ \bibnamefont {Fulling}}, \ and\ \bibinfo {author} {\bibfnamefont {D.~N.}\ \bibnamefont {Page}},\ }\href {\doibase 10.1103/PhysRevLett.121.071301} {\bibfield  {journal} {\bibinfo  {journal} {Phys. Rev. Lett.}\ }\textbf {\bibinfo {volume} {121}},\ \bibinfo {pages} {071301} (\bibinfo {year} {2018})}\BibitemShut {NoStop}%
\bibitem [{\citenamefont {Svidzinsky}(2019)}]{Svidzinsky:2019jqr}%
  \BibitemOpen
  \bibfield  {author} {\bibinfo {author} {\bibfnamefont {A.~A.}\ \bibnamefont {Svidzinsky}},\ }\href {\doibase 10.1103/physrevresearch.1.033027} {\bibfield  {journal} {\bibinfo  {journal} {Phys. Rev. Res.}\ }\textbf {\bibinfo {volume} {1}},\ \bibinfo {pages} {033027} (\bibinfo {year} {2019})}\BibitemShut {NoStop}%
\bibitem [{\citenamefont {Fulling}\ and\ \citenamefont {Wilson}(2019)}]{Fulling:2018lez}%
  \BibitemOpen
  \bibfield  {author} {\bibinfo {author} {\bibfnamefont {S.~A.}\ \bibnamefont {Fulling}}\ and\ \bibinfo {author} {\bibfnamefont {J.~H.}\ \bibnamefont {Wilson}},\ }\href {\doibase 10.1088/1402-4896/aaecaa} {\bibfield  {journal} {\bibinfo  {journal} {Phys. Scripta}\ }\textbf {\bibinfo {volume} {94}},\ \bibinfo {pages} {014004} (\bibinfo {year} {2019})},\ \Eprint {http://arxiv.org/abs/1805.01013} {arXiv:1805.01013 [quant-ph]} \BibitemShut {NoStop}%
\bibitem [{\citenamefont {Good}(2011)}]{Good:2011nue}%
  \BibitemOpen
  \bibfield  {author} {\bibinfo {author} {\bibfnamefont {M.~R.~R.}\ \bibnamefont {Good}},\ }\emph {\bibinfo {title} {{Quantized scalar fields under the influence of moving mirror and anisotropic curved spacetime}}},\ \href {\doibase 10.17615/a075-pg88} {Ph.D. thesis},\ \bibinfo  {school} {North Carolina U.} (\bibinfo {year} {2011})\BibitemShut {NoStop}%
\bibitem [{\citenamefont {Carlitz}\ and\ \citenamefont {Willey}(1987)}]{Carlitz:1986nh}%
  \BibitemOpen
  \bibfield  {author} {\bibinfo {author} {\bibfnamefont {R.~D.}\ \bibnamefont {Carlitz}}\ and\ \bibinfo {author} {\bibfnamefont {R.~S.}\ \bibnamefont {Willey}},\ }\href {\doibase 10.1103/PhysRevD.36.2327} {\bibfield  {journal} {\bibinfo  {journal} {Phys. Rev. D}\ }\textbf {\bibinfo {volume} {36}},\ \bibinfo {pages} {2327} (\bibinfo {year} {1987})}\BibitemShut {NoStop}%
\bibitem [{\citenamefont {Haro}\ and\ \citenamefont {Elizalde}(2008)}]{Haro:2008zza}%
  \BibitemOpen
  \bibfield  {author} {\bibinfo {author} {\bibfnamefont {J.}~\bibnamefont {Haro}}\ and\ \bibinfo {author} {\bibfnamefont {E.}~\bibnamefont {Elizalde}},\ }\href {\doibase 10.1103/PhysRevD.77.045011} {\bibfield  {journal} {\bibinfo  {journal} {Phys. Rev. D}\ }\textbf {\bibinfo {volume} {77}},\ \bibinfo {pages} {045011} (\bibinfo {year} {2008})},\ \Eprint {http://arxiv.org/abs/0712.4141} {arXiv:0712.4141 [quant-ph]} \BibitemShut {NoStop}%
\bibitem [{\citenamefont {Nicolaevici}(2009)}]{Nicolaevici:2009zz}%
  \BibitemOpen
  \bibfield  {author} {\bibinfo {author} {\bibfnamefont {N.}~\bibnamefont {Nicolaevici}},\ }\href {\doibase 10.1103/PhysRevD.80.125003} {\bibfield  {journal} {\bibinfo  {journal} {Phys. Rev. D}\ }\textbf {\bibinfo {volume} {80}},\ \bibinfo {pages} {125003} (\bibinfo {year} {2009})}\BibitemShut {NoStop}%
\bibitem [{\citenamefont {{Walker}}\ and\ \citenamefont {{Davies}}(1982)}]{1982JPhA...15L.477W}%
  \BibitemOpen
  \bibfield  {author} {\bibinfo {author} {\bibfnamefont {W.~R.}\ \bibnamefont {{Walker}}}\ and\ \bibinfo {author} {\bibfnamefont {P.~C.~W.}\ \bibnamefont {{Davies}}},\ }\href {\doibase 10.1088/0305-4470/15/9/008} {\bibfield  {journal} {\bibinfo  {journal} {Journal of Physics A Mathematical General}\ }\textbf {\bibinfo {volume} {15}},\ \bibinfo {pages} {L477} (\bibinfo {year} {1982})}\BibitemShut {NoStop}%
\bibitem [{\citenamefont {Good}\ \emph {et~al.}(2013)\citenamefont {Good}, \citenamefont {Anderson},\ and\ \citenamefont {Evans}}]{Good:2013lca}%
  \BibitemOpen
  \bibfield  {author} {\bibinfo {author} {\bibfnamefont {M.~R.~R.}\ \bibnamefont {Good}}, \bibinfo {author} {\bibfnamefont {P.~R.}\ \bibnamefont {Anderson}}, \ and\ \bibinfo {author} {\bibfnamefont {C.~R.}\ \bibnamefont {Evans}},\ }\href {\doibase 10.1103/PhysRevD.88.025023} {\bibfield  {journal} {\bibinfo  {journal} {Phys. Rev. D}\ }\textbf {\bibinfo {volume} {88}},\ \bibinfo {pages} {025023} (\bibinfo {year} {2013})},\ \Eprint {http://arxiv.org/abs/1303.6756} {arXiv:1303.6756 [gr-qc]} \BibitemShut {NoStop}%
\bibitem [{\citenamefont {Good}\ \emph {et~al.}(2016)\citenamefont {Good}, \citenamefont {Anderson},\ and\ \citenamefont {Evans}}]{Good:2016oey}%
  \BibitemOpen
  \bibfield  {author} {\bibinfo {author} {\bibfnamefont {M.~R.~R.}\ \bibnamefont {Good}}, \bibinfo {author} {\bibfnamefont {P.~R.}\ \bibnamefont {Anderson}}, \ and\ \bibinfo {author} {\bibfnamefont {C.~R.}\ \bibnamefont {Evans}},\ }\href {\doibase 10.1103/PhysRevD.94.065010} {\bibfield  {journal} {\bibinfo  {journal} {Phys. Rev. D}\ }\textbf {\bibinfo {volume} {94}},\ \bibinfo {pages} {065010} (\bibinfo {year} {2016})},\ \Eprint {http://arxiv.org/abs/1605.06635} {arXiv:1605.06635 [gr-qc]} \BibitemShut {NoStop}%
\bibitem [{\citenamefont {Good}\ and\ \citenamefont {Linder}(2017)}]{Good:2017kjr}%
  \BibitemOpen
  \bibfield  {author} {\bibinfo {author} {\bibfnamefont {M.~R.~R.}\ \bibnamefont {Good}}\ and\ \bibinfo {author} {\bibfnamefont {E.~V.}\ \bibnamefont {Linder}},\ }\href {\doibase 10.1103/PhysRevD.96.125010} {\bibfield  {journal} {\bibinfo  {journal} {Phys. Rev. D}\ }\textbf {\bibinfo {volume} {96}},\ \bibinfo {pages} {125010} (\bibinfo {year} {2017})},\ \Eprint {http://arxiv.org/abs/1707.03670} {arXiv:1707.03670 [gr-qc]} \BibitemShut {NoStop}%
\bibitem [{\citenamefont {Good}\ \emph {et~al.}(2020{\natexlab{a}})\citenamefont {Good}, \citenamefont {Linder},\ and\ \citenamefont {Wilczek}}]{Good:2020rmk}%
  \BibitemOpen
  \bibfield  {author} {\bibinfo {author} {\bibfnamefont {M.~R.~R.}\ \bibnamefont {Good}}, \bibinfo {author} {\bibfnamefont {E.~V.}\ \bibnamefont {Linder}}, \ and\ \bibinfo {author} {\bibfnamefont {F.}~\bibnamefont {Wilczek}},\ }\href {\doibase 10.1142/S0217732320400064} {\bibfield  {journal} {\bibinfo  {journal} {Mod. Phys. Lett. A}\ }\textbf {\bibinfo {volume} {35}},\ \bibinfo {pages} {2040006} (\bibinfo {year} {2020}{\natexlab{a}})},\ \Eprint {http://arxiv.org/abs/2108.11188} {arXiv:2108.11188 [quant-ph]} \BibitemShut {NoStop}%
\bibitem [{\citenamefont {Mintz}\ \emph {et~al.}(2006)\citenamefont {Mintz}, \citenamefont {Farina}, \citenamefont {Maia~Neto},\ and\ \citenamefont {Rodrigues}}]{Mintz:2006yz}%
  \BibitemOpen
  \bibfield  {author} {\bibinfo {author} {\bibfnamefont {B.}~\bibnamefont {Mintz}}, \bibinfo {author} {\bibfnamefont {C.}~\bibnamefont {Farina}}, \bibinfo {author} {\bibfnamefont {P.~A.}\ \bibnamefont {Maia~Neto}}, \ and\ \bibinfo {author} {\bibfnamefont {R.~B.}\ \bibnamefont {Rodrigues}},\ }\href {\doibase 10.1088/0305-4470/39/36/013} {\bibfield  {journal} {\bibinfo  {journal} {J. Phys. A}\ }\textbf {\bibinfo {volume} {39}},\ \bibinfo {pages} {11325} (\bibinfo {year} {2006})},\ \Eprint {http://arxiv.org/abs/hep-th/0605221} {arXiv:hep-th/0605221} \BibitemShut {NoStop}%
\bibitem [{\citenamefont {Barton}\ and\ \citenamefont {Calogeracos}(1995)}]{Barton:1995he}%
  \BibitemOpen
  \bibfield  {author} {\bibinfo {author} {\bibfnamefont {G.}~\bibnamefont {Barton}}\ and\ \bibinfo {author} {\bibfnamefont {A.}~\bibnamefont {Calogeracos}},\ }\href {\doibase 10.1006/aphy.1995.1021} {\bibfield  {journal} {\bibinfo  {journal} {Annals Phys.}\ }\textbf {\bibinfo {volume} {238}},\ \bibinfo {pages} {227} (\bibinfo {year} {1995})}\BibitemShut {NoStop}%
\bibitem [{\citenamefont {Calogeracos}\ and\ \citenamefont {Barton}(1995)}]{Calogeracos:1995he}%
  \BibitemOpen
  \bibfield  {author} {\bibinfo {author} {\bibfnamefont {A.}~\bibnamefont {Calogeracos}}\ and\ \bibinfo {author} {\bibfnamefont {G.}~\bibnamefont {Barton}},\ }\href {\doibase 10.1006/aphy.1995.1022} {\bibfield  {journal} {\bibinfo  {journal} {Annals Phys.}\ }\textbf {\bibinfo {volume} {238}},\ \bibinfo {pages} {268} (\bibinfo {year} {1995})}\BibitemShut {NoStop}%
\bibitem [{\citenamefont {Golestanian}\ and\ \citenamefont {Kardar}(1997)}]{Golestanian:1997ks}%
  \BibitemOpen
  \bibfield  {author} {\bibinfo {author} {\bibfnamefont {R.}~\bibnamefont {Golestanian}}\ and\ \bibinfo {author} {\bibfnamefont {M.}~\bibnamefont {Kardar}},\ }\href {\doibase 10.1103/PhysRevLett.78.3421} {\bibfield  {journal} {\bibinfo  {journal} {Phys. Rev. Lett.}\ }\textbf {\bibinfo {volume} {78}},\ \bibinfo {pages} {3421} (\bibinfo {year} {1997})},\ \Eprint {http://arxiv.org/abs/quant-ph/9701005} {arXiv:quant-ph/9701005} \BibitemShut {NoStop}%
\bibitem [{\citenamefont {Golestanian}\ and\ \citenamefont {Kardar}(1998)}]{Golestanian:1998bx}%
  \BibitemOpen
  \bibfield  {author} {\bibinfo {author} {\bibfnamefont {R.}~\bibnamefont {Golestanian}}\ and\ \bibinfo {author} {\bibfnamefont {M.}~\bibnamefont {Kardar}},\ }\href {\doibase 10.1103/PhysRevA.58.1713} {\bibfield  {journal} {\bibinfo  {journal} {Phys. Rev. A}\ }\textbf {\bibinfo {volume} {58}},\ \bibinfo {pages} {1713} (\bibinfo {year} {1998})},\ \Eprint {http://arxiv.org/abs/quant-ph/9802017} {arXiv:quant-ph/9802017} \BibitemShut {NoStop}%
\bibitem [{\citenamefont {Sopova}\ and\ \citenamefont {Ford}(2002)}]{Sopova:2002cs}%
  \BibitemOpen
  \bibfield  {author} {\bibinfo {author} {\bibfnamefont {V.}~\bibnamefont {Sopova}}\ and\ \bibinfo {author} {\bibfnamefont {L.~H.}\ \bibnamefont {Ford}},\ }\href {\doibase 10.1103/PhysRevD.66.045026} {\bibfield  {journal} {\bibinfo  {journal} {Phys. Rev. D}\ }\textbf {\bibinfo {volume} {66}},\ \bibinfo {pages} {045026} (\bibinfo {year} {2002})},\ \Eprint {http://arxiv.org/abs/quant-ph/0204125} {arXiv:quant-ph/0204125} \BibitemShut {NoStop}%
\bibitem [{\citenamefont {Galley}\ \emph {et~al.}(2013)\citenamefont {Galley}, \citenamefont {Behunin},\ and\ \citenamefont {Hu}}]{Galley:2012qz}%
  \BibitemOpen
  \bibfield  {author} {\bibinfo {author} {\bibfnamefont {C.~R.}\ \bibnamefont {Galley}}, \bibinfo {author} {\bibfnamefont {R.~O.}\ \bibnamefont {Behunin}}, \ and\ \bibinfo {author} {\bibfnamefont {B.~L.}\ \bibnamefont {Hu}},\ }\href {\doibase 10.1103/PhysRevA.87.043832} {\bibfield  {journal} {\bibinfo  {journal} {Phys. Rev. A}\ }\textbf {\bibinfo {volume} {87}},\ \bibinfo {pages} {043832} (\bibinfo {year} {2013})},\ \Eprint {http://arxiv.org/abs/1204.2569} {arXiv:1204.2569 [quant-ph]} \BibitemShut {NoStop}%
\bibitem [{\citenamefont {Wang}\ and\ \citenamefont {Unruh}(2014)}]{Wang:2013lex}%
  \BibitemOpen
  \bibfield  {author} {\bibinfo {author} {\bibfnamefont {Q.}~\bibnamefont {Wang}}\ and\ \bibinfo {author} {\bibfnamefont {W.~G.}\ \bibnamefont {Unruh}},\ }\href {\doibase 10.1103/PhysRevD.89.085009} {\bibfield  {journal} {\bibinfo  {journal} {Phys. Rev. D}\ }\textbf {\bibinfo {volume} {89}},\ \bibinfo {pages} {085009} (\bibinfo {year} {2014})},\ \Eprint {http://arxiv.org/abs/1312.4591} {arXiv:1312.4591 [gr-qc]} \BibitemShut {NoStop}%
\bibitem [{\citenamefont {Wang}\ and\ \citenamefont {Unruh}(2015)}]{Wang:2015axa}%
  \BibitemOpen
  \bibfield  {author} {\bibinfo {author} {\bibfnamefont {Q.}~\bibnamefont {Wang}}\ and\ \bibinfo {author} {\bibfnamefont {W.~G.}\ \bibnamefont {Unruh}},\ }\href {\doibase 10.1103/PhysRevD.92.063520} {\bibfield  {journal} {\bibinfo  {journal} {Phys. Rev. D}\ }\textbf {\bibinfo {volume} {92}},\ \bibinfo {pages} {063520} (\bibinfo {year} {2015})},\ \Eprint {http://arxiv.org/abs/1506.05531} {arXiv:1506.05531 [gr-qc]} \BibitemShut {NoStop}%
\bibitem [{\citenamefont {Walker}(1985)}]{Walker:1984vj}%
  \BibitemOpen
  \bibfield  {author} {\bibinfo {author} {\bibfnamefont {W.~R.}\ \bibnamefont {Walker}},\ }\href {\doibase 10.1103/PhysRevD.31.767} {\bibfield  {journal} {\bibinfo  {journal} {Phys. Rev. D}\ }\textbf {\bibinfo {volume} {31}},\ \bibinfo {pages} {767} (\bibinfo {year} {1985})}\BibitemShut {NoStop}%
\bibitem [{\citenamefont {Fabbri}\ and\ \citenamefont {Navarro-Salas}(2005)}]{Fabbri:2005mw}%
  \BibitemOpen
  \bibfield  {author} {\bibinfo {author} {\bibfnamefont {A.}~\bibnamefont {Fabbri}}\ and\ \bibinfo {author} {\bibfnamefont {J.}~\bibnamefont {Navarro-Salas}},\ }\href@noop {} {\emph {\bibinfo {title} {{Modeling black hole evaporation}}}}\ (\bibinfo {year} {2005})\BibitemShut {NoStop}%
\bibitem [{\citenamefont {Sorge}(2005)}]{Sorge:2005ed}%
  \BibitemOpen
  \bibfield  {author} {\bibinfo {author} {\bibfnamefont {F.}~\bibnamefont {Sorge}},\ }\href {\doibase 10.1088/0264-9381/22/23/012} {\bibfield  {journal} {\bibinfo  {journal} {Class. Quant. Grav.}\ }\textbf {\bibinfo {volume} {22}},\ \bibinfo {pages} {5109} (\bibinfo {year} {2005})}\BibitemShut {NoStop}%
\bibitem [{\citenamefont {Sorge}(2019)}]{Sorge:2019ldb}%
  \BibitemOpen
  \bibfield  {author} {\bibinfo {author} {\bibfnamefont {F.}~\bibnamefont {Sorge}},\ }\href {\doibase 10.1088/1361-6382/ab4def} {\bibfield  {journal} {\bibinfo  {journal} {Class. Quant. Grav.}\ }\textbf {\bibinfo {volume} {36}},\ \bibinfo {pages} {235006} (\bibinfo {year} {2019})}\BibitemShut {NoStop}%
\bibitem [{\citenamefont {Celeri}\ \emph {et~al.}(2009)\citenamefont {Celeri}, \citenamefont {Pascoal},\ and\ \citenamefont {Moussa}}]{Celeri:2008ui}%
  \BibitemOpen
  \bibfield  {author} {\bibinfo {author} {\bibfnamefont {L.~C.}\ \bibnamefont {Celeri}}, \bibinfo {author} {\bibfnamefont {F.}~\bibnamefont {Pascoal}}, \ and\ \bibinfo {author} {\bibfnamefont {M.~H.~Y.}\ \bibnamefont {Moussa}},\ }\href {\doibase 10.1088/0264-9381/26/10/105014} {\bibfield  {journal} {\bibinfo  {journal} {Class. Quant. Grav.}\ }\textbf {\bibinfo {volume} {26}},\ \bibinfo {pages} {105014} (\bibinfo {year} {2009})},\ \Eprint {http://arxiv.org/abs/0809.3706} {arXiv:0809.3706 [quant-ph]} \BibitemShut {NoStop}%
\bibitem [{\citenamefont {R\"atzel}\ \emph {et~al.}(2018{\natexlab{b}})\citenamefont {R\"atzel}, \citenamefont {Schneiter}, \citenamefont {Braun}, \citenamefont {Bravo}, \citenamefont {Howl}, \citenamefont {Lock},\ and\ \citenamefont {Fuentes}}]{Ratzel:2017etl}%
  \BibitemOpen
  \bibfield  {author} {\bibinfo {author} {\bibfnamefont {D.}~\bibnamefont {R\"atzel}}, \bibinfo {author} {\bibfnamefont {F.}~\bibnamefont {Schneiter}}, \bibinfo {author} {\bibfnamefont {D.}~\bibnamefont {Braun}}, \bibinfo {author} {\bibfnamefont {T.}~\bibnamefont {Bravo}}, \bibinfo {author} {\bibfnamefont {R.}~\bibnamefont {Howl}}, \bibinfo {author} {\bibfnamefont {M.~P.~E.}\ \bibnamefont {Lock}}, \ and\ \bibinfo {author} {\bibfnamefont {I.}~\bibnamefont {Fuentes}},\ }\href {\doibase 10.1088/1367-2630/aac0ac} {\bibfield  {journal} {\bibinfo  {journal} {New J. Phys.}\ }\textbf {\bibinfo {volume} {20}},\ \bibinfo {pages} {053046} (\bibinfo {year} {2018}{\natexlab{b}})},\ \Eprint {http://arxiv.org/abs/1711.11320} {arXiv:1711.11320 [gr-qc]} \BibitemShut {NoStop}%
\bibitem [{\citenamefont {Sorge}\ and\ \citenamefont {Wilson}(2019)}]{Sorge:2019ecb}%
  \BibitemOpen
  \bibfield  {author} {\bibinfo {author} {\bibfnamefont {F.}~\bibnamefont {Sorge}}\ and\ \bibinfo {author} {\bibfnamefont {J.~H.}\ \bibnamefont {Wilson}},\ }\href {\doibase 10.1103/PhysRevD.100.105007} {\bibfield  {journal} {\bibinfo  {journal} {Phys. Rev. D}\ }\textbf {\bibinfo {volume} {100}},\ \bibinfo {pages} {105007} (\bibinfo {year} {2019})},\ \Eprint {http://arxiv.org/abs/1909.07357} {arXiv:1909.07357 [gr-qc]} \BibitemShut {NoStop}%
\bibitem [{\citenamefont {Wilson}\ \emph {et~al.}(2020)\citenamefont {Wilson}, \citenamefont {Sorge},\ and\ \citenamefont {Fulling}}]{Wilson:2019ago}%
  \BibitemOpen
  \bibfield  {author} {\bibinfo {author} {\bibfnamefont {J.~H.}\ \bibnamefont {Wilson}}, \bibinfo {author} {\bibfnamefont {F.}~\bibnamefont {Sorge}}, \ and\ \bibinfo {author} {\bibfnamefont {S.~A.}\ \bibnamefont {Fulling}},\ }\href {\doibase 10.1103/PhysRevD.101.065007} {\bibfield  {journal} {\bibinfo  {journal} {Phys. Rev. D}\ }\textbf {\bibinfo {volume} {101}},\ \bibinfo {pages} {065007} (\bibinfo {year} {2020})},\ \Eprint {http://arxiv.org/abs/1911.04492} {arXiv:1911.04492 [hep-th]} \BibitemShut {NoStop}%
\bibitem [{\citenamefont {Fagnocchi}\ \emph {et~al.}(2010)\citenamefont {Fagnocchi}, \citenamefont {Finazzi}, \citenamefont {Liberati}, \citenamefont {Kormos},\ and\ \citenamefont {Trombettoni}}]{Fagnocchi:2010sn}%
  \BibitemOpen
  \bibfield  {author} {\bibinfo {author} {\bibfnamefont {S.}~\bibnamefont {Fagnocchi}}, \bibinfo {author} {\bibfnamefont {S.}~\bibnamefont {Finazzi}}, \bibinfo {author} {\bibfnamefont {S.}~\bibnamefont {Liberati}}, \bibinfo {author} {\bibfnamefont {M.}~\bibnamefont {Kormos}}, \ and\ \bibinfo {author} {\bibfnamefont {A.}~\bibnamefont {Trombettoni}},\ }\href {\doibase 10.1088/1367-2630/12/9/095012} {\bibfield  {journal} {\bibinfo  {journal} {New J. Phys.}\ }\textbf {\bibinfo {volume} {12}},\ \bibinfo {pages} {095012} (\bibinfo {year} {2010})},\ \Eprint {http://arxiv.org/abs/1001.1044} {arXiv:1001.1044 [gr-qc]} \BibitemShut {NoStop}%
\bibitem [{\citenamefont {Friis}\ \emph {et~al.}(2013)\citenamefont {Friis}, \citenamefont {Lee},\ and\ \citenamefont {Louko}}]{Friis:2013eva}%
  \BibitemOpen
  \bibfield  {author} {\bibinfo {author} {\bibfnamefont {N.}~\bibnamefont {Friis}}, \bibinfo {author} {\bibfnamefont {A.~R.}\ \bibnamefont {Lee}}, \ and\ \bibinfo {author} {\bibfnamefont {J.}~\bibnamefont {Louko}},\ }\href {\doibase 10.1103/PhysRevD.88.064028} {\bibfield  {journal} {\bibinfo  {journal} {Phys. Rev. D}\ }\textbf {\bibinfo {volume} {88}},\ \bibinfo {pages} {064028} (\bibinfo {year} {2013})},\ \Eprint {http://arxiv.org/abs/1307.1631} {arXiv:1307.1631 [quant-ph]} \BibitemShut {NoStop}%
\bibitem [{\citenamefont {Lima}\ \emph {et~al.}(2019)\citenamefont {Lima}, \citenamefont {Alencar}, \citenamefont {Muniz},\ and\ \citenamefont {Landim}}]{Lima:2019pbo}%
  \BibitemOpen
  \bibfield  {author} {\bibinfo {author} {\bibfnamefont {A.~P. C.~M.}\ \bibnamefont {Lima}}, \bibinfo {author} {\bibfnamefont {G.}~\bibnamefont {Alencar}}, \bibinfo {author} {\bibfnamefont {C.~R.}\ \bibnamefont {Muniz}}, \ and\ \bibinfo {author} {\bibfnamefont {R.~R.}\ \bibnamefont {Landim}},\ }\href {\doibase 10.1088/1475-7516/2019/07/011} {\bibfield  {journal} {\bibinfo  {journal} {JCAP}\ }\textbf {\bibinfo {volume} {07}},\ \bibinfo {pages} {011} (\bibinfo {year} {2019})},\ \Eprint {http://arxiv.org/abs/1903.00512} {arXiv:1903.00512 [hep-th]} \BibitemShut {NoStop}%
\bibitem [{\citenamefont {Scully}\ \emph {et~al.}(2003)\citenamefont {Scully}, \citenamefont {Kocharovsky}, \citenamefont {Belyanin}, \citenamefont {Fry},\ and\ \citenamefont {Capasso}}]{Scully:2003zz}%
  \BibitemOpen
  \bibfield  {author} {\bibinfo {author} {\bibfnamefont {M.~O.}\ \bibnamefont {Scully}}, \bibinfo {author} {\bibfnamefont {V.~V.}\ \bibnamefont {Kocharovsky}}, \bibinfo {author} {\bibfnamefont {A.}~\bibnamefont {Belyanin}}, \bibinfo {author} {\bibfnamefont {E.}~\bibnamefont {Fry}}, \ and\ \bibinfo {author} {\bibfnamefont {F.}~\bibnamefont {Capasso}},\ }\href {\doibase 10.1103/PhysRevLett.91.243004} {\bibfield  {journal} {\bibinfo  {journal} {Phys. Rev. Lett.}\ }\textbf {\bibinfo {volume} {91}},\ \bibinfo {pages} {243004} (\bibinfo {year} {2003})},\ \Eprint {http://arxiv.org/abs/quant-ph/0305178} {arXiv:quant-ph/0305178} \BibitemShut {NoStop}%
\bibitem [{\citenamefont {Dolan}\ \emph {et~al.}(2020)\citenamefont {Dolan}, \citenamefont {Hunter-McCabe},\ and\ \citenamefont {Twamley}}]{Dolan:2020hzm}%
  \BibitemOpen
  \bibfield  {author} {\bibinfo {author} {\bibfnamefont {B.~P.}\ \bibnamefont {Dolan}}, \bibinfo {author} {\bibfnamefont {A.}~\bibnamefont {Hunter-McCabe}}, \ and\ \bibinfo {author} {\bibfnamefont {J.}~\bibnamefont {Twamley}},\ }\href {\doibase 10.1088/1367-2630/ab7bd5} {\bibfield  {journal} {\bibinfo  {journal} {New J. Phys.}\ }\textbf {\bibinfo {volume} {22}},\ \bibinfo {pages} {033026} (\bibinfo {year} {2020})},\ \Eprint {http://arxiv.org/abs/2003.02258} {arXiv:2003.02258 [quant-ph]} \BibitemShut {NoStop}%
\bibitem [{\citenamefont {Scully}\ \emph {et~al.}(2019)\citenamefont {Scully}, \citenamefont {Svidzinsky},\ and\ \citenamefont {Unruh}}]{PhysRevResearch.1.033115}%
  \BibitemOpen
  \bibfield  {author} {\bibinfo {author} {\bibfnamefont {M.~O.}\ \bibnamefont {Scully}}, \bibinfo {author} {\bibfnamefont {A.~A.}\ \bibnamefont {Svidzinsky}}, \ and\ \bibinfo {author} {\bibfnamefont {W.}~\bibnamefont {Unruh}},\ }\href {\doibase 10.1103/PhysRevResearch.1.033115} {\bibfield  {journal} {\bibinfo  {journal} {Phys. Rev. Res.}\ }\textbf {\bibinfo {volume} {1}},\ \bibinfo {pages} {033115} (\bibinfo {year} {2019})}\BibitemShut {NoStop}%
\bibitem [{\citenamefont {Good}(2016)}]{Good:2016bsq}%
  \BibitemOpen
  \bibfield  {author} {\bibinfo {author} {\bibfnamefont {M.~R.~R.}\ \bibnamefont {Good}},\ }in\ \href {\doibase 10.1142/9789813203952_0078} {\emph {\bibinfo {booktitle} {{2nd LeCosPA Symposium}: {Everything about Gravity, Celebrating the Centenary of Einstein's General Relativity}}}}\ (\bibinfo {year} {2016})\ \Eprint {http://arxiv.org/abs/1602.00683} {arXiv:1602.00683 [gr-qc]} \BibitemShut {NoStop}%
\bibitem [{\citenamefont {Good}(2020)}]{Good:2020nmz}%
  \BibitemOpen
  \bibfield  {author} {\bibinfo {author} {\bibfnamefont {M.~R.~R.}\ \bibnamefont {Good}},\ }\href {\doibase 10.1103/PhysRevD.101.104050} {\bibfield  {journal} {\bibinfo  {journal} {Phys. Rev. D}\ }\textbf {\bibinfo {volume} {101}},\ \bibinfo {pages} {104050} (\bibinfo {year} {2020})},\ \Eprint {http://arxiv.org/abs/2003.07016} {arXiv:2003.07016 [gr-qc]} \BibitemShut {NoStop}%
\bibitem [{\citenamefont {Good}\ \emph {et~al.}(2020{\natexlab{b}})\citenamefont {Good}, \citenamefont {Zhakenuly},\ and\ \citenamefont {Linder}}]{Good:2020byh}%
  \BibitemOpen
  \bibfield  {author} {\bibinfo {author} {\bibfnamefont {M.~R.~R.}\ \bibnamefont {Good}}, \bibinfo {author} {\bibfnamefont {A.}~\bibnamefont {Zhakenuly}}, \ and\ \bibinfo {author} {\bibfnamefont {E.~V.}\ \bibnamefont {Linder}},\ }\href {\doibase 10.1103/PhysRevD.102.045020} {\bibfield  {journal} {\bibinfo  {journal} {Phys. Rev. D}\ }\textbf {\bibinfo {volume} {102}},\ \bibinfo {pages} {045020} (\bibinfo {year} {2020}{\natexlab{b}})},\ \Eprint {http://arxiv.org/abs/2005.03850} {arXiv:2005.03850 [gr-qc]} \BibitemShut {NoStop}%
\bibitem [{\citenamefont {Bekenstein}(1973)}]{Bekenstein:1973ur}%
  \BibitemOpen
  \bibfield  {author} {\bibinfo {author} {\bibfnamefont {J.~D.}\ \bibnamefont {Bekenstein}},\ }\href {\doibase 10.1103/PhysRevD.7.2333} {\bibfield  {journal} {\bibinfo  {journal} {Phys. Rev. D}\ }\textbf {\bibinfo {volume} {7}},\ \bibinfo {pages} {2333} (\bibinfo {year} {1973})}\BibitemShut {NoStop}%
\bibitem [{\citenamefont {{Alfaro}}\ \emph {et~al.}(1976)\citenamefont {{Alfaro}}, \citenamefont {{Fubini}},\ and\ \citenamefont {{Furlan}}}]{1976NCimA..34..569A}%
  \BibitemOpen
  \bibfield  {author} {\bibinfo {author} {\bibfnamefont {V.}~\bibnamefont {{Alfaro}}}, \bibinfo {author} {\bibfnamefont {S.}~\bibnamefont {{Fubini}}}, \ and\ \bibinfo {author} {\bibfnamefont {G.}~\bibnamefont {{Furlan}}},\ }\href {\doibase 10.1007/BF02785666} {\bibfield  {journal} {\bibinfo  {journal} {Nuovo Cimento A Serie}\ }\textbf {\bibinfo {volume} {34}},\ \bibinfo {pages} {569} (\bibinfo {year} {1976})}\BibitemShut {NoStop}%
\bibitem [{\citenamefont {Camblong}\ and\ \citenamefont {Ordonez}(2005{\natexlab{a}})}]{Camblong:2004ye}%
  \BibitemOpen
  \bibfield  {author} {\bibinfo {author} {\bibfnamefont {H.~E.}\ \bibnamefont {Camblong}}\ and\ \bibinfo {author} {\bibfnamefont {C.~R.}\ \bibnamefont {Ordonez}},\ }\href {\doibase 10.1103/PhysRevD.71.104029} {\bibfield  {journal} {\bibinfo  {journal} {Phys. Rev. D}\ }\textbf {\bibinfo {volume} {71}},\ \bibinfo {pages} {104029} (\bibinfo {year} {2005}{\natexlab{a}})},\ \Eprint {http://arxiv.org/abs/hep-th/0411008} {arXiv:hep-th/0411008} \BibitemShut {NoStop}%
\bibitem [{\citenamefont {Camblong}\ and\ \citenamefont {Ordonez}(2005{\natexlab{b}})}]{Camblong:2004ec}%
  \BibitemOpen
  \bibfield  {author} {\bibinfo {author} {\bibfnamefont {H.~E.}\ \bibnamefont {Camblong}}\ and\ \bibinfo {author} {\bibfnamefont {C.~R.}\ \bibnamefont {Ordonez}},\ }\href {\doibase 10.1103/PhysRevD.71.124040} {\bibfield  {journal} {\bibinfo  {journal} {Phys. Rev. D}\ }\textbf {\bibinfo {volume} {71}},\ \bibinfo {pages} {124040} (\bibinfo {year} {2005}{\natexlab{b}})},\ \Eprint {http://arxiv.org/abs/hep-th/0412309} {arXiv:hep-th/0412309} \BibitemShut {NoStop}%
\bibitem [{\citenamefont {Camblong}\ \emph {et~al.}(2020)\citenamefont {Camblong}, \citenamefont {Chakraborty},\ and\ \citenamefont {Ordonez}}]{Camblong:2020pme}%
  \BibitemOpen
  \bibfield  {author} {\bibinfo {author} {\bibfnamefont {H.~E.}\ \bibnamefont {Camblong}}, \bibinfo {author} {\bibfnamefont {A.}~\bibnamefont {Chakraborty}}, \ and\ \bibinfo {author} {\bibfnamefont {C.~R.}\ \bibnamefont {Ordonez}},\ }\href {\doibase 10.1103/PhysRevD.102.085010} {\bibfield  {journal} {\bibinfo  {journal} {Phys. Rev. D}\ }\textbf {\bibinfo {volume} {102}},\ \bibinfo {pages} {085010} (\bibinfo {year} {2020})},\ \Eprint {http://arxiv.org/abs/2009.06580} {arXiv:2009.06580 [gr-qc]} \BibitemShut {NoStop}%
\bibitem [{\citenamefont {Azizi}\ \emph {et~al.}(2021{\natexlab{a}})\citenamefont {Azizi}, \citenamefont {Camblong}, \citenamefont {Chakraborty}, \citenamefont {Ordonez},\ and\ \citenamefont {Scully}}]{Azizi:2021qcu}%
  \BibitemOpen
  \bibfield  {author} {\bibinfo {author} {\bibfnamefont {A.}~\bibnamefont {Azizi}}, \bibinfo {author} {\bibfnamefont {H.~E.}\ \bibnamefont {Camblong}}, \bibinfo {author} {\bibfnamefont {A.}~\bibnamefont {Chakraborty}}, \bibinfo {author} {\bibfnamefont {C.~R.}\ \bibnamefont {Ordonez}}, \ and\ \bibinfo {author} {\bibfnamefont {M.~O.}\ \bibnamefont {Scully}},\ }\href {\doibase 10.1103/PhysRevD.104.084086} {\bibfield  {journal} {\bibinfo  {journal} {Phys. Rev. D}\ }\textbf {\bibinfo {volume} {104}} (\bibinfo {year} {2021}{\natexlab{a}}),\ 10.1103/PhysRevD.104.084086},\ \Eprint {http://arxiv.org/abs/2108.07570} {arXiv:2108.07570 [gr-qc]} \BibitemShut {NoStop}%
\bibitem [{\citenamefont {Azizi}\ \emph {et~al.}(2021{\natexlab{b}})\citenamefont {Azizi}, \citenamefont {Camblong}, \citenamefont {Chakraborty}, \citenamefont {Ordonez},\ and\ \citenamefont {Scully}}]{Azizi:2021yto}%
  \BibitemOpen
  \bibfield  {author} {\bibinfo {author} {\bibfnamefont {A.}~\bibnamefont {Azizi}}, \bibinfo {author} {\bibfnamefont {H.~E.}\ \bibnamefont {Camblong}}, \bibinfo {author} {\bibfnamefont {A.}~\bibnamefont {Chakraborty}}, \bibinfo {author} {\bibfnamefont {C.~R.}\ \bibnamefont {Ordonez}}, \ and\ \bibinfo {author} {\bibfnamefont {M.~O.}\ \bibnamefont {Scully}},\ }\href {\doibase 10.1103/PhysRevD.104.084085} {\bibfield  {journal} {\bibinfo  {journal} {Phys. Rev. D}\ }\textbf {\bibinfo {volume} {104}} (\bibinfo {year} {2021}{\natexlab{b}}),\ 10.1103/PhysRevD.104.084085},\ \Eprint {http://arxiv.org/abs/2108.07572} {arXiv:2108.07572 [gr-qc]} \BibitemShut {NoStop}%
\bibitem [{\citenamefont {Maldacena}\ and\ \citenamefont {Seiberg}(2005)}]{Maldacena:2005he}%
  \BibitemOpen
  \bibfield  {author} {\bibinfo {author} {\bibfnamefont {J.~M.}\ \bibnamefont {Maldacena}}\ and\ \bibinfo {author} {\bibfnamefont {N.}~\bibnamefont {Seiberg}},\ }\href {\doibase 10.1088/1126-6708/2005/09/077} {\bibfield  {journal} {\bibinfo  {journal} {JHEP}\ }\textbf {\bibinfo {volume} {09}},\ \bibinfo {pages} {077} (\bibinfo {year} {2005})},\ \Eprint {http://arxiv.org/abs/hep-th/0506141} {arXiv:hep-th/0506141} \BibitemShut {NoStop}%
\bibitem [{\citenamefont {Morita}(2019)}]{Morita:2019bfr}%
  \BibitemOpen
  \bibfield  {author} {\bibinfo {author} {\bibfnamefont {T.}~\bibnamefont {Morita}},\ }\href {\doibase 10.1103/PhysRevLett.122.101603} {\bibfield  {journal} {\bibinfo  {journal} {Phys. Rev. Lett.}\ }\textbf {\bibinfo {volume} {122}},\ \bibinfo {pages} {101603} (\bibinfo {year} {2019})},\ \Eprint {http://arxiv.org/abs/1902.06940} {arXiv:1902.06940 [hep-th]} \BibitemShut {NoStop}%
\bibitem [{\citenamefont {Maitra}\ \emph {et~al.}(2020)\citenamefont {Maitra}, \citenamefont {Maity},\ and\ \citenamefont {Majhi}}]{Maitra:2019eix}%
  \BibitemOpen
  \bibfield  {author} {\bibinfo {author} {\bibfnamefont {M.}~\bibnamefont {Maitra}}, \bibinfo {author} {\bibfnamefont {D.}~\bibnamefont {Maity}}, \ and\ \bibinfo {author} {\bibfnamefont {B.~R.}\ \bibnamefont {Majhi}},\ }\href {\doibase 10.1140/epjp/s13360-020-00451-3} {\bibfield  {journal} {\bibinfo  {journal} {Eur. Phys. J. Plus}\ }\textbf {\bibinfo {volume} {135}},\ \bibinfo {pages} {483} (\bibinfo {year} {2020})},\ \Eprint {http://arxiv.org/abs/1906.04489} {arXiv:1906.04489 [hep-th]} \BibitemShut {NoStop}%
\bibitem [{\citenamefont {Dalui}\ and\ \citenamefont {Majhi}(2020)}]{Dalui:2020qpt}%
  \BibitemOpen
  \bibfield  {author} {\bibinfo {author} {\bibfnamefont {S.}~\bibnamefont {Dalui}}\ and\ \bibinfo {author} {\bibfnamefont {B.~R.}\ \bibnamefont {Majhi}},\ }\href {\doibase 10.1103/PhysRevD.102.124047} {\bibfield  {journal} {\bibinfo  {journal} {Phys. Rev. D}\ }\textbf {\bibinfo {volume} {102}},\ \bibinfo {pages} {124047} (\bibinfo {year} {2020})},\ \Eprint {http://arxiv.org/abs/2007.14312} {arXiv:2007.14312 [gr-qc]} \BibitemShut {NoStop}%
\bibitem [{\citenamefont {Dalui}\ \emph {et~al.}(2020)\citenamefont {Dalui}, \citenamefont {Majhi},\ and\ \citenamefont {Mishra}}]{Dalui:2019esx}%
  \BibitemOpen
  \bibfield  {author} {\bibinfo {author} {\bibfnamefont {S.}~\bibnamefont {Dalui}}, \bibinfo {author} {\bibfnamefont {B.~R.}\ \bibnamefont {Majhi}}, \ and\ \bibinfo {author} {\bibfnamefont {P.}~\bibnamefont {Mishra}},\ }\href {\doibase 10.1103/PhysRevD.102.044006} {\bibfield  {journal} {\bibinfo  {journal} {Phys. Rev. D}\ }\textbf {\bibinfo {volume} {102}},\ \bibinfo {pages} {044006} (\bibinfo {year} {2020})},\ \Eprint {http://arxiv.org/abs/1910.07989} {arXiv:1910.07989 [gr-qc]} \BibitemShut {NoStop}%
\bibitem [{\citenamefont {Dalui}\ and\ \citenamefont {Majhi}(2022)}]{Dalui:2021tvy}%
  \BibitemOpen
  \bibfield  {author} {\bibinfo {author} {\bibfnamefont {S.}~\bibnamefont {Dalui}}\ and\ \bibinfo {author} {\bibfnamefont {B.~R.}\ \bibnamefont {Majhi}},\ }\href {\doibase 10.1016/j.physletb.2022.136899} {\bibfield  {journal} {\bibinfo  {journal} {Phys. Lett. B}\ }\textbf {\bibinfo {volume} {826}},\ \bibinfo {pages} {136899} (\bibinfo {year} {2022})},\ \Eprint {http://arxiv.org/abs/2103.11613} {arXiv:2103.11613 [gr-qc]} \BibitemShut {NoStop}%
\bibitem [{\citenamefont {Dalui}\ \emph {et~al.}(2021)\citenamefont {Dalui}, \citenamefont {Majhi},\ and\ \citenamefont {Padmanabhan}}]{Dalui:2021sme}%
  \BibitemOpen
  \bibfield  {author} {\bibinfo {author} {\bibfnamefont {S.}~\bibnamefont {Dalui}}, \bibinfo {author} {\bibfnamefont {B.~R.}\ \bibnamefont {Majhi}}, \ and\ \bibinfo {author} {\bibfnamefont {T.}~\bibnamefont {Padmanabhan}},\ }\href {\doibase 10.1103/PhysRevD.104.124080} {\bibfield  {journal} {\bibinfo  {journal} {Phys. Rev. D}\ }\textbf {\bibinfo {volume} {104}},\ \bibinfo {pages} {124080} (\bibinfo {year} {2021})},\ \Eprint {http://arxiv.org/abs/2110.12665} {arXiv:2110.12665 [gr-qc]} \BibitemShut {NoStop}%
\bibitem [{\citenamefont {Kane}\ and\ \citenamefont {Majhi}(2022)}]{Kane:2022zcg}%
  \BibitemOpen
  \bibfield  {author} {\bibinfo {author} {\bibfnamefont {G.~R.}\ \bibnamefont {Kane}}\ and\ \bibinfo {author} {\bibfnamefont {B.~R.}\ \bibnamefont {Majhi}},\ }\href@noop {} {\  (\bibinfo {year} {2022})},\ \Eprint {http://arxiv.org/abs/2210.04056} {arXiv:2210.04056 [gr-qc]} \BibitemShut {NoStop}%
\bibitem [{\citenamefont {Chatterjee}\ \emph {et~al.}(2021)\citenamefont {Chatterjee}, \citenamefont {Gangopadhyay},\ and\ \citenamefont {Majumdar}}]{Chatterjee:2021fue}%
  \BibitemOpen
  \bibfield  {author} {\bibinfo {author} {\bibfnamefont {R.}~\bibnamefont {Chatterjee}}, \bibinfo {author} {\bibfnamefont {S.}~\bibnamefont {Gangopadhyay}}, \ and\ \bibinfo {author} {\bibfnamefont {A.~S.}\ \bibnamefont {Majumdar}},\ }\href {\doibase 10.1103/PhysRevD.104.124001} {\bibfield  {journal} {\bibinfo  {journal} {Phys. Rev. D}\ }\textbf {\bibinfo {volume} {104}},\ \bibinfo {pages} {124001} (\bibinfo {year} {2021})},\ \Eprint {http://arxiv.org/abs/2104.10531} {arXiv:2104.10531 [quant-ph]} \BibitemShut {NoStop}%
\bibitem [{\citenamefont {Sen}\ \emph {et~al.}(2022)\citenamefont {Sen}, \citenamefont {Mandal},\ and\ \citenamefont {Gangopadhyay}}]{Sen:2022tru}%
  \BibitemOpen
  \bibfield  {author} {\bibinfo {author} {\bibfnamefont {S.}~\bibnamefont {Sen}}, \bibinfo {author} {\bibfnamefont {R.}~\bibnamefont {Mandal}}, \ and\ \bibinfo {author} {\bibfnamefont {S.}~\bibnamefont {Gangopadhyay}},\ }\href {\doibase 10.1103/PhysRevD.105.085007} {\bibfield  {journal} {\bibinfo  {journal} {Phys. Rev. D}\ }\textbf {\bibinfo {volume} {105}},\ \bibinfo {pages} {085007} (\bibinfo {year} {2022})},\ \Eprint {http://arxiv.org/abs/2202.00671} {arXiv:2202.00671 [hep-th]} \BibitemShut {NoStop}%
\bibitem [{\citenamefont {Chakraborty}\ and\ \citenamefont {Majhi}(2019)}]{Chakraborty:2019ltu}%
  \BibitemOpen
  \bibfield  {author} {\bibinfo {author} {\bibfnamefont {K.}~\bibnamefont {Chakraborty}}\ and\ \bibinfo {author} {\bibfnamefont {B.~R.}\ \bibnamefont {Majhi}},\ }\href {\doibase 10.1103/PhysRevD.100.045004} {\bibfield  {journal} {\bibinfo  {journal} {Phys. Rev. D}\ }\textbf {\bibinfo {volume} {100}},\ \bibinfo {pages} {045004} (\bibinfo {year} {2019})},\ \Eprint {http://arxiv.org/abs/1905.10554} {arXiv:1905.10554 [gr-qc]} \BibitemShut {NoStop}%
\bibitem [{\citenamefont {Bukhari}\ and\ \citenamefont {Wang}(2023)}]{Bukhari:2023yuy}%
  \BibitemOpen
  \bibfield  {author} {\bibinfo {author} {\bibfnamefont {S.~M. A.~S.}\ \bibnamefont {Bukhari}}\ and\ \bibinfo {author} {\bibfnamefont {L.-G.}\ \bibnamefont {Wang}},\ }\href@noop {} {\  (\bibinfo {year} {2023})},\ \Eprint {http://arxiv.org/abs/2309.11958} {arXiv:2309.11958 [gr-qc]} \BibitemShut {NoStop}%
\bibitem [{\citenamefont {Parikh}\ and\ \citenamefont {Wilczek}(2000)}]{Parikh:1999mf}%
  \BibitemOpen
  \bibfield  {author} {\bibinfo {author} {\bibfnamefont {M.~K.}\ \bibnamefont {Parikh}}\ and\ \bibinfo {author} {\bibfnamefont {F.}~\bibnamefont {Wilczek}},\ }\href {\doibase 10.1103/PhysRevLett.85.5042} {\bibfield  {journal} {\bibinfo  {journal} {Phys. Rev. Lett.}\ }\textbf {\bibinfo {volume} {85}},\ \bibinfo {pages} {5042} (\bibinfo {year} {2000})},\ \Eprint {http://arxiv.org/abs/hep-th/9907001} {arXiv:hep-th/9907001} \BibitemShut {NoStop}%
\bibitem [{\citenamefont {Visser}(2015)}]{Visser:2014ypa}%
  \BibitemOpen
  \bibfield  {author} {\bibinfo {author} {\bibfnamefont {M.}~\bibnamefont {Visser}},\ }\href {\doibase 10.1007/JHEP07(2015)009} {\bibfield  {journal} {\bibinfo  {journal} {JHEP}\ }\textbf {\bibinfo {volume} {07}},\ \bibinfo {pages} {009} (\bibinfo {year} {2015})},\ \Eprint {http://arxiv.org/abs/1409.7754} {arXiv:1409.7754 [gr-qc]} \BibitemShut {NoStop}%
\bibitem [{\citenamefont {Ma}\ \emph {et~al.}(2018)\citenamefont {Ma}, \citenamefont {Cai}, \citenamefont {Dong},\ and\ \citenamefont {Sun}}]{Ma:2017odv}%
  \BibitemOpen
  \bibfield  {author} {\bibinfo {author} {\bibfnamefont {Y.-H.}\ \bibnamefont {Ma}}, \bibinfo {author} {\bibfnamefont {Q.-Y.}\ \bibnamefont {Cai}}, \bibinfo {author} {\bibfnamefont {H.}~\bibnamefont {Dong}}, \ and\ \bibinfo {author} {\bibfnamefont {C.-P.}\ \bibnamefont {Sun}},\ }\href {\doibase 10.1209/0295-5075/122/30001} {\bibfield  {journal} {\bibinfo  {journal} {EPL}\ }\textbf {\bibinfo {volume} {122}},\ \bibinfo {pages} {30001} (\bibinfo {year} {2018})},\ \Eprint {http://arxiv.org/abs/1711.10704} {arXiv:1711.10704 [quant-ph]} \BibitemShut {NoStop}%
\bibitem [{\citenamefont {Kastor}\ and\ \citenamefont {Traschen}(1996)}]{Kastor:1993mj}%
  \BibitemOpen
  \bibfield  {author} {\bibinfo {author} {\bibfnamefont {D.}~\bibnamefont {Kastor}}\ and\ \bibinfo {author} {\bibfnamefont {J.~H.}\ \bibnamefont {Traschen}},\ }\href {\doibase 10.1088/0264-9381/13/10/013} {\bibfield  {journal} {\bibinfo  {journal} {Class. Quant. Grav.}\ }\textbf {\bibinfo {volume} {13}},\ \bibinfo {pages} {2753} (\bibinfo {year} {1996})},\ \Eprint {http://arxiv.org/abs/gr-qc/9311025} {arXiv:gr-qc/9311025} \BibitemShut {NoStop}%
\bibitem [{\citenamefont {Bhattacharya}(2018)}]{Bhattacharya:2018ltm}%
  \BibitemOpen
  \bibfield  {author} {\bibinfo {author} {\bibfnamefont {S.}~\bibnamefont {Bhattacharya}},\ }\href {\doibase 10.1103/PhysRevD.98.125013} {\bibfield  {journal} {\bibinfo  {journal} {Phys. Rev. D}\ }\textbf {\bibinfo {volume} {98}},\ \bibinfo {pages} {125013} (\bibinfo {year} {2018})},\ \Eprint {http://arxiv.org/abs/1810.13260} {arXiv:1810.13260 [gr-qc]} \BibitemShut {NoStop}%
\bibitem [{\citenamefont {Qiu}\ and\ \citenamefont {Traschen}(2020)}]{Qiu:2019qgp}%
  \BibitemOpen
  \bibfield  {author} {\bibinfo {author} {\bibfnamefont {Y.}~\bibnamefont {Qiu}}\ and\ \bibinfo {author} {\bibfnamefont {J.}~\bibnamefont {Traschen}},\ }\href {\doibase 10.1088/1361-6382/ab8bba} {\bibfield  {journal} {\bibinfo  {journal} {Class. Quant. Grav.}\ }\textbf {\bibinfo {volume} {37}},\ \bibinfo {pages} {135012} (\bibinfo {year} {2020})},\ \Eprint {http://arxiv.org/abs/1908.02737} {arXiv:1908.02737 [hep-th]} \BibitemShut {NoStop}%
\bibitem [{\citenamefont {Bukhari}\ \emph {et~al.}(2023)\citenamefont {Bukhari}, \citenamefont {Bhat}, \citenamefont {Xu},\ and\ \citenamefont {Wang}}]{Bukhari:2022wyx}%
  \BibitemOpen
  \bibfield  {author} {\bibinfo {author} {\bibfnamefont {S.~M. A.~S.}\ \bibnamefont {Bukhari}}, \bibinfo {author} {\bibfnamefont {I.~A.}\ \bibnamefont {Bhat}}, \bibinfo {author} {\bibfnamefont {C.}~\bibnamefont {Xu}}, \ and\ \bibinfo {author} {\bibfnamefont {L.-G.}\ \bibnamefont {Wang}},\ }\href {\doibase 10.1103/PhysRevD.107.105017} {\bibfield  {journal} {\bibinfo  {journal} {Phys. Rev. D}\ }\textbf {\bibinfo {volume} {107}},\ \bibinfo {pages} {105017} (\bibinfo {year} {2023})},\ \Eprint {http://arxiv.org/abs/2211.08793} {arXiv:2211.08793 [gr-qc]} \BibitemShut {NoStop}%
\bibitem [{\citenamefont {Bartlett}\ \emph {et~al.}(2007)\citenamefont {Bartlett}, \citenamefont {Rudolph},\ and\ \citenamefont {Spekkens}}]{Bartlett:2006tzx}%
  \BibitemOpen
  \bibfield  {author} {\bibinfo {author} {\bibfnamefont {S.~D.}\ \bibnamefont {Bartlett}}, \bibinfo {author} {\bibfnamefont {T.}~\bibnamefont {Rudolph}}, \ and\ \bibinfo {author} {\bibfnamefont {R.~W.}\ \bibnamefont {Spekkens}},\ }\href {\doibase 10.1103/RevModPhys.79.555} {\bibfield  {journal} {\bibinfo  {journal} {Rev. Mod. Phys.}\ }\textbf {\bibinfo {volume} {79}},\ \bibinfo {pages} {555} (\bibinfo {year} {2007})},\ \Eprint {http://arxiv.org/abs/quant-ph/0610030} {arXiv:quant-ph/0610030} \BibitemShut {NoStop}%
\bibitem [{\citenamefont {Ralph}\ and\ \citenamefont {Downes}(2012)}]{Ralph:2011hp}%
  \BibitemOpen
  \bibfield  {author} {\bibinfo {author} {\bibfnamefont {T.~C.}\ \bibnamefont {Ralph}}\ and\ \bibinfo {author} {\bibfnamefont {T.~G.}\ \bibnamefont {Downes}},\ }\href {\doibase 10.1080/00107514.2011.640146} {\bibfield  {journal} {\bibinfo  {journal} {Contemp. Phys.}\ }\textbf {\bibinfo {volume} {53}},\ \bibinfo {pages} {1} (\bibinfo {year} {2012})},\ \Eprint {http://arxiv.org/abs/1111.2648} {arXiv:1111.2648 [quant-ph]} \BibitemShut {NoStop}%
\bibitem [{\citenamefont {Summers}\ and\ \citenamefont {Werner}(1987)}]{Summers:1987ze}%
  \BibitemOpen
  \bibfield  {author} {\bibinfo {author} {\bibfnamefont {S.~J.}\ \bibnamefont {Summers}}\ and\ \bibinfo {author} {\bibfnamefont {R.}~\bibnamefont {Werner}},\ }\href {\doibase 10.1007/BF01207366} {\bibfield  {journal} {\bibinfo  {journal} {Commun. Math. Phys.}\ }\textbf {\bibinfo {volume} {110}},\ \bibinfo {pages} {247} (\bibinfo {year} {1987})}\BibitemShut {NoStop}%
\bibitem [{\citenamefont {Reznik}\ \emph {et~al.}(2005)\citenamefont {Reznik}, \citenamefont {Retzker},\ and\ \citenamefont {Silman}}]{Reznik:2003mnx}%
  \BibitemOpen
  \bibfield  {author} {\bibinfo {author} {\bibfnamefont {B.}~\bibnamefont {Reznik}}, \bibinfo {author} {\bibfnamefont {A.}~\bibnamefont {Retzker}}, \ and\ \bibinfo {author} {\bibfnamefont {J.}~\bibnamefont {Silman}},\ }\href {\doibase 10.1103/PhysRevA.71.042104} {\bibfield  {journal} {\bibinfo  {journal} {Phys. Rev. A}\ }\textbf {\bibinfo {volume} {71}},\ \bibinfo {pages} {042104} (\bibinfo {year} {2005})},\ \Eprint {http://arxiv.org/abs/quant-ph/0310058} {arXiv:quant-ph/0310058} \BibitemShut {NoStop}%
\bibitem [{\citenamefont {Salton}\ \emph {et~al.}(2015)\citenamefont {Salton}, \citenamefont {Mann},\ and\ \citenamefont {Menicucci}}]{Salton:2014jaa}%
  \BibitemOpen
  \bibfield  {author} {\bibinfo {author} {\bibfnamefont {G.}~\bibnamefont {Salton}}, \bibinfo {author} {\bibfnamefont {R.~B.}\ \bibnamefont {Mann}}, \ and\ \bibinfo {author} {\bibfnamefont {N.~C.}\ \bibnamefont {Menicucci}},\ }\href {\doibase 10.1088/1367-2630/17/3/035001} {\bibfield  {journal} {\bibinfo  {journal} {New J. Phys.}\ }\textbf {\bibinfo {volume} {17}},\ \bibinfo {pages} {035001} (\bibinfo {year} {2015})},\ \Eprint {http://arxiv.org/abs/1408.1395} {arXiv:1408.1395 [quant-ph]} \BibitemShut {NoStop}%
\bibitem [{\citenamefont {Pozas-Kerstjens}\ and\ \citenamefont {Martin-Martinez}(2015)}]{Pozas-Kerstjens:2015gta}%
  \BibitemOpen
  \bibfield  {author} {\bibinfo {author} {\bibfnamefont {A.}~\bibnamefont {Pozas-Kerstjens}}\ and\ \bibinfo {author} {\bibfnamefont {E.}~\bibnamefont {Martin-Martinez}},\ }\href {\doibase 10.1103/PhysRevD.92.064042} {\bibfield  {journal} {\bibinfo  {journal} {Phys. Rev. D}\ }\textbf {\bibinfo {volume} {92}},\ \bibinfo {pages} {064042} (\bibinfo {year} {2015})},\ \Eprint {http://arxiv.org/abs/1506.03081} {arXiv:1506.03081 [quant-ph]} \BibitemShut {NoStop}%
\bibitem [{\citenamefont {Zhou}\ \emph {et~al.}(2022)\citenamefont {Zhou}, \citenamefont {Hu},\ and\ \citenamefont {Yu}}]{Zhou:2022nur}%
  \BibitemOpen
  \bibfield  {author} {\bibinfo {author} {\bibfnamefont {Y.}~\bibnamefont {Zhou}}, \bibinfo {author} {\bibfnamefont {J.}~\bibnamefont {Hu}}, \ and\ \bibinfo {author} {\bibfnamefont {H.}~\bibnamefont {Yu}},\ }\href {\doibase 10.1103/PhysRevD.106.105028} {\bibfield  {journal} {\bibinfo  {journal} {Phys. Rev. D}\ }\textbf {\bibinfo {volume} {106}},\ \bibinfo {pages} {105028} (\bibinfo {year} {2022})}\BibitemShut {NoStop}%
\bibitem [{\citenamefont {Liu}\ \emph {et~al.}(2023{\natexlab{b}})\citenamefont {Liu}, \citenamefont {Zhang},\ and\ \citenamefont {Yu}}]{Liu:2022uhf}%
  \BibitemOpen
  \bibfield  {author} {\bibinfo {author} {\bibfnamefont {Z.}~\bibnamefont {Liu}}, \bibinfo {author} {\bibfnamefont {J.}~\bibnamefont {Zhang}}, \ and\ \bibinfo {author} {\bibfnamefont {H.}~\bibnamefont {Yu}},\ }\href {\doibase 10.1103/PhysRevD.107.045010} {\bibfield  {journal} {\bibinfo  {journal} {Phys. Rev. D}\ }\textbf {\bibinfo {volume} {107}},\ \bibinfo {pages} {045010} (\bibinfo {year} {2023}{\natexlab{b}})},\ \Eprint {http://arxiv.org/abs/2208.14825} {arXiv:2208.14825 [quant-ph]} \BibitemShut {NoStop}%
\bibitem [{\citenamefont {Bozanic}\ \emph {et~al.}(2023)\citenamefont {Bozanic}, \citenamefont {Naeem}, \citenamefont {Gallock-Yoshimura},\ and\ \citenamefont {Mann}}]{Bozanic:2023okm}%
  \BibitemOpen
  \bibfield  {author} {\bibinfo {author} {\bibfnamefont {L.}~\bibnamefont {Bozanic}}, \bibinfo {author} {\bibfnamefont {M.}~\bibnamefont {Naeem}}, \bibinfo {author} {\bibfnamefont {K.}~\bibnamefont {Gallock-Yoshimura}}, \ and\ \bibinfo {author} {\bibfnamefont {R.~B.}\ \bibnamefont {Mann}},\ }\href {\doibase 10.1103/PhysRevD.108.105017} {\bibfield  {journal} {\bibinfo  {journal} {Phys. Rev. D}\ }\textbf {\bibinfo {volume} {108}},\ \bibinfo {pages} {105017} (\bibinfo {year} {2023})},\ \Eprint {http://arxiv.org/abs/2308.06329} {arXiv:2308.06329 [quant-ph]} \BibitemShut {NoStop}%
\bibitem [{\citenamefont {Zhang}\ and\ \citenamefont {Yu}(2020)}]{Zhang:2020xvo}%
  \BibitemOpen
  \bibfield  {author} {\bibinfo {author} {\bibfnamefont {J.}~\bibnamefont {Zhang}}\ and\ \bibinfo {author} {\bibfnamefont {H.}~\bibnamefont {Yu}},\ }\href {\doibase 10.1103/PhysRevD.102.065013} {\bibfield  {journal} {\bibinfo  {journal} {Phys. Rev. D}\ }\textbf {\bibinfo {volume} {102}},\ \bibinfo {pages} {065013} (\bibinfo {year} {2020})},\ \Eprint {http://arxiv.org/abs/2008.07980} {arXiv:2008.07980 [quant-ph]} \BibitemShut {NoStop}%
\bibitem [{\citenamefont {Liu}\ \emph {et~al.}(2021)\citenamefont {Liu}, \citenamefont {Zhang},\ and\ \citenamefont {Yu}}]{Liu:2020jaj}%
  \BibitemOpen
  \bibfield  {author} {\bibinfo {author} {\bibfnamefont {Z.}~\bibnamefont {Liu}}, \bibinfo {author} {\bibfnamefont {J.}~\bibnamefont {Zhang}}, \ and\ \bibinfo {author} {\bibfnamefont {H.}~\bibnamefont {Yu}},\ }\href {\doibase 10.1007/JHEP08(2021)020} {\bibfield  {journal} {\bibinfo  {journal} {JHEP}\ }\textbf {\bibinfo {volume} {08}},\ \bibinfo {pages} {020} (\bibinfo {year} {2021})},\ \Eprint {http://arxiv.org/abs/2101.00114} {arXiv:2101.00114 [quant-ph]} \BibitemShut {NoStop}%
\bibitem [{\citenamefont {Ye}\ \emph {et~al.}(2021)\citenamefont {Ye}, \citenamefont {Yu},\ and\ \citenamefont {Hu}}]{Ye:2021muj}%
  \BibitemOpen
  \bibfield  {author} {\bibinfo {author} {\bibfnamefont {Y.}~\bibnamefont {Ye}}, \bibinfo {author} {\bibfnamefont {H.}~\bibnamefont {Yu}}, \ and\ \bibinfo {author} {\bibfnamefont {J.}~\bibnamefont {Hu}},\ }\href {\doibase 10.1088/1572-9494/abf03d} {\bibfield  {journal} {\bibinfo  {journal} {Commun. Theor. Phys.}\ }\textbf {\bibinfo {volume} {73}},\ \bibinfo {pages} {065104} (\bibinfo {year} {2021})}\BibitemShut {NoStop}%
\bibitem [{\citenamefont {Liu}\ \emph {et~al.}(2023{\natexlab{c}})\citenamefont {Liu}, \citenamefont {Zhang},\ and\ \citenamefont {Yu}}]{Liu:2023zro}%
  \BibitemOpen
  \bibfield  {author} {\bibinfo {author} {\bibfnamefont {Z.}~\bibnamefont {Liu}}, \bibinfo {author} {\bibfnamefont {J.}~\bibnamefont {Zhang}}, \ and\ \bibinfo {author} {\bibfnamefont {H.}~\bibnamefont {Yu}},\ }\href {\doibase 10.1007/JHEP11(2023)184} {\bibfield  {journal} {\bibinfo  {journal} {JHEP}\ }\textbf {\bibinfo {volume} {11}},\ \bibinfo {pages} {184} (\bibinfo {year} {2023}{\natexlab{c}})},\ \Eprint {http://arxiv.org/abs/2310.07164} {arXiv:2310.07164 [quant-ph]} \BibitemShut {NoStop}%
\bibitem [{\citenamefont {Li}\ and\ \citenamefont {Zhao}(2024)}]{Li:2024dvs}%
  \BibitemOpen
  \bibfield  {author} {\bibinfo {author} {\bibfnamefont {R.}~\bibnamefont {Li}}\ and\ \bibinfo {author} {\bibfnamefont {Z.}~\bibnamefont {Zhao}},\ }\href@noop {} {\  (\bibinfo {year} {2024})},\ \Eprint {http://arxiv.org/abs/2401.16018} {arXiv:2401.16018 [quant-ph]} \BibitemShut {NoStop}%
\bibitem [{\citenamefont {Barman}\ and\ \citenamefont {Majhi}(2023)}]{Barman:2023wkr}%
  \BibitemOpen
  \bibfield  {author} {\bibinfo {author} {\bibfnamefont {D.}~\bibnamefont {Barman}}\ and\ \bibinfo {author} {\bibfnamefont {B.~R.}\ \bibnamefont {Majhi}},\ }\href {\doibase 10.1103/PhysRevD.108.085007} {\bibfield  {journal} {\bibinfo  {journal} {Phys. Rev. D}\ }\textbf {\bibinfo {volume} {108}},\ \bibinfo {pages} {085007} (\bibinfo {year} {2023})},\ \Eprint {http://arxiv.org/abs/2306.09943} {arXiv:2306.09943 [gr-qc]} \BibitemShut {NoStop}%
\bibitem [{\citenamefont {Ji}\ \emph {et~al.}(2024)\citenamefont {Ji}, \citenamefont {Zhang},\ and\ \citenamefont {Yu}}]{Ji:2024fcq}%
  \BibitemOpen
  \bibfield  {author} {\bibinfo {author} {\bibfnamefont {Y.}~\bibnamefont {Ji}}, \bibinfo {author} {\bibfnamefont {J.}~\bibnamefont {Zhang}}, \ and\ \bibinfo {author} {\bibfnamefont {H.}~\bibnamefont {Yu}},\ }\href@noop {} {\  (\bibinfo {year} {2024})},\ \Eprint {http://arxiv.org/abs/2401.13406} {arXiv:2401.13406 [quant-ph]} \BibitemShut {NoStop}%
\bibitem [{\citenamefont {Martin-Martinez}\ \emph {et~al.}(2016)\citenamefont {Martin-Martinez}, \citenamefont {Smith},\ and\ \citenamefont {Terno}}]{Martin-Martinez:2015qwa}%
  \BibitemOpen
  \bibfield  {author} {\bibinfo {author} {\bibfnamefont {E.}~\bibnamefont {Martin-Martinez}}, \bibinfo {author} {\bibfnamefont {A.~R.~H.}\ \bibnamefont {Smith}}, \ and\ \bibinfo {author} {\bibfnamefont {D.~R.}\ \bibnamefont {Terno}},\ }\href {\doibase 10.1103/PhysRevD.93.044001} {\bibfield  {journal} {\bibinfo  {journal} {Phys. Rev. D}\ }\textbf {\bibinfo {volume} {93}},\ \bibinfo {pages} {044001} (\bibinfo {year} {2016})},\ \Eprint {http://arxiv.org/abs/1507.02688} {arXiv:1507.02688 [quant-ph]} \BibitemShut {NoStop}%
\bibitem [{\citenamefont {Hu}\ \emph {et~al.}(2022)\citenamefont {Hu}, \citenamefont {Zhang},\ and\ \citenamefont {Yu}}]{Hu:2022nxc}%
  \BibitemOpen
  \bibfield  {author} {\bibinfo {author} {\bibfnamefont {H.}~\bibnamefont {Hu}}, \bibinfo {author} {\bibfnamefont {J.}~\bibnamefont {Zhang}}, \ and\ \bibinfo {author} {\bibfnamefont {H.}~\bibnamefont {Yu}},\ }\href {\doibase 10.1007/JHEP05(2022)112} {\bibfield  {journal} {\bibinfo  {journal} {JHEP}\ }\textbf {\bibinfo {volume} {05}},\ \bibinfo {pages} {112} (\bibinfo {year} {2022})},\ \Eprint {http://arxiv.org/abs/2204.01219} {arXiv:2204.01219 [quant-ph]} \BibitemShut {NoStop}%
\bibitem [{\citenamefont {Cong}\ \emph {et~al.}(2020)\citenamefont {Cong}, \citenamefont {Qian}, \citenamefont {Good},\ and\ \citenamefont {Mann}}]{Cong:2020nec}%
  \BibitemOpen
  \bibfield  {author} {\bibinfo {author} {\bibfnamefont {W.}~\bibnamefont {Cong}}, \bibinfo {author} {\bibfnamefont {C.}~\bibnamefont {Qian}}, \bibinfo {author} {\bibfnamefont {M.~R.~R.}\ \bibnamefont {Good}}, \ and\ \bibinfo {author} {\bibfnamefont {R.~B.}\ \bibnamefont {Mann}},\ }\href {\doibase 10.1007/JHEP10(2020)067} {\bibfield  {journal} {\bibinfo  {journal} {JHEP}\ }\textbf {\bibinfo {volume} {10}},\ \bibinfo {pages} {067} (\bibinfo {year} {2020})},\ \Eprint {http://arxiv.org/abs/2006.01720} {arXiv:2006.01720 [gr-qc]} \BibitemShut {NoStop}%
\bibitem [{\citenamefont {Henderson}\ \emph {et~al.}(2018)\citenamefont {Henderson}, \citenamefont {Hennigar}, \citenamefont {Mann}, \citenamefont {Smith},\ and\ \citenamefont {Zhang}}]{Henderson:2017yuv}%
  \BibitemOpen
  \bibfield  {author} {\bibinfo {author} {\bibfnamefont {L.~J.}\ \bibnamefont {Henderson}}, \bibinfo {author} {\bibfnamefont {R.~A.}\ \bibnamefont {Hennigar}}, \bibinfo {author} {\bibfnamefont {R.~B.}\ \bibnamefont {Mann}}, \bibinfo {author} {\bibfnamefont {A.~R.~H.}\ \bibnamefont {Smith}}, \ and\ \bibinfo {author} {\bibfnamefont {J.}~\bibnamefont {Zhang}},\ }\href {\doibase 10.1088/1361-6382/aae27e} {\bibfield  {journal} {\bibinfo  {journal} {Class. Quant. Grav.}\ }\textbf {\bibinfo {volume} {35}},\ \bibinfo {pages} {21LT02} (\bibinfo {year} {2018})},\ \Eprint {http://arxiv.org/abs/1712.10018} {arXiv:1712.10018 [quant-ph]} \BibitemShut {NoStop}%
\bibitem [{\citenamefont {Carib\'e}\ \emph {et~al.}(2023)\citenamefont {Carib\'e}, \citenamefont {Jonsson}, \citenamefont {Casals}, \citenamefont {Kempf},\ and\ \citenamefont {Mart\'\i{}n-Mart\'\i{}nez}}]{Caribe:2023fhr}%
  \BibitemOpen
  \bibfield  {author} {\bibinfo {author} {\bibfnamefont {J.~a. G.~A.}\ \bibnamefont {Carib\'e}}, \bibinfo {author} {\bibfnamefont {R.~H.}\ \bibnamefont {Jonsson}}, \bibinfo {author} {\bibfnamefont {M.}~\bibnamefont {Casals}}, \bibinfo {author} {\bibfnamefont {A.}~\bibnamefont {Kempf}}, \ and\ \bibinfo {author} {\bibfnamefont {E.}~\bibnamefont {Mart\'\i{}n-Mart\'\i{}nez}},\ }\href {\doibase 10.1103/PhysRevD.108.025016} {\bibfield  {journal} {\bibinfo  {journal} {Phys. Rev. D}\ }\textbf {\bibinfo {volume} {108}},\ \bibinfo {pages} {025016} (\bibinfo {year} {2023})},\ \Eprint {http://arxiv.org/abs/2303.01402} {arXiv:2303.01402 [quant-ph]} \BibitemShut {NoStop}%
\bibitem [{\citenamefont {Hu}\ and\ \citenamefont {Yu}(2013)}]{Hu:2013ypa}%
  \BibitemOpen
  \bibfield  {author} {\bibinfo {author} {\bibfnamefont {J.}~\bibnamefont {Hu}}\ and\ \bibinfo {author} {\bibfnamefont {H.}~\bibnamefont {Yu}},\ }\href {\doibase 10.1103/PhysRevD.88.104003} {\bibfield  {journal} {\bibinfo  {journal} {Phys. Rev. D}\ }\textbf {\bibinfo {volume} {88}},\ \bibinfo {pages} {104003} (\bibinfo {year} {2013})},\ \Eprint {http://arxiv.org/abs/1310.7650} {arXiv:1310.7650 [gr-qc]} \BibitemShut {NoStop}%
\bibitem [{\citenamefont {Bueley}\ \emph {et~al.}(2022)\citenamefont {Bueley}, \citenamefont {Huang}, \citenamefont {Gallock-Yoshimura},\ and\ \citenamefont {Mann}}]{Bueley:2022ple}%
  \BibitemOpen
  \bibfield  {author} {\bibinfo {author} {\bibfnamefont {K.}~\bibnamefont {Bueley}}, \bibinfo {author} {\bibfnamefont {L.}~\bibnamefont {Huang}}, \bibinfo {author} {\bibfnamefont {K.}~\bibnamefont {Gallock-Yoshimura}}, \ and\ \bibinfo {author} {\bibfnamefont {R.~B.}\ \bibnamefont {Mann}},\ }\href {\doibase 10.1103/PhysRevD.106.025010} {\bibfield  {journal} {\bibinfo  {journal} {Phys. Rev. D}\ }\textbf {\bibinfo {volume} {106}},\ \bibinfo {pages} {025010} (\bibinfo {year} {2022})},\ \Eprint {http://arxiv.org/abs/2205.07891} {arXiv:2205.07891 [quant-ph]} \BibitemShut {NoStop}%
\bibitem [{\citenamefont {Gallock-Yoshimura}\ \emph {et~al.}(2021)\citenamefont {Gallock-Yoshimura}, \citenamefont {Tjoa},\ and\ \citenamefont {Mann}}]{Gallock-Yoshimura:2021yok}%
  \BibitemOpen
  \bibfield  {author} {\bibinfo {author} {\bibfnamefont {K.}~\bibnamefont {Gallock-Yoshimura}}, \bibinfo {author} {\bibfnamefont {E.}~\bibnamefont {Tjoa}}, \ and\ \bibinfo {author} {\bibfnamefont {R.~B.}\ \bibnamefont {Mann}},\ }\href {\doibase 10.1103/PhysRevD.104.025001} {\bibfield  {journal} {\bibinfo  {journal} {Phys. Rev. D}\ }\textbf {\bibinfo {volume} {104}},\ \bibinfo {pages} {025001} (\bibinfo {year} {2021})},\ \Eprint {http://arxiv.org/abs/2102.09573} {arXiv:2102.09573 [quant-ph]} \BibitemShut {NoStop}%
\end{thebibliography}%
\end{document}